\documentclass[11pt]{article}
\pdfoutput=1
\usepackage{jcapmod}

\usepackage{booktabs}
\usepackage[english]{babel}
\usepackage{amsmath,amssymb,amsbsy,amstext, amsthm, simplewick, amsfonts}
\usepackage{hyperref}
\usepackage{graphicx}
\usepackage[small]{caption}
\usepackage{siunitx}
\usepackage{upgreek}
\usepackage{framed}
\usepackage{wrapfig}
\usepackage{multirow}
\usepackage{bbm}
\usepackage[svgnames,dvipsnames,x11names]{xcolor}
\usepackage[utf8x]{inputenc}
\usepackage{selinput}
\usepackage{bm}
\usepackage{float}
\usepackage{geometry}
\usepackage{yfonts}
\usepackage{caption}
\usepackage{subcaption}
\usepackage{sidecap}
\usepackage{longtable}
\usepackage{anyfontsize}
\usepackage[hang,flushmargin]{footmisc} 

\usepackage[framemethod=default]{mdframed}
\newmdenv[skipabove=7pt,
skipbelow=7pt,
rightline=false,
leftline=false,
topline=false,
bottomline=false,
backgroundcolor=gray!10,
linecolor=gray,
innerleftmargin=5pt,
innerrightmargin=5pt,
innertopmargin=5pt,
innerbottommargin=5pt,
leftmargin=0cm,
rightmargin=0cm,
linewidth=4pt]{eBox}
\newmdenv[skipabove=7pt,
skipbelow=7pt,
rightline=false,
leftline=false,
topline=false,
bottomline=false,
backgroundcolor=gray!10,
linecolor=gray,
innerleftmargin=5pt,
innerrightmargin=5pt,
innertopmargin=-5pt,
innerbottommargin=5pt,
leftmargin=0cm,
rightmargin=0cm,
linewidth=4pt]{eBox2}

\definecolor{blue3}{RGB}{31, 119, 180}
\definecolor{red3}{RGB}{	214, 39, 40}
\definecolor{orange3}{RGB}{255, 127, 14}
\definecolor{green3}{RGB}{44, 160, 44}

\newcommand{\blue}[1]{\textcolor{blue3}{#1}}
\newcommand{\green}[1]{\textcolor{green3}{#1}}
\newcommand{\red}[1]{\textcolor{red3}{#1}}

\newcommand{\orange}[1]{\textcolor{orange3}{#1}}

\usepackage{colortbl}
\definecolor{lightgreen}{cmyk}{0.2, 0, 0.2, 0.2}
\definecolor{lightgray}{cmyk}{0.1,0.2,0,0.1}
\definecolor{lightgray2}{cmyk}{0.1,0.1,0,0.1}

\makeatletter
\newlength{\apb@width}
\newcommand{\autoparbox}[2][c]{\settowidth{\apb@width}{#2}\parbox[#1]{\apb@width}{#2}}

\makeatother


\def\d{{\rm d}}

\def\k{{\bf k}}

\def\q{{\bf q}}

\def\s{{\bf s}}

\def\x{{\bf x}}
\def\y{{\bf y}}
\def\z{{\bf z}}

\def\A{{\cal A}}

\def\C{{\cal C}}

\def\F{{\cal F}}

\def\K{{\cal K}}

\def\nn{\nonumber\\}

\def\hs{\hskip 1pt}

\def\beq{\begin{equation}}
\def\eeq{\end{equation}}

\allowdisplaybreaks[1]
\begin{document}



\begin{titlepage}
\setcounter{page}{1} \baselineskip=15.5pt 
\thispagestyle{empty}

\begin{center}
{\fontsize{20}{18} \bf The Cosmological Bootstrap:}\\[14pt]
{\fontsize{15.3}{18} \bf  Inflationary Correlators from Symmetries and Singularities}
\end{center}

\vskip 20pt
\begin{center}
\noindent
{\fontsize{12}{18}\selectfont Nima Arkani-Hamed,$^1$ Daniel Baumann,$^2$ Hayden Lee,$^{3}$ and Guilherme L.~Pimentel$^2$}
\end{center}

\begin{center}
  \vskip 8pt
\textit{$^1$ School of Natural Sciences, Institute for Advanced Study,  Princeton, NJ 08540, USA} 

  \vskip8pt
\textit{$^2$ Institute of Physics, University of Amsterdam, Amsterdam, 1098 XH, The Netherlands}

\vskip 8pt
\textit{$^3$  Department of Physics, Harvard University, Cambridge, MA 02138, USA}
\end{center}

\vspace{0.4cm}
 \begin{center}{\bf Abstract}
 \end{center}
 
 \noindent
Scattering amplitudes at weak coupling are highly constrained by Lorentz invariance, locality and unitarity, and depend on model details only through coupling constants and the particle content of the theory.
For example, four-particle amplitudes are analytic for contact interactions and have simple poles with appropriately positive residues for tree-level exchange. In this paper, we develop an understanding of 
inflationary correlators which parallels that of flat-space scattering amplitudes. Specifically, we study slow-roll inflation with weak couplings to extra massive particles, for which all correlation functions are controlled by an approximate conformal symmetry on the boundary of the spacetime.  
After systematically classifying all possible contact terms in de Sitter space, we derive an analytic expression for the four-point function of conformally coupled scalars mediated by the tree-level exchange of massive scalars. Conformal symmetry implies that the correlator satisfies a pair of differential equations with respect to spatial momenta, encoding bulk time evolution in purely boundary terms. The absence of unphysical singularities (and the correct normalization of physical ones) completely fixes this correlator. 
Moreover, a ``spin-raising'' operator relates it to the correlators associated with the exchange of particles with spin, while ``weight-shifting'' operators map it to the four-point function of massless scalars. We explain how these de Sitter four-point functions can be perturbed to obtain inflationary three-point functions.
Using our formalism, we reproduce many classic results in the literature, such as the three-point function of slow-roll inflation, and provide a complete classification of all inflationary three- and four-point functions arising from weakly broken conformal symmetry. Remarkably, the inflationary bispectrum associated with the exchange of particles with arbitrary spin is completely characterized by the soft limit of the simplest scalar-exchange four-point function of conformally coupled scalars and a series of contact terms. 
Finally, we demonstrate that the inflationary correlators contain flat-space scattering amplitudes via a suitable analytic continuation of the external momenta, which can also be directly connected with the signals for particle production seen in the squeezed limit.

\end{titlepage}

\restoregeometry

\newpage
\setcounter{tocdepth}{2}
\tableofcontents

\newpage
\section{Time Without Time}

Cosmology is famously an observational rather than an experimental science. No experimentalists were present in the early universe, and the experiment of the birth and subsequent evolution of the universe cannot be repeated. Instead, we can only measure the spatial correlations between cosmological structures at late times.   The central challenge of modern cosmology is to construct a consistent ``history" of the universe that explains these correlations.  This cosmological history is a narrative, a story we tell to give a rational accounting of the patterns that we see in the cosmological correlations. 

\begin{figure}[b!]
\centering
      \includegraphics[scale=0.6]{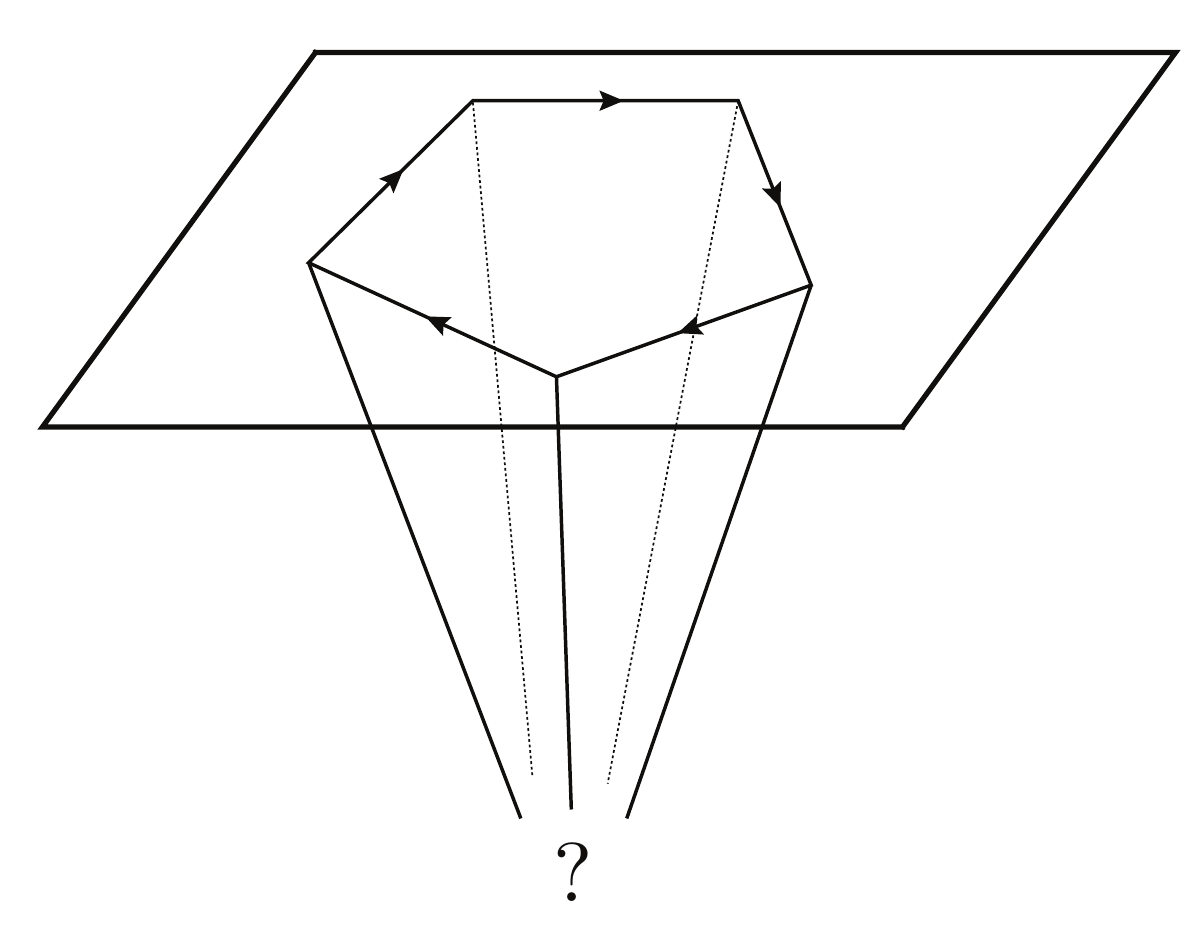}
           \caption{Cosmological observations can be traced back to the end of inflation where they become spatial correlations on the boundary of the approximate de Sitter spacetime.} 
    \label{fig:dS}
\end{figure}

\vskip 4pt
In inflationary cosmology~\cite{Guth:1980zm, Linde:1981mu, Albrecht:1982wi, TASI2008}, all cosmological correlations can be traced back to the origin of the hot big bang, or the end of inflation, where they reside on the ``boundary" of an approximate de Sitter (dS) spacetime (see Fig.~\ref{fig:dS}). Given that 
cosmological observations are firmly anchored to the spatial slice at future infinity, it is natural to ask whether we can reproduce all of these spatial correlations in a radically different way, without making explicit reference to the time evolution in the bulk spacetime. This is far from an academic issue, 
since we suspect that amongst other things the notion of time itself must break down in the initial big bang singularity. It is possible that we will eventually be forced to replace cosmological time evolution with something else. How can this be done? Or, borrowing a Wheeler-ism, how can we have ``time without time"?

\vskip 4pt
There are good reasons to suspect that the only well-defined quantum mechanical observables in a theory of quantum gravity must live at infinity. Only at infinity can we have infinitely massive measuring apparatuses, separated by infinite distances,  capable of making infinitely many measurements, with all these infinities needed to yield quantum-mechanically precise results unpolluted by gravitational effects. Thus, quantum gravity forces us to study boundary observables, and is therefore ``holographic" in nature. The past few decades have seen an intensive focus on various sorts of holographic theories for boundary observables. The most precise of these theories have been formulated in asymptotically anti-de Sitter (AdS) space~\cite{Maldacena:1997re, Witten:1998qj, Gubser:1998bc, Aharony:1999ti}.  In that case, the boundary is an ordinary flat spacetime of one lower dimension. The boundary observables are correlation functions that can be measured by ``pinging" the boundary, and are functions of the spacetime points on the boundary. The gauge-gravity duality then provides a precise boundary theory to compute the boundary observables. A crucial feature that makes the identification of this dual description possible is that, while space and gravity are emergent from the dynamics of the boundary gauge theory, the boundary theory is still an ordinary physical system with a standard notion of locality and of time. Moreover, time flows on the boundary the same way it does in the interior. 

\vskip 4pt
The situation is much murkier in asymptotically flat spacetimes, where the observable is the $S$-matrix, and the boundary does not have standard notions of either locality or of time evolution. In that case, it is less clear what the rules should be that govern a potential boundary theory of scattering amplitudes. Indeed, even some first hints for such a description, seen in perturbation theory, involve much more alien combinatorial, geometric and number-theoretic ideas \cite{Arkani-Hamed:2013jha,Arkani-Hamed:2017mur,ArkaniHamed:2012nw, Arkani-Hamed:2017vfh, Cheung:2016iub, Pasterski:2016qvg,Pasterski:2017ylz,Strominger:2017zoo}, from which the physics of spacetime and quantum mechanics---locality and unitarity---emerge as derivative notions, with a huge amount left to be understood. 

\vskip 4pt
A holographic description of cosmology~\cite{Maldacena:2002vr, Strominger:2001pn,Witten:2001kn,Larsen:2003pf,van2004inflationary, McFadden:2009fg,Antoniadis:2011ib,Maldacena:2011nz, Bzowski:2011ab, Creminelli:2011mw, Anninos:2011ui,Mata:2012bx,Raju:2012zs,Coriano:2012hd,Schalm:2012pi,Bzowski:2013sza,Ghosh:2014kba,Pajer:2016ieg, Anninos:2017eib} is even more confusing, since the boundary geometry is a Euclidean manifold, and there is no notion of boundary time. We thus face a rich irony: The cosmological spacetimes for which the notion of ``emergent spacetime" is of the most pressing significance, are the ones where we have the least clue of how to make a meaningful start!

\vskip 4pt
However, before getting too far ahead of ourselves with such lofty questions and ambitions, it behooves us to ask a much simpler question. Suppose we were handed a candidate set of cosmological correlations. How would we check if they are right or wrong? How could we tell whether they arise from a consistent picture of causal time evolution? Precisely the same question arises for AdS boundary correlators and the flat-space $S$-matrix. In AdS, it has a well-defined answer, since the thorny issue of ``time" can be sidestepped altogether by defining boundary correlators in Euclidean space satisfying sharply defined consistency conditions of Euclidean conformal field theories (CFTs) with a unitary spectrum and a consistent operator product expansion (OPE).  
In flat space, on the other hand, the question of how causal time evolution is imprinted in the $S$-matrix is much more difficult, and cannot be shunted to Euclidean space. There is clearly a sense in which causality is reflected in the analytic structure of the $S$-matrix, and while the $S$-matrix program of the 1960's hoped to derive these analytic properties from first principles, to this day, we do not know what they are, even in perturbation theory. Finally, the consistency conditions on cosmological correlators are the least understood. It is disturbing that our understanding of these issues becomes more and more primitive the closer and closer we get to the real world.

\vskip 4pt
The explosion of progress in the understanding of scattering amplitudes over the past few decades has been fueled by a more pragmatic attitude, exploiting various situations where the analytic structure is understood well enough to make progress, as in \cite{Bern:1994cg, Britto:2005fq} (see~\cite{Elvang:2015rqa, Cheung:2017pzi, Carrasco:2015iwa, Benincasa:2013faa, Weinzierl:2016bus, Burger:2017yod} for recent reviews). For instance, consistent tree amplitudes have an obvious analytic structure reflecting locality and unitarity---they must have poles where the sums of the external momenta corresponding to an internal propagator go on-shell (locality), and they must factorize into products of lower-point amplitudes on the residues of these poles (unitarity). Beyond tree level, amplitudes have branch cuts, with a completely understood analytic structure at one loop and a steadily growing control of this structure at higher loop orders. Moreover, when the notion of ``the integrand" of the multi-loop scattering amplitudes is  available, the analytic structure is again nearly as simple as at tree level---we have rational functions of external and loop momenta, with the locations of poles and the factorization on residues dictated by the cutting rules reflecting locality and unitarity. All of this has allowed, in a huge number of examples, the direct determination of scattering amplitudes from first principles, eschewing the crippling complexity of the Lagrangian formalism and Feynman diagrams. Aside from its utility in making predictions for collider experiments, these computations have generated an ocean of ``theoretical data", from which the outlines of the more radical theories replacing locality and unitarity with new mathematical and physical structures can more easily be seen.

\vskip 4pt
Returning to cosmology, our present understanding of the way consistent time evolution is encoded in cosmological correlators is still in its infancy, matching not even the level of understanding for tree-level scattering amplitudes. As a simple and startling illustration of this fact, we currently don't even have good analytic control for the four-point function of massless scalars in de Sitter space, mediated by the exchange of a massive scalar! At the same time, there are many indications that a similarly rich and deep structure controlling cosmological correlators {\it must} exist---not least because, as we will review, these correlators contain and generalize flat-space scattering amplitudes in a beautiful way~\cite{Raju:2012zr, Maldacena:2011nz}.

\vskip 4pt
In this paper, we initiate a systematic exploration of de Sitter and inflationary correlators, from the point of view of ``time without time" or, equivalently, that of the ``cosmological bootstrap". Indeed, it is useful to phrase our goals in the language of various ``bootstrap" programs that have been undertaken over the past fifty years. As we have already alluded to, the earliest attempt to bootstrap physical observables by directly imposing physical principles such as unitary, Lorentz invariance and causality, was for the $S$-matrix. This endeavor was stymied by the fundamental difficulty of not knowing the precise rules for encoding causality in the analytic structure of the $S$-matrix~\cite{Eden:1966dnq}. The major success of this program---the discovery of string theory \cite{Veneziano:1968yb}---was made possible by restricting attention to tree amplitudes with only poles as singularities, where the rules are well-defined. The more recent wave of advances in the field has used the bootstrap philosophy in combination with perturbation theory, which restricts the analytic structure of the functions that can appear in the final results in a controllable way.   The story is rather different for the ``conformal bootstrap" program for CFTs, where by focusing on Euclidean correlation functions, the rules are completely, and even nonperturbatively, well-defined, so a systematic exploration from numerical and analytical points of view becomes possible~\cite{Rychkov:2016iqz,Simmons-Duffin:2016gjk}. The character of the ``cosmological bootstrap" that we pursue in this paper is closer in spirit to its modern incarnation in scattering amplitudes, striving to use a simplified analytic structure for correlators in perturbation theory, together with symmetries and singularities, to fully determine the final answer without reference to bulk time evolution.

\vskip 4pt
We will study the fundamentals of this physics, focusing our attention on the simplest case of four-point correlators at tree level. One important motivation for doing this comes from experiment: We wish to give a completely invariant and physical characterization of the way in which the exchange of particles of general masses and spins can be extracted from cosmological probes of non-Gaussianity in the coming decades. This completes the dictionary of ``cosmological collider physics"~\cite{Chen:2009zp, Baumann:2011nk, Assassi:2012zq, Chen:2012ge, Pi:2012gf, Noumi:2012vr, Baumann:2012bc, Assassi:2013gxa, Gong:2013sma, Arkani-Hamed:2015bza, Lee:2016vti, Kehagias:2017cym, Kumar:2017ecc, An:2017hlx, An:2017rwo, Baumann:2017jvh}, giving physically motivated templates for comparison with observational data. Another motivation is more theoretical: We strongly believe that, as with the exploration of scattering amplitudes, the ability to systematically compute cosmological correlators will generate a wealth of theoretical data that might stimulate the possible discovery of deep new physical and mathematical structures underlying this physics. Finally, as an incidental by-product of wider interest,   our investigations involve a detailed study of the constraints of conformal symmetry in momentum space, which appear to have some unfamiliar and beautiful properties that deserve further exploration in their own right.

\begin{figure}[h!]
    \centering
      \includegraphics[scale=1]{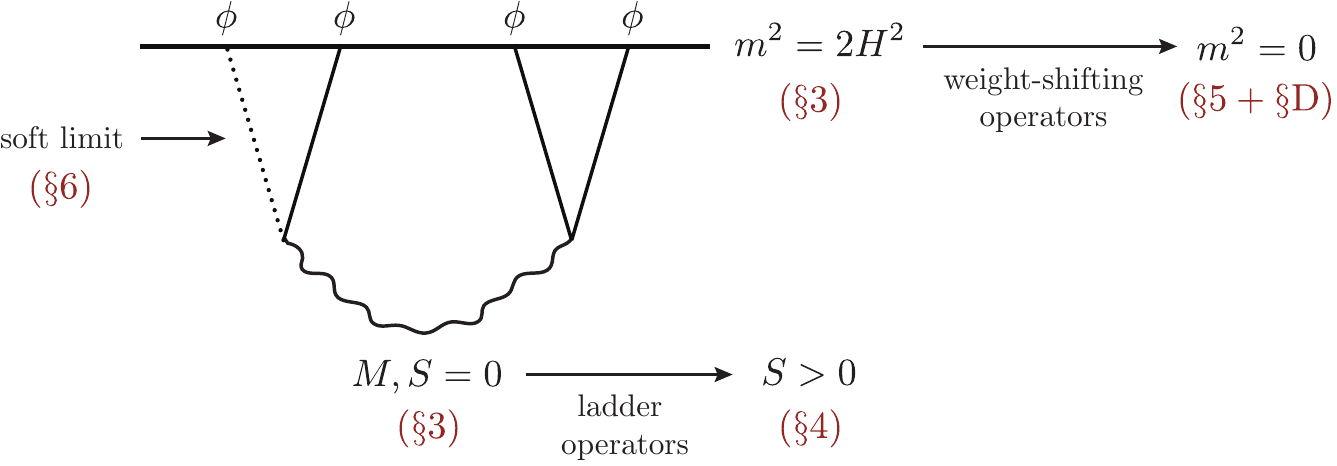}
           \caption{Schematic illustration of the logical connections between the different parts of the paper.}
    \label{fig:RoadMap}
\end{figure}
  
\paragraph{Outline} Figure~\ref{fig:RoadMap} provides a road map through this paper. In Section~\ref{sec:sym}, we first review how the structure of scattering amplitudes in flat space is fixed by symmetries and singularities. We then sketch that a similar logic can be used to determine four-point functions in de Sitter space.
In Section~\ref{sec:dS-4pt}, we use this approach to derive an analytic solution for the four-point function of {\it conformally coupled} scalars in de Sitter space, mediated by the tree-level exchange of massive scalars.  In Section~\ref{sec:spin}, we show that a simple spin-raising operator maps this solution to the solutions with {\it spinning} internal particles. 
In Section~\ref{sec:Massless}, we introduce weight-shifting operators that relate the solutions for conformally coupled scalars to correlators with {\it massless} external fields, which is the case relevant to inflation.   Our derivation of the spin-raising and weight-shifting operators involves a mix of bulk and boundary intuition. In an upcoming paper~\cite{CosmoBoot2}, we will present a more systematic derivation of these operators (and their generalizations) using tools of conformal field theory~\cite{Dirac:1936fq, Costa:2011mg, Karateev:2017jgd}.
In Section~\ref{sec:inflation}, we show how the de Sitter four-point functions can be perturbed to obtain inflationary three-point functions~\cite{Kundu:2014gxa,Kundu:2015xta}. We reproduce many classic results in the literature and provide a systematic way to obtain new results.  In Section~\ref{equ:pheno}, we comment briefly on a few phenomenological consequences of our results.
Our conclusions are summarized in Section~\ref{sec:conclusions}. 

\vskip 4pt
A number of appendices contain additional technical details and derivations.
In Appendix~\ref{app:ward}, we derive the conformal symmetry constraints on three- and four-point functions in de Sitter space.  In Appendix~\ref{app:bulk}, we discuss the singularities of our solutions from both the bulk and boundary perspectives.
In Appendix~\ref{app:hyper}, we present the solution of Section~\ref{sec:dS-4pt} in hypergeometric form, and analyze various limits of the result.  In Appendix~\ref{app:D3}, we provide further details on the solutions for massless external fields presented in Section~\ref{sec:Massless}. In Appendix~\ref{sec:ladder}, we introduce a set of weight-shifting operators that allow us the bootstrap the solutions for massless external fields from those for conformally coupled fields. Appendix~\ref{app:identities} contains a few useful identities for the hypergeometric functions, and Appendix~\ref{app:notation} collects important variables used in the paper.

\paragraph{Reading guide} Since the paper is quite long, it may be helpful to provide a short reading guide. 
We recommend beginning with a careful reading of Sections~\ref{sec:sym} and~\ref{sec:dS-4pt}, which first explain the bootstrap philosophy and then apply it to the specific example of scalar exchange. The following Sections~\ref{sec:spin} and~\ref{sec:Massless}, on the other hand, can be skipped on a first reading. In particular, the rest of the paper can be read without having absorbed the details of this part of the paper. The reader can therefore continue directly with Sections~\ref{sec:inflation} and~\ref{equ:pheno}, where we apply our formalism to inflationary bispectra and their phenomenology. All appendices are for aficionados.

\paragraph{Main results} The main results of this paper are highlighted by gray boxes: 
\begin{itemize}
\item Equation (\ref{equ:SeriesSol}) is the solution for the four-point function of conformally coupled scalars arising from the exchange of a massive scalar. This provides the fundamental building block from which all other correlators are derived by spin-raising and weight-shifting operators. 
\item Equations (\ref{equ:Ansatz}) and (\ref{equ:coeff}) display the solution for massive spin-exchange. \item Equations (\ref{Fdelta3S0}), (\ref{equ:F111}) and (\ref{equ:Fspin2}) are the four-point functions of massless scalars arising from the exchange of massive scalars, spin-1 particles and spin-2 particles, respectively.
\item Equation (\ref{equ:BINF}) is the inflationary bispectrum for arbitrary spin-exchange, which becomes (\ref{equ:thrptscalar}) for scalar exchange and (\ref{equ:3SR}) for graviton exchange. 
\item Equations~(\ref{equ:bispectrum}) and (\ref{equ:Bfinal}) characterize all inflationary three-point functions arising from interactions that only weakly break conformal invariance. 
\end{itemize}

\paragraph{Notation and conventions} Our metric signature is $({-}\hs{+}\hs{+}\hs{+})$, with apologies to our particle physics friends.
Throughout the paper, we use natural units, $\hbar = c \equiv 1$. 
Generic scalar operators (of dimension $\Delta$) will be denoted by $O$.
We will use $\varphi$ and $\phi$ for operators with $\Delta=2$ and $\Delta=3$, respectively. When we need to refer to the corresponding bulk fields, we will use $\upvarphi$ and~$\upphi$.
We use Greek letters for spacetime indices, $\mu =0,1,2,3$, and Latin letters for spatial indices, $i=1,2,3$. Three-dimensional vectors will be denoted in boldface (${\bf k}$) or with Latin superscripts~($k^i$). The magnitude of vectors is defined as $k = |\k|$ and unit vectors are written as $\hat \k = \k/k$. The momentum of the $n$-th leg of a correlation function is denoted by $\k_n$ and its magnitude is $k_n \equiv |\k_n|$. Our conventions for scattering amplitudes are the same as in~\cite{Elvang:2015rqa, Cheung:2017pzi}. We will use the following Mandelstam variables $s_{nm} \equiv -(p_n+p_m)^2$, where $s \equiv s_{12}$, $t \equiv s_{23}$ and $u \equiv s_{24}$.  We will often denote them by $s_{\rm flat}$ and $t_{\rm flat}$ to avoid confusion with $s \equiv |\k_1+\k_2|$ and $t \equiv |\k_2+\k_3|$, which we employ for the exchange momenta in cosmological correlators.

\newpage
\section{Amplitudes Meet Cosmology}
\label{sec:sym}

We will begin with a general discussion of the power of symmetries to constrain the structure of scattering amplitudes in flat space (\S\ref{sec:FlatSpace}) and correlation functions in de Sitter space (\S\ref{sec:dS}).

\subsection{Amplitudes in Flat Space}
\label{sec:FlatSpace}

Consider a theory of scalars of mass $m$ in $d$ spatial dimensions. All other particles are taken to be significantly heavier than $m$. 
We now review how basic physical requirements such as Lorentz invariance, locality and unitarity severely restrict the analytic structure of the four-particle scattering amplitude $A_4$ at tree level.

\vskip 4pt
By {\it Lorentz invariance}, the scattering amplitude is a function of the Mandelstam variables~$s,t,u$. 
At low energies, the theory is described by contact interactions, and $A_4$ is a purely analytic function of $s,t,u$, which can be written as a power series for small $s,t,u$. At higher energies, we become sensitive to the exchange of massive particles. For example, at tree level, the existence of a particle of mass~$M$ leads to poles as $s,t,u \to M^2$. As a consequence of {\it locality}, these poles must be simple poles.  Moreover, the residues of the poles are fixed by kinematics.
In particular, near the $s$-channel pole, the amplitude can be written as\footnote{Note that our convention for the overall sign of the scattering amplitude is the opposite of that used in most textbooks,~e.g.~\cite{Peskin:1995ev, Srednicki:2007qs, Schwartz:2013pla}, but is common in the modern literature on scattering amplitudes, e.g.~\cite{Elvang:2015rqa, Cheung:2017pzi, Carrasco:2015iwa}. Of course, this sign is physically irrelevant.}
\beq
{A}_4(s,t,u) \xrightarrow{\ s \to M^2\ } \frac{1}{s -M^2} \sum_\lambda {A}_3^\lambda(p_1,p_2,p_I) \hs {A}_3^{-\lambda}(p_3,p_4,-p_I)\, . \label{equ:A4}
\eeq
We see that the residue of the pole factorizes into a product of on-shell three-particle amplitudes~$A_3^\lambda$ involving two scalars and a massive spin-$S$ particle with helicity $\lambda$.
These three-particle amplitudes are simple because there are no kinematical invariants analogous to $s,t,u$ for three-particle scattering. To see this, note that 
$p_n^2=-m^2$, implies $(p_1+p_2)^2=(-p_3)^2 = -m^2$, etc. 
As we show in the insert below, the allowed three-particle amplitudes depend on the spin of the exchanged particle.
Summing over the different helicity contributions, we get
\beq
\begin{aligned}
{A}_4(s,t,u) \xrightarrow{\ s \to M^2\ } 
 & \ g^2\frac{(4m^2-M^2)^S}{s-M^2}P_{d,S}\bigg(1 + \frac{2t}{M^2-4m^2} \bigg)\, ,
 \end{aligned}
\label{equ:A4final}
\eeq
where $g$ is a coupling constant and $P_{d,S}$ is the $d$-dimensional Gegenbauer polynomial. For $d=2$ and $d=3$, the Gegenbauer polynomials become the Chebyshev and Legendre polynomials, respectively.  Note that the coefficient of the Gegenbauer polynomial is positive; this is a consequence of~{\it unitarity}.

\begin{framed}
{\small \noindent {\it Derivation.}---In this insert, we derive (\ref{equ:A4final}) for exchange particles with different spin.
\begin{itemize}
\item ${\boldsymbol{S=0}}$: If the internal state is a massive scalar, then the three-particle amplitude ${A}_3$ is just a constant,
\beq
{A}_3(p_1,p_2,p_3) = g\, ,
\eeq
where $g$ determines 
 the interaction strength.
\item ${\boldsymbol{S=1}}$:  If the internal state is a massive vector, then ${A}_3^\lambda$ must depend on the polarization vector~$\epsilon_\mu^\lambda$. Since $p^\mu \epsilon_\mu^\lambda =0$, the most general amplitude is of the form
\beq
{A}^\lambda_3(p_1,p_2,p_3) = g\, (p_1^\mu - p^\mu_2)\epsilon_\mu^\lambda(p_3)\, .
\eeq
\item ${\boldsymbol{S\ge 1}}$:  If the internal state is a massive particle of general spin $S$, then we have
\beq
{A}^\lambda_3(p_1,p_2,p_3) = g\, (p_1 - p_2)^{\mu_1} \cdots (p_1 - p_2)^{\mu_S} \epsilon_{\mu_1 \ldots \mu_S}^\lambda(p_3)\, .
\eeq
\end{itemize}
The polarization sum in the four-point amplitude (\ref{equ:A4}) then reads
\beq
P_{d,S}(k,q) \equiv \sum_\lambda  k^{\mu_1} \cdots k^{\mu_S}\, q^{\nu_1} \cdots q^{\nu_S}\,  \epsilon_{\mu_1 \ldots \mu_S}^\lambda\,\epsilon_{\nu_1 \ldots \nu_S}^{-\lambda}\, ,
\eeq
where $k\equiv p_1-p_2$ and $q \equiv p_3-p_4$.  This sum is evaluated most easily in  the rest frame of the massive particle, where only the spatial components of the polarization tensors are non-vanishing. 
Switching  to real space variables for the sake of notational familiarity, we then have
\beq
P_{d,S}(\x,\y) = \sum_\lambda x^{i_1} \cdots x^{i_S} \,y^{j_1} \cdots y^{j_S} \epsilon_{i_1\ldots i_S}^\lambda \epsilon_{j_1\ldots j_S}^{-\lambda}  \equiv |\x|^S |\y|^S P_{d,S}(\cos \theta)\, ,
\eeq
where $\cos \theta \equiv \x\cdot \y/|\x||\y|$.
The sum over the polatization tensors must be made out of Kronecker delta's, be symmetric in all $i$'s and $j$'s and traceless in the $i$'s and $j$'s separately. 
The tracelessness condition implies
\beq
\nabla_\x^2 P_{d,S}(\x,\y) = 0\, .
\eeq
The solution to the $d$-dimensional Laplace equation is
\begin{align}
 \frac{1}{|\x-\y|^{d-2}} &= \frac{1}{(|\x|^2 -2 |\x||\y| \cos\theta + |\y|^2)^{d/2-1}} =  \frac{1}{|\y|^{d-2}} \sum_{S=0}^\infty \left(\frac{|\x|}{|\y|}\right)^S C_{S}^{(d/2-1)}(\cos \theta)\, ,
\end{align}
where we have introduced the $d$-dimensional Gegenbauer polynomials.
From the term proportional to $|\x|^S$, we read off $P_{d,S}(\cos \theta)= C_{S}^{(d/2-1)}(\cos \theta)$. Writing the amplitude in a Lorentz-invariant form gives \eqref{equ:A4final}. }
\end{framed}

\noindent
 The most general four-particle amplitude arising from tree-level exchange therefore takes the form 
 \beq
A_4(s,t,u) = g^2\frac{(4m^2-M^2)^S}{s-M^2} P_{d,S}\bigg(1 + \frac{2t}{M^2-4m^2} \bigg) 
\,+\, \text{$t$- and $u$-channels} \,+\, 
\text{analytic}\, .
 \label{equ:M4-2}
\eeq
Symmetries and kinematics have fixed the structure of the amplitude near physical poles. The rest of the amplitude is analytic. Note that the separation of the amplitude into ``pole" and ``polynomial" parts is not unique; it is only the residues on the poles that have unambiguous meaning. 

\vskip 4pt
A similar argument applies to the exchange of massless particles.\footnote{In the case of massless particles, the requirement of consistent factorization of four-particle amplitudes is highly restrictive, and makes almost
all theories, other than the familiar gauge theories and gravity, inconsistent~\cite{Weinberg:1964ew,Coleman:1967ad,Weinberg:1980kq,Benincasa:2007xk,Bekaert:2010hw,McGady:2013sga,Rahman:2015pzl,Arkani-Hamed:2017jhn}.} The structure of the three-particle amplitude is the same but with the replacement $P_{d,S}(\cos \theta) \to P_{d-1,S}(\cos \theta)$.  This is a consequence of the little group reducing from $SO(d)$ to $SO(d-1)$ for massless particles. 
Note that for scalars and vectors the Gegenbauer polynomials are independent of dimension, namely $P_{d,S=0}(\cos \theta)=1$ and $P_{d,S=1}(\cos \theta) = \cos \theta$ for all $d$. This implies that the amplitudes have a smooth limit as $M \to 0$.
Starting at $S=2$, however, the Gegenbauer polynomials do depend on the dimension; for example, $P_{d,S=2}(\cos \theta)= d \cos^2 \theta - 1$.
The amplitudes for particles with $S\ge 2$ therefore have a discontinuity in the massless limit. For spin two, this is the famous vDVZ discontinuity of massive gravity~\cite{vanDam:1970vg, Zakharov:1970cc}. Finally, we note that in the limit $M,m\to 0$ the amplitude simplifies to $t^S/s$.

\subsection{Correlators in de Sitter Space}
\label{sec:dS}

We have seen that the four-scalar scattering amplitude was determined by Poincar\'e invariance, locality, and unitarity, together with an ansatz giving the amplitude the ``simplest possible" analytic structure: being analytic in the case of contact interactions, and having simple poles for tree-level particle exchange. In this paper, we will show that a similar logic constrains the structure of conformally-invariant\footnote{Strictly speaking, the correlation functions in question are {\it covariant} under the conformal symmetry. For simplicity, we will not make this distinction and use conformal covariance and invariance synonymously throughout the paper.} four-point functions in de Sitter space. 

\vskip 4pt
We will first review the de Sitter isometries and show how they completely fix two- and three-point functions. 
After that, we will set up the problem of determining the structure of four-point functions both for contact interactions and for tree-level exchange. 

\subsubsection{Boundary Perspective}

In most of this paper, we will study quantum fields on a fixed four-dimensional de Sitter background.  In flat slicing, the metric of the dS spacetime can be written as
\beq
\d s^2 = \frac{- \d \eta^2 + \d \x^2}{(H\eta)^2}\, ,
\eeq
where $H$ is the Hubble scale and $\eta$ is conformal time.  The line element is manifestly invariant under spatial translations and rotations. Less obvious isometries are dilatation and special conformal transformations (SCTs), whose associated Killing vectors are
\begin{align}
{\cal D} &\equiv - \eta \partial_\eta - x^i \partial_{x^i}\,, \label{equ:D}\\
{\cal K}^i &\equiv 2x^i \eta \partial_\eta  +  2 x^i x^j \partial_{x^j} +(\eta^2- |\x|^2)\partial_{x_i}\, . \label{equ:SCT}
\end{align}
Correlators of quantum fields in dS must be 
invariant under the action of these isometries.
At late times, $\eta \to 0$, a massive scalar field $\sigma(\eta,\x)$ behaves as
\beq
\lim_{\eta \to 0} \sigma(\eta,\x) = O^+(\x)\, \eta^{\Delta^+} + O^-(\x) \,\eta^{\Delta^-}\, , \label{equ:dual}
\eeq
where the scaling dimensions are
\beq\label{equ:mdimsc}
\Delta^\pm = \frac{3}{2}\pm i \mu\, , \quad \mu \equiv \sqrt{\frac{m^2}{H^2} - \frac{9}{4}}\, .
\eeq 
We will mostly be interested in the correlation functions of conformally coupled scalars $\varphi$ and massless scalars $\phi$, for which $\Delta^-=2$ and $3$, respectively. The action of the generators (\ref{equ:D}) and (\ref{equ:SCT}) on the boundary operators $O^\pm(\x)$ becomes 
\begin{align}
{\cal D} &\equiv - \Delta^\pm - x^j \partial_{x^j}\,, \label{equ:D2}\\
{\cal K}^i &\equiv 2\Delta^\pm\, x^i +  2 x^i x^j \partial_{x^j} - |\x|^2 \,\partial_{x_i}\, , \label{equ:SCT2}
\end{align}
which are the generators of the conformal group in three dimensions. Because of translational invariance, cosmological correlators are usually studied in Fourier space.
The Fourier transforms of the operators (\ref{equ:D2}) and (\ref{equ:SCT2}) are
\begin{align}
{\cal D} &\equiv - (\Delta^\pm-3) + k^j \partial_{k^j}\,, \label{equ:D3}\\
{\cal K}^i &\equiv 2(\Delta^\pm-3) \partial_{k_i} -  2 k^j \partial_{k^j} \partial_{k_i}  + k^i \partial_{k^j}\partial_{k_j}\, , \label{equ:SCT3}
\end{align}
which are the generators acting on $O^\pm(\k)$.\footnote{For spinning operators, the action of the conformal group is more complicated, as SCTs also rotate the indices of the operator.}  Unless stated otherwise, we will present the correlation functions of $O^-$, but drop the superscript to avoid clutter. The correlation functions of $O^+$ are related to those of $O^-$ by simple momentum-dependent rescalings.

\paragraph{Two-point functions} The conformal group constrains the functional form of two-point functions to be  
\beq
\langle {O}_1  {O}_2 \rangle = \begin{cases} c_{O_1} k_1^{2 \Delta_1 - 3} \times (2\pi)^3 \delta^3(\k_1+\k_2)\,, & \Delta_1=  \Delta_2 \, , \\ 0\,, & \Delta_1\ne  \Delta_2 \, , 
\end{cases}
\eeq 
where $O_n$ is shorthand for $O_n(\k_n)$. 
Dilatation symmetry fixes the overall momentum scaling, while SCTs only allow nonzero two-point functions for $\Delta_1=  \Delta_2$. The only freedom left is in the overall size of the correlation function, set by the constant $c_{O_1}$. 

\paragraph{Three-point functions} Rotations and translations fix the form of generic three-point functions to be
\beq
\langle {O}_1 {O}_2{O}_3\rangle = B(k_1,k_2,k_3) \times (2\pi)^3 \delta^3 (\k_1 +\k_2+\k_3)\, . \label{equ:scalarB}
\eeq
Just like Poincar\'e symmetry completely fixes three-particle scattering amplitudes, conformal symmetry determines the functional form of three-point functions \cite{Polyakov:1970xd,Osborn:1993cr} (see \cite{Bzowski:2013sza} for a detailed analysis in momentum space). During inflation, the breaking of the exact conformal symmetry allows for more freedom. As we will show in Section~\ref{sec:inflation}, if the breaking is sufficiently weak, these three-point correlators can be derived from the soft limits of conformally-invariant four-point functions (see also \cite{Kundu:2014gxa, Arkani-Hamed:2015bza}).

\paragraph{Four-point functions} 
It is well known that four-point functions in conformal field theories are less constrained kinematically. In position space, they are given by an arbitrary function of two conformally-invariant cross-ratios. In this paper, we will study the kinematic constraints due to SCTs in momentum space, where the four-point function of scalar operators takes the form
\beq
\langle {O}_1{O}_2 {O}_3{O}_4 \rangle = F(k_1,k_2,k_3,k_4,s,t) \times (2\pi)^3 \delta^3 (\k_1 +\cdots+\k_4)\, . \label{equ:scalarF}
\eeq
Momentum conservation and rotational invariance imply that the four-point function $F$ depends on six independent variables before imposing conformal symmetry.  It is convenient to take these variables to be the magnitudes $k_n\equiv |\k_n|$, and Mandelstam-like variables\hs\footnote{From now on, we will denote the flat space, four-dimensional Mandelstam variables by $s_{\rm flat}$ and $t_{\rm flat}$.} $s\equiv |\k_1+\k_2|$ and $t\equiv |\k_2+\k_3|$. In the inflationary literature, $k_I$ is often used in place of $s$, and the sum of the energies $k_n$ is written as $k_t \equiv \sum_n k_n$. Sometimes, we will trade $t$ for $\tau \equiv (\k_1-\k_2)\cdot(\k_3-\k_4)$.  
Constraints from dilatation symmetry and SCTs reduce these six independent variables to just two, which in position space are the conformally-invariant cross-ratios.
In momentum space, we will use 
\beq
u \equiv \frac{s}{k_1+k_2} \, , \quad v \equiv \frac{s}{k_3+k_4} \, .\label{equ:p}
\eeq
Invariance under (\ref{equ:D3}) and (\ref{equ:SCT3}) impose the constraints  $(-3 + \sum {\cal D}_n) F =0$ and $\sum {\cal K}^i_n F = 0$ on the four-point function: 
\begin{align}
	\left[9-\sum_{n=1}^4 \left(\Delta_n - k_n^j\frac{\partial}{\partial k_n^j}\right)\right] F&= 0 \, ,\label{WI_D} \\
	\sum_{n=1}^{4}\left[k_n^i\frac{\partial^2}{\partial k_n^j\partial k_n^j}-2k_n^j\frac{\partial^2}{\partial k_n^j\partial k_n^i}+2(\Delta_n-3)\frac{\partial}{\partial k_n^i}\right] F &= 0\, .\label{WI_SCT}
\end{align}
Dilatation symmetry is trivially reflected in the overall scaling dimension of the correlator: for a general $N$-point function, the correlator must have scaling dimension $\Delta_t \equiv \sum_n \Delta_n$ in position space, and thus dimension 
$ \Delta_t - 3N$ in momentum space. Stripping off the momentum-conserving delta function leaves us with a function of dimension $\Delta_t- 3(N-1)$. To make the dilatation symmetry of the four-point function manifest, it will be convenient to define $F = s^{\Delta_t - 9} \hat F$, where the form of the dimensionless function $\hat F$ will be dictated by special conformal invariance. 

\vskip 4pt
Invariance under SCTs implies three differential equations that must be satisfied by the correlators.
A bit of tedious algebra turns (\ref{WI_SCT}) into
\begin{align}
	\sum_{n=1}^4\, k_n^i\hs D_n F =0\, ,\label{WI_SCT2}
\end{align}
where we have defined 
\begin{align}
	D_1F &\equiv \bigg[\frac{\partial^2}{\partial k_1^2} + \frac{1}{s}\frac{\partial}{\partial s}\left(k_1\frac{\partial }{\partial k_1}+k_2\frac{\partial }{\partial k_2} \right)+\frac{1}{t}\frac{\partial}{\partial t}\left(k_1\frac{\partial }{\partial k_1}+k_4\frac{\partial }{\partial k_4} \right)-\frac{k_3^2}{st}\frac{\partial^2 }{\partial s \partial t}\nn[4pt]
	& \hspace{0.5cm}- \frac{2(\Delta_1-2)}{k_1}\frac{\partial }{\partial k_1} +\frac{\Delta_1+\Delta_2}{s}\frac{\partial }{\partial s}+\frac{\Delta_1+\Delta_4}{t}\frac{\partial }{\partial t} \bigg] F\, ,\label{W1F}
\end{align}
and the rest are given by the cyclic permutation of the indices (remembering that $t \to s$ under a cyclic shift). The operator $D_n$ is a combination of the SCT and dilatation operators, whose derivation can be found in Appendix~\ref{app:ward}. The only nontrivial way to satisfy \eqref{WI_SCT2} is to demand that all $D_nF$ are equal to each other, so that $\sum_n k_n^i\hs D_n F$ vanishes as a consequence of momentum conservation.\footnote{If all momenta are collinear, then perhaps other possibilities are allowed. However, in that case the four-point function would have to vanish in any non-collinear configuration, which we rule out on the basis of continuity.}
To satisfy the SCT constraint, we can therefore pick any three of the six conditions
\beq
(D_n  - D_m) F=0\, , \label{equ:conformal}
\eeq
 for $n,m = 1,\ldots, 4$.  
 
\paragraph{Outline and strategy} In this paper, we will give a systematic classification of the solutions to (\ref{equ:conformal}). Since the details are rather technical, we will begin with a rough sketch of our general strategy for solving these equations.

\vskip 4pt
As we have alluded to above (see Fig.~\ref{fig:RoadMap}), an object of particular interest is the four-point function of conformally coupled scalars, mediated by the tree-level exchange of massive scalars.  In that case, the $s$-channel contribution\footnote{It suffices to impose conformal invariance of a single channel to fix the whole four-point function. To see this, note that the correlator $\langle{O}{O} {O}^\prime {O}^\prime \rangle$ with $\Delta_{O}=\Delta_{O^\prime}$, which is conformally 
invariant, coincides with the $s$-channel of~$\langle{O}{O}{O}{O}\rangle$. The other channels can be included by cyclic permutations, e.g.~by replacing $u$ with $|\k_2+\k_3|/(k_2+k_3)$ and $|\k_2+\k_4|/(k_2+k_4)$ for the $t$- and $u$-channels, respectively. Finally, for contact interactions, ``$s$-channel" refers to a specific permutation of the external momenta.} can be written as
 $F = s^{-1} \hat F(u,v)$, an ansatz which automatically satisfies the equations $D_{12}F=0$ and  $D_{34}F=0$, where $D_{nm}\equiv D_n-D_m$.
The remaining conformal invariance equation $D_{13}F=0$ becomes
\begin{align}
	(\Delta_u-\Delta_{v}) \hat F =0\, , \label{contactgeneral} 
\end{align}
where we have introduced the differential operator 
\beq
\Delta_u \equiv u^2(1-u^2)\partial_u^2 - 2u^3\partial_u\,. \label{equ:Du}
\eeq  
The simplest solutions to this equation correspond to the four-point functions arising from {\it contact interactions} (see~\S\ref{sec:contact}). These solutions, which we will denote by $\hat C(u,v)$, are characterized by the simplest singularity structure possible. For four-particle scattering amplitudes, the simplest analytic structure we could possibly have corresponded to polynomials in the Mandelstam variables. But simply by scaling, this is impossible for our correlators, since even if they are rational functions, they must have some sort of poles. 
The simplest choice corresponding to ``contact" terms in the bulk is one where the correlator has poles in the ``total energy" variable $k_t \equiv \sum k_n$. As we will review in greater detail below, this pole reflects a universal singularity of the correlator associated with bulk time integrals where all the times head off to the infinite past, and the residue of this singularity is related to the flat-space scattering amplitude. The contact terms can be classified by the order of the pole.  For example, the simplest solution corresponding to the bulk $\upvarphi^4$ interaction is 
\beq
\hat C_0 = \frac{s}{k_t} = \frac{uv}{u+v}\, .  \label{equ:C00}
\eeq
A tower of higher-derivative contact interactions is created by repeated application of $\Delta_u$:\hskip 1pt\footnote{The basis (\ref{equ:Cn}) corresponds only to a subset of all possible contact terms, namely those arising from integrating out scalar particles. To generate contact terms coming from the exchange of massive particles with spin, we must feed the $\hat C_n$ into the spin-exchange ansatz of \S\ref{sec:results}--- cf.~(\ref{equ:Ansatz}) and (\ref{equ:coeff})---and sum over permutations. The resulting basis will be over-complete, but will encompass all possible scalar contact interactions. We thank Scott Melville and the anonymous referee for discussions on this point.} 
\beq
\hat C_n \equiv \Delta_u^n \hat C_0 = \left(\frac{s}{k_t}\right)^{2n+1} \hat f_n(u,v)\, , \label{equ:Cn}
\eeq
where the functional form of $\hat f_n(u,v)$ is fixed by conformal invariance.

\begin{figure}[t!]
    \centering
      \includegraphics[scale=1.0]{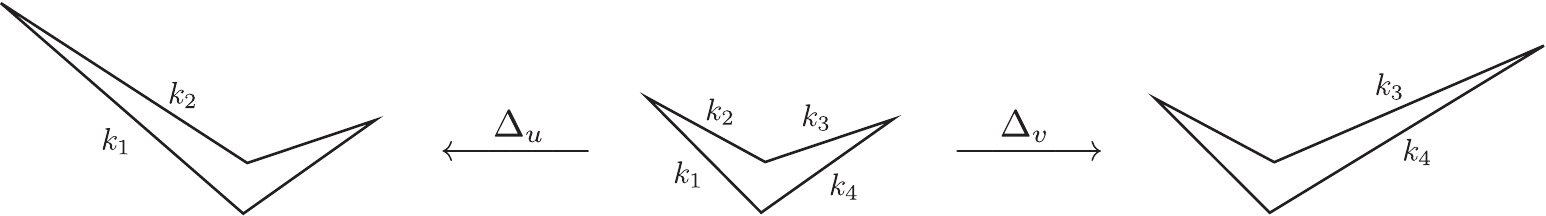}
           \caption{The four-point point function arising from tree exchange in the $s$-channel satisfy
a pair of ordinary differential equations (\ref{equ:exchange}), that determine the dependence as $u \propto (k_1 + k_2)^{-1}$ is varied, or as $v \propto (k_3 + k_4)^{-1}$ is varied. These correspond to two different ``holographic" pictures for time evolution, which are nontrivially mutually consistent.}
    \label{fig:shape}
\end{figure}

\vskip 4pt
For general {\it tree exchange},  we can write (\ref{contactgeneral}) as {\it ordinary} differential equations in $u$ and $v$ separately:
\begin{align}
	(\Delta_u+M^2)\hat F=(\Delta_{v}+ M^2)\hat F= \hat C(u,v)\, , \label{equ:exchange}
\end{align}
where the function $\hat C(u,v)$ must satisfy (\ref{contactgeneral}) and the parameter $M^2 \equiv \mu^2 + \frac{1}{4}$ is fixed in terms of the mass of the exchange particle.  The operator $\Delta_u$ determines how the four-point function changes as we scale  $k_1+k_2$, while keeping $k_3$ and $k_4$ fixed. Similarly, $\Delta_{v}$ describes the change when varying $k_3+k_4$, for fixed $k_1$ and $k_2$ (see Fig.~\ref{fig:shape}). In the limit $\mu \to \infty$, the differential operators $\Delta_u$ and $\Delta_{v}$ become irrelevant, and the solution reduces to the contact interactions obtained previously, $\hat F_c \equiv \mu^{-2} \hat C$.  This makes sense, since in this limit the exchange particle can be integrated out, and the theory should reduce to pure contact terms.  The four-point function for general tree exchanges is then obtained by solving the differential equations (\ref{equ:exchange}) with source terms given by the allowed conformally-invariant contact terms. 

\vskip 4pt
 A formal solution of (\ref{equ:exchange}) is 
\beq
\hat F = \sum_n \left(- \frac{\Delta_u}{M^2}\right)^n \frac{\hat C}{M^2}\, .\label{formalsol}
\eeq 
This is the {\it effective field theory} (EFT) expansion of the correlation function. We see that the solution is a sum over the contact terms $\Delta_u^n \hat C$, organized in powers of $M^{-2}$, i.e.~as an expansion in the inverse mass of the exchange particle is units of the Hubble parameter. 
Moreover, the solution is analytic to all orders in $M^{-2}$. For scattering amplitudes this statement would be exact, but in a time-dependent background we expect nonperturbative corrections due to spontaneous particle production. This effect scales as $e^{-\pi \mu}$ and  is encoded in additional homogeneous solutions to~(\ref{equ:exchange}).

\vskip 4pt
To identify the effects of {\it particle production}, it is useful to write the solution of the inhomogeneous equation (\ref{equ:exchange}) as two separate series expansions around $u=0$ and $u=\infty$. Unlike the EFT expansion, which is formal, these expansions will have a well-defined radius of convergence. Matching these two series expansions at $|u|=|v|$ will force us to add a specific homogeneous solution. This extra piece captures the effect of spontaneous particle production, but here arises simply from the wish to define the solution for all $u$ and $v$. 
The freedom to add further homogeneous terms to the solution is removed by requiring the solution only to have the physically expected singularities.

\begin{figure}[b!]
     \centering
     \begin{subfigure}[b]{0.35\textwidth}
         \centering
         \includegraphics[width=\textwidth]{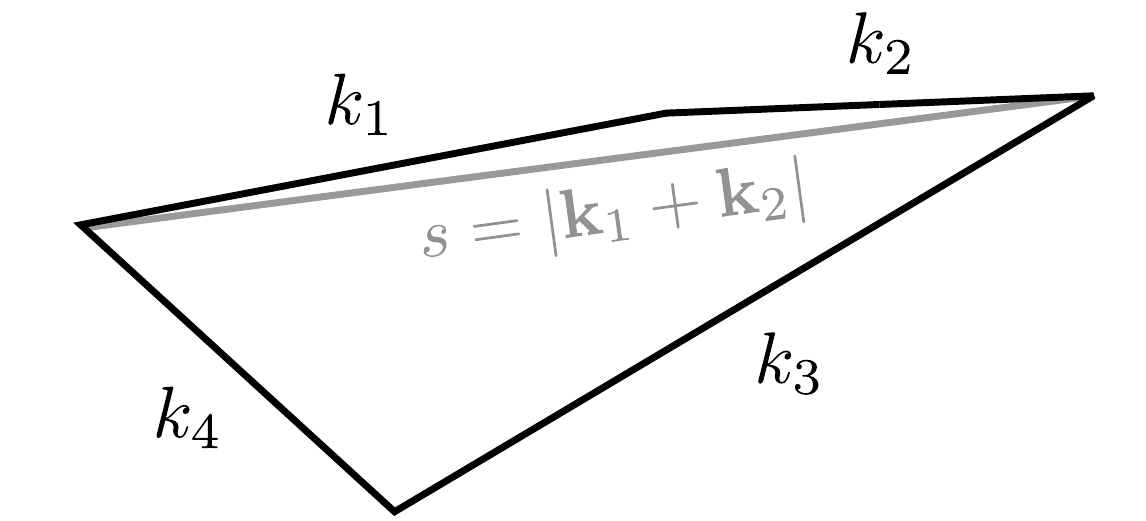}
         \caption{Collinear limit: $u \to +1$}
         \label{fig:colinear}
     \end{subfigure}
     \hspace{1.5cm}
     \begin{subfigure}[b]{0.35\textwidth}
         \centering
      \includegraphics[width=\textwidth]{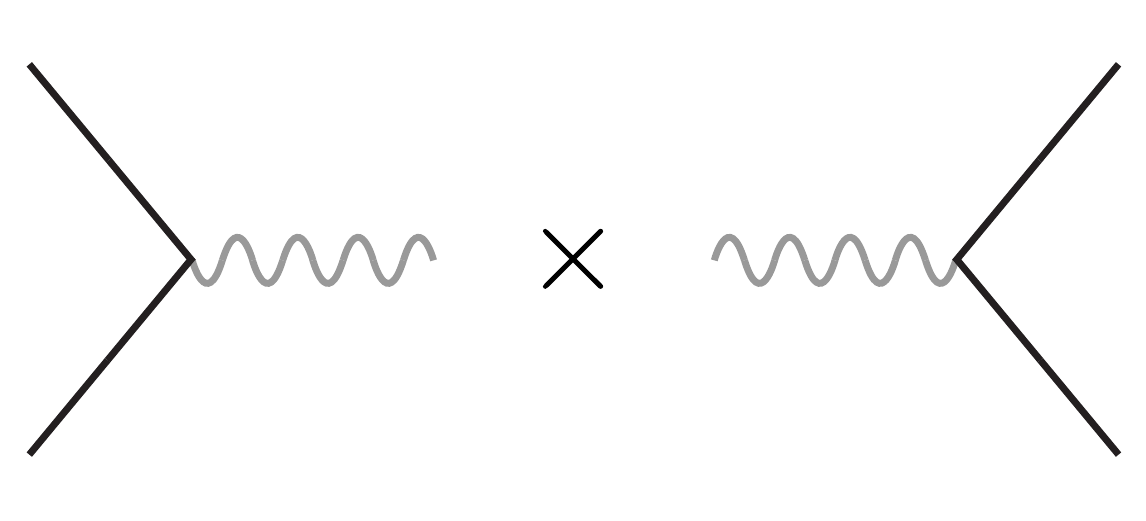}
         \caption{Factorization channel: $u,v \to -1$}
         \label{fig:factorization}
     \end{subfigure}
        \caption{Illustration of two important singularities of the solution $\hat F(u,v)$. The singularity in the collinear limit should be absent for the adiabatic vacuum. The singularity in the factorization channel is an avatar of the standard factorization of the scattering amplitude and therefore expected to be present.}
        \label{fig:singular}
\end{figure}

\vskip 4pt
Inspection of the operator $\Delta_u$ shows that the general solution of (\ref{equ:exchange}) has the following two singularities (see also Fig.~\ref{fig:singular}):
\begin{align}
\lim_{u \to +1} \hat F &\propto \log(1-u)\, , \label{equ:unphysical}\\
\lim_{u,v \to -1} \hat F &\propto \log(1+u)\log(1+v)\, . \label{equ:physical}
\end{align}
The former should be absent for the standard adiabatic vacuum, while the latter corresponds to the standard factorization of the flat-space amplitude and is therefore expected to be there. 
Requiring the absence of the unphysical singularity (\ref{equ:unphysical}) and correctly normalizing the physical singularity~(\ref{equ:physical}) completely fixes the solution  (see \S\ref{sec:exchange}).  
For small~$u$ and $v$, with $u\ll v$, we find 
\beq
\hat F =  \sum_n \frac{(-1)^n}{(n+\frac{1}{2})^2 + \mu^2} \left(\frac{u}{v}\right)^{n+1} + \frac{\pi}{2\cosh \pi \mu} \left(\frac{u}{v}\right)^{\frac{1}{2}} \frac{\sin(\mu \log u/v)}{\mu}\, ,
\eeq
where the first term is analytic and corresponds to the EFT expansion, while the second term is the non-analytic correction due to particle production.
The solution for general values of $u$ and $v$ takes a similar form and will be presented in Section~\ref{sec:dS-4pt}.
We see that the nonperturbative correction oscillates in the collapsed limit $u \to 0$, with a frequency set by the mass of the exchange particle. This characteristic signature of massive particle exchange during inflation was first highlighted in~\cite{Arkani-Hamed:2015bza}.
 Solutions corresponding to the exchange of particles with spin will be given in Section~\ref{sec:spin}.
Finally, in Section~\ref{sec:Massless}, we will show that these solutions can be mapped to the corresponding solutions for massless external fields.

\vskip 4pt
Note that this derivation, mirroring that of the construction of tree-level scattering amplitudes, reflects the ability to encode ``time without time", reproducing what is normally thought of as an intrinsically ``time-dependent" phenomenon from a purely boundary perspective.

\subsubsection{Bulk Perspective}

It is instructive to re-derive the conformal constraint equations from the bulk integrals defining the correlators in the in-in formalism~\cite{Maldacena:2002vr, Weinberg:2005vy}.
From the bulk perspective, the tree-exchange diagram involves integrating over two different
times $\eta$ and $\eta'$ associated with the three-point interactions.
A differential operator in $k_1+k_2$ can be applied to collapse the internal propagator, giving the equation we have found
above. 

\vskip 4pt
For concreteness, let us assume a three-point  vertex of the form $g\hs\upvarphi^2 \sigma$, where $\sigma$ is the massive particle being exchanged and $g$ is a coupling constant.
The corresponding four-point function can schematically be written as (see \S\ref{sec:bulk} for a more precise discussion)
\beq
\hat F =  -  \frac{g^2}{s^2} \int \frac{\d \eta}{\eta^2} \frac{\d \eta'}{\eta'{}^2} \, e^{ik_{12}\eta} e^{ik_{34} \eta'} \hat G(s\eta, s \eta')\, ,
\eeq
where $\hat G(s\eta,s\eta')\equiv s^3G(s,\eta,\eta')$ and we have set $H \equiv 1$. Instead of trying to compute the integral, we note that bulk-to-bulk propagator satisfies
\beq
\left(\eta^2 \partial_\eta^2 - 2 \eta \partial_\eta + s^2 \eta^2 + m^2\right)G(s,\eta,\eta') =  -i\eta^2 \eta'{}^2 \, \delta(\eta-\eta')\, . \label{propagator}
\eeq
Since $\hat G$ depends on $\eta,\eta'$ only in the combinations $s \eta$ and  $s \eta'$, it is easy to trade $\eta$-derivatives with $s$-derivatives. 
To have the derivatives act only on the first argument of the Green's function, we rescale $\eta' \to \eta'/s$. The function $s\hat F$ then satisfies the differential equation
\beq
\frac{1}{s}\left(s^2 \partial_s^2 - 2 s \partial_s - s^2 \partial_{k_{12}}^2 + m^2\right)\big(s  \hat F \big) = i\hs g^2s  \int \d \eta\, e^{i k_t \eta} = g^2 \frac{s}{k_t}\, .
\eeq
Letting $\hat F = \hat F(u,v)$, and using (\ref{equ:mdimsc}), we find 
\beq
	(\Delta_u+M^2)\hat F = g^2 \frac{uv}{u+v}\, ,
\label{equ:34}
\eeq
where $M^2 \equiv m^2 -2$.
This is precisely our previous result (\ref{equ:exchange}) for the lowest-order contact term~(\ref{equ:C00}). Permutation symmetry implies a second equation with $u \leftrightarrow v$. We can think of (\ref{equ:34}) as tracking the evolution in $\eta$ purely in boundary terms, while the corresponding equation in terms of $v$ tracks the evolution in $\eta'$. The fact
that these two histories are consistent, and give the same four-point function, is made manifest in the bulk picture, but it is nonetheless a nontrivial property of the solution.

\vskip 6pt
 A particularly interesting limit of the correlator is $k_t \to 0$. Of course, this limit cannot be reached for physical momenta with positive magnitudes, but requires an analytic continuation of the momenta.  
In this limit, we expect the correlator to have a singularity with a coefficient that is related to the flat-space scattering amplitude~\cite{Maldacena:2011nz,Raju:2012zr}. Rather remarkably, cosmological correlators therefore contain in them flat-space scattering amplitudes. This provides a strong consistency condition on the structure of cosmological correlators.

\vskip 4pt
This feature of the correlator has a simple bulk interpretation. The four-point function coming from a contact term involves an integral of the form
\beq
F \sim \int_{-\infty}^{0} \d \eta\, \eta^{p-1} e^{i k_t \eta}  A(k_1,k_2,k_3,k_4)  \,\to\, \frac{A_{\rm flat}}{k_t^p} \, ,
\eeq
where $A_{\rm flat}$ is the flat-space amplitude in the high-energy limit upon a suitable analytic continuation of the energies. 
The singularity for $k_t \to 0$ arises when all bulk interactions are pushed to very early times $\eta \to - \infty$.  Since all interactions are then far from the future boundary of the de Sitter spacetime, we expect to recover flat-space results. For instance, taking the flat-space limit of the contact terms in (\ref{equ:Cn}), we find
\beq
\lim_{k_t \to 0} C_n = (2n)!\, \frac{s_{\rm flat}^n}{k_t^{2n+1}}\, , \label{equ:flatC}
\eeq
where $s_{\rm flat} = (k_1+k_2)^2 - |\k_1+\k_2|^2$ is the four-dimensional Mandelstam invariant. We recognize the numerator as the derivative expansion of the flat-space amplitude.  
Similarly, the flat-space limit of the scalar-exchange solution turns out to be (see~\S\ref{sec:FSL}) 
\beq
\lim_{k_t \to 0} F = \frac{1}{s_{\rm flat}}\, (-k_t \log k_t) \, .
\eeq
We see that the discontinuity at $k_t=0$ is related to the high-energy limit of the scattering amplitude, $A_{\rm flat} = 1/s_{\rm flat}$.

\vskip 4pt
Finally, let us show that conformal invariance of the correlator implies Lorentz invariance of the amplitude.
To see this, consider the SCT constraint (\ref{WI_SCT}) for the ansatz $F = A_{\rm flat} k_t^{-p}$.
The most singular contribution, proportional to $k_t^{-p-2}$, arises from terms with two derivatives in the generator. However, the result is proportional to the sum of spatial momenta and therefore vanishes due to momentum conservation.  The next most singular piece, proportional to $k_t^{-p-1}$, is more interesting and leads to
\begin{equation}
		\sum_n k_n b^j \partial_{k_n^j} A_{\rm flat}=0\, . \label{equ:LIaflat}
\end{equation}
If we consider the ``four-momentum" $k^\mu\equiv (k,\k)^\mu$, with $k =|\k|$, then (\ref{equ:LIaflat}) implies that $A_{\rm flat}$ is a Lorentz-invariant function of the null momenta $k^\mu$, with Lorentz transformations given by $\delta k^0 = b^jk^j$ and $\delta k^j=b^jk^0$.

\subsection{Symmetries and Singularities}

In this section, we have shown how symmetries and singularities constrain the structure of correlation functions in de Sitter space in a way that is completely analogous to the 
bootstrapping of amplitudes in flat space. Let us summarize the many parallels and a few small differences:
\begin{itemize}
\item In flat space, Lorentz invariance demands that scattering amplitudes are functions of the Mandelstam invariants, $A_4(s,t)$, while in de Sitter space, conformal invariance imposes the constraint $(\Delta_u-\Delta_v) \hat F=0$ on the dimensionless four-point function $\hat F$.
\item While contact terms in flat space correspond to purely analytic terms in the amplitude, in de Sitter space contact terms have poles at $k_t  = 0$. Higher-order poles correspond to higher-derivative interactions.
\item By locality, the only allowed singularities of tree-level amplitudes are simple poles, 
i.e.~$(s-M^2) A_4 = \text{analytic}$,
corresponding to the exchange of massive particles. In de Sitter space, tree exchange is described by a pair of differential equations
\beq
\begin{aligned}
(\Delta_u + M^2) \hat F= \hat C\,, \\
(\Delta_v + M^2) \hat F= \hat C\,, 
\end{aligned}
\eeq
 where $\hat C$ is one of the contact solutions.
\item On the poles, $s \to M^2$, four-particle amplitudes factorize into products of three-particle amplitudes with positive coefficients. Similarly, de Sitter four-point functions factorize into products of three-particle 	amplitudes in the limit $u,v \to -1$. The correct normalization of this limit is an important condition in determining the solution.
\item The de Sitter correlators can have a singularity as $u \to 1$, corresponding to the 	flattened momentum configuration shown in Fig.~\ref{fig:singular}. This singularity specifies the initial state and has no analog in the amplitudes bootstrap. We expect no such singularity if the initial state is the standard adiabatic vacuum. 
Demanding the absence of this singularity, together with the correct normalization of the factorization channel, completely fixes the solution. 
\end{itemize}
In the rest of the paper, we will apply this formalism to derive analytic solutions for de Sitter four-point functions and inflationary three-point functions.

\section{De Sitter Four-Point Functions}
\label{sec:dS-4pt}

We begin our exploration with the four-point function of scalar fields in de Sitter space. We will consider both contact interactions and tree-level exchange of massive scalars. We will take the external fields to be conformally coupled scalars $\varphi$, with scaling dimension $\Delta=2$ (corresponding to bulk scalars with $m = \sqrt{2} H$). A spin-raising operator will relate the solutions that we will obtain in this section to solutions for general spin exchange~(see~\S\ref{sec:spin}), while a weight-shifting operator will map the solutions to four-point functions of external scalars~$\phi$, with  $\Delta=3$ (corresponding to massless scalars in the bulk)~(see \S\ref{sec:Massless}+\S\ref{sec:ladder}).

\vskip 4pt
In \S\ref{sec:contact}, we derive the simplest solutions to the conformal invariance equations (\ref{equ:conformal}), corresponding to contact interactions in the bulk. We show that the solutions can be organized by inverse powers of $k_t = \sum k_n$. In \S\ref{sec:exchange}, we study tree-level exchange for which the conformal invariance equations separate into a pair of ordinary differential equations. We first consider a simple limit of these equations, where the dynamics reduces to that of a forced harmonic oscillator. We identify the oscillatory part of the solution with the effects of particle production in the expanding spacetime. Finding the general solution is more involved, but the structure of the answer is the same as in the harmonic oscillator limit. In \S\ref{sec:FSL}, we explicitly confirm the expectation that the correlator contains the scattering amplitude in the limit $k_t \to 0$. Moreover, we relate the discontinuity at $k_t=0$  to the effects of particle production, providing an interesting link between scattering in flat space and particle production in curved space. Finally, in \S\ref{sec:UV}, we speculate about the fate of the $k_t$-singularity in gravitational UV completions.

\subsection{Contact Interactions}
\label{sec:contact}

We first consider the four-point functions associated with contact interactions.
Up to six derivatives, the independent bulk interactions are $\upvarphi^4$, $\upvarphi^2 \upvarphi_{;\mu \nu}\upvarphi^{;\mu \nu}$ and $\upvarphi^2 \upvarphi_{;\mu \nu \rho}\upvarphi^{;\mu \nu \rho}$, using integration by parts and equations of motion.\footnote{Counting such independent operators up to total derivatives and equations of motion is equivalent to counting
flat-space $S$-matrices; see e.g.~the discussion in \S4.1~of~\cite{Heemskerk:2009pn}.} In the following, we will reproduce this fact from the boundary perspective and determine the corresponding four-point functions.

\vskip 4pt
First, we assume that the four-point function $F$ depends only on the magnitudes~$k_n$. In that case, the $s$ and $t$ dependences drop out in \eqref{W1F}, and the constraint equations (\ref{equ:conformal}) simply become  
\beq
(\partial_{k_n}^2-\partial_{k_m}^2)F=0\, .
\eeq
 These wave equations are solved by $F(k_n) =F_c(\pm k_1\pm k_2\pm k_3\pm k_4)$, where $F_c$ is an arbitrary function. The absence of any singularities in the physical region $k_n >0$, and the fact that $F_c$ has mass dimension $-1$, fixes the solution to be 
\beq
F_c(k_n)  = \frac{c_0}{k_t}\, , \label{equ:Fkt}
\eeq 
where $c_0$ is an arbitrary coupling constant. This is the four-point function due to the bulk interaction $\upvarphi^4$. 

\vskip 4pt
Next, we allow $F$ to depend on $s$ and $t$. It is easy to see from the form of the conformal invariance equation that the dependence on  $s$ and $t$ must be a polynomial dependence on $s^2$ and~$t^2$. We therefore try the ansatz 
\begin{align}
	F_c(k_n,s,t)=g_s(k_n)\hskip 1pt s^2+g_t(k_n) \hskip 1pt t^2+h(k_n)\, ,
\end{align}
assuming higher orders of $s^2$ and $t^2$ to be absent. The conformal invariance equation then implies  that the coefficient functions satisfy $(\partial_{k_n}^2-\partial_{k_m}^2)g_{s,t}=0$, which must take the form $g_{s,t}=c_{s,t} k_t^{-3}$ because $F_c$ has mass dimension $-1$. The only singularities of the function $h(k_n)$ are also of the form $k_t^{-3}$, and the general solution is
\begin{align}
F_c(k_n,s,t)=	\frac{c_s[s^2+(k_1+k_2)(k_3+k_4)]+c_t[t^2+(k_2+k_3)(k_1+k_4)]}{k_t^3}+\frac{c_0}{k_t}\, . \label{equ:Fck}
\end{align}
A symmetry under the exchange $k_1 \leftrightarrow k_3$ would require the coefficients $c_s$ and $c_t$ to be equal to each other, while a symmetry under $k_1 \leftrightarrow k_2$ would enforce $c_s=c_t = 0$. The four-point function of identical scalars therefore has no nontrivial dependence on $s^2$ and $t^2$ at this order. 
This has a simple bulk interpretation: The interaction $\upvarphi^2 (\partial_\mu\upvarphi)^2$ does not give rise to a new shape, since it is identical to $\upvarphi^4$ after integration by parts; i.e.~$\upvarphi^2 (\partial_\mu\upvarphi)^2 \sim \upvarphi^3\Box\upvarphi =2H^2\upvarphi^4$ on-shell.

\vskip 4pt
To derive the most general form of the four-point function due to contact interactions, we now solve the conformal invariance equation systematically.
Consider the ansatz $F=s^{-1} \hat F(u,v)$, where $\hat F$ satisfies (\ref{contactgeneral}):
\beq
(\Delta_u - \Delta_v) \hat F = 0\, , \label{equ:DuvF}
\eeq
where the differential operator $\Delta_u$ was defined in (\ref{equ:Du}). 
The simplest solution of this equation is given by  (\ref{equ:Fkt}), which we can write as
\beq
\hat F_c(u,v) = c_0 \frac{uv}{u+v}  \equiv c_0\hs \hat C_0(u,v) \, .
\eeq 
All higher-derivative contact interactions coming from the integrating out of massive scalars can be generated by acting with $\Delta_u$ on $\hat C_0$, i.e.~$\hat C_n(u,v) \equiv \Delta_u^n \hat C_0(u,v)$, 
which trivially satisfy the constraint $(\Delta_u -\Delta_{v})\hat C_n=0$. 
A general contact solution is a linear combination of these solutions
\begin{align}
	\hat F_c(u,v) 
	&\ =\ \sum_{n=0}^\infty c_n \Delta_u^n \hat C_0(u,v)  \nonumber \\[4pt]
	&\ =\ c_0 \frac{uv}{u+v}-2\hs c_1\left( \frac{uv}{u+v}\right)^3 \frac{1+uv}{uv} \nonumber \\[2pt]
	&\qquad  - 4\hs c_2 \left( \frac{uv}{u+v}\right)^5 \frac{u^2+v^2 +uv(3u^2+3v^2-4) - 6(uv)^2-6(uv)^3}{(uv)^3}+\cdots\, ,\label{generalcontact}
\end{align}
where the dimensionless parameters $c_n$ are couplings constants.\footnote{In a bulk theory with a derivative expansion, i.e. a weakly coupled EFT of a self-interacting scalar field with a cutoff $\Lambda$, we expect the expansion coefficients to satisfy the scaling $c_n\sim (H/\Lambda)^{2 n}$. Even though the contact terms appear to be on an equal footing in (\ref{generalcontact}), the $\hat C_n$'s therefore become less relevant for increasing $n$.}  The solution has a few interesting features:
\begin{itemize}
\item The expansion is organized in powers of  
$uv/(u+v) = k_t^{-1}$, multiplied by functions whose form is dictated by conformal symmetry. 
Already at this early stage, we can appreciate the power of this boundary perspective on de Sitter correlators. Even these contact interactions are relatively intricate functions of the momenta, but are fully controlled by symmetries and singularities without any reference to a Lagrangian description.

\item The shapes produced by different bulk interactions correspond to linear combinations of the contact terms $\hat C_n$; for example  $\{c_0,c_1,c_2\} =\{1,0,0\}$ for $\upvarphi^4$ and $\{1,1,1/4\}$ for $(\partial_\mu\upvarphi)^4$. 
Here, the coefficients $c_n$ were determined by looking at a single arrangement of the external legs, and the final four-point function is obtained by summing over all permutations. For identical fields, the contact term $\hat C_1$ then won't contribute to the four-point function; see the discussion below~(\ref{equ:Fck}).

\item The solution is symmetric under the exchange $u \leftrightarrow v$.  In fact, this is a general feature of all conformally-invariant four-point interactions. This is manifest in (\ref{generalcontact}) because $\hat C_0(u,v)$ is symmetric and $\hat C_1 = \Delta_u \hat C_0 =   \Delta_v \hat C_0$, etc.~(by conformal symmetry). 
\item Finally, the flat-space limit $k_t \to 0$, or $u\to - v$, of the individial contact terms is
\beq
\lim_{k_t \to 0} C_n = (2n)!  \, \frac{s_{\rm flat}^n}{k_t^{2n+1}}\, . \label{equ:Cnflat}
\eeq
We see that each contact term has a pole in $k_t$, whose residue is a positive power of the Mandelstam invariant~$s_{\rm flat}$. 
This confirms the expected relation between the singularity of the correlator at $k_t=0$ and the scattering amplitude for contact interactions. When the correlator is a sum of contact terms, the coefficients of the expansion should inherit positivity from the positivity of the corresponding parameters in the low-energy limit of the scattering amplitude~\cite{Adams:2006sv}.
\end{itemize}

\subsection{Tree-Level Exchange}
\label{sec:exchange}

Next, we consider the four-point function involving the tree-level exchange of a scalar particle. 
We will perform the analysis in several steps of increasing level of generality. 

\vskip 4pt
In the operator product expansion (OPE) limit, we expect the four-point function to factorize into a product of three-point functions. A simple product of conformally invariant three-point functions should therefore produce a conformally invariant four-point function. To see this, consider the ansatz
\begin{align}
	\hat F(u,v) = \hat I(u) \hat J(v)\, , \label{equ:FIJ}
\end{align}
where $\hat I(u)$ and $\hat J(v)$ satisfy the correct constraint equation for a three-point function involving a massive particle 
(see \S\ref{app:3pt}): 
\beq
\begin{aligned}
	\left[\Delta_u+\left(\mu^2+\frac{1}{4}\right)\right]\hat I(u) &=0\, , \\
	\left[\Delta_v+\left(\mu^2+\frac{1}{4}\right)\right]\hat J(v) &=0\, .
\end{aligned}
\eeq
It is then trivial to see that (\ref{equ:FIJ}) solves the conformal invariance condition (\ref{equ:DuvF}).

\vskip 4pt 
This motivates us to find a more general solution of conformal invariance, not of the factorized form, but satisfying $(\Delta_u-\Delta_{v})\hat F=0$ by taking 
\begin{align}
	\left[\Delta_u+\left(\mu^2+\frac{1}{4}\right)\right]\hat F=\left[\Delta_{v}+\left(\mu^2+\frac{1}{4}\right)\right]\hat F= \hat C(u,v)\, .
	\label{InhomogeneousFu}
\end{align}
Note that these are now {\it ordinary} differential equations in $u$ and $v$ separately.  The solution corresponding to a product of three-point functions corresponds to homogeneous solutions to these equations, which explains the introduction of the $\mu$-dependent terms, to dictate the mass of the exchanged particle. For consistency, the source functions $\hat C(u,v)$ must satisfy $(\Delta_u -\Delta_{v})\hat C(u,v)=0$ and are thus themselves conformally invariant. 
For large $\mu$, the interaction effectively reduces to a contact interaction, $\hat F(u,v)\to \hat F_c(u,v) = \mu^{-2}\hat C(u,v)$. Given our classification of the contact solutions in \S\ref{sec:contact}, we can therefore classify the corresponding exchange solutions.

\vskip 4pt
The differential operator $\Delta_u$ has three singularities at $u \to 0$ and $u \to \pm 1$, and thus the homogeneous solutions of (\ref{InhomogeneousFu})  can be expressed as hypergeometric functions, which we will denote by $\hat F_+$ and $\hat F_-$.
The Wronskian $W[\hat F_+,\hat F_-]= \hat F_+ \hat F_-^{\hskip 1pt \prime}- \hat F_-\hat F_+^{\hskip 1pt \prime}$ satisfies $\partial_u W= 2u/(1-u^2)\hskip 1ptW$, so it is natural to normalize the solutions so that $W=1/(1-u^2)$.\footnote{This choice of normalization avoids unnecessary factors of $\mu$ in the intermediate steps, but the final solution will of course not depend on it. } 
The homogeneous solutions are then
\begin{align}
	\hat F_\pm (u) &= \left(\frac{i u}{2\mu}\right)^{\frac{1}{2} \pm i\mu }  {}_2F_1\Bigg[\begin{array}{c} \frac{1}{4}\pm\frac{i\mu}{2},\hs \frac{3}{4}\pm\frac{i\mu}{2}\\[2pt] 1\pm i\mu \end{array}\Bigg|\, u^2\Bigg]\, .\label{equ:homo}
\end{align}
From the differential equation directly, we can see that the solution has logarithmic singularities for $u \to \pm 1$. This is reflected in the fact that the sum of the first two parameters of the hypergeometric function equals the third, which implies that there is a logarithmic singularity as we take its argument $u^2\to 1$ (see Appendix~\ref{app:identities}). In the limit $u \to 1$, the solution has the following singular behavior 
\begin{align}
\lim_{u \to 1} \hat F_\pm(u) &= \alpha_\pm \log(1-u) \, ,\quad {\rm where} \quad \alpha_\pm \equiv -\left(\frac{i}{2\mu}\right)^{\frac{1}{2}\pm i\mu}\frac{\Gamma(1\pm i\mu)}{\Gamma(\frac{1}{4}\pm\frac{i\mu}{2})\Gamma(\frac{3}{4}\pm\frac{i\mu}{2})}\, .\label{equ:Shomo}
\end{align} 
As we explained in \S\ref{sec:dS}, this corresponds to a singularity in the collinear limit, which should be absent in the standard adiabatic vacuum. Removing this singularity will be important for determining the physically relevant solution.

\vskip 4pt
We now wish to find the particular solution to the inhomogeneous equation \eqref{InhomogeneousFu}. For concreteness, we will choose the simplest contact term $\hat C_0$ as a source:\footnote{As we will see below, the exchange solutions with higher-order contact terms as sources are related in a very simple way to the solution derived from $\hat C_0$.}
\begin{align}
	\left[\Delta_u+\left(\mu^2+\frac{1}{4}\right)\right]\hat F= g^2 \frac{uv}{u+v}\, .
	\label{InhomogeneousFu2}
\end{align}
The coupling constant $g^2$ can be absorbed into the normalization of $\hat F$, so we will set $g^2 \equiv 1$.
We will first study this equation in a simple limit and then for the general case. 

\paragraph{Harmonic oscillator}
Consider the limit $v\to 0$. Writing $u=\xi v$ and $\hat F=v\tilde F$, the differential equation (\ref{InhomogeneousFu2}) then takes the form
\begin{align}
	\left[\xi^2\partial_\xi^2+\left(\mu^2+\frac{1}{4}\right)\right] \tilde F = \frac{\xi}{1+\xi}\, .\label{InhomogeneousFx}
\end{align}
Defining $\tilde F = \sqrt{\xi} \,q$ and $\xi = e^\rho$, this becomes the equation of a {\it forced harmonic oscillator}
\beq
	(\partial_\rho^2 + \mu^2)\hs q= \frac{1}{2\cosh(\frac{1}{2}\rho)}\, . \label{equ:forced}
\eeq
The homogenous solutions of this equation are $e^{\pm i \mu \rho} = \xi^{\pm i \mu}$. The particular solution that vanishes at early ``times", i.e.~when $\xi 
\to 0$, can be written as a power series in $\xi$:
\beq
q_<(\xi) = \sum_{n=0}^\infty (-1)^n \frac{\xi^{n+1/2}}{(n+\frac{1}{2})^2 + \mu^2}\, .
\eeq
Modulo the overall factor of $\sqrt{\xi}$, this solution is analytic around $\xi=0$.
It is convergent for $\xi \le 1$ and divergent for $\xi>1$.
We are interested in the analytic continuation of the solution for $\xi > 1$. In this simple case, we could recognize the solution as a hypergeometric function 
\begin{align}
	q_<(\xi) = \sum_\pm \frac{\pm i\sqrt{\xi}}{\mu(1\pm 2i\mu)}\,{}_2F_1\Bigg[\begin{array}{c} \frac{1}{2}\pm i\mu,\hs 1\\[2pt] \frac{3}{2}\pm i\mu \end{array}\Bigg|\, -\xi\Bigg]\, ,\label{Fsol1}
\end{align}
and rely on the known analytic structure of the hypergeometric functions. 
To make contact with the general case, however, it will be more useful to understand the regime $\xi > 1$  directly from the properties of the differential equation and the series expansion itself.

\vskip 4pt
Note that another solution of the differential equation, which is analytic around $\xi \to \infty$, can be written as a power series in $1/\xi$:
\beq
q_>(\xi) =  \sum_{n=0}^\infty (-1)^n \frac{\xi^{-n-1/2}}{(n+\frac{1}{2})^2 + \mu^2}\, .
\eeq
Since $q_<(\xi)$ and $q_>(\xi)$ satisfy the same differential equation, the difference $q_<(\xi)-q_>(\xi)$ must be a solution of the homogeneous equation, i.e.
\beq
q_<(\xi)-q_>(\xi) = \sum_{\pm }A_\pm \xi^{\pm i\mu} \, .
\eeq
Continuity
at $\xi=1$ then implies that $A_+=-A_- \equiv A$ and
\begin{align}
2i\mu A&= \sum_n^\infty (-1)^n \frac{(2n+1)}{(n+\frac{1}{2})^2+\mu^2} = \frac{\pi}{\cosh\pi \mu} \, ,
\end{align}
where the second equality can be established by showing that the residues at  $\mu=\pm i(k+\tfrac{1}{2})$ match on both sides. We have therefore obtained an explicit form for the solution of the differential equation which is analytic around the origin: 
\beq
\tilde F_<(\xi) = \begin{cases}  \displaystyle \ \sum_{n=0}^\infty (-1)^n \frac{\xi^{n+1}}{(n+\frac{1}{2})^2 + \mu^2} & \text{for $\xi \le 1$}\, , \\[20pt]
 \displaystyle \ \sum_{n=0}^\infty (-1)^n \frac{\xi^{-n}}{(n+\frac{1}{2})^2 + \mu^2}  + \frac{\pi}{\cosh \pi \mu} \frac{\xi^{\frac{1}{2}-i\mu} -\xi^{\frac{1}{2}+i\mu}}{2i\mu}& \text{for $\xi \ge 1$}\, . \end{cases} 
 \label{equ:f0}
\eeq
We see that it is impossible for the solution which is analytic around $\xi = 0$  to be analytic around $\xi = \infty$. It is also notable that for large $\mu$, we have an
expansion for $\tilde F_<(\xi)$ in powers of $1/\mu$, and each term in this ``effective field theory expansion" is analytic both around $\xi \to 0$ and $\xi \to \infty$; but there is a nonperturbative correction of order $e^{-\pi\mu}$ which spoils this property. In the oscillator analog, we begin with a ball at rest at early times, and ``kick" it with the forcing term, ending up with an oscillating ball at late times. In the cosmological context, the presence of these oscillatory terms can physically be attributed to particle production by the time-dependent inflationary background. It is striking that an effect  we normally so vividly ascribe to ``time-dependence", follows directly from consideration of the conformal invariance equations, which are formulated purely on the future boundary of the spacetime and make no direct reference to ``time evolution" whatsoever.

\paragraph{General case} Having studied the solution in the limit $v\to 0$, let us return to the full differential equation \eqref{InhomogeneousFu}. We will again find it easier to understand the analytic properties of the inhomogeneous solution directly from a series expansion. The solution written in canonical hypergeometric form can be found in Appendix~\ref{app:hyper}.

\vskip 4pt
We again have two solutions,  $\hat F_<(u,v)$ and $\hat F_>(u,v)$, which are analytic for $u \to 0$ and $u \to \infty$, repsectively.
We will give a convergent series expansion for $\hat F_<(u,v)$ when $u<v$, and another for $\hat F_>(u,v)$ when $u>v$. As before, we will match the solutions at $u=v$, so that $\hat F_<(u,v)$ can be extended to $u>v$.
The explicit solution for $\hat F_<$ is
\beq
\hat F_<(u,v) = \begin{cases}  \displaystyle \ \sum_{m,n=0}^\infty c_{mn}u^{2m+1}(u/v)^{n} & \text{$u \le v$}\, , \\[20pt]
 \displaystyle \ \sum_{m,n=0}^\infty c_{mn}v^{2m+1}(v/u)^{n} + \frac{\pi}{\cosh \pi \mu}  \left(\hat{F}_+(v) \hat {F}_-(u)  -  \hat{F}_-(v) \hat{F}_+(u)  \right) & \text{$u \ge v$}\, , \end{cases} \label{equ:SOL1}
\eeq
where the series coefficients are given by
\beq
c_{mn}  = \frac{(-1)^n(n+1)(n+2)\cdots(n+2m)}{[(n+\frac{1}{2})^2+\mu^2][(n+\frac{5}{2})^2+\mu^2]\cdots [(n+\frac{1}{2}+2m)^2+\mu^2]}\, .\label{cmn}
\eeq
The details of the derivation can be found in the following insert.

\begin{framed}
\noindent
{\small {\it Derivation.}---For $u < v$, we seek a power series solution of the form
\begin{align}
	\hat F_<(u,v) = \sum_{m,n=0}^\infty c_{mn}u^{m+1}(u/v)^{n}\, .
\end{align}
This form of the series solution is motivated by the series expansion of the source term $uv/(u+v)$.
Plugging this ansatz into the differential equation, we find that the series coefficients obey the following recursive relation:
\begin{align}
	c_{0n} = \frac{(-1)^n}{(n+\frac{1}{2})^2+\mu^2}\, , \quad c_{1n} = 0\, , \quad c_{m+2,n} = \frac{(m+n+2)(m+n+1)}{(m+n+\tfrac{5}{2})^2+\mu^2}\, c_{mn}\, ,
\end{align}
where the condition $c_{1n}=0$ implies that $c_{mn}$ vanishes for all odd $m$. Redefining $2m\to m$ and solving the recursive relation, we find
\beq
	\hat F_<(u,v)= \sum_{m,n=0}^\infty c_{mn}\,u^{2m+1}(u/v)^{n}\, ,\label{fuv}
\eeq
with 
\beq
c_{mn}  = \frac{(-1)^n(n+1)(n+2)\cdots(n+2m)}{[(n+\frac{1}{2})^2+\mu^2][(n+\frac{5}{2})^2+\mu^2]\cdots [(n+\frac{1}{2}+2m)^2+\mu^2]}\, .\label{cmn}
\eeq
Note that the series solution (\ref{fuv}) is the unique particular solution that is regular at the origin, since both homogeneous solutions are non-analytic at $u=0$. In contrast, regularity around $u=\infty$ does not uniquely fix $\hat F_>(u,v)$. 
Instead, we will demand that the full particular solution is symmetric under the exchange $u \leftrightarrow v$.
For $u > v$, the solution therefore is
\beq
\hat F_>(u,v) = \hat F_<(v,u) =  \sum_{m,n=0}^\infty c_{mn}\,v^{2m+1}(v/u)^{n}\, , \label{equ:F+}
\eeq
where $c_{mn}$ are same coefficients as in (\ref{cmn}).
It is straightforward to check that (\ref{equ:F+}) solves~(\ref{InhomogeneousFu}).

\vskip 4pt
As before, the difference between the two particular solutions must be a homogeneous solution, so we can write 
\beq
\hat F_<(u,v) - \hat F_>(u,v)= \sum_\pm A_\pm(v;\mu) \hat {F}_\pm(u) \,,
\eeq
where the functions $\hat {F}_\pm$ are given by~(\ref{equ:homo}). Evaluating this at $u= v$ gives $A_\pm (v;\mu)= \mp a(v;\mu) \hat {F}_\mp (v)$, for some function $a(v;\mu)$. Matching the $u$-derivative at $u = v$ fixes the function in terms of the Wronskian of the homogeneous solutions, namely
\begin{align}
\left.\left(\partial_u \hat F_< - \partial_u \hat F_>\right)\right|_{u \to v} &= a(v;\mu) \left(\hat {F}_+(v) \hat {F}_-^{\hskip 1pt \prime}(v) - \hat {F}_+^{\hskip 1pt \prime}(v)  \hat {F}_-(v) \right) \nonumber \\
&= \frac{a(v;\mu)}{1-v^2} = a(v;\mu) \times \sum_{m=0}^\infty v^{2m}\, .
\end{align}
It remains to evaluate the left-hand side
\begin{align}
\left.\left(\partial_u \hat F_< - \partial_u \hat F_>\right)\right|_{u \to v} = \sum_{m,n=0}^\infty (2m+2n+1)c_{mn}v^{2m}\, .
\end{align}
Somewhat remarkably, the sum over $n$ above is $m$-independent; we have the identity
\begin{align}
	\sum_{n=0}^\infty (2m+2n+1)c_{mn} = \sum_{n=0}^\infty\frac{(-1)^n (2m+2n+1)(n+1)\cdots(n+2m)}{[(n+\frac{1}{2})^2+\mu^2]\cdots [(n+\frac{1}{2}+2m)^2+\mu^2]}=\frac{\pi}{\cosh\pi\mu}\, .
\end{align}
Once again, this identity can be established by matching the residues on both sides. For instance, for $m=1$, only the terms in the sum with $n=k-2,k$ have a residue as $\mu\to \pm i(k+\frac{1}{2})$; the residue is
\begin{align}
	-\frac{(-1)^k(1+k)(2+k)}{2(1+2k)}+\frac{(-1)^k(k-1)k}{2(1+2k)}=(-1)^k\, ,
\end{align}
which matches that of the right-hand side. This allows us to fix $a(v;\mu)={\pi/\cosh\pi\mu}$ independent of~$v$. }
\end{framed}

\noindent
The solution (\ref{equ:SOL1}) still has two deficiencies: 
\begin{itemize}
\item First, it isn't symmetric in $u \leftrightarrow v$, as required by consistency of the bulk evolution (and the symmetry of the conformally-invariant contact terms). In particular, the nonperturbative correction is absent as $u\to 0$, and as a result the solution is analytic in this limit.

\item Second, it has a singularity at $u = 1$: 
\beq
\lim_{u \to 1}\hat F_<(u,v) = \frac{\pi}{\cosh \pi \mu} \Big(\alpha_+ \hat {F}_-(v) - \alpha_- \hat {F}_+(v)\Big) \log(1-u)\, ,
\eeq
where the constants $\alpha_\pm$ were defined in (\ref{equ:Shomo}). This singularity in the folded configuration, $k_1+k_2 = |\k_1+\k_2|$, is a signature of excited initial states~\cite{Holman:2007na,Flauger:2013hra}  and should be absent in the standard Bunch-Davies vacuum. Removing this singularity further restricts the solution.
\end{itemize}

\noindent
In order to find a solution that is symmetric in $u \leftrightarrow v$ and regular at $u = 1$, we add appropriate homogeneous solutions. A residual freedom in adding further homogeneous solutions is fixed by imposing a boundary condition in the limit $u,v \to -1$. In this limit, the four-point function factorizes into a product of three-particle amplitudes. Correctly normalizing this limit uniquely fixes the solution to be
\begin{eBox}
\beq
\hat F(u,v) = \begin{cases}  \displaystyle \ \sum_{m,n=0}^\infty c_{mn}u^{2m+1}(u/v)^{n} + \frac{\pi}{2\cosh\pi \mu} \, \hat g(u,v)& \text{$u \le v$}\, , \\[20pt]
 \displaystyle \ \sum_{m,n=0}^\infty c_{mn}v^{2m+1}(v/u)^{n} + \frac{\pi}{2\cosh \pi \mu} \, \hat g(v,u)& \text{$u \ge v$}\,, \end{cases} \label{equ:SeriesSol}
\eeq
\end{eBox}
where we have defined 
\begin{align}
	\hat g(u,v) &\equiv \hat {F}_+(u)\hat{F}_-(v) - \hat {F}_-(u) \hat{F}_+(v) \,-\,\frac{\alpha_-}{\alpha_+} (\beta_0+1) \hat{F}_+(u)  \hat{F}_+(v) \,-\,\frac{\alpha_+}{\alpha_-}  (\beta_0-1) \hat{F}_-(u)  \hat{F}_-(v)\nn
&\quad +\beta_0\Big[\hat {F}_-(u)\hat{F}_+(v) + \hat {F}_-(v) \hat{F}_+(u)\Big]\, ,\label{guv} 
\end{align}
with
\beq
\beta_0 \equiv \frac{1}{i\sinh\pi\mu}\, .
\eeq
It is interesting that the only solutions that are symmetric and free of spurious singularities are necessarily non-analytic around $u\to 0$ and $v \to 0$. This is how particle production in the time-dependent bulk spacetime is encoded in the boundary correlators.

\begin{figure}[h!]
\centering
\includegraphics[scale=0.5]{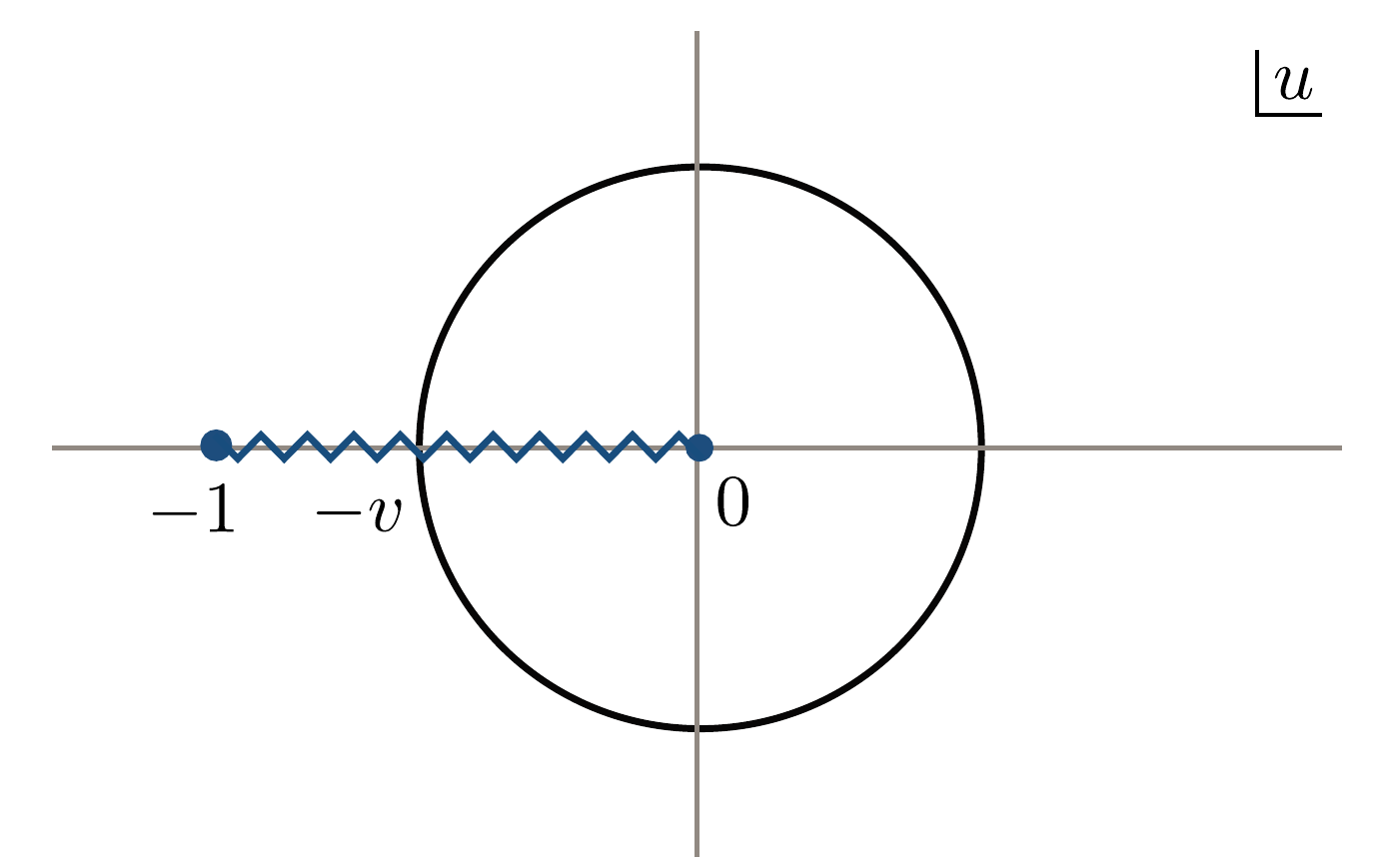}
\caption{Illustration of the analytic structure of the function $\hat F(u,v)$. } 
\label{fig:analytic}
\end{figure}

\begin{framed}
\noindent
{\small {\it Derivation.}---The most general solution of the differential equation that is symmetric in $u\leftrightarrow v$ is 
\begin{align}
	\hat F(u,v) &\,=\,  \hat F_<(u,v)+\frac{\pi}{2\cosh \pi\mu}\Big\{\big[\hat {F}_+(u) \hat{F}_-(v) - \hat{F}_-(u)  \hat{F}_+(v) \big]\nn[4pt]
	&\quad\ +\Big[ \beta_+\hat F_+(u)\hat F_+(v)+\beta_-\hat F_-(u)\hat F_-(v)+\beta_0\big[\hat F_-(u)\hat F_+(v)+\hat F_+(u)\hat F_-(v)\big]\Big]\Big\}\, .\label{generalF}
\end{align}
Demanding the absence of the spurious singularity at $u=1$ implies $\beta_\pm = -(\beta_0\mp 1)\alpha_\mp/\alpha_\pm$, but still leaves $\beta_0$ unfixed. To fix $\beta_0$, we take the limit $u=v \to -1$. 
Since the function $\hat F(u,v)$ has a branch point at $u=v=-1$ (see Fig.~\ref{fig:analytic}), the limit will depend on how this point is approached. In particular, for $u=v=-1^+ \equiv e^{i\pi}$ and $u=v=-1^- \equiv e^{-i\pi}$, we get
\begin{align}
	\lim_{u,v\to -1^+}\hat F(u,v) &\,=\,\frac{i}{2} \big(\beta_0 \sinh \pi\mu \,-\, \cosh \pi\mu \big)\log(1+u)\log(1+v)\, , \\
	\lim_{u,v\to -1^-}\hat F(u,v) &\,=\,\frac{i}{2} \big(\beta_0 \sinh \pi\mu \,+\, \cosh \pi\mu \big)\log(1+u)\log(1+v)\, . 
\end{align}
We see that these limits have a universal $\beta_0\sinh \pi\mu$ contribution and ambiguous $\cosh\pi\mu$ terms. To isolate the physically relevant contribution, we define the $u=v \to -1$ limit as 
\begin{align}
\lim_{u,v\to -1} \hat F(u,v) &\,\equiv\,	\frac{1}{2}\left(\lim_{u,v\to -1^+} + \lim_{u,v\to -1^-} \right)\hat F(u,v) \nonumber \\[4pt]
 &\,=\, \frac{i\beta_0 \sinh\pi\mu}{2}\log(1+u)\log(1+v)\, . \label{uvminus1}
\end{align}
Restoring the coupling constant $g^2$, the coefficient of this singularity takes the form of a product of three-particle amplitudes.
The correct normalization of the singularity that follows from a simple bulk computation is $\frac{1}{2}$ (see Appendix~\ref{app:bulk}), which we get by setting
\beq
\beta_0= \frac{1}{i\sinh\pi\mu}\, .
\eeq
As we show in Appendix~\ref{app:bulk}, the same value of $\beta_0$ can also be found purely from the boundary point of view. This can be done by imposing the correct normalization of the disconnected contribution to the four-point function. Finally, another way to determine $\beta_0$ is by comparing the $u \to -1$ limit, for general $0<v<1$, with the corresponding limit of the bulk calculation. In that limit, the four-point function factorizes into the product of a three-point correlation function and a three-particle amplitude. This limit has the advantage that it doesn't depend on the way the branch point is approached.}
\end{framed}

\paragraph{Collapsed limit} The effects of particle production are most manifest in the collapsed limit,  $u,v \to 0$, in which the homogeneous part of the solution (\ref{equ:SeriesSol}) dominates. Keeping only the terms which are non-analytic in $s$, we get
\begin{align}\label{ccexchangecollpased}
	\lim_{u,v\to 0} 
	\hat F(u,v) \,&=\, \left(\frac{u v}{4}\right)^{\frac{1}{2}+i\mu} (1+i\sinh \pi\mu) \frac{\Gamma(\tfrac{1}{2}+i\mu)^2\Gamma(-i\mu)^2}{2\pi}+c.c.\, ,
\end{align}
which agrees with equation (5.98) in \cite{Arkani-Hamed:2015bza}. The overall amplitude scales as $e^{-\pi\mu}$ for large $\mu$. This leading part captures the interference effect between physically producing and not producing a pair of massive particles. 
Restoring the coupling constant $g^2$, we see that the overall coefficient of the collapsed limit is manifestly positive. This is a consequence of {\it bulk unitarity} and is related to the fact that going to the collapsed limit is equivalent to taking the OPE limit, for which the four-point function factorizes and becomes proportional to the two-point function of the intermediate particle.  As we will see in Section~\ref{sec:spin}, the positivity of the collapsed limit is also true for spin exchanges, for which there are nontrivial angular dependences with positive coefficients. This can be viewed as an analog of the positivity of the coefficients of the Gegenbauer polynomials on the $s$-channel pole for amplitudes (see \S\ref{sec:FlatSpace}). 

\vskip 4pt
So far, we have presented results for the exchange of particles belonging to the {\it principal series}, for which $m > \frac{3}{2}H$ or $\mu> 0$, and the signal in the collapsed limit displayed distinct oscillatory features. 
However, our results are also valid for the exchange of lighter particles belonging to the {\it complementary series}, with masses $0< m< \frac{3}{2}H$, corresponding to $\mu \to i\nu$, with $\nu\in (0,\frac{3}{2})$.\footnote{Except when $m=0$ or $\sqrt{2}H$, for which our solution becomes singular. These special cases will be treated separately in \S\ref{sec:PM}.} In this case, the nonperturbative contribution is no longer suppressed relative to the EFT part. Instead of giving oscillations, the intermediate particle now leads to a smooth, but still non-analytic, scaling in the collapsed limit, $(uv)^{1/2-\nu}$. This scenario is often dubbed {\it quasi-single-field inflation}, and has been studied extensively e.g.~in~\cite{Chen:2009zp, Baumann:2011nk, Assassi:2012zq, Noumi:2012vr}.

\vskip 4pt
The $\mu\to 0$ limit of the result, which lies on the boundary of the complementary and principal series,  is also interesting. 
Although the formulas above naively blow up, taking the limit carefully one can show that the leading $\mu$-independent behavior is
\begin{align}
	\lim_{\mu\to 0}\lim_{u,v\to 0}\hat F(u,v) = \sqrt{uv}\hs \log u\log v\, .\label{collapsedmu0}
\end{align} 
Because of the logarithms, some of the terms that were analytic in $s$ and therefore neglected in \eqref{ccexchangecollpased} now also contribute in the collapsed limit. This behavior is consistent with the bulk expectation, since massive particles with $m = \frac{3}{2}H$ scale as $\eta^{3/2}\log\eta$ at late times. As we show in Appendix~\ref{app:hyper}, it also agrees precisely with an explicit calculation for $\mu=0$.

\begin{figure}[h!]
    \centering
      \includegraphics[height=5.45cm]{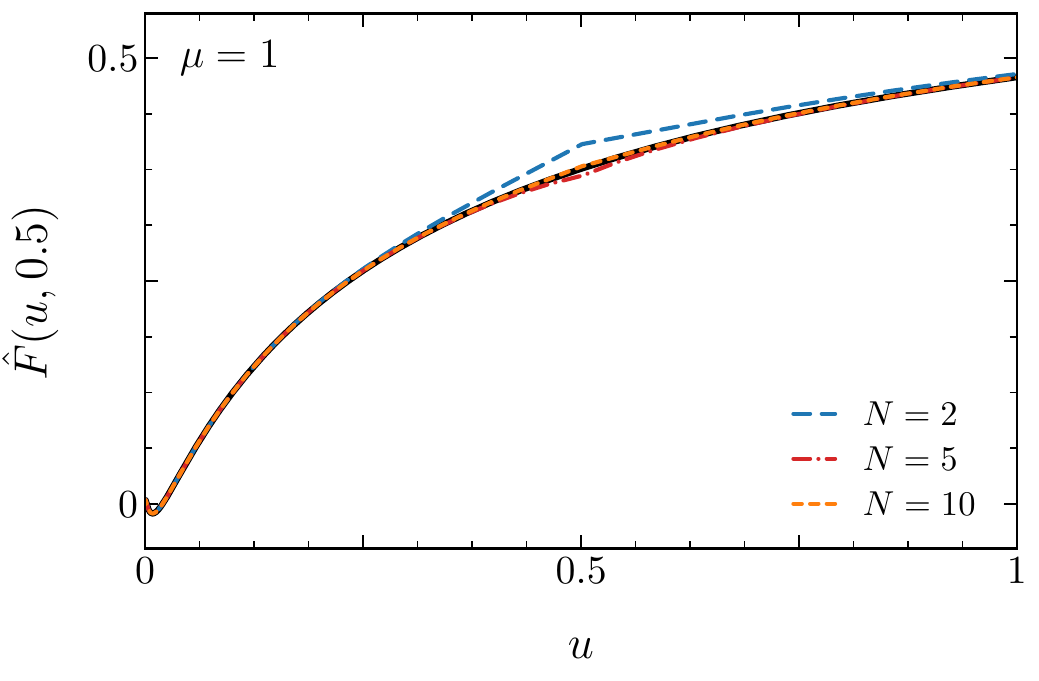}
         \includegraphics[height=5.45cm]{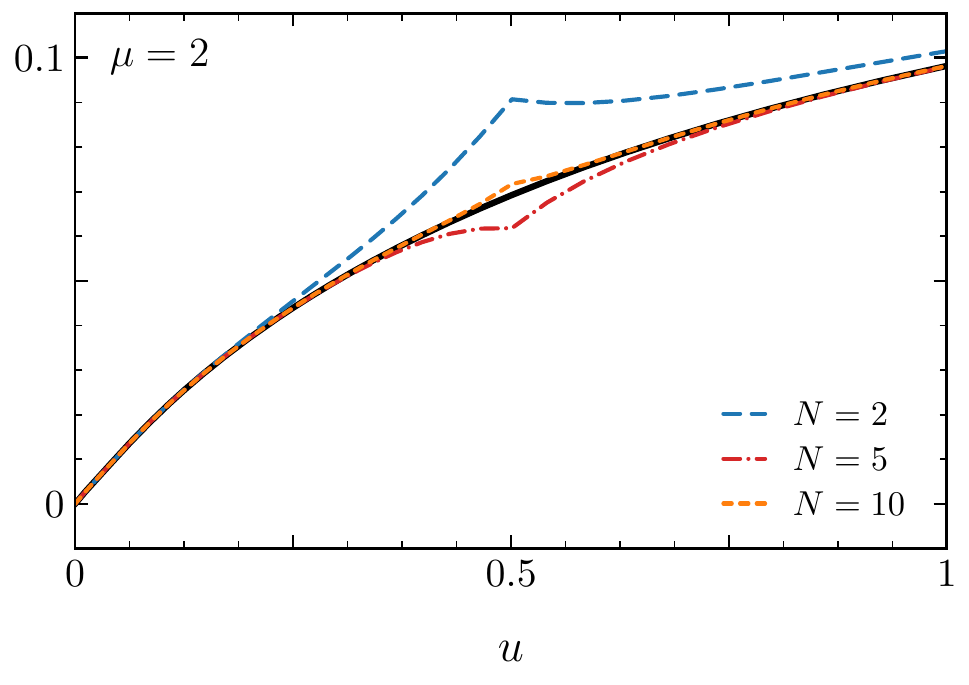}
         \\
    \caption{Comparison of the analytic expression for $\hat F(u,v)$ keeping only terms from $m,n=0,\ldots, N$ to the numerical solution of the differential equation (solid black lines). We have chosen $v=0.5$. The left and right plots are for $\mu=1$ and $\mu=2$, respectively.
    }
    \label{fig:comparison}
\end{figure}

\paragraph{Convergence}
Figure~\ref{fig:comparison} shows a comparison between the analytic solution (\ref{equ:SeriesSol}) and the numerical solution obtained by directly integrating (\ref{InhomogeneousFu}). As we can see, the expression \eqref{equ:SeriesSol} is both exact and practical, since the series is highly convergent around $u=0$, and we get a good approximation by keeping only a few terms. The convergence is particularly fast for small values of $\mu$ because in this limit the nonperturbative part plays a dominant role in determining the overall shape of the solution. For larger $\mu$, we see that more terms need to be kept to achieve convergence. In particular, 	when full convergence has not been reached, a kink appears at $u=v$. This is a consequence of the fact that, although the full solution is smooth everywhere in the physical domain, the individual perturbative and nonperturbative parts are not; for instance, the first derivatives of $\hat g(u,v)$ and $\hat g(v,u)$ do not agree at $u=v$. The reason for the slow convergence near $u=v$ is that this point lies precisely on the boundary of the two disks of convergence. A better convergence behavior could then be achieved e.g.~by gluing \eqref{equ:SeriesSol} with the series expansion around $u=v$. In this case, the different series expansions would have overlapping disks of convergence, resulting in a smoother transition. 

\paragraph{Higher-derivative interactions} So far, we have presented a systematic classification of all contact terms, as well as derived the explicit solution for the simplest tree-level exchange, corresponding to the vertex $\upvarphi^2 \sigma$ in the bulk. It remains to determine the exchange solutions associated with higher-derivative interactions in the bulk, such as $(\partial_\mu \upvarphi)^2 \sigma$. As we will now show, these solutions can be expressed in terms of the simplest exchange solution and a sum of contact terms. 

\vskip 4pt
Consider the generalized inhomogeneous equation 
\beq
	(\Delta_u +M^2) \hat F_n = (-1)^n\hs\hat C_n\, ,\label{Fnrecursive}
\eeq
where the source term on the right-hand side is a higher-order contact term, $\hat C_n \equiv \Delta_u^n \hat C_0$, and we have defined $M^2 \equiv \mu^2 + \frac{1}{4}$.  We have chosen the alternating signs for the source term so that the solution to (\ref{Fnrecursive}) can be written in the following recursive form:
\beq
\hat F_{n} =  M^2 \hat F_{n-1} - \hat C_{n-1} \, .
\eeq
Applying this relation iteratively, the $n$-th order solution $\hat F_n$ can be written in terms of the $0$-th order solution $\hat F_0$ and a sum of contact terms:
\beq
\hat F_{n} = M^{2n} \hat F_{0} \,-\, \sum_{m=0}^{n-1}  M^{2(n-1-m)} \hat C_m \, .\label{highercontact}
\eeq
In this way, all solutions to (\ref{Fnrecursive}) can be related to the solutions $\hat F_0$ and $\hat C_m$ that we obtained before.
The fact that the higher-derivative exchange solutions can be reduced to the lowest-derivative exchange solution and a series of contact terms also has a close analog in the treatment of scattering amplitudes. Indeed, a general exchange amplitude will have the form
\beq
A =\sum_n\frac{ a_n s^n}{s-M^2}\,=\,\frac{b_{-1}}{s-M^2} +\sum_n b_n s^n\, ,
\eeq
where the coefficients $b_n$ are easily determined from the coefficients $a_n$ by matching residues. This shows that the exchange of a new particle can always be represented as $(s-M^2)^{-1}$, and higher-derivative interactions can be reinterpreted as contact terms in the EFT expansion.

\subsection{Flat-Space Limit}
\label{sec:FSL}

Having the full solution for the correlators allows us to analytically continue them in the complex plane. A particular interesting limit is 
\beq
k_t \equiv \sum_n |{\bf k}_n| \to 0\qquad \text{or} \qquad u \to - v\, , 
\eeq
where we expect to recover flat-space results~\cite{Maldacena:2011nz, Raju:2012zr}. In the following, we explicitly verify that our solution indeed has a discontinuity at $k_t = 0$ whose coefficient is given by the flat-space scattering amplitude. Moreover, we show that this discontinuity is related to the homogeneous part of the solution, which, as we have seen, characterizes the effects of particle production. This provides a fascinating link between scattering in flat space and particle production in curved space.

\vskip 4pt
To study the $k_t \to 0$ limit, we go back to the differential equation~\eqref{InhomogeneousFu2}. For $u \to -v$, the most singular piece of the equation is
\begin{align}
	\lim_{u\to -v} \frac{\partial^2\hat F}{\partial u^2} = -\frac{1}{1-v^2}\frac{1}{u+v} \, . \label{equ:lim}
\end{align}
Integrating this, we get\footnote{This scaling is true for $\hat C_0$ as the source term in \eqref{InhomogeneousFu}. For higher-order contact terms $\hat C_n$ as source terms, we must first subtract the more divergent terms $\hat C_{m}$, with $m<n$, 
that contribute to the four-point function. See the discussion on ``higher-derivative interactions" at the end of the previous section.} 
\beq
\lim_{u\to -v} F \equiv \frac{1}{s}\, \lim_{u\to -v}\hat F =  -\frac{1}{s(1-v^2)} (u+v)\log(u+v)  = \frac{1}{s_{\rm flat}}(- k_t \log k_t) \, , 
\label{equ:Flog}
\eeq
where we have used the definitions of $v$ and $s_{\rm flat}$ in the second equality. We see that the solution has a discontinuity at $k_t =0$ whose coefficient is given by the high-energy limit of the flat-space amplitude, $A_{\rm flat} = 1/s_{\rm flat}$. Next, we show that analyticity arguments allow us to relate ${A}_{\rm flat}$ to the discontinuity of the {\it homogeneous} solution of the differential equation (or more accurately the discontinuity of its first derivative), which controls the nonperturbative piece of the four-point function, and is intimately tied to  particle production. 

\paragraph{Harmonic oscillator}  We begin by going back to the limit $v \to 0$ and  $u = \xi v \to 0$, for which the conformal invariance equation took the form of the equation of motion of a forced harmonic oscillator.  We found series solutions to this equation, $\tilde F_<(\xi)$ and $\tilde F_>(\xi)$, which are regular around $\xi=0$ and $\xi=\infty$, respectively. 
 By construction, the function $\tilde F_<(\xi)$ is analytic for $|\xi| \le 1$, but has a branch cut for $|\xi| > 1$. On the other hand, by looking at the large $n$ behavior of the series coefficients, we see that the first derivative $\tilde F_<^{\hskip 1pt \prime}(\xi)$ fails to converge at $\xi=-1$. To understand the behavior at $\xi=-1$, we let $\xi \to -1 + \delta$, for infinitesimal $\delta > 0$.
Using the equation of motion, together with the fact that $\tilde F_<$ is finite at $\xi=-1$, we find
\beq
\lim_{\xi \to -1+\delta}\frac{\d^2 \tilde F_<}{\d\xi^2}=-\frac{1}{\delta} \quad \Rightarrow\quad \lim_{\delta\to 0} \tilde F_<(-1+\delta) = {A}_{\rm flat}(-\delta \log \delta)\, ,  \label{equ:log}
\eeq 
where the ellipses denote less singular terms, and we have introduced ${A}_{\rm flat}=1$ to keep track of the expected dependence on the scattering amplitude.\footnote{Note that we would have gotten ${A}_{\rm flat}=g^2$ if we had kept the coupling constant $g^2$ in the source term in (\ref{InhomogeneousFu2}).}
The first derivative of (\ref{equ:log}) is $\tilde F_<^{\hskip 1pt\prime} =- {A}_{\rm flat} \log\delta + \cdots$, which at $\xi =-1$, has the following discontinuity
\beq
\lim_{\delta \to 0^+}{\rm Disc}[\tilde F_<^{\hskip 1pt\prime}(-1+\delta)] = -2\pi i\times {A}_{\rm flat}\, , \label{equ:Disc}
\eeq
where ${\rm Disc}[f(\xi)] \equiv f(\xi+i0)-f(\xi-i0)$. We see that the discontinuity is related to the scattering amplitude. 

\begin{figure}
\centering
\includegraphics[scale=0.5]{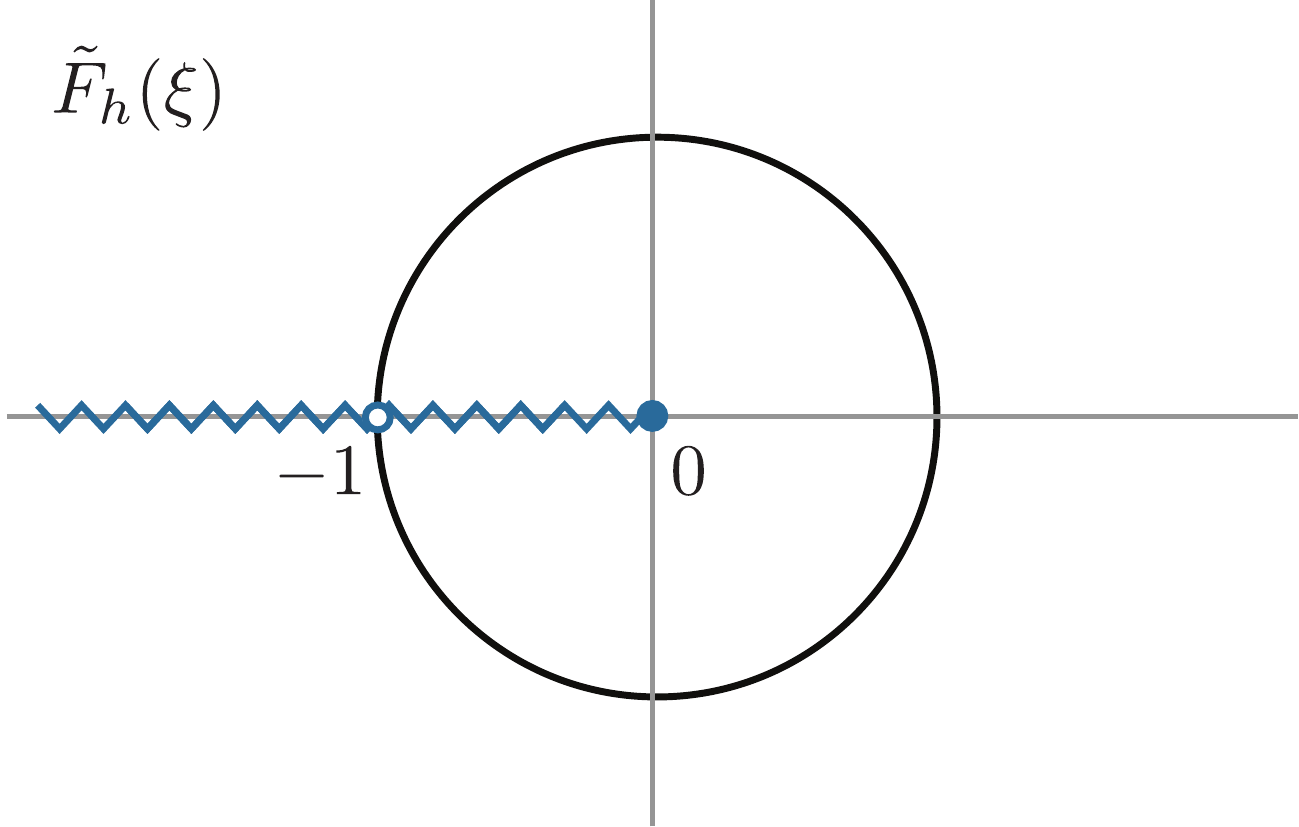} \hspace{0.5cm}
\includegraphics[scale=0.5]{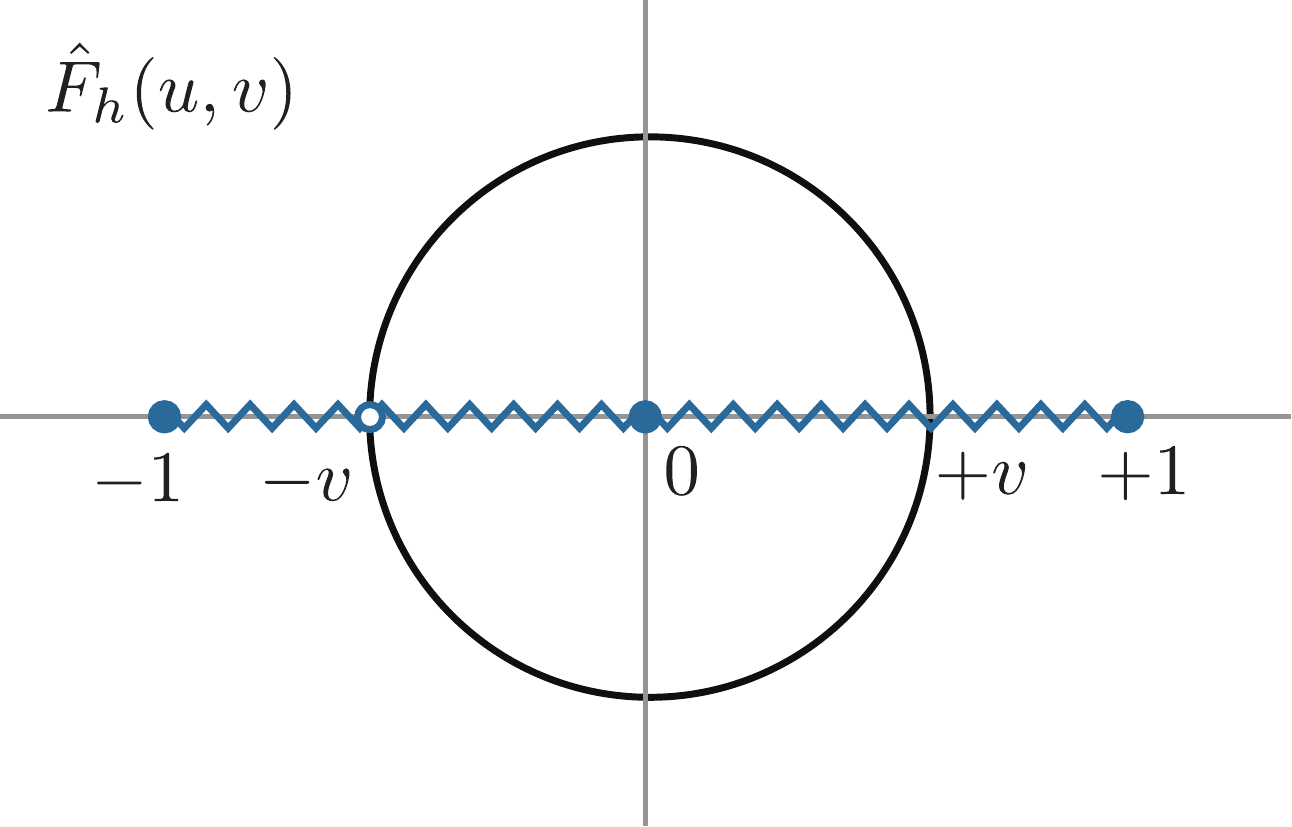}
\caption{ Illustration of the analytic structures of the functions $\tilde F_h(\xi)$ ({\it left})  and $\hat F_h(u,v)$ ({\it right}).} 
\label{fig:analytic2}
\end{figure}

\vskip 4pt
An alternative way to compute the discontinuity at $\xi =-1$ is to first recall that 
\beq
\tilde F_<(\xi) - \tilde F_>(\xi) = \frac{\pi}{\cosh \pi \mu} \frac{\xi^{\frac{1}{2}+i\mu} -\xi^{\frac{1}{2}-i\mu}}{2i\mu} \equiv \tilde F_h(\xi)\, ,
\eeq
where $\tilde F_h(\xi)$ is the homogeneous solution whose analytic structure is illustrated in the left panel of Fig.~\ref{fig:analytic2}.  Since $\tilde F_>^{\hskip 1pt \prime}(\xi)$ is analytic for $|\xi|>1$, we obtain
\begin{align}
\lim_{\delta \to 0^+}{\rm Disc}[\tilde F_<^{\hskip 1pt\prime}(-1 +\delta)]  =  {\rm Disc}[\tilde F_h^{\hskip 1pt\prime}(-1)]  =-2\pi i\, .
\end{align}
Comparing this to (\ref{equ:Disc}), we find
\beq
{A}_{\rm flat} = -\frac{{\rm Disc}[\tilde F_h^{\hskip 1pt\prime}(-1)] }{2\pi i} = 1\,,
\eeq
which is the expected amplitude.  
Although this example looks somewhat trivial since the scattering amplitude was just a constant,we will see below the same relation between the flat-space amplitude and the discontinuity of the homogeneous solution holds in the general case.

\paragraph{General case}
By construction, the function $\hat F_<(u,v)$ is analytic within the disk of convergence, $|u|<|v|$, but becomes non-analytic outside the disk. 
Conversely, the function $\hat F_>(u,v)$ is analytic for $|u| > |v|$.  
 The difference between the two solutions is 
 \begin{align}
	\hat F_<(u,v)-  \hat F_>(u,v)=  \frac{\pi}{\cosh\pi\mu}\left(\hat{F}_+(u) \hat {F}_-(v)  -  \hat{F}_-(u) \hat{F}_+(v)  \right) \equiv \hat F_h(u,v)\, , \label{equ:Hom}
\end{align}
where the analytic structure of $\hat F_h(u,v)$ is illustrated in the right panel of Fig.~\ref{fig:analytic2}. 
The discontinuity of $\hat F_<^{\hskip 1pt\prime}$ at $u=-v$ can be defined as the limit  
\beq
\lim_{\delta \to 0^+} {\rm Disc}[\hat F_<^{\hskip 1pt \prime}(-v +\delta, v)] = -2\pi i\times \frac{s^2}{v^2}\,  {A}_{\rm flat}\, , \label{equ:Disc2}
\eeq
where the equality follows from (\ref{equ:Flog}).
Since $\hat F_>^{\hskip 1pt \prime}$ is analytic at $u = -v$, we get
\beq
{A}_{\rm flat} = -\frac{v^2}{s^2} \frac{{\rm Disc}[\hat F_h^{\hskip 1pt \prime}(u=-v, v)] }{2\pi i}\, ,
\eeq
which, as before, relates the scattering amplitude in flat space to the homogeneous solution associated with particle production in curved space. Substituting (\ref{equ:Hom}), we find 
\beq
{A}_{\rm flat} = \frac{v^2}{s^2} \left[\hat {F}_+(v)\hat {F}_-^{\hskip 1pt \prime}(v)- \hat {F}_-(v)\hat {F}_+^{\hskip 1pt \prime}(v)\right] \equiv \frac{v^2}{s^2} \,W[\hat F_+,\hat F_-] \, , \label{equ:relation}
\eeq
where $W = 1/(1-v^2)$ is the Wronksian of the solution.  Using the definitions of $v$ and $s$, this becomes
\beq
{A}_{\rm flat} = \frac{1}{(k_1+k_2)^2-(\k_1+\k_2)^2}= \frac{1}{s_{\rm flat}}\, ,
\eeq
as expected. 

\begin{framed}
{\small \noindent {\it Conformal invariance.}---In (\ref{equ:relation}), we have established a precise relation between the scattering amplitude and the Wronskian of the homogenous solutions of the conformal invariance equation.  As we will show in this insert, this relationship can be inverted: Requiring the Wronskian of the homogeneous solutions to a general second-order differential equation to match the flat-space amplitude fixes the form of the equation to be of the form of the conformal invariance constraint.

\vskip 4pt
Suppose that we are looking for a second-order differential equation of the form
\begin{align}
	\big(p(u)\partial_u^2+q(u)\partial_u+m^2\big)\hat F=\frac{uv}{u+v}\, ,\label{diffgeneral}
\end{align}
and a similar equation for $v$. Symmetry in $u\leftrightarrow v$ demands that
\beq
	 \big(p(u)\partial_u^2+q(u)\partial_u-p(v)\partial_v^2-q(v)\partial_v\big)\frac{uv}{u+v}=0 \, ,
	 \eeq
	 and hence
	\beq 
	2\big(u^2p(v)-v^2q(u)\big)-(u+v)\big(u^2q(v)-v^2q(u)\big) =0\, .
\eeq
It is easy to see that the general solution to this equation is
\beq
\begin{aligned}
	p(u)&=a u^4+b u^2\, ,\\ 
	q(u)&=2au^3+cu^2\, ,
\end{aligned}
\label{pqfunc}
\eeq
where $a,b,c$ are arbitrary constants.
The special case of the conformal differential equations corresponds to the choice $a= -  b = -1$ and $c=0$.
We will now show that the same choice of parameters follows from demanding that the singularity at $k_t=0$ has the right flat-space scattering amplitude as its discontinuity.

\vskip 4pt
Above, we saw that the scattering amplitude matches the Wronskian of the differential equation. Given \eqref{diffgeneral} and \eqref{pqfunc}, the Wronskian is
\begin{align}
	W(u) = \frac{d}{b+au^2}\left(\frac{\sqrt{b}-i\sqrt{a}u}{\sqrt{b}+i\sqrt{a}u}\right)^{-\frac{ic}{2\sqrt{ab}}}\, ,
\end{align}
where $d$ is an additional constant. Demanding that this correctly reproduces the flat-space scattering amplitude, i.e.~$W(u) = \lambda (1-u^2)^{-1}$ fixes the parameters to be $a=-b$, $c=0$. The ratio $d/b$ can then be interpreted as the coupling constant.}
\end{framed}

\subsection{Ultraviolet Completion}
\label{sec:UV}

As we have just seen, the coefficient of the leading divergence of the correlator as $k_t \to 0$ is related to the {\it high-energy} limit of the flat-space scattering amplitude.
It therefore probes the {\it ultraviolet completion} of the physics. Note in particular that taking this limit does not commute with the effective field theory expansion. For physical momenta, we can imagine adding higher-dimension operators to the theory, and these make progressively smaller contributions to the correlators. However, near $k_t \to 0$ the behavior of the singularity will be dominated by the operator with the highest dimension, which means that the EFT approximation is not useful in this limit, and one really needs a UV-complete theory for high-energy scattering to correctly compute the behavior of the singularity near $k_t \to 0$.

\vskip 4pt
It is interesting to note that we expect radically different behavior as $k_t \to 0$, for ``field
theoretic" vs.~``stringy/gravitational" UV completions. In a field-theoretic UV completion,
amplitudes die as powers of energy, which translates into a suppression of the bulk integrals for cosmological correlators at early times. For example, we expect
\beq
F \sim \int \d \eta\,\eta^{-p} \,e^{ik_t \eta} \sim k_t^{p-1} \log k_t\, ,
\eeq
which still has a branch-cut singularity near $k_t \to 0$.\footnote{There is a singularity as $k_t \to 0$ because $k_t$ only regulates the oscillatory integral when it is a nonzero real number.  
When $k_t$ has a negative imaginary part, the $e^{ik_t \eta}$ factor is exponentially growing. For this reason, we cannot have analyticity in $k_t$ as $k_t \to 0$.} Note that the presence of the singularity in the complex $k_t$-plane can be detected far from the origin: a contour integral with large radius, circling around the origin, will give a nonzero result.  Indeed, we might imagine that the low-energy observer only saw a contact $\upvarphi^4$ interaction, and inferred a $1/k_t$ singularity. Approaching $k_t\to 0$, this singularity could be softened to $k_t \log k_t$, if  the $\upvarphi^4$ interaction is seen to arise from integrating out a massive field with a $\upvarphi^2 \sigma$ coupling. However, the presence of some kind of singularity as $k_t\to 0$ is correctly captured by the effective theory, and in particular, the contour integral with large radius around $k_t=0$ is correctly computed by the $\upvarphi^4$ approximation.

\vskip 4pt
The presence of singularities as $k_t \to 0$ is a sharp feature of local quantum field theory in de Sitter space. Of course, we know that, in the presence of dynamical gravity, we cannot have such field-theoretic UV completions, and the behavior of the high-energy scattering amplitude drops faster than exponentially at high energies. For example, in a weakly coupled string theory the high-energy amplitude scales as $e^{-E \log E}$, while in any consistent theory of gravity, we expect it to behave like $e^{-S_{\rm BH}(E)}  = e^{-E^2/M_{\rm pl}^2}$. It is then easy to see that
\beq
F \sim \int \d \eta \, e^{-c\eta^2}\, e^{ik_t \eta} \ \xrightarrow{\ k_t \to 0\ }\ \text{analytic}\, . 
\eeq
The integral is convergent, in every direction of
complex $k_t$-space, as long as the UV-softening factor is stronger than $e^{-E}$ at high energies.
If the same effective $\upvarphi^4$ contact interaction is UV completed into a gravitational theory,
the local effective field theory expectation of a singularity as $k_t \to 0$ is therefore completely incorrect,
and the contour integral around $k_t \to 0$ even with large radius is dramatically different
than the effective-field-theoretic expectation. If there is any weakly coupled stringy description for
computing the correlators, then in the same approximation, the 
flat-space amplitude still falls fast enough that the same conclusion holds. This is how ``stringy UV completeness"
is encoded in the structure of the correlators. 

\vskip 4pt
This discussion also holds in AdS space, where it is simply the momentum-space avatar of the absence of the ``bulk-point singularity" for boundary CFT correlators \cite{Maldacena:2015iua}. Having said that, the situation in dS is even more interesting. We are motivated to look for some ``Veneziano correlator" in de Sitter space, with the magical property of having the oscillatory features in the squeezed limit corresponding the particle production of string excitations,  with appropriately positive weights, while not leaving a singular mark of any sort as $k_t \to 0$. This poses a concrete (if difficult!) challenge for an inroad into whatever ``string theory in de Sitter space" should mean, which can be sought even in the absence of any sort of standard worldsheet description.

\section{Exchange of Spinning Particles}
\label{sec:spin}

In this section, we generalize our results for the de Sitter four-point functions with scalar exchange to tree exchange of particles with spin, again with the external fields being conformally coupled scalars. We show that these solutions can be obtained from our previous results for the scalar exchange through the action of suitable ladder operators. We will determine these operators through a combination of bulk reasoning, boundary considerations, and educated guesswork. In a future publication~\cite{CosmoBoot2}, we will show  how to obtain them in a more systematic way, using tools from conformal field theory~\cite{Dirac:1936fq, Costa:2011mg, Karateev:2017jgd}.

\vskip 4pt
In \S\ref{sec:prelim}, we introduce an ansatz for the four-point function associated with the exchange of particles with spin, and derive the conformal invariance constraints satisfied by each of its helicity components. In \S\ref{sec:results}, we solve these equations, first for spins 1 and 2, and then for arbitrary spins. In \S\ref{sec:PM}, we specialize to the exchange of (partially) massless particles. 

\subsection{Polarization Basis}\label{sec:prelim}

In the case of scalar exchange, and for $\Delta=2$, the four-point function depended on the following variables 
\beq
x\equiv k_1+k_2\,, \quad y\equiv k_3+k_4\, , \quad s= |\k_1+\k_2|\, , \label{equ:xy}
\eeq 
and we have usually worked with the dimensionless combinations $u = s/x$ and $v=s/y$. 
As in the case of flat-space scattering amplitudes (cf.~\S\ref{sec:FlatSpace}), intermediate particles with spin lead to a characteristic dependence of the four-point function on the transverse momentum $t = |\k_2+\k_3|$. 
More generally, the contractions of the polarization tensors of the intermediate spinning particles with the external momenta lead to a dependence on the following variables
\beq
\alpha\equiv k_1-k_2\,, \quad \beta\equiv k_3-k_4\, , \quad \tau \equiv (\k_1-\k_2)\cdot(\k_3-\k_4)\, ,
\label{equ:tau}
\eeq
where $\tau$ is related to $t$ via
\beq
\tau = \frac{1}{2}\left(\alpha^2 + \beta^2 + x^2 + y^2\right) -s^2-2t^2 \, .
\eeq
We showed, in \S\ref{sec:FlatSpace}, how the angular dependence of higher-spin exchange in flat space is  encapsulated in the Gegenbauer polynomial of order $S$, the spin of the exchange particle. In that case, a boost symmetry allowed us to go to the center-of-mass frame, in which there is a single angle controlling the scattering and the polarization sums could be determined in a simple way.
The situation is a bit different in de Sitter space. Unlike in flat space, different helicity\footnote{By helicity, we mean the helicity of the representation with respect to the spatial slices of de Sitter space, and not the helicities of representations of the Poincar\'e group.} states now have different amplitudes, which are fixed by imposing (special) conformal symmetry of the full answer. This implies that the four-point function can naively depend on more angles.  A second difficulty is the absence of a boost symmetry, so that it is significantly more complicated to determine this angular dependence compared to flat space.

\vskip 4pt
We will show that the conformal invariance constraints are solved by an ansatz of the form

\begin{eBox}
\beq
F_S=\sum_{m=0}^S  \bar \Pi_{m} \tilde \Pi_{S,m}\hs A_{S,m} \,, \quad {\rm with} \quad A_{S,m}=
\frac{1}{s}\hs\hat A_{S,m}(u,v)\, ,
\label{equ:Ansatz}
\eeq
\end{eBox}
where $\bar \Pi_{m}(\tau,\alpha,\beta)$ and $\tilde \Pi_{S,m}(\alpha,\beta)$ are the polarization sums of the transverse and longitudinal modes, respectively. In the following, we will first determine the functional forms of the polarization sums, and then derive the conformal invariance constraints satisfied by the coefficient functions.

\paragraph{Transverse polarizations} 
We first consider the contractions of the external momenta with internal polarization tensors $\bar\epsilon^{\lambda}_{i_1 \cdots i_S}$ that are traceless and transverse in all of their indices. 
By the same reasoning as in \S\ref{sec:FlatSpace}, the unique polarization sum is 
\beq
\bar\Pi_{S} \equiv  {\alpha}_{i_1} \cdots {\alpha}_{i_S} {\beta}_{j_1} \cdots {\beta}_{j_S} \underbrace{\sum_{\lambda=\pm} \bar \epsilon^{\lambda}_{i_1 \cdots i_S}(\hat\s)\bar \epsilon^{\hskip 1pt -\lambda}_{j_1 \cdots j_S}(\hat\s)}_{\displaystyle \equiv P^{j_1\cdots j_S}_{i_1\cdots i_S}(\hat \s)}\ , \label{equ:pol1}
\eeq
where ${\boldsymbol\alpha}\equiv \k_1-\k_2$ and ${\boldsymbol\beta}\equiv \k_3-\k_4$.\footnote{Warning: Note that $|\boldsymbol\alpha|\ne \alpha$ and $|{\boldsymbol\beta}|\ne \beta$.} 
The tensor $P^{j_1\cdots j_S}_{i_1\cdots i_S}(\hat \s)$ projects all momenta orthogonal to the direction of the internal momentum $\s \equiv \k_1+\k_2$.
For example, the spin-1 projection tensor is
\beq
P_{ij}(\hat\s)  = \delta_{ij} - \hat s_i \hat s_j\, . \label{equ:S1P}
\eeq 
All higher-spin projection tensors can be built out of this tensor (see e.g.~\cite{Chandrasekaran:2018qmx}):\footnote{The indices of the projection tensors are raised and lowered with $\delta_{ij}$.}
\beq
\begin{aligned}
	P_{i_1\cdots i_S}^{j_1\cdots j_S} &=\sum_{m=0}^{\lfloor S/2\rfloor}C(S,m)\,P_{(i_1i_2}\cdots P_{i_{2m-1}i_{2m}}P^{(j_1j_2}\cdots P^{j_{2m-1}j_{2m}} P_{i_{2m+1}}^{j_{2m+1}}\cdots P_{i_S)}^{j_S)}\, ,\\
	C(S,m) &\equiv \frac{(-1/2)^m S!}{m!(S-2m)!}\frac{(2S-2m-2)!!}{(2S-2)!!}\, .
\end{aligned}
\label{projtensor}
\eeq 
For example, the spin-2 and spin-3 projection tensors are
\begin{align}
	P_{i_1i_2}^{j_1j_2} &=P_{(i_1}^{(j_1}P_{i_2)}^{j_2)}-\frac{1}{2}P_{i_1i_2}P^{j_1j_2}\, ,\\
	P_{i_1i_2i_3}^{j_1j_2j_3} &=P_{(i_1}^{(j_1}P_{i_2}^{j_2}P_{i_3)}^{j_3)}-\frac{3}{4}P_{(i_1i_2}P^{(j_1j_2}P_{i_3)}^{j_3)}\, . 
\end{align}
Since the external momenta are contracted with the projection tensor $P_{ij}$ in two different ways, the transverse polarization sums $\bar\Pi_S$ are functions of the following two dimensionless variables: 
\begin{align}
\hat T &\equiv  \frac{\alpha_i P^{ij}\beta_j}{s^2}= \frac{s^2\hs\tau + x y \alpha\beta}{s^4}\, , \label{equ:T}\\
\hat L^2 &\equiv \frac{\alpha_i P^{ij}\alpha_j \, \beta_k P^{kl}\beta_l}{s^4}= \frac{(s^2-x^2)(s^2-\alpha^2)(s^2-y^2)(s^2-\beta^2)}{s^8}\, . \label{equ:Z}
\end{align}
Given the projection tensors defined in (\ref{projtensor}), we find
\beq
\bar\Pi_S = 2^{S}\sum_{m=0}^{\lfloor S/2\rfloor}C(S,m) \,\hat T^{S-2m} \hat L^{2m}\, , \label{equ:trans}
\eeq
with $\bar\Pi_0\equiv 1$, where the normalization was chosen for later convenience. Explicitly, the first few terms are $\bar\Pi_1 =2 \hat T$, $\bar\Pi_2 =4 \hat T^2 - 2 \hat L^2$, and $\bar\Pi_3 = 8\hat T^3-6 \hat T \hat L^2$. 
It will be important, in Section~\ref{sec:inflation} and Appendix~\ref{app:D3}, that these functions are homogeneous in $\hat T$ and $\hat L$, and that, in the soft limit $k_4\to0$, both $\hat T$ and $\hat L$ vanish. 

\paragraph{Longitudinal polarizations} 
To describe the lower-helicity contributions for fixed spin, we will need to use polarization tensors that carry some transverse and some longitudinal indices. 
 A generic polarization tensor can then be expressed as a symmetrized product of transverse and longitudinal tensors 
\begin{align}
	\epsilon_{i_1\cdots i_S}^{\lambda,m}(\hat \s) = \bar \epsilon_{(i_1\cdots i_m}^\lambda(\hat \s)\, \tilde\epsilon_{i_{m+1}\cdots i_S)}(\hat \s)\, , \label{equ:pol}
\end{align}
where $\bar \epsilon$ and $\tilde\epsilon$ denote the transverse and longitudinal parts of the polarization tensor,
respectively. The polarization tensor is labeled by its total spin $S$, its helicity $\lambda$ and the number of transverse indices $m$. The longitudinal piece $\tilde\epsilon$ is built out of the unit vector $\hat \s$ and Kronecker delta's, and needs to be constructed in such a way that the complete tensor $\epsilon^{\lambda,m}_S$ is symmetric and traceless.
A key property of the longitudinal piece of the polarization tensor is 
\begin{align}\label{contrlong}
	 \hat q_{i_{m+1}}\cdots \hat q_{i_{S}}\tilde\epsilon_{i_{m+1}\cdots i_{S}}(\hat\s) &= \tilde P_S^m(\hat\q\cdot\hat\s)\, , 
\end{align}
where $\tilde P_S^m(x) \equiv (1-x^2)^{-|m|/2} P_S^m(x)$, with $P_S^m(x)$ the associated Legendre polynomial.
  Notice that \eqref{contrlong} depends not only on the number of longitudinal indices, but also on the total spin.

\vskip 4pt
This time, however, it is not a priori  clear which combination of $\k_1$, $\k_2$ and $\s$ will be contracted with the longitudinal polarization tensor; we have not derived the longitudinal polarization sum from first principles. 
However, by a combination of educated guesswork and trial-and-error, we have found\footnote{It is worth noting that $s^{-1}\tilde\Pi_{S,m}$ solves the conformal constraints  by itself.}
\beq
	\tilde \Pi_{S,m} \equiv  (-1)^m \tilde P_S^m(\alpha/s)\hskip 1pt \tilde P_S^{-m}(\beta/s)\, .\label{equ:long}
\eeq
As we will see, using this form of the polarization sum in the ansatz (\ref{equ:Ansatz}) allows us to solve the conformal invariance conditions. Note that (\ref{equ:long}) reduces to $\tilde\Pi_{S,S} = \Gamma(\frac{1}{2}+S)/(\sqrt\pi S!)$ for $m=S$. 
This completes our discussion of the polarization basis.

\paragraph{Coefficient functions}  
The functional form of the coefficient functions $\hat A_{S,m}$ is determined by solving the differential equations $D_{ij}F_S=0$. 
It is slightly easier to first write these equations in terms of $(x,y)$ rather than~$(u,v)$.
For $\Delta=2$, the equations $D_{12}F_S=0$
 and $D_{34}F_S=0$ then are 
 \begin{align}
\bigg[\partial_{x\alpha} +\partial_\tau\Big(y \partial_\beta+\beta\partial_y+\frac{y \beta}{s}\partial_s\Big)-\alpha x \partial^2_\tau\bigg] F_S &= 0\, ,\\
\bigg[\partial_{y\beta}+\partial_\tau\Big(x \partial_\alpha+\alpha\partial_x+\frac{x \alpha}{s}\partial_s\Big)-\beta y \partial^2_\tau \bigg] F_S&= 0\, ,
\end{align}
while $D_{13}F_S=0$ reads
\begin{align}
	&\bigg[\partial_x^2-\partial_y^2+\partial_\alpha^2-\partial_\beta^2-2\Big((\alpha\partial_\tau-\partial_\alpha)\partial_x-(\beta\partial_\tau-\partial_\beta)\partial_y+(x\partial_\alpha-y\partial_\beta)\partial_\tau\Big)\\
	&+\frac{1}{s}\partial_s\Big(x\partial_x-y\partial_y+\alpha\partial_\alpha-\beta\partial_\beta +2(y\beta-x\alpha)\partial_\tau\Big) +(x+y-\alpha-\beta)(x-y-\alpha+\beta)\,\partial_\tau^2\bigg]F_S=0\, . \nonumber
\end{align}
Substituting the ansatz (\ref{equ:Ansatz}), we find
\beq\label{equ:spinSconsttop}
(\Delta_{m,u}-\Delta_{m,v})\hat A_{S,m}=0\, ,
\eeq 
where we introduced the following differential operator for each helicity component
\beq
\Delta_{m,u}\equiv u^2(1-u^2)\partial^2_u -2u(u^2+m)\partial_u \,. \label{equ:Dum}
\eeq This equation follows from looking at the coefficient of the various powers of $\tau$ in the $D_{12}$ and $D_{34}$ equations. These will determine certain cross derivatives of the coefficient functions. Demanding consistency of those equations forces
 $\hat A_{S,m}$ to satisfy \eqref{equ:spinSconsttop}.

\vskip 4pt
We see that the functions $\hat A_{S,m}$ obey an equation that is very similar to the scalar equation. The only difference lies in the modified differential operator $\Delta_{m,u}$ instead of $\Delta_{0,u}\equiv \Delta_u$. 
To relate $\hat A_{S,m}$ to a solution of the scalar equation, we introduce the following {\it spin-raising operator} 
\beq
D_{uv}\equiv (u v)^2 \partial_u\partial_v\, . \label{equ:SpinRaising}
\eeq
This operator
satisfies the very useful relation
\beq
\Delta_{m,u} D_{uv} =D_{uv}(\Delta_{m-1,u}-2m) \, ,
\eeq
when acting on any function of $(u,v)$.  This means that acting with the operator $D^m_{uv}$ on a function turns the action of $\Delta_{m,u}$ into the action of $\Delta_u$, the scalar differential operator. In other words, a general solution of \eqref{equ:spinSconsttop} can be written in the form 
\beq
\hat A_{S,m}(u,v)=D^m_{uv} \hat f_m(u,v)\, ,
\eeq
where $\hat f_m(u,v)$ are solutions of the scalar constraint equation $(\Delta_u-\Delta_v)\hat f_m=0$. 
Moreover, as we will see below, starting from the highest-helicity component $\hat A_{S,S}=D^S_{uv}\hat f$, all lower-helicity components $\hat A_{S,m < S}$ can be written as suitable powers of $D_{uv}$ and $\Delta_u$ acting on the {\it same} function~$\hat f$. 
This means that the full answer in the spinning case is determined in terms of the scalar-exchange solutions obtained in \S\ref{sec:exchange}.

\subsection{Results for Spin Exchange}
\label{sec:results}

Before presenting general results for arbitrary spin, we will provide explicit results for spins 1 and 2. 
The pattern that emerges from these examples is then easily generalized.

\paragraph{Spin-1 exchange} 

For the case of spin-1 exchange, the ansatz (\ref{equ:Ansatz}) takes the following simple form\hs\footnote{Notice that this is odd under $k_1\leftrightarrow k_2$ or $k_3\leftrightarrow k_4$, which implies that the four-point function vanishes if the external scalars are all identical. This is true for all odd-spin exchange. In these cases, we must assume that the external fields are described by complex or non-identical scalars.}
\beq
F_1=\frac{1}{s}\left[ \bar \Pi_1\tilde\Pi_{1,1}D_{uv} \hs \hat f +  \tilde\Pi_{1,0}\hs \hat A_{1,0}\right] . 
\eeq
It can be shown that this solves the $D_{12}$ and $D_{34}$ equations if 
\beq
\hat A_{1,0} = \Delta_u \hat f\,.
\eeq 
The four-point function corresponding to spin-1 exchange can therefore  be written as
\beq
F_1=\frac{1}{s}\Big[ \bar \Pi_1\tilde\Pi_{1,1}\hs D_{uv} +  \tilde\Pi_{1,0}\hs \Delta_u \Big]\hat f\, ,\label{equ:spin1}
\eeq
where $\hat f$ is any solution of the scalar constraint equation, $(\Delta_u - \Delta_v)\hat f=0$. 
Substituting the scalar solution (\ref{equ:SeriesSol}) therefore gives the four-point function arising from spin-1 exchange.

\paragraph{Spin-2 exchange} 
For the case of spin-2 exchange, the ansatz becomes a bit more complicated
\beq\label{equ:spin2d2}
F_2=\frac{1}{s}\left[  \bar\Pi_2\tilde\Pi_{2,2}\hs D_{uv}^2+ \bar \Pi_1 \tilde \Pi_{2,1}\hs \hat A_{2,1}+  \tilde\Pi_{2,0}\hs \hat A_{2,0}\right] ,
\eeq
which solves the $D_{12}$ and $D_{34}$ equations if 
\begin{align}
\hat A_{2,1} &= D_{uv}(\Delta_u-2) \hat f \, , \\
\hat A_{2,0}&=\Delta_u(\Delta_u-2) \hat f \, .
\end{align}
The four-point function corresponding to spin-2 exchange hence is
\beq
F_2=\frac{1}{s}\Big[ \bar \Pi_2\tilde\Pi_{2,2}\hs D^2_{uv}+ \bar \Pi_1 \tilde \Pi_{2,1}\hs D_{uv}(\Delta_u-2) +  \tilde\Pi_{2,0}\hs \Delta_u(\Delta_u-2)\Big]\hat f \, , \label{equ:spin2}
\eeq
where $(\Delta_u - \Delta_v)\hat f=0$. Substituting (\ref{equ:SeriesSol}) for $\hat f$ completes the analysis.

\paragraph{Spin-S exchange}
Inspection of the spin-1 and spin-2 solutions (\ref{equ:spin1}) and (\ref{equ:spin2}) suggests a systematic procedure to determine the coefficient functions for general spin. Indeed, we will now show how all lower-helicity components are related to highest-helicity component $\hat A_{S,S} = D_{uv}^S \hat f$, and hence to the solution $\hat f$ for scalar exchange.

\vskip 4pt
To obtain the lower-helicity components $\hat A_{S,m}$, we replace the outermost operator of $\hat A_{S,m+1}$ by $\Delta_{m,u}$, and then commute it through the ladder operators $m$ times all the way to the right. 
For example, the lower-helicity components for spin-4 exchange are obtained by
\begin{align}
	\hat  A_{4,3}\ &=\ \Delta_{3,u}D_{uv}^3 \hat f &&\hskip -10pt\to\quad D_{uv}^3(\Delta_u-12)\hat f\, ,\\[4pt]
	\hat A_{4,2}\ &=\ \Delta_{2,u}D_{uv}^2(\Delta_u-12)\hat f&&\hskip -10pt\to\quad D_{uv}^2(\Delta_u-6)(\Delta_u-12)\hat f\, ,\\[4pt]
	\hat A_{4,1}\ &=\ \Delta_{1,u} D_{uv}(\Delta_u-6)(\Delta_u-12)\hat f &&\hskip -10pt\to\quad D_{uv}(\Delta_u-2)(\Delta_u-6)(\Delta_u-12)\hat f\, , \\[4pt]
	\hat A_{4,0} \ &=\ \Delta_u(\Delta_u-2)(\Delta_u-6)(\Delta_u-12)\hat f\, .
\end{align}
Generalizing this pattern to arbitrary spin, we find
\begin{eBox}
\beq
	\hat A_{S,m} = D_{uv}^{m}\prod_{j=1}^{S-m} \big(\Delta_u-(S-j)(S-j+1)\big)\hat f\, . \label{equ:coeff}
\eeq
\end{eBox}
 This solves all the constraints as long as $(\Delta_u - \Delta_v ) \hat f=0$. This is quite remarkable because it means that we can bootstrap our way up to all higher spins by using our previous solutions for scalar exchange. 
To derive the contact interactions arising from integrating out massive particles with spin, we therefore substitute (\ref{generalcontact}) into (\ref{equ:coeff}). Similarly, for exchange interactions we substitute the scalar solution (\ref{equ:SeriesSol}). In the following, we will show that our solution has the correct behavior in the collapsed and flat-space limits.

\paragraph{Collapsed limit}  It is straightforward to verify that our solution has the correct limiting behavior for $u,v \to 0$. To see this, first note that $\Delta_u \to u^2\partial_u^2$ as $u\to 0$. Since $\hat f \equiv \hat f^+ + \hat f^-$, with $ f^\pm \propto (uv)^{\frac{1}{2}\pm i\mu}$ in the same limit, each action of $\Delta_u$ on $\hat f$ brings down a factor of $-(\mu^2+\frac{1}{4})$. On the other hand, it is easy to verify that applying $D_{uv}^m$ to $\hat f$ gives $D_{uv}^m \hat f^\pm = (uv)^{m}[(\frac{1}{2}\pm i\mu)_{m}]^2\hat f^\pm$, where $(\cdot)_n$ is the Pochhammer symbol. Combining these facts, we find
\begin{align}
	\lim_{u,v\to 0}\hat A_{S,m}^\pm  &=  (-1)^S(uv)^m\,\frac{\Gamma(\frac{1}{2}+m\pm i\mu)\Gamma(\frac{1}{2}-m\pm i\mu)}{\Gamma(\frac{1}{2}-S+i\mu)\Gamma(\frac{1}{2}-S-i\mu)}\frac{\Gamma(\frac{1}{2}\mp i\mu)}{\Gamma(\frac{1}{2}\pm 	i\mu)}\lim_{u,v\to 0}\hat f^\pm \, \nn[5pt]
	&= (-uv)^{m-S} I_2^\pm (S,m)\, \lim_{u,v\to 0}\hat A_{S,S}^\pm \, ,
\end{align}
where 
we have defined
\begin{align}
	I_2^\pm (S,m)\equiv \frac{\Gamma(\frac{1}{2}+m\pm i\mu)\Gamma(\frac{1}{2}+S\mp i\mu)}{\Gamma(\frac{1}{2}+m\mp i\mu)\Gamma(\frac{1}{2}+S\pm i\mu)}\, .\label{I2}
\end{align}
As we mentioned before, in the collapsed limit the four-point function becomes proportional to the two-point function of the intermediate particle, whose normalization is given by $I_2^\pm$ (see Appendix~\ref{app:ward}). 

\begin{figure}[t!]
   \centering
      \includegraphics[scale =.9]{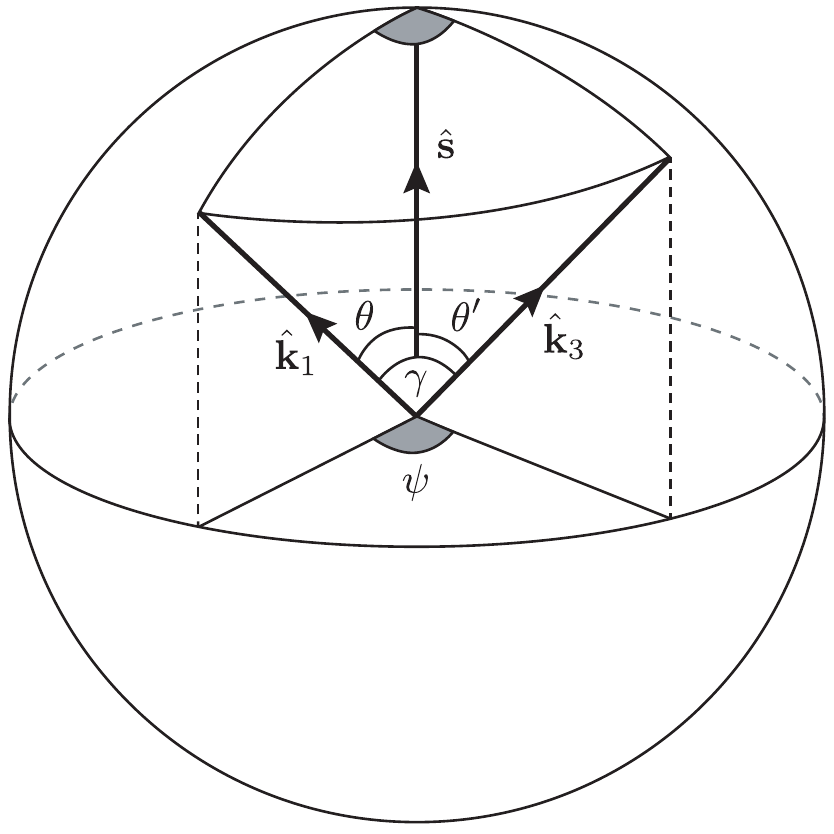}
      \caption{Illustration of the angles used in (\ref{equ:angles1}) and (\ref{equ:angles2}).}
      \label{fig:angles}
\end{figure}

\vskip 4pt
Next, we analyze the behavior of the polarization structure in the collapsed limit. Taking $u,v\to 0$, the angular variables become 
\beq
\begin{aligned}
\tau &\to 4\,\k_1\cdot\k_3\equiv 4k_1k_3\cos\gamma\,, \\[4pt]
	\frac{\alpha}{u} &\to 2k_1(\hat\k_1\cdot\hat\s)\equiv 2k_1\cos\theta\, , \quad \frac{\beta}{v} \to -2k_3(\hat\k_3\cdot\hat\s)\equiv -2k_3\cos\theta'\, ,
\end{aligned} 
\label{equ:angles1}
\eeq
and the variables defined in (\ref{equ:T}) and (\ref{equ:Z}) can be written as
\begin{align}
	\hat T &\to \frac{4k_1k_3}{s^2}(\cos\gamma-\cos\theta\cos\theta')\, ,\\[4pt]
	 \hat L &\to \frac{4k_1k_3}{s^2}\sin\theta\sin\theta'\, .
\end{align}
Using the trigonometric identity
\begin{align}
	\cos\gamma   &= \cos\theta\cos\theta'+\sin\theta\sin\theta'\cos\psi\, ,
	\label{equ:angles2}
	\end{align}
where $\psi$ is the angle between $\hat\k_1$ and $\hat\k_3$ projected onto the plane perpendicular to $\hat\s$ (see Fig.~\ref{fig:angles}), the polarization sums in (\ref{equ:trans}) and (\ref{equ:long}) become 
\begin{align}
	\bar \Pi_{m} &\to (2-\delta_{0m})(-uv)^{-m}(\sin \theta\sin \theta')^m\cos(m\psi)\, ,\\
	\tilde\Pi_{S,m} &\to (-1)^m \tilde P^m_S(\cos\theta)\tilde P^{-m}_S(\cos\theta')\, .
\end{align}
We find that the contribution from each helicity component becomes 
\begin{align}
	\hat A_{S,m}^\pm \bar \Pi_{m}\tilde\Pi_{S,m} \to\ &(-uv)^{-S}\hat A_{S,S}^\pm (2-\delta_{0m})  (-1)^m I_2^\pm (S,m)\hs\cos(m\psi)P^m_S(\cos\theta)P^{-m}_S(\cos\theta') \, .\label{angdep}
\end{align} 
This means that different helicity components  scale in the same way as a function of $u$ and $v$, and only differ by the angular dependence.  Substituting the behavior of the highest-helicity component in the collapsed limit 
 \begin{align}
	(uv)^{-S}\hat A_{S,S}(u,v) \,&\to\, \left(\frac{u v}{4}\right)^{\frac{1}{2}+i\mu} \frac{(1+i\sinh\pi\mu) \Gamma(\tfrac{1}{2}+S+i\mu)^2\Gamma(-i\mu)^2}{2\pi}+c.c.\label{collapsedcoeff}
\end{align}
we reproduce equation (5.120) in \cite{Arkani-Hamed:2015bza}.

\vskip 4pt
For contact diagrams, all helicity components contribute equally and we get
\begin{align}
	F_S \to  \frac{(2S)!}{2}\frac{(k_1k_3)^S}{(k_1+k_3)^{2S+1}}P_S(\cos\gamma)\, .
\end{align}
However, the appearance of the Legendre polynomial is an artifact of looking only at the $s$-channel contribution. Summing over all channels, the spin-$S$ contact contribution will just be a linear combination of the scalar contact terms $\hat C_n$ that we have classified in~\S\ref{sec:contact}.

\paragraph{Flat-space limit}
Another useful consistency check is the flat-space limit, where we expect to reproduce the angular dependence of the flat-space scattering amplitude. 
For this purpose, it suffices to determine the leading behavior of the solution as $u+v\to 0$. Since all helicity components follow from the scalar solution, let us examine
\begin{align}
	\left[u^2(1-u^2)\partial_u^2-2u^3\partial_u + \left(\mu^2+\frac{1}{4}\right)\right]\hat f = \frac{uv}{u+v}\, .
\end{align}
The most singular piece in the flat-space limit is given by 
\begin{align}
	\lim_{u\to -v} \partial_u^2\hat f = -\frac{1}{1-v^2}\frac{1}{u+v} \quad\Rightarrow\quad \lim_{u\to -v}\hat f =-\frac{(u+v)\log(u+v)}{1-v^2}\, .
\end{align}
Substituting this into \eqref{equ:coeff}, the limiting behavior of the coefficient functions becomes 
\begin{align}
	\lim_{u\to -v}\frac{1}{(2S)!}\hat A_{S,m} = \frac{v^{4S-2}}{(u+v)^{2S+1}}\left(\frac{1-v^2}{v^2}\right)^{S-m-1}\, .
\end{align}
Since the polarization sums do not diverge in the flat-space limit, all the helicity coefficients contribute in this limit. 
Switching to the variables $x$ and $y$, we find  
\beq
		\lim_{x+y\to 0} \frac{(-1)^S}{(2S)!}(x+y)^{2S-1}\,F_S \,=\, s^{S-1}_{\rm flat}\times P_S\bigg(1+2\frac{t_{\rm flat}}{s_{\rm flat}}\bigg) \,\propto\, \frac{t_{\rm flat}^S}{s_{\rm flat}}\,+\,{\rm contact~terms}\, ,
\eeq
where have introduce the flat-space Mandelstam variables
\begin{align}
	s_{\rm flat} &= (k_1+k_2)^2 - |\k_1+\k_2|^2 \,\to\, y^2-s^2\, ,\\
	t_{\rm flat} &= (k_2+k_3)^2 - |\k_2+\k_3|^2 \,\to\, \frac{1}{2v^2}\big((\tau-\alpha\beta)v^2-s^2(1-v^2)\big)\, .
\end{align}
This shows that the leading behavior of the four-point function in the flat-space limit gives the scattering amplitude (\ref{equ:M4-2}) (in the high-energy, massless limit).

\subsection{(Partially) Massless Exchange} 
\label{sec:PM}

The conformal dimension of a field $\sigma$ with spin $S$ and mass $M_\sigma$ is 
\begin{align}
	\Delta_\sigma = \frac{3}{2} + \sqrt{\left(S-\frac{1}{2}\right)^2 - \frac{M_\sigma^2}{H^2}}\, , \label{equ:Deltas}
\end{align}
for $S\ge 1$, while the formula for $S=0$ is given by shifting the mass $M_\sigma^2\to M_\sigma^2-2H^2$. 
So far, we have been working with intermediate particles in the principal and complementary series. Another interesting case is the {\it discrete series}, for which $\Delta_\sigma$ takes integer values, corresponding to massless and {partially massless} particles~\cite{Deser:2003gw}. The latter are an interesting generalization of massless particles in (anti-)de Sitter space, for which there is no analog in flat space. Naively extrapolating our previous results to these mass values would lead to divergences, so we will have to treat these cases separately. As was shown by explicit bulk calculations in~\cite{Baumann:2017jvh}, these correlation functions turn out to be rational functions. 

\paragraph{$\boldsymbol{\Delta_\sigma=2}$ exchange} 
Consider the exchange of a conformally coupled scalar field with  $\Delta_\sigma=2$. At leading order in derivatives, the exchange contribution is given by $\Delta_u\hat F = \hat C_0$, and its 
explicit solution  is~\cite{Arkani-Hamed:2015bza}
 \begin{align}
	\hat F= \frac{1}{2}\bigg[{\rm Li}_2\bigg(\frac{x-s}{x+y}\bigg)+{\rm Li}_2\bigg(\frac{y-s}{x+y}\bigg)+\log\bigg(\frac{x+s}{x+y}\bigg)\log\bigg(\frac{y+s}{x+y}\bigg)+\frac{\pi^2}{3}\bigg]\, \equiv \hat F_{\Delta_\sigma=2}\, ,\label{ccscalarexchange}
\end{align}
where ${\rm Li}_2$ denotes the dilogarithm. Using the formulas in the previous section, this can be used to obtain all solutions for spin exchange with $\Delta_\sigma=2$.

\vskip 4pt 
Before presenting a few explicit examples, let us make an important comment on the distinction between the wavefunction coefficients and correlation functions, which is particularly relevant for exchange particles in the discrete series (see also Appendix~\ref{app:ward}). First of all, our proposed ansatz in terms of a scalar solution of the conformal constraints guarantees that the four-point function we obtain will be conformally 
invariant. However, for particles in the discrete series, the disconnected contribution to the correlation function---the product of two three-point functions of two external scalars and the spinning particle---is not exactly conformally 
invariant. This is because particles in the discrete series satisfy Ward identities related to their gauge symmetries. In practice, this means that the three-point function, instead of satisfying a homogeneous differential equation, satisfies an inhomogeneous equation, where the source term is local and enforced by the Ward identity. In other words, the disconnected four-point function satisfies an ordinary differential equation with local source terms that are controlled by symmetry.
 These contributions are not captured by our ansatz, but can be accounted for by adding the known expressions in the literature~\cite{Mata:2012bx,Bzowski:2013sza}.\footnote{However, as far as we know, the form of $\langle \varphi \varphi \sigma \rangle$ for an arbitrary member of the discrete series is not known.}

\vskip 4pt
Substituting the scalar solution \eqref{ccscalarexchange} into \eqref{equ:spin1}, we get the four-point function for massless spin-1 exchange
\beq
\hat F_1 = \frac{s^3}{(x+y)(x+s)(y+s)}\,\bar\Pi_1 -  \frac{s}{x+y}\,\tilde\Pi_{1,0}\, . 
\eeq
Since gauge fields are conformally coupled in $\text{dS}_4$, the result is the same as in flat space. We see that the helicity-1 component has singularities not only at $k_t \equiv x+y \to 0$, but also at $x+s\to 0$ and $y+s\to 0$. These new singularities aren't present in  the longitudinal part, because it is a pure constraint. We also see that the result is a rational function, unlike the case of scalar exchange.

\vskip 4pt
Similar to the spin-1 example, the result for the {\it partially massless} spin-2 \cite{Deser:2001us,Deser:2013uy,Joung:2012rv, deRham:2012kf} exchange turns out to be a rational function. 
However, since a partially massless spin-2 field is not conformal, it has a more complicated structure than the spin-1 example. 
Substituting \eqref{ccscalarexchange} into (\ref{equ:spin2}), we find
\begin{equation}
	\hat F_2 = \frac{s^5(f_2+2s(x+y))}{(x+y)^3 (x+s)^2(y+s)^2}\,\bar\Pi_2\tilde\Pi_{2,2} -\frac{s^3(f_2+s(x+y))}{(x+y)^3 (x+s)(y+s)}\,\bar\Pi_1\tilde\Pi_{2,1}+\frac{s f_2}{(x+y)^3}\,\tilde\Pi_{2,0}\, ,
\end{equation}
where $f_2\equiv s^2+xy+(x+y)^2$. Again, we see that the longitudinal part, being a constraint, does not have a singularity as $x+s\to 0$ and $y+s\to 0$. However, the helicity-1 and 2 components have nontrivial singularities, which indicate that they carry information about propagating degrees of freedom.

\paragraph{$\boldsymbol{\Delta_\sigma>2}$ exchange} Beyond $\Delta_\sigma=2$, the discrete series consists of a whole tower of particles with integer conformal weights $\Delta_\sigma=3,4,\cdots$. The corresponding higher-spin exchange solutions can be obtained once the scalar-exchange solution of the same conformal dimension is known. For example, (partially) massless spin-$S$ exchange with $\Delta_\sigma=3$ can be obtained from the massless scalar exchange solution, while for $\Delta_\sigma>3$ we need solutions for tachyonic scalars. For generic~$\Delta_\sigma$, the scalar solution satisfies 
\begin{align}
	\big[\Delta_u-(\Delta_\sigma-1)(\Delta_\sigma-2)\big] \hat F_{\Delta_\sigma} = \hat C_0\, .\label{discreteeq}
\end{align}
Remarkably, the solution for every integer $\Delta_\sigma$ can be obtained from the $\Delta_\sigma=2$ solution \eqref{ccscalarexchange}. To see this, consider the operator 
\begin{align}
	M_{uv,\Delta_\sigma}(\cdot) \equiv -\frac{1}{(\Delta_\sigma-2)^2}\bigg[\frac{(u(1-u^2)\partial_u-\Delta_\sigma+2)(v(1-v^2)\partial_v-\Delta_\sigma+2)(\cdot)}{uv} +\frac{uv+1}{u+v}\bigg]\, ,\label{Muv}
\end{align}
which shifts the weight of the internal field. This operator can be found by inspection of the homogeneous solutions of the exchange equation (\ref{discreteeq}). The first term in $M_{uv,\sigma}$ relates the solutions of homogeneous equations with different $\Delta_\sigma$, while the second, multiplicative term, is necessary to relate the inhomogeneous solutions of the exchange equation.\footnote{Note that this operator only relates the tree-exchange solutions with the contact term $\hat C_0$ as the source term. As we have seen in Section~\ref{sec:dS-4pt}, however, this is sufficient to generate other solutions with higher-order source terms.} For example, the four-point function associated with the exchange of a massless scalar is given by 
\begin{align}
	\hat F_{\Delta_\sigma=3} &= M_{uv,3} \hat F_{\Delta_\sigma=2}\nn
	& = -\frac{1}{uv}\bigg[\hat F_{\Delta_\sigma=2}-u\log\bigg(\frac{u(1+v)}{u+v}\bigg)-v\log\bigg(\frac{v(1+u)}{u+v}\bigg)+uv\bigg]\, .\label{Fmasslessscalar}
\end{align}
It is easy to check that this satisfies the equation $(\Delta_u-2) \hat F_{\Delta_\sigma=3}=\hat C_0$. 
\vskip 4pt
The solution for general $\Delta_\sigma$ is then given by
\begin{align}
	\hat F_{\Delta_\sigma=2+n} = M_{uv,2+n}\cdots M_{uv,3} \hat F_{\Delta_\sigma=2}\, .
\end{align}
In this way, we can determine all solutions for the exchange of particles belonging to the discrete series from that of a conformally coupled scalar.\footnote{Note that, for $\Delta_\sigma > 3$, the other members of the scalar discrete series are all tachyonic, but their correlation functions can be used in the ansatz for the spinning four-point function to generate the non-tachyonic members of the discrete series at high enough spin, since the number of partially massless fields increases with spin.}

\vskip 4pt
 Using the solution for massless scalar exchange above, we can obtain the graviton-exchange four-point function of conformally coupled external scalars. It is given by
\begin{equation}
	\hat F_{2,\Delta_\sigma=3} = \frac{s^5(s^2+xy+2s(x+y))}{(x+y)^3 (x+s)^2(y+s)^2}\,\bar\Pi_2\tilde\Pi_{2,2} -\frac{s^3}{(x+y)^3}\,\bar\Pi_1\tilde\Pi_{2,1}+\frac{s (s^2+xy)}{(x+y)^3}\,\tilde\Pi_{2,0}\, .
\end{equation}
An important feature of this solution is that it is not purely transverse, but also includes lower helicity components as constraints. For example, the partially massless graviton has nontrivial poles in its helicity-1 components, unlike the ordinary graviton. Note also that this four-point function automatically eliminates the lower-helicity modes without reference to a gauge-fixed propagator. 
Demanding consistent propagation of a massless spin-2 field in de Sitter therefore automatically ensures that only the helicity-2 degree of freedom is propagating.

\section{Massless External Fields}
\label{sec:Massless}

In Sections~\ref{sec:dS-4pt} and \ref{sec:spin}, we presented results for the four-point functions of conformally coupled scalars, while for inflation we need the corresponding results for (nearly) massless scalar fields. In this section, we introduce a set of weight-shifting operators that map the four-point functions of conformally coupled scalars $\varphi$~($\Delta=2$) to the corresponding solutions with massless scalars~$\phi$ ($\Delta=3$). We will present the explicit weight-shifting operators for exchange particles of spin $0, 1, 2$ and sketch a procedure to determine the operators for general spin.
In Appendix~\ref{app:D3}, we show how the conformal invariance constraints on the four-point functions of massless scalars can be solved directly, and confirm that both approaches give consistent results.

\subsection[{Conformal Invariance for $\Delta=3$}]{Conformal Invariance for $\boldsymbol{\Delta=3}$}

Modulo one technical detail, the analysis of the four-point functions of massless scalars is very similar to that for conformally coupled scalars.  We will briefly review the relevant conformal invariance equations, but leave their detailed analysis to  Appendix~\ref{app:D3}.

\vskip 4pt
For notational clarity, we denote the four-point functions of massless scalars by ${\cal F}$. 
In the case of scalar exchange, these are functions only of $(k_n,s)$. Substituting $\Delta_n =3$ into (\ref{W1F}), we find that $(D_1-D_2)\F=0$ implies
\beq
\left(\partial_{k_1}^2 - \frac{2}{k_1} \partial_{k_1}\right) {\cal F} \,-\, \left(\partial_{k_2}^2 - \frac{2}{k_2} \partial_{k_2}\right) {\cal F} \,=\,0\,. \label{equ:350}
\eeq
A similar equation follows from $(D_3-D_4)\F=0$. 
Both equations are solved by the ansatz 
\beq\label{equ:anssol1234}
{\cal F} = s^3 O_{12} O_{34} \,\hat {\cal F}(u, v)\, ,
\eeq
where $\hat {\cal F}$ is (so far) an arbitrary function of ($u$, $v$), and
 $O_{12}$ and  $O_{34}$ are the following differential operators
\beq
O_{nm} \equiv 1 - \frac{k_n k_m}{k_{nm}} \, \partial_{k_{nm}}\, . \label{equ:Oij2}
\eeq
The ansatz (\ref{equ:anssol1234}) works because commuting $D_n$ with $O_{nm}$ gives a result that is symmetric in the two momenta; for instance,
\beq
\left(\partial_{k_1}^2 - \frac{2}{k_1} \partial_{k_1}\right) O_{12}\, h(k_1+k_2) = O_{12} \left(\partial_{k_{12}}^2 - \frac{2}{k_{12}} \partial_{k_{12}}\right) h(k_1+k_2)\, .
\eeq
We see that the introduction of the operators $O_{12}$ and $O_{34}$ has trivialized the 12- and 34-equations.
The dimensionless function $\hat{\cal F}(u,v)$ is then determined by the remaining constraint equation, $(D_1-D_3)\F=0$, which becomes 
\beq
\big(\tilde\Delta_u -\tilde \Delta_{v}\big) \hat {\cal F}= 0\, , 
\label{equ:d3cons}
\eeq
where we have introduced the differential operator\hs\footnote{Note that $u^{-3}$ is a zero mode of the operator $\tilde\Delta_u$. This implies that ${\hat \F}_{\rm loc} \equiv u^{-3}+v^{-3}$ is a trivial solution of (\ref{equ:d3cons}) and can be added (with arbitrary coefficient) to any of the solutions studied below. This extra term corresponds to ``local non-Gaussianity", and can be removed by a field redefinition.}
\beq
\tilde\Delta_u \equiv u^2(1-u^2)\partial^2_u+4u(1-u^2)\partial_u\, .
\label{equ:tildeD}
\eeq
Equation (\ref{equ:d3cons}) is very similar to equation (\ref{contactgeneral}) for the four-point function of conformally coupled scalars.  In Appendix~\ref{app:D3},  we will present the explicit solutions of (\ref{equ:d3cons}) corresponding to contact terms and the exchange of massive particles.

\subsection[{Mapping $\Delta=2$ to $\Delta=3$}]{Mapping $\boldsymbol{\Delta=2}$ to $\boldsymbol{\Delta=3}$}
\label{sec:mapping}

 Instead of embarking on a detailed analysis of the solutions to the $\Delta=3$ constraint equations, we will now show that the same solutions can be obtained directly by applying certain differential operators to the four-point functions with external $\Delta=2$ legs. These operators depend on the spin of the exchanged particle. We analyze explicitly the cases of spins $0$, $1$, $2$, and briefly describe the generalization to general spin exchange.
 
\paragraph{Spin-0 exchange}  We first consider scalar exchange, and take inspiration from the bulk calculation. Although we will analyze a specific three-point vertex for the bulk four-point function, conformal symmetry guarantees that our result will apply more generally. This can be confirmed explicitly by checking that the ansatz derived below solves the conformal constraint equations.

\vskip 4pt
Let us write the mode functions of massless and conformally coupled scalar fields as
\beq
\begin{aligned}
\upphi_k &\equiv(1+i k\eta) h_k\,,\\
 \upvarphi_k &\equiv \eta h_k\,, 
\end{aligned}
\label{equ:modes}
\eeq
where $h_k(\eta) \equiv e^{i k_\mu x^\mu}=e^{-ik \eta+i \k\cdot\x}$. The product $\upvarphi_{k_1}\upvarphi_{k_2}$ is a function of $k_1+k_2$ only, which allowed us to write the four-point function of conformally coupled scalars as $F = s^{-1}\hat F(u,v)$, where $\hat F$ is a function of $(u,v)$ only. The situation for massless scalars is slightly more complicated.  However, after a bit of work, we can show that
\beq
\upphi_{k_1} \upphi_{k_2} =  O_{12}\,\upphi_{k_1+k_2}\,, \label{equ:relationOmn}
\eeq
where $O_{12}$ is the same operator as defined above.
The relation (\ref{equ:relationOmn}) is the bulk reason why ${\cal F}$ can be written in the form (\ref{equ:anssol1234}).

\vskip 4pt
If the inflaton couples in a shift-symmetric fashion, say through the coupling $(\nabla \upphi)^2\sigma$, then its four-point functions can be related directly to that of conformally coupled scalars, with coupling $\upvarphi^2 \sigma$, 
by introducing differential operators constructed out of the operators $O_{12}$ and $O_{34}$.  In the insert below we derive the following relation between the $\upphi$ and $\upvarphi$ mode functions:
\beq
\nabla_\mu \upphi_{k_1}\nabla^\mu \upphi_{k_2} =  s^2\, U_{12}(\upvarphi_{k_1} \upvarphi_{k_2})\,, \label{equ:relation}
\eeq
where we have defined the following operator 
\beq
U_{12} (\cdot)\equiv \frac{1}{2}\,O_{12}\bigg[\frac{1-u^2}{u^2}\partial_u (u \, \cdot) \bigg] \, . \label{equ:U12}
\eeq
The relation (\ref{equ:relation}) maps bulk interactions of conformally coupled scalars to those of massless scalars with two additional derivatives per vertex. Although we utilized particular bulk vertices to derive $U_{12}$, it is easily confirmed that $U_{12}U_{34}$ applied to {\it any} solution of the $\Delta=2$ constraints solves the conformal invariance constraints for  $\Delta=3$. This implies that the four-point functions of $\varphi$ and $\phi$ are related as follows
\begin{eBox}
\beq
{\cal F}_{0}  =s^3\hs U_{12} U_{34}\hs \hat f (u,v)\, ,\label{Fdelta3S0}
\eeq
\vskip 4pt
\end{eBox}
where $\hat f \equiv \hat F_0$ is any solution of the scalar $\Delta=2$ equation (\ref{contactgeneral}). 
Equation (\ref{Fdelta3S0}) allows us to bootstrap the four-point functions of massless scalars from the solutions obtained in Section~\ref{sec:dS-4pt}.

\vskip 4pt 
Of course, this prescription is not exclusive to exchange diagrams, but also applies to contact interactions. For instance, applying $U_{12}U_{34}$ to the contact term~$\hat C_0$ and symmetrizing, we obtain the four-point function corresponding to the bulk interaction~$(\partial_\mu\upphi)^4$, i.e.~we reproduce precisely equation (17) of \cite{Creminelli:2011mw}. 

\begin{framed}
{\small \noindent {\it Derivation.}---From (\ref{equ:modes}), we have 
\begin{align}
g^{\mu\nu}\nabla_\mu \upphi_{k_1} \nabla_\nu \upphi_{k_2} &=-(1-k_1\partial_{k_1})(1-k_2\partial_{k_2}) (k_1 \cdot k_{2} \, \upvarphi_{k_1} \upvarphi_{k_2}) \, ,
\end{align}
where we have used that the derivatives pass through the momentum operators, so that $\nabla_\mu \upphi_k = (1-k\partial_k)(i k_\mu h_k)$, with $k^\mu\equiv(k,\k)$ and $k_1\cdot k_2\equiv \eta_{\mu\nu} k_1^\mu k_2^\nu$. The factors of conformal time in $g^{\mu\nu}$ have turned $h_{k_1}h_{k_2}$ into $\upvarphi_{k_1} \upvarphi_{k_2}$. Because $\upvarphi_{k_1}\upvarphi_{k_2}$ depends on the energy of the modes $k_1$ and $k_2$ only through their sum, we have $\upvarphi_{k_1}\upvarphi_{k_2}= h(k_1+k_2)$. Including the action of the derivatives on $k_1 \cdot k_2 \equiv -k_1 k_2 + \k_1\cdot \k_2$, we find
\beq
(1-k_1\partial_{k_1})(1-k_2\partial_{k_2}) (k_1 \cdot k_{2} \, h(k_1+k_2))= \left[ \k_1\cdot\k_2\, (1-x \partial_x)+k_1 k_2 (k_1 \cdot k_{2}) \,\partial^2_x \right] h\, ,
\eeq
which implies the result (\ref{equ:relation}).}
\end{framed}

In the following, we will determine the corresponding operators $U_{12}^{S,m}$ and $U_{34}^{S,m}$ for spin-$S$ exchange. The action of these operators on the helicity-$m$ components of the $\Delta = 2$ solutions gives the corresponding $\Delta=3$ solutions, i.e. 
\beq
{\cal F}_S  = \sum_m  s^4 \hs U_{12}^{S,m} U_{34}^{S,m} F_{S,m}\, , \label{equ:USm}
\eeq
where we have extracted a factor of $s^4$, so that the $U$-operators are dimensionless. Since these operators shift the weights (masses) of the external fields, we will call them the {\it weight-shifting} operators.\footnote{Other  weight-shifting operators are defined in Appendix~\ref{app:D3}, which may be useful for other purposes. Moreover, an operator that shifts the weight of the internal field was introduced in \eqref{Muv}. Similar weight-shifting operators for conformal correlators in position space have recently been discovered in~\cite{Karateev:2017jgd, Costa:2018mcg}.} We will illustrate the procedure explicitly up to spin-2 exchange.

\paragraph{Spin-1 exchange} 
We begin with the example of spin-1 exchange. For concreteness, we will consider the four-point function involving conformally coupled scalars interacting with a spin-1 field through the minimal number of derivatives. In other words, we will study a three-point vertex of the form $J_\alpha^\varphi \sigma^\alpha$, with
the current given by  
\beq
J_\alpha^{\upvarphi} \equiv \upvarphi_{k_1}\nabla_\alpha \upvarphi^*_{k_2} - (k_1\leftrightarrow k_2)\,. \label{J1}
\eeq 
As in the scalar example, we wish to relate $F_1$ to ${\cal F}_1$ obtained by the substitution $J_\alpha^\upvarphi\to J_\alpha^{\upphi}$, for a suitable choice of $J_\alpha^{\upphi}$. Following the scalar example, the current involving massless scalars must have two additional derivatives per vertex. We choose 
\beq
J_\alpha^\upphi \equiv \nabla_\mu \upphi_{k_1} \nabla^\mu\nabla_{\alpha} \upphi_{k_2}^* - (k_1\leftrightarrow k_2)\,. \label{J2}
\eeq
As we prove in the insert below, the two currents are related as follows \begin{align}
J^\upphi_i &\approx U^{1,1}_{12} J^\upvarphi_i \, ,  \label{equ:Jveceqs}\\
J^\upphi_0&= U^{1,0}_{12} J^\upvarphi_0 \, , 
\end{align}
where we have introduced distinct operators for the different helicity components:
\begin{align}
U^{1,1}_{12}(\cdot) &\equiv \frac{1}{2}\,O_{12}\bigg[\frac{1-u^2}{u^2}\partial_u(u\,\cdot)-\frac{2}{u^2} (\cdot)\bigg]\, , \label{equ:U121}\\
U^{1,0}_{12}(\cdot) &\equiv \frac{1}{2u}\,O_{12}\bigg[(1-u^2)\partial_u(\cdot) \bigg]\, . \label{equ:U122}
\end{align}
Acting on the second vertex, these operators become $U^{1,m}_{34}$, which follow from (\ref{equ:U121}) and (\ref{equ:U122}) with the substitutions $u\leftrightarrow v$ and $\alpha \leftrightarrow \beta$. 
The $\approx$ symbol in (\ref{equ:Jveceqs})
 means that the relation between the currents is only true when they are contracted with the propagator of a spin-$1$ particle. In other words, in (\ref{equ:Jveceqs}), we dropped terms that are longitudinal with respect to the momentum  of the exchange particle.
 
 \vskip 4pt
Given the $\Delta=2$ solution (\ref{equ:spin1}), 
\beq
F_{1}=\frac{1}{s}\Big[\bar \Pi_1 \tilde \Pi_{1,1}  D_{uv} \, + \tilde \Pi_{1,0} \Delta_u \Big]  \hat f\, ,
\eeq 
we therefore obtain the following $\Delta=3$ solution 
\begin{eBox}
\beq
{\cal F}_{1}= s^3 \Big[  \bar \Pi_1 \tilde \Pi_{1,1} \hs U^{1,1}_{12}U^{1,1}_{34} D_{uv} + \tilde \Pi_{1,0} \hs U^{1,0}_{12}U^{1,0}_{34}\Delta_u \Big] \hat f\, .  \label{equ:F111}
\eeq 
\end{eBox}
It is straightforward to verify that the function ${\cal F}_{1}$ solves the conformal constraints, for any solution $\hat f$ of the scalar $\Delta=2$ equation. 
Note that the polarization sums in (\ref{equ:F111}) are not acted on with the differential operators. As we will see below for spin 2, this is not generally the case. 

\begin{framed}
{\small \noindent {\it Derivation.}---First, we work out the two-index object $\nabla_\mu\nabla_\alpha \upphi_k$. 
A straightforward computation yields
\begin{align}
\nabla_\mu\nabla_\alpha \upphi_k= (1-k \partial_k)\left[-k_\mu k_\alpha+ \frac{i}{\eta}(k_\mu \delta_\alpha^0+k_\alpha \delta_\mu^0-\eta_{\alpha\mu}k) \right] h_k\, .
\end{align}
We thus find 
\begin{align}
\nabla^\mu \upphi_{k_1}\nabla_{\mu}\nabla_\alpha \upphi_{k_2}^* = &\ (1-k_1\partial_{k_1})(1-k_2\partial_{k_2})\Big[k_{1} \cdot k_2 \left(k_{2,\alpha} \partial_{k_2} - \delta_\alpha^0\right)   + \nonumber\\[2pt] &+ (k_{1,\alpha} k_2 - k_{2,\alpha} k_1)   \Big] \upvarphi_{k_1} h_{k_2}^* \, ,
\end{align}
where the second line vanishes for $\alpha=0$, since $k_{n,0}=-k_n$. The first line can be written in terms of $\upvarphi_{k_1}\nabla_\alpha \upvarphi_{k_2}^*$, because 
\beq
\upvarphi_{k_1}\nabla_\alpha \upvarphi_{k_2}^* =-(k_{2,\alpha} \partial_{k_2}-\delta_\alpha^0) \upvarphi_{k_1} h_{k_2}^* \, .
\eeq
We therefore get 
\begin{align}
\nabla^\mu \upphi_{k_1}\nabla_{\mu}\nabla_\alpha \upphi_{k_2}^* = \ &-(1-k_1\partial_{k_1})(1-k_2\partial_{k_2})\big[k_{1} \cdot k_2\,(\upvarphi_{k_1}\nabla_\alpha \upvarphi_{k_2}^*) \big] +\nonumber\\ &+(1-k_1\partial_{k_1})(1-k_2\partial_{k_2})\big[(k_{1,\alpha} k_2- k_{2,\alpha} k_1)\, \eta\hs h_{k_1} h_{k_2}^* \big]\, ,
\end{align} 
where the second line vanishes for $\alpha = 0$.
This implies 
\begin{align}
J_i^\upphi &\approx -(1-k_1\partial_{k_1})(1-k_2\partial_{k_2})\left[k_{1}\cdot k_2\,J_i^\upvarphi \right] -k_1^2(1-k_2\partial_{k_2}) J_i^\upvarphi -k_2^2(1-k_1\partial_{k_1}) J_i^\upvarphi  \, , \label{spatspin1}\\
J_0^\upphi &= -(1-k_1\partial_{k_1})(1-k_2\partial_{k_2})\left[k_{1}\cdot k_2 \, J_0^\upvarphi \right]  \, . \label{timespin1}
\end{align}
To obtain the relation between the spatial components, we dropped a term proportional to $s_i$. This is because $J^\upphi_i$ and $J^\upvarphi_i$ are contracted with a propagator for the spinning field with momentum $s_i$, and this contribution vanishes. We must act with the operator \eqref{spatspin1} on the transverse piece of the four-point function and with the operator \eqref{timespin1} on the longitudinal piece. This is because the longitudinal part of the spatial components of the exchanged particle can be converted into time components through the constraint equations. In other words, once we substitute $J^\upvarphi_\alpha=(k_{1,\alpha}-k_{2,\alpha}) h(k_1+k_2)$ on the right-hand sides of \eqref{spatspin1} and \eqref{timespin1}, we obtain two operators that act on the transverse and longitudinal pieces of the four-point function.}
\end{framed}

\paragraph{Spin-2 exchange}  
The analysis of spin-2 exchange is a bit more subtle. 
We wish to relate couplings like $T_{\alpha \beta}^\upvarphi\hs \sigma^{\alpha \beta}$ and $T_{\alpha \beta}^\upphi\hs \sigma^{\alpha \beta}$, where the tensors $T_{\alpha \beta}^\upvarphi$ and $T_{\alpha \beta}^\upphi$ are quadratic in the fields $\upvarphi$ and~$\upphi$. However, the naive higher-derivative object,  $\nabla_\alpha \nabla^\mu \upphi_{k_1} \nabla_\beta \nabla_{\mu} \upphi_{k_2}$, is not related in a simple way to $\nabla_\alpha \upvarphi_{k_1} \nabla_\beta \upvarphi_{k_2}$.  
Since the exchanged particle is a symmetric traceless spin-2 field, we must instead contract it with a spin-2 tensor that is (spacetime) transverse and traceless.
Like for the spatial component of the spin-1 example, the equivalence between the conformally coupled spin-2 tensor and the massless spin-2 tensor only needs to be true up to terms that are longitudinal with respect to the exchanged momentum, i.e.~terms proportional to $s_i s_j$. 
 We are also free to discard pieces in the tensors proportional to the metric, as these will produce traces of the exchanged field, which are zero.

\vskip 4pt
As the computations are more tedious, we state the results without derivation, but they are straightforward to check. We will consider the map between the following two transverse tensors:
\begin{align}
T^\upvarphi_{\alpha\beta} &\equiv \upvarphi_{k_1} \nabla_\alpha \nabla_\beta \upvarphi_{k_2}-2\nabla_\alpha \upvarphi_{k_1}\nabla_\beta \upvarphi_{k_2} +(k_1\leftrightarrow k_2) \, , \label{equ:tensors0}\\
T^\upphi_{\alpha\beta} &\equiv\nabla_\alpha\nabla_\beta \nabla_\mu \upphi_{k_1}\nabla^\mu \upphi_{k_2}-2 \nabla_\alpha\nabla_\mu \upphi_{k_1}\nabla_\beta\nabla^\mu \upphi_{k_2}-3 \nabla_\alpha \upphi_{k_1} \nabla_\beta \upphi_{k_2}+(k_1\leftrightarrow k_2)\, , \label{equ:tensors}
\end{align}
where we have dropped terms proportional to $g_{\alpha \beta}$.
Explicitly, we have 
\begin{align}
T^\upphi_{ij} &\approx - \bigg[2 (k_1+k_2)^2 -(k_1+k_2)\,\k_1\cdot\k_2\, (\partial_{k_1}+\partial_{k_2}) +\partial_{k_1}\partial_{k_2}(k_1 k_2\, k_{1}\cdot k_2)\bigg]\,T^\upvarphi_{ij}\, , \label{equ:Tij} \\
T^\upphi_{0i} &\approx -\bigg[(k_1^2+k_1 k_2+k_2^2)- (k_1 k_2+ \k_1 \hskip -1pt\cdot \hskip -1pt \k_2)\frac{k_1+k_2}{2}(\partial_{k_1}+\partial_{k_2})+(k_1k_2\, k_{1}\hskip -1pt\cdot \hskip -1pt k_2)\partial_{k_1 k_2}\bigg]\, T^\upvarphi_{0i}\, , \label{equ:T0i}\\
T^\upphi_{00} &\approx - \bigg[(1-k_1\partial_{k_1})(1-k_2\partial_{k_2}) \,k_{1}\cdot k_2  -\frac{(k_1 k_2)^2}{k_1-k_2} (\partial_{k_1}-\partial_{k_2}) \bigg] \,T^\upvarphi_{00} \, ,  \label{equ:T00}
\end{align}
where $\approx$ means 
that we dropped terms proportional to $s_i$ and $g_{\alpha \beta}$. 
Like in the spin-1 case, the operator that relates $T^\upphi_{00}$ and $T^\upvarphi_{00}$ acts on the longitudinal component of the four-point function, the operator that relates $T^\upphi_{0i}$ and $T^\upvarphi_{0i}$ acts on the helicity-one component, and the operator that relates $T^\upphi_{ij}$ and $T^\upvarphi_{ij}$ on the helicity-two component. The reason is that the longitudinal parts of the spatial components of the exchanged field can always be turned into temporal components via the constraint equations.

\vskip 4pt
Acting with (\ref{equ:Tij}) and (\ref{equ:T0i}) on the top and middle helicity components of $F_2$, we find that the operators commute with the polarization sums, and so we can act on the coefficient functions with 
\begin{align}
U^{2,2}_{12}(\cdot)&=\frac{1}{2} \,O_{12}\bigg[\frac{1-u^2}{u^2}\partial_u (u\,\cdot)-\frac{4}{u^2}(\cdot)\bigg] , \\
U^{2,1}_{12}(\cdot) &=\frac{1}{2u}\,O_{12}\bigg[(1-u^2)\partial_u(\cdot) -\frac{2}{u}(\cdot) \bigg] .
\end{align}
The operator relating $T_{00}^\upphi$ and $T_{00}^\upvarphi$ in (\ref{equ:T00}) acts on the longitudinal solution of the spin-2 equation, which is of the form $P_2(\alpha/s) F(u)$ (where we suppressed the $\beta$, $v$ dependence). This time, however, we cannot act directly on the coefficient function and pull out the polarization sum from the $U^{2,0}_{12}$ operator. Using (\ref{equ:T00}), we obtain  
\beq\label{sp2long}
U_{12}^{2,0}(\cdot) = U_{12}^{1,0}(\cdot) +\left[\frac{\hat\alpha^2}{P_2(\hat \alpha)}-\frac{1+u^2}{2u^2}\right](\cdot)\, .
\eeq
where $\hat\alpha\equiv\alpha/s$ and $U_{12}^{1,0}$ was defined in \eqref{equ:U122}. This expression makes it manifest that the longitudinal component vanishes in the soft limit. Taking, say $k_2\to0$, we have 
$\hat\alpha\to 1$ and $u\to 1$, so that 
$U_{12}^{1,0}$ vanishes and the two terms in the brackets cancel against each other.

\vskip 4pt

The corresponding $U^{2,m}_{34}$ operators are obtained by the substitutions $u \leftrightarrow v$ and $\alpha\leftrightarrow \beta$. Given the solution $F_{2}$ in (\ref{equ:spin2d2}), we then obtain 
\begin{eBox2}
\begin{equation}
	{\cal F}_{2}= s^3\bigg[ \bar\Pi_{2}\tilde\Pi_{2,2}\hs U^{2,2}_{12}U^{2,2}_{34}D_{uv}^2   + \bar \Pi_{1} \tilde\Pi_{2,1}\hs U^{2,1}_{12}U^{2,1}_{34} D_{uv}(\Delta_u-2)  + \tilde\Pi_{2,0} \hskip 1pt U^{2,0}_{12}U_{34}^{2,0}\Delta_u(\Delta_u-2) \bigg] \hat f\hskip 1pt .\label{equ:Fspin2}
\end{equation}
\end{eBox2}
This solves the constraints for $\Delta=3$, if $\hat f$ is any scalar solution of the $\Delta=2$ constraint. Notice that the polarization sum $\tilde\Pi_{2,0}$ appears inside $U_{12}^{2,0} U_{34}^{2,0}$, cf.~\eqref{sp2long}, rather than as a prefactor, since it does not commute with the operator that relates $T^\upphi_{00}$ to $T^\upvarphi_{00}$.
Also note that, although the coefficient functions for the helicity-2 and 1 component do not vanish in the soft limit, the polarization sums do. This implies that the four-point function vanishes in the soft limit, as expected for derivatively coupled interactions.

\paragraph{Graviton exchange}  The leading four-point function in single-field slow-roll inflation arises from graviton exchange~\cite{Seery:2008ax} (see also~\cite{Seery:2006vu, Arroja:2008ga, Ghosh:2014kba}).
Equipped with our ansatz for spin-2 exchange, we can now show how this famous result is reproduced in our formalism.

\vskip 4pt
The inflationary trispectrum, denoted by $ {\cal F}_{\rm inf}$, consists of two parts: a contribution from the exchange of the transverse-traceless graviton $\gamma_{ij}$ (GE) and a contribution from contact terms~(CT):
\begin{align}
	 {\cal F}_{\rm inf} =  {\cal F}_{\rm GE} +  {\cal F}_{\rm CT}\, ,\label{inf4pt}
\end{align}
where $ {\cal F}_{\rm GE}$ and $ {\cal F}_{\rm CT}$ are given by equations (2.33) and (4.1) in \cite{Seery:2008ax} (after stripping off the overall factors of $\prod_n k_n^3$). It is useful to split the GE contribution into its  connected and disconnected parts,
which we denote by $ {\cal F}_{\rm GE, c}$ and $ {\cal F}_{\rm GE, d}$.\footnote{By the disconnected part, we mean the piece that looks like a product of two three-point functions (see \S\ref{sec:Psi}), not the Gaussian contribution to the four-point function $\langle \phi^2\rangle^2$.}

\vskip 4pt
The connected piece is conformally invariant and given by 
\begin{align}
	{\cal F}_{\rm GE,c} + {\cal F}_{\rm CT} =\frac{1}{6}\left(  {\cal F}_{2}[\hat f_{-1}] - 4{\cal F}_{0}[\hat C_0] \right)+ \text{perms}\, ,\label{ConnectedPart}
\end{align}
where ${\cal F}_{2}[\hat f_{-1}]$ and ${\cal F}_{0}[\hat C_0]$ were defined in \eqref{equ:Fspin2} and \eqref{Fdelta3S0}, respectively.  Note that our spin-2 ansatz has eight derivatives acting on the scalar solution $\hat f_{-1}$.  In order to obtain the graviton exchange from a scalar exchange solution, we must pick $\hat f_{-1}$ to be a solution of the equation for massless scalar exchange. To obtain the four-point function that arises from the interaction $\partial^\mu\phi\partial^\nu\phi \hskip 1pt \sigma_{\mu\nu}$, we use as an input the function $\hat f_{-1}$ that solves the massless scalar exchange equation $(\Delta_u-2)\hat f_{-1}=\hat C_{-1}$ with the source satisfying $\Delta_u\hat C_{-1} = \hat C_0$.\footnote{The solution to this exchange equation always captures the physical exchange of the graviton irrespective of the type of contact term used as the source. Different solutions are then simply related by a shift of lower-order contact terms. Our choice of $\hat f_{-1}$ is convenient because it requires the minimal number of contact terms to capture the graviton exchange piece. Explicitly, the solution can be written as $\hat f_{-1}=\frac{1}{2}(\hat F_{\Delta_\sigma=3}-\hat F_{\Delta_\sigma=2})$.} 
\vskip 4pt
The disconnected part, on the other hand, breaks the conformal symmetry in a way that is dictated by Ward identities (see \S\ref{sec:PM} and ref.~\cite{Ghosh:2014kba}): 
\begin{align}
	{\cal F}_{\rm GE,d} 
 	&= \frac{1}{12} B_{OOT}(k_1,k_2,s)B_{OOT}(k_3,k_4,s)\, \bar\Pi_2\tilde\Pi_{2,2}+ \text{perms}\, , \label{DisconnectedPart}
\end{align}
where $B_{OOT}$ is given by (\ref{equ:BOOT}) in Appendix~\ref{app:ward}; see also \cite{Bzowski:2013sza, Mata:2012bx}. 
Since ${\cal F}_{\rm GE,d}$ breaks the conformal symmetry, it cannot be captured by a suitable choice of $\Delta=2$ input function, which was obtained as a solution of the conformal invariance equations. 
The coefficient of the disconnected piece can be determined in a similar fashion as the parameter $\beta_0$ was determined for the massive exchange solutions discussed earlier; see Appendix~\ref{app:bulk} for details. The resulting expression for ${\cal F}_{\rm inf}$ agrees with equation (4.7) in~\cite{Seery:2008ax}.

\paragraph{General spin}  In this section, we presented explicit results for the four-point functions of massless scalars due to the exchange of particles with spin 0, 1, and 2. We showed how transverse traceless tensors that are bilinear in $\upphi$ can be written in terms of tensors that are bilinear in~$\upvarphi$. The tensors in $\varphi$ had the minimum number of derivatives, while the tensor in $\upphi$ was slightly non-minimal, having two additional derivatives. The extension of these results to higher-spin exchange in conceptually straightforward, but technically cumbersome.  We therefore only sketch the overall strategy for obtaining the weight-shifting operators associated with the exchange of higher-spin particles.

\vskip 4pt
The operators relating the spin-$S$ tensors $T^\upphi_{\mu_1 ... \mu_S}$ to $T^\upvarphi_{\mu_1 ... \mu_S}$ have at most two derivatives, and the coefficient functions are rational functions of $(k_1,k_2,\k_1\cdot \k_2)$ of homogeneity $+2$ under rescaling of the momenta, cf.~Eqs.~(\ref{equ:Tij}) to (\ref{equ:T00}). This suggests an algorithm to determine the operators $U_{12}^{S,m}$ for general spin $S$: First, we write general functions of $(k_1, k_2, \k_1\cdot\k_2)$ with the right homogeneity properties, times a differential operator that is at most second order in $(k_1,k_2)$. Then, we apply such an operator to the known $\Delta=2$ solution, being careful not to act with the derivatives on the transverse polarization sums $\bar \Pi_{S,m}$. This gives a candidate solution of the constraint equations. Finally, we find the unknown functions inside the differential operator order by order in $\tau$, thus determining the operators $U^{S,m}_{12}$ that act on each helicity mode separately. 

\vskip 4pt
In an upcoming paper~\cite{CosmoBoot2}, we will provide a more systematic derivation of the weight-shifting operators for general spin following a more group-theoretical approach, without resorting to bulk kinematics. For the longitudinal component, $m=0$, this analysis gives
\begin{align}
U_{12}^{S,0} &\equiv U_{12}^{1,0}+\frac{1}{2}\left(\frac{(S-1)(S-2) u^2 \hat \alpha^2 - (S+2)(S-1)}{4u^2}+\hat \alpha S\, \frac{P_{S-1}(\hat \alpha)}{P_S(\hat \alpha)}-1\right)   .  \label{equ:U12L}
\end{align}
As we will show in the next section, only the longitudinal part of the spin-exchange four-point functions survives in inflationary three-point functions, so the result \eqref{equ:U12L} will be sufficient for that purpose.

\section{Inflationary Three-Point Functions}
\label{sec:inflation}

We have seen that tree-level correlation functions in de Sitter space are completely fixed by symmetries and the absence of unphysical singularities.
However, the time dependence of inflation breaks scale and special conformal invariance, so it needs to be discussed how much of this structure survives in the inflationary context. In this section, we will show that the small breaking of the de Sitter symmetries due to the inflationary background can be accounted for systematically. 
As an illustration, we will reproduce Maldacena's famous result for the three-point function of slow-roll inflation~\cite{Maldacena:2002vr} from a simple deformation of the de Sitter four-point function due to graviton exchange (see also~\cite{Kundu:2015xta}). We will also provide a systematic classification of the inflationary three-point functions arising from massive particle exchange.

\begin{figure}[h!]
    \centering
      \includegraphics[scale=0.85]{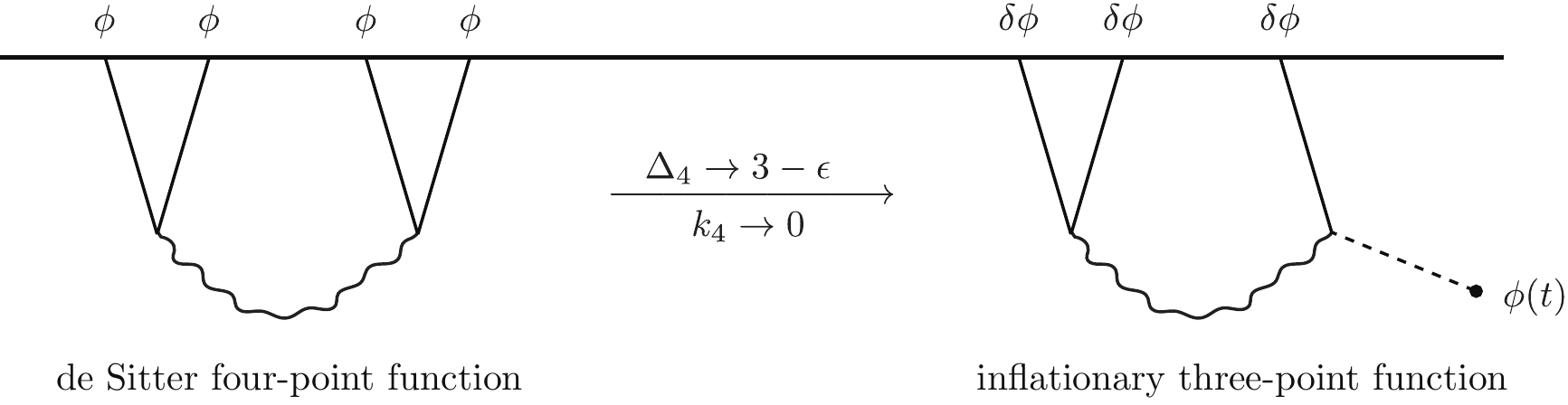}
           \caption{Inflationary three-point functions can be derived from de Sitter four-point functions by evaluating one of the legs on the time-dependent background.}
    \label{fig:3pt}
\end{figure}

\subsection{Perturbed de Sitter}

We assume a standard slow-roll scenario in which the breaking of the de Sitter isometries is controlled by the slow-roll parameter
\beq
\epsilon \equiv \frac{\dot {\upphi}^2}{2M_{\rm pl}^2 H^2} \ll 1\, .\label{epsilon}
\eeq
At leading order, the four-point functions we studied above don't feel the effects of the weak symmetry breaking and our results therefore also apply for inflation (after mapping the results for conformally coupled external fields to massless fields). On the other hand, the possibilities for inflationary three-point functions that are exactly de Sitter invariant are very limited.\footnote{For scalar fields, a de Sitter-invariant bispectrum corresponds to a $\upphi^3$ interaction in the bulk. This interaction breaks the shift symmetry but can be naturally small in models of slow-roll inflation, $\alpha$-attractors or Starobinsky inflation~\cite{Pajer:2016ieg}.} 
However, if the breaking of the conformal symmetry is weak, then inflationary three-point functions can also be obtained by taking one of the legs of the de Sitter four-point functions to have a soft momentum and a small mass~\cite{Creminelli:2003iq, Kundu:2015xta, Arkani-Hamed:2015bza} (see Fig.~\ref{fig:3pt}). 
This corresponding to studying the conformal correlation function for $\Delta=3 - \epsilon$, where $\epsilon$ plays the role of the slow-roll parameter in the bulk. For some deformations of the boundary theory, the perturbation to the operator dimension may be a combination of slow-roll parameters in the bulk, not just the parameter $\epsilon$ defined in (\ref{epsilon}). This gives us a purely boundary way of thinking about slow-roll deviations from pure de Sitter space in the inflationary correlators.\footnote{From the boundary perspective, slow-roll parameters are usually defined as deformations either of the conformal field theory or of the conformal dimension of the operator dual to the inflaton~\cite{Maldacena:2002vr, Larsen:2003pf, van2004inflationary, Bzowski:2012ih, McFadden:2013ria, Ghosh:2014kba,  Baumann:2015xxa, Isono:2016yyj}.} As we will show below, we only need to consider the soft leg to have dimension $\Delta=3 - \epsilon$, and can keep the dimensions of the remaining fields at $\Delta=3$. 
We will choose the last leg of the correlator to be soft, i.e.~$k_4\to0$, and determine the correction to the correlators in this limit at leading order in~$\epsilon$. This turns out to give the leading contribution to the inflationary three-point function.

\subsection{Inflationary Bispectra}
\label{sec:InfBi}

To obtain inflationary three-point functions from the massless de Sitter four-point functions of Section~\ref{sec:Massless}, we must evaluate one leg on the background.  Say that we single out the fourth leg and take $k_4 \to 0$. For massless fields interacting in a shift-symmetric fashion, this soft limit of the four-point function will be zero. A nonzero result is only obtained if we take into account the small inflaton mass proportional to the slow-roll parameter $\epsilon$, so that the mode function associated with the soft leg is 
\beq
\upphi_{k_4, \epsilon}(x) =\bigg((1+i k_4 \eta)+  \frac{\epsilon}{2} \log (-k_4 \eta)+\cdots\bigg)\,e^{i k_4\cdot x}\, .\label{equ:PM}
\eeq
The four-point function will then have a nontrivial soft limit proportional to the slow-roll parameter~$\epsilon$, and the ellipsis represents slow-roll corrections to the mode function that are irrelevant in the $k_4\to0$ limit.  

\vskip 4pt
We expect only the longitudinal component of the four-point function to contribute to the inflationary three-point function. From the bulk, it is easy to see why. The parts of the four-point function with helicities greater than zero involve the contraction of a polarization tensor for the exchanged particle, with intermediate momentum $s_i$, with the external momenta. In particular, we will always have at least one contraction of the form $(k_3-k_4)_i\,\epsilon_{i j}(s_i)$.  By transversality of the polarization tensor, this contraction vanishes in the soft limit $k_4 \to 0$.  This can be checked explicitly by setting $k_4\to0$, $\hat T\to0$ and $\hat L\to 0$,  in the polarization sums \eqref{equ:T} and \eqref{equ:Z}.

\vskip 4pt
Taking the soft limit of the longitudinal part of the four-point function, $s^3\, \tilde\Pi_{S,0}  U_{12}^{S,0} U_{34}^{S,0} \hat A_L$, we get an inflationary three-point function of the form
\begin{eBox}
\beq
B(k_1,k_2,k_3) = \frac{\epsilon}{2} \hs k_3^3\hs P_S(\alpha/s) \hs U_{12}^{S,0}\hs \hat b_S(u) \, +\, \text{perms}\, ,  \label{equ:BINF}
\eeq
\vskip 4pt
\end{eBox}
where $U_{12}^{S,0}$ is given by (\ref{equ:U12L}) and we have introduced the {\it source function}
\beq
\hat b_S(u) \,\equiv\, \lim_{v \to 1} \hat A_L(u,v) 
 \,=\, \prod_{j=1}^{S} \big(\Delta_u-(S-j)(S-j+1)\big)\hat f(u,1)\, . \label{equ:bs}
\eeq
We see that inflationary bispectra are determined by the $v \to 1$ limit of the four-point function of conformally coupled scalars in de Sitter space, $\hat f(u,v)$.
In the following, we will show that this reproduces classic results from the inflationary literature, as well as providing an elegant way to classify all bispectra due to the exchange of particles with spin.

\paragraph{Scalar exchange/contact} 
Consider the soft limit $k_4 \to 0$ of the operator $\nabla_\mu \upphi_{k_3} \nabla^\mu \upphi_{k_4,\epsilon}$. In terms of the mode functions of conformally coupled scalars, this can be written as 
\beq\label{equ:soflimsc}
\lim_{k_4\to 0}\nabla_\mu \upphi_{k_3} \nabla^\mu \upphi_{k_4,\epsilon} =  \frac{\epsilon}{2} \hs k_3^2 \lim_{k_4\to 0} (\upvarphi_{k_3}\upvarphi_{k_4})\, .
\eeq
Letting $s^2 U_{34} \to (\epsilon/2)\hs k_3^2$ in the scalar exchange four-point function (\ref{Fdelta3S0}) and taking the limit $k_4 \to 0$, we get
\begin{eBox}
\beq\label{equ:thrptscalar}
B = \frac{\epsilon}{2} \hs k_3^3\hs U_{12} \hat f(u,1) + \text{perms}  \, ,
\eeq
\vskip 4pt
\end{eBox}
where $u=k_3/k_{12}$ for $v \to 1$. 

\vskip 4pt
As a concrete example, let us take the simplest contact term for the $\Delta=2$ solution, namely
\beq
\label{equ:solC0}
\hat f = \hat C_0 = \frac{uv}{u+v}  \ \xrightarrow{\ v  \to 1 \ } \ \frac{u}{u+1} \, .
\eeq 
Substituting this into (\ref{equ:thrptscalar}), we get 
\beq
B =  -\frac{\epsilon}{2}\,k_3^2\left(\frac{k_3}{2}-\frac{k_1^2+k_1 k_2+k_2^2}{k_t}+\frac{k_1k_2k_3}{k_t^2}\right)+\text{perms}\, .\label{Bcontact}
\eeq
Symmetrizing the momenta, we get
\begin{align}
	B\equiv -\frac{\epsilon}{4k_t^2} \left[\sum_n k_n^5+\sum_{n\ne m}(2k_n^4k_m^{\phantom{4}} -3k_n^3k_m^2)+\sum_{n\ne m\ne l}(k_n^3k_m^{\phantom{3}} k_l^{\phantom{3}} -4k_n^2k_m^2k_l^{\phantom{2}})\right] , \label{equ:equil}
\end{align}
which is precisely the famous equilateral bispectrum arising from the bulk interaction $(\partial_\mu \upphi)^4$; cf.~equation~(14) in~\cite{Creminelli:2003iq}.

\paragraph{Massless scalar exchange} 
The case of massless exchange is particularly interesting. A common diagnosis for other light fields during inflation is the appearance of local non-Gaussianity, which is absent in single-field inflation \cite{Creminelli:2004yq}. If this source of non-Gaussianity is generated during inflation,\footnote{This excludes scenarios like the curvaton mechanism~\cite{Enqvist:2001zp,Lyth:2001nq}, where non-Gaussianities are sourced after inflation.} then the relevant shape comes from massless exchange. As we now show, the bispectrum due to the exchange of a massless scalar does leave an imprint in the squeezed limit, although the shape is not the same as the local one, even in the squeezed limit. Substituting $\hat F_{\Delta_\sigma=3}$, given in (\ref{Fmasslessscalar}), into (\ref{equ:thrptscalar}), we obtain 
\beq\label{bimassscex}
B=\frac{\epsilon}{4}\left(k_3^3 \log k_t-\frac{k_3^2(k_1^2+k_1 k_2+k_2^2-k_3^2)}{k_t}\right)+{\rm perms}\, ,
\eeq
where we discarded terms of the form $\sum_n k_n^3 \log k_n$ which are artifacts of imposing conformal invariance of the answer. The logarithm can be regulated by introducing an IR cutoff.  
Note that the pole in $k_t$ in (\ref{bimassscex}) cancels after summing over permutations.
Finally, the soft behavior of the bispectrum is 
\beq
\lim_{k_3\to 0} B= \frac{\epsilon}{2}\hskip 1pt k_1^3\log(2k_1)\, ,
\eeq 
which has a small, logarithmic deviation from the local shape. 
We show in Appendix~\ref{app:D3} that the same three-point function is also given by a local self-interaction of the inflaton, as long as we allow for self-interactions which break shift symmetry. In other words, this shape can be interpreted in two different ways: either as an indication of a new massless scalar state in the spectrum, or as a local breakdown of shift symmetry of the inflaton. The zero mode of the inflaton (or of the extra massless scalar) has a long range effect, and is ultimately responsible for this slow decay of the squeezed limit. 

\paragraph{Graviton exchange} Another interesting example is the bispectrum associated with graviton exchange in the four-point function. This is expected to lead to the three-point function of slow-roll inflation~\cite{Maldacena:2002vr, Kundu:2014gxa}. We will now show that this is indeed the case.

\vskip 4pt
Consider the tensors $T^\upphi_{\mu \nu}$ and $T^\upvarphi_{\mu \nu}$ defined in (\ref{equ:tensors}), but with the perturbed mode function (\ref{equ:PM}) in $T^\upphi_{\mu \nu}$. In the soft limit $k_4 \to 0$, the $00$-component of the tensors are related as 
\beq
\lim_{k_4 \to 0}T^{\upphi,\varepsilon}_{00}  
= \frac{\epsilon}{2}\hskip 1pt k_3^2 \lim_{k_4 \to 0}T^\upvarphi_{00} \, ,
\eeq
where we had to adjust the trace part of $T^\upphi_{\mu \nu}$ to get a precise matching with $T^\upvarphi_{\mu \nu}$.
Letting $s^2 U_{34}^{2,0} \to (\epsilon/2) \hskip 1pt k_3^2$ in the longitudinal part of the four-point function (\ref{equ:Fspin2}), and taking the soft limit, we get 
\begin{eBox}
\beq
B= \frac{\epsilon}{2}\hskip 1pt k_3^3\, P_2(\alpha/s)\, U_{12}^{2,0} \big(\Delta_u(\Delta_u-2)\hat f(u,1)\big)  + \text{perms} \, . \label{equ:3SR}
\eeq
\vskip 1pt
\end{eBox}

\vskip 4pt
\noindent
To determine the bispectrum due to graviton exchange, we must substitute the scalar solution associated with massless exchange, $(\Delta_u-2)\hat f = \hat C(u,v)$. Because the operator $U_{12}$ raises the number of derivatives of the effective interaction, we use the unphysical ``contact term" $\hat C_{-1}$, defined implicitly via $\Delta_u \hat C_{-1}= \hat C_0$.\footnote{The term $\hat C_{-1}$ has a physical interpretation as the four-point function due to the exchange of a conformally coupled scalar. Applying $\Delta_u$ to it collapses its internal propagator and outputs the lowest-derivative contact term~$\hat C_0$.} 
This implies
\begin{align}
\Delta_u(\Delta_u-2)\hat f &= \Delta_u \hat C_{-1} \nonumber  \\
&= \hat C_0= \frac{u v}{(u+v)} \ \xrightarrow{\ v  \to 1 \ } \ \frac{u}{u+1} \, .
\end{align} 
Substituting this into (\ref{equ:3SR}), we obtain 
\begin{align}
	B[\hat f_{-1}] 
=\frac{\epsilon}{8}\bigg(k_t^3-6\sum_{n>m}k_n^{\phantom{2}}k^2_m-4k_1k_2k_3+\frac{8(\sum_{n>m}k_nk_m)^2}{k_t}+\frac{32k_1k_2k_3\sum_{n>m}k_nk_m}{k_t^2}\bigg)\, .
\end{align}
In Section~\ref{sec:Massless}, we saw that the four-point function corresponding to physical graviton exchange involves the combination ${\cal F}_{2}[\hat f_{-1}] - 4{\cal F}_{0}[\hat C_0]$. This suggests that we should subtract the contribution found in~\eqref{Bcontact}, which we denote by $B_{(\partial \upphi)^4}$. We therefore find
\begin{align}
	B_{\rm inf} &\,\equiv\, B[\hat f_{-1}] - 4B_{(\partial \upphi)^4} \nonumber \\ 
	&\,=\,  -\frac{3}{8} \epsilon\left[\sum_{n\ne m}k_n k_m^2+\frac{8}{k_t} \sum_{n>m}k_n^2k_m^2 -3\sum_n k_n^3 \right] . \label{equ:Binf}
\end{align}
Up to a local contribution, this is the classic three-point function of slow-roll inflation~\cite{Maldacena:2002vr}. 
Note that we have computed the bispectrum of inflaton fluctuations, $\delta \phi$. 
To perform the precise comparison with the result in~\cite{Maldacena:2002vr}, the bispectrum must be written in comoving gauge, in which the scalar fluctuation is proportional to the trace of the metric, usually referred as the comoving curvature perturbation $\zeta$. From the boundary perspective, computing correlators of the metric is equivalent to computing correlators of the trace of the stress tensor. Since the stress tensor must satisfy a boundary Ward identity \cite{Pimentel:2013gza, Goldberger:2013rsa, Berezhiani:2013ewa}, we must add an extra piece to the inflationary bispectrum. This extra piece in the three-point function of $\zeta$ is proportional to $B_{\rm loc}\equiv \sum_n k_n^3$, with coefficient determined by the tilt of the scalar power spectrum, $n_s-1$. It is the only non-vanishing piece of the bispectrum in the squeezed limit and, from the bulk perspective, follows from consistency conditions \cite{Maldacena:2002vr, Creminelli:2004yq, Assassi:2012zq} that relate a long-wavelength metric fluctuation to a change of coordinates. This extra contribution to the squeezed bispectrum does not generate observable effects in the late universe \cite{Pajer:2013ana}.

\paragraph{Massive spin-2 exchange} Finally, the result (\ref{equ:3SR}) straightforwardly includes the bispectra associated with the exchange of massive particles.  We will verify this explicitly for massive spin-2 exchange. Again, we will feed in the scalar solution $\hat f_{-1}$, which now satisfies $(\Delta_u+\mu^2+\frac{1}{4})\hat f_{-1}=\hat C_{-1}$.
This implies that
\begin{align}
	\Delta_u(\Delta_u-2)\hat f_{-1} = (\Delta_u-2) \hat f_0\, ,
\end{align}
where $\hat f_0$ is the solution given in \eqref{equ:SeriesSol}. Substituting this into \eqref{equ:3SR}, we get
\begin{align}
	B &= \frac{\epsilon}{2}\hskip 1pt k_3^3\, P_2(\alpha/s)\, U_{12}^{2,0}\Bigg[\sum_{m,n=0}^\infty c_{mn}(2m+n+2)\,u^{2m+n+1}\big(2m+n-1-(2m+n+1)u^2\big) \nn
	&\qquad\qquad\quad -\frac{\pi^{5/2}(\frac{9}{4}+\mu^2)(\frac{1}{4}+\mu^2)(\tanh \pi\mu -1)}{\sinh(2\pi\mu)}\, g_2(u)\Bigg]+\text{perms}\, ,
\end{align}
where we have defined the function 
\begin{align}
	g_2(u) &\equiv \frac{(e^{\pi\mu}-i)^2}{1+2i\mu}\frac{\Gamma(-\frac{1}{2}+i\mu)}{\Gamma(1+i\mu)}\left(\frac{u}{2}\right)^{\frac{1}{2}+i\mu}{}_2 F_1\Bigg[\begin{array}{c} \frac{1}{4}+ \frac{i\mu}{2},\hs \frac{3}{4}+ \frac{i\mu}{2} \\[2pt] 1+i\mu\end{array}\Bigg|\, u^2\Bigg]+c.c.
\end{align}
In the squeezed limit, only the homogeneous solution in one of the channels dominates, and we find
\begin{align}
	\lim_{k_3\to 0} B = \epsilon\hskip 1pt k_1^3\left[ \left(\frac{k_3}{4k_1}\right)^{\frac{3}{2}+i\mu} a_2(\mu) + c.c.\right] P_2(\cos\theta) \, ,\label{massive3pt}
\end{align}
where $\cos\theta=\hat\k_1\cdot\hat\k_3$ and 
\begin{align}
	a_2(\mu) \equiv\, & \frac{\pi(\frac{1}{4}+\mu^2)}{\cosh\pi\mu}\frac{\Gamma(\frac{5}{2}+i\mu)\Gamma(\frac{5}{2}-i\mu)}{128\sqrt\pi} (1+i\sinh\pi\mu)\frac{\frac{9}{2}+i\mu}{\frac{1}{2}+i\mu}\frac{\Gamma(-i\mu)}{\Gamma(\frac{1}{2}-i\mu)}\, .\label{Jfunc}
\end{align}
The result (\ref{massive3pt}) agrees with equation (6.142) in \cite{Arkani-Hamed:2015bza}, with the Legendre polynomial indicating that we are exchanging a massive spin-2 particle.\footnote{An extra factor of $\mu^2+\frac{1}{4}$ in \eqref{Jfunc} compared to (6.144) in \cite{Arkani-Hamed:2015bza} is due to the fact that we have used the solution with a higher-derivative source term as the input function. Again, the difference is  given by a contact term, and the extra prefactor can simply be absorbed in the coupling constant.}

\newpage
\section{Comments on Phenomenology}
\label{equ:pheno}

\begin{figure}[h!]
   \centering
      \includegraphics[scale =0.6]{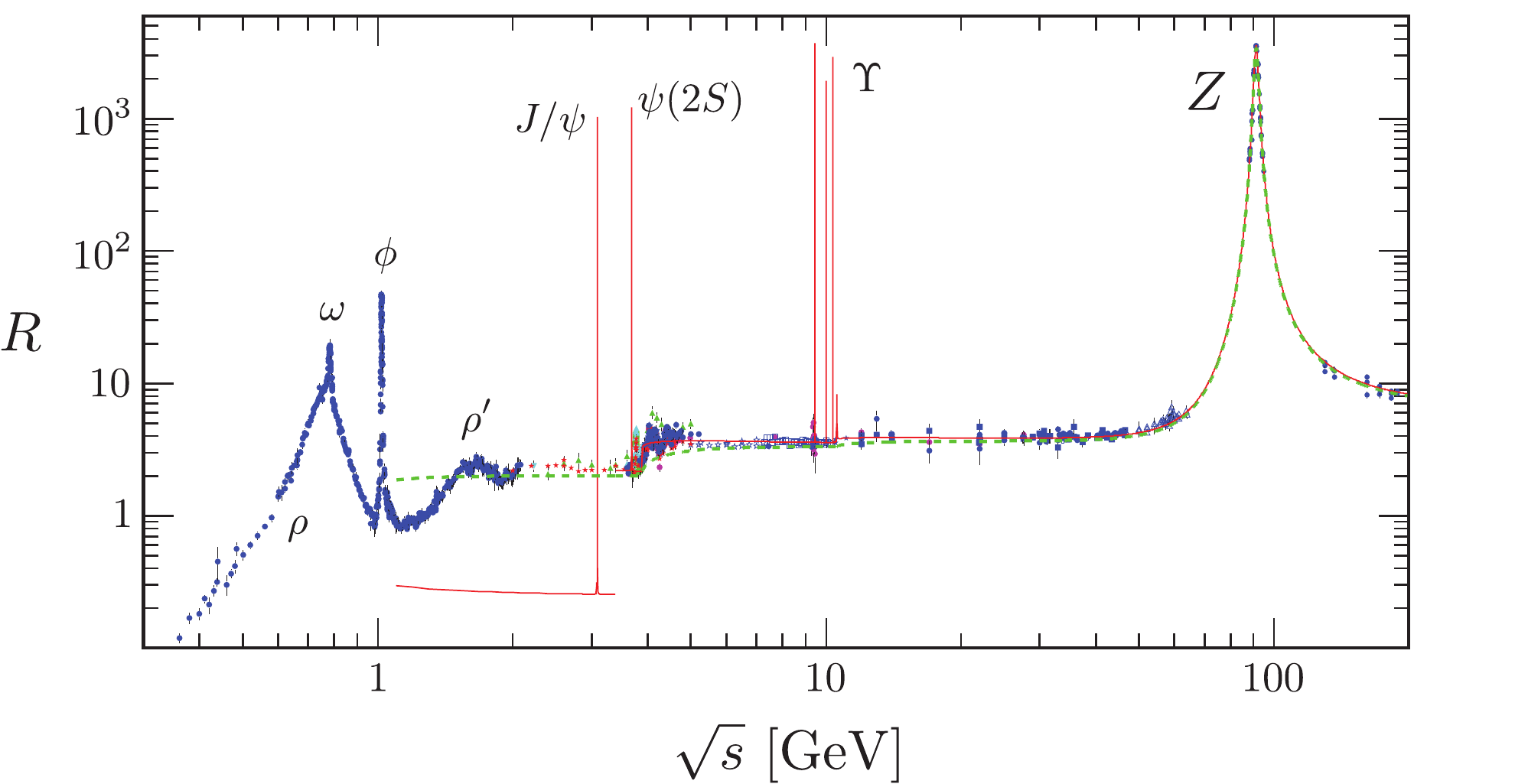}
      \caption{Plot of $R \equiv \sigma(e^+ e^- \to {\rm hadrons})/\sigma(e^+ e^- \to \mu^+ \mu^-)$ as a function of the center-of-mass energy (figure adapted from~\cite{Patrignani:2016xqp}).}
      \label{fig:resonance}
\end{figure}

Figure~\ref{fig:resonance} shows the cross section for $e^+ e^- \to {\rm hadrons}$ as a function of the center-of-mass energy.
The different resonance peaks, such as the famous $Z$ resonance near $100$ GeV, prove the existence of new particles and determines their properties. For example, the position of a peak measures the mass of the particle, while its height and width probe the lifetime of the particle and hence its couplings to lighter degrees of freedom in the Standard Model.
The angular dependence of the decay products puts constraints on the spin of the intermediate particle.
In this section, we will discuss how similar spectroscopic information is encoded in the structure of inflationary correlators.  We will also present a new physically-motivated basis of shapes for inflationary three-point functions with weakly broken conformal symmetry.

\subsection{Cosmological Collider Physics}

\begin{figure}[t!]
   \centering
      \includegraphics[scale =0.4]{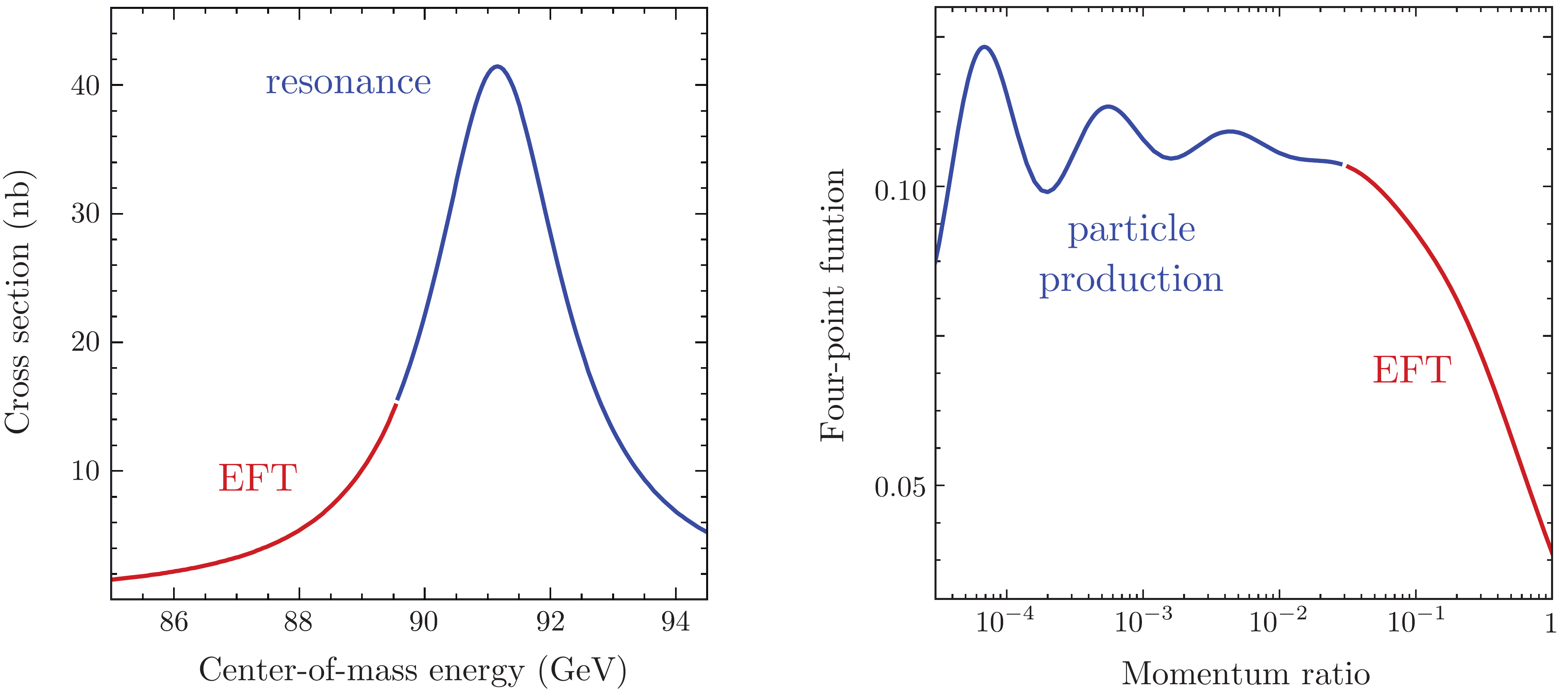} 
      \caption{{\it Left panel:} Shape of the $Z$ resonance as measured by LEP. {\it Right panel:} Example of scalar exchange, $u^{-1} \hat F(u,0.5)$, for external particles with $\Delta=2$ and an internal particle with~$\mu=3$. Note that the four-point function has been rescaled by $u^{-1}$ which visually enhances the effect of the oscillations. In practice, the particle production effect will be harder to observe than the EFT contribution.}
      \label{fig:resonance2}
\end{figure}

The right panel in Figure~\ref{fig:resonance2} displays our solution for the exchange of a massive scalar particle, $\hat F(u,v)$, for fixed $v=0.5$.
We see that the signal in the collapsed limit, $u \to 0$, oscillates with a frequency that is set by the mass of the exchange particle. Measuring these oscillations is the analog of measuring the position of a resonance peak in collider physics. It would prove the existence of new particles and determine their masses.  Going away from the squeezed limit, the particular solution will start to dominate over the homogeneous solution.
This provides a smooth contribution to the four-point function, whose shape will also be determined by the mass of the exchange particle.
This is the analog of going off resonance and measuring the shape of the resonance peak in collider physics.
Measuring both the oscillations and the smooth shape provides an important consistency check for the signal.

\vskip 4pt
In colliders, we begin with low-energy measurements where all interactions are pure contact interactions. For example, at low energies the electroweak theory is approximated by the four-point interaction of Fermi theory. In the latter case, the energy dependence of the interaction hints at a violation of perturbative unitarity at a higher scale.  This suggests the existence of new particles (in the case of the electroweak theory, $W$ bosons) to improve the UV behavior of the effective theory.  Going to higher energies, colliders may start producing these particles as resonances.  Predicting the shape of the resonance is essential for extracting the detailed properties of the new particles.  It also provides the opportunity to identify additional new physics.  For example, any unexplained excess in the cross section may be due to additional particle exchanges.

\vskip 4pt
In cosmology, we first expect to observe signals in the limit of relatively large momenta. This is where the signals are strongest and the observations are most sensitive. Initially, we would see the shape of a pure contact interaction.  With increased sensitivity we may then be able to observe a small deviation from the pure contact shape (see Fig.~\ref{fig:source} in \S\ref{sec:shapes}).\footnote{In practice, it will be hard to reliably extract the precise shape of the smooth part of the signal from large-scale structure observations because late-time nonlinearities produce non-Gaussianities of a similar form. Although the oscillatory part of the signal is smaller, it is more distinctive and cannot be mimicked by late-time effects.} Using the hypothesis of the exchange of a single massive particle to fit the smooth part of the signal would then allow us to predict the amplitude and frequency of the characteristic oscillations in the soft limit. 
Finding consistency between the smooth and oscillating parts of the signal would be essential to establish the underlying physics. 
Any discrepancies between the two components may be a signal of additional new physics.

\subsection{Challenges and Opportunities}
\label{sec:shapes}

In this paper, we have worked under the lamppost of weakly broken conformal symmetry.
This has allowed us to derive particularly clean insights into the analytic structure of inflationary correlators.
However, it also restricts the strength of the couplings between the inflaton and additional massive fields.
This makes the observational challenge to detect these effects enormous.
To achieve larger levels of non-Gaussianity, $f_{\rm NL} > 1$, requires interactions that break the conformal symmetry more strongly, as in models with a reduced sound speed of the inflationary fluctuations~\cite{Cheung:2007st, Senatore:2010wk, Lee:2016vti, Silverstein:2003hf}, or strongly coupled fluctuations, as in some holographic models of dS/CFT~\cite{Afshordi:2016dvb}.
While the main observational signatures of massive particles---oscillations and a distinct angular dependence in the squeezed limit---are preserved~\cite{Lee:2016vti}, the details of the non-Gaussian shapes will be modified.\footnote{It is conceivable that our approach can be generalized to cases with nonlinearly conformal symmetry or interactions constrained by the non-relativistic conformal group (see e.g.~\cite{Son:2008ye}). We will explore this in future work.} Having made this important qualifier, we will nevertheless present a systematic classification of inflationary three-point functions arising from weakly broken conformal symmetry providing a physically-motivated basis of templates in the search for primordial non-Gaussianity.  This may be viewed as an ultimate target for future generations of cosmological observations~\cite{Meerburg:2016zdz, MoradinezhadDizgah:2017szk, MoradinezhadDizgah:2018ssw}.

\vskip 4pt
In Section~\ref{sec:inflation}, we showed that all inflationary three-point functions arising from interactions that only weakly break conformal invariance can be written as follows

\vskip 4pt
\begin{eBox}
\vskip 5pt
\beq
B(k_1,k_2,k_3)  \,=\,\epsilon\, k_3^3 \sum_{S,n} \, P_S(\alpha/s)\, U_{12}^{S,0}\Big(c_{S,n}\hat b_{S,n}(u)+c_n \hat C_n(u,1)\Big)  \,+\, {\rm perms}\, , \label{equ:bispectrum}
\eeq
\vskip 3pt
\end{eBox}

\begin{figure}[t!]
   \centering
      \includegraphics[height =7.5cm]{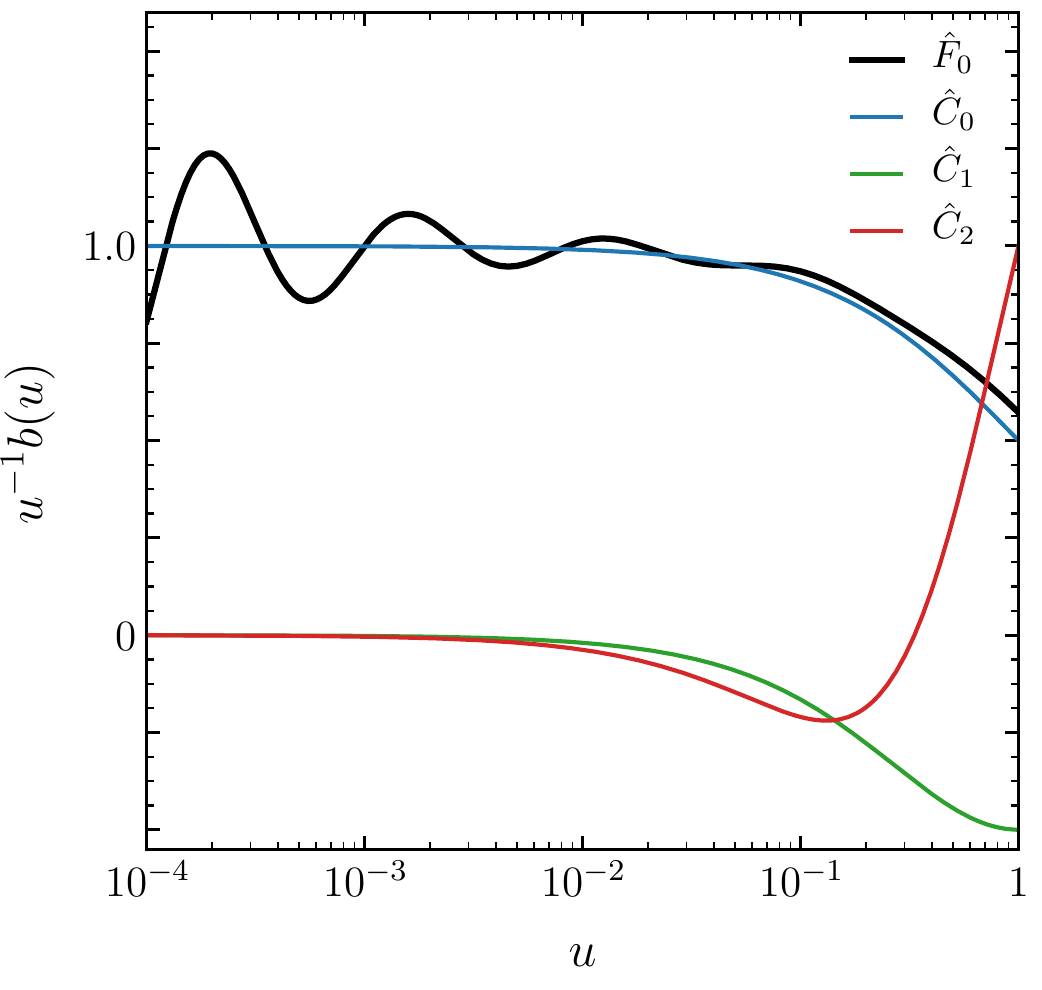}  \hspace{0.2cm}
\includegraphics[height =7.5cm]{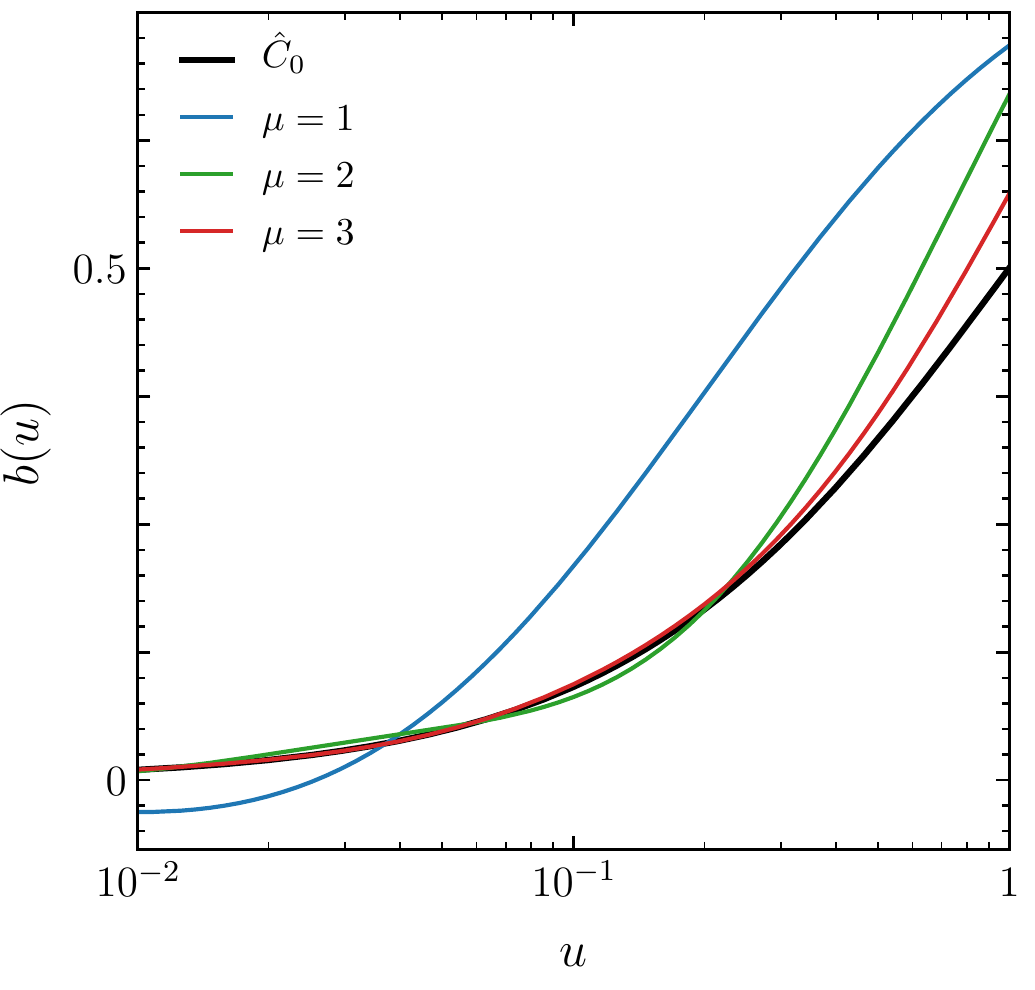} 
      \caption{{\it Left panel:} Comparison of the source functions $b(u)$ as a function of $u$ between the scalar exchange solution $(\mu^2+\frac{1}{4})\hat F_0$ for $\mu=3$ and the contact terms $\hat C_n$ for $n=0,1,2$. {\it Right panel:} Comparison between the leading contact term $\hat C_0$ and $(\mu^2+\frac{1}{4})\hat F_0$ for $\mu =1,2,3$.} 
      \label{fig:source}
\end{figure}

\vskip 4pt
\noindent
where
\beq
\hat b_{S,n}(u) \equiv \prod_{j=1}^{S} \big(\Delta_u-(S-j)(S-j+1)\big)\,\hat F_n(u,1)\, ,
\eeq
with $\hat F_n(u,v)$ being a solution of the de Sitter four-point function of conformally coupled scalars induced by the tree-level exchange of a massive scalar.
Using $[\Delta_u + (\mu^2+\frac{1}{4})] \hat F_n = \hat C_n$ and $\hat C_n = \Delta_u^n \hat C_0$, we can write the source function in terms of the simplest exchange solution $\hat F_0$ and a sum over contact terms:
\beq
\hat b_{S,n}(u) = \lim_{v\to 1} \left[a_{S,n}(\mu)\, \hat F_0 + \sum_{m=0}^{S-1} a_{S,nm}(\mu)\, \hat C_m \right] , \label{equ:bSn}
\eeq
where the coefficients are known functions of $\mu$, but their explicit forms won't be needed.
After summing over permutations, the contact contributions in (\ref{equ:bSn}) can be absorbed into the contact contributions in (\ref{equ:bispectrum}). 
Up to local terms that can be removed by field redefinitions, the inflationary bispectrum can then be written as\hs\footnote{This basis covers all shift-symmetric contact interactions of massless scalar $\upphi$. A violation of the shift symmetry can be interpreted either as adding local terms without derivatives in the inflaton Lagrangian, or as an indication of an extra massless scalar in the spectrum.}
\begin{eBox}
\vskip 5pt
\beq
B(k_1,k_2,k_3) \,=\, \epsilon\, k_3^3  \sum_{S} \,P_S(\alpha/s)\, U_{12}^{S,0} \bigg(d_S \hat F_0(u,1) \, +\, \sum_{n} e_n \hat C_n(u,1)\bigg)  \, +\, {\rm perms}\, ,
\label{equ:Bfinal}
\eeq
\vskip 4pt
\end{eBox}
where $d_S$ and $e_n$ are constants. Rather remarkably, the inflationary bispectrum arising from arbitrary spin-exchange is completely described by the soft limit of the simplest scalar-exchange four-point function of conformally coupled scalars, $\hat F_0$, and a series of contact terms, $\hat C_n$.  This physically-motivated basis of shape functions is illustrated in the left panel of Fig.~\ref{fig:source}.
We see that the higher-order contact terms $\hat C_{n>0}$ are suppressed away from the equilateral limit $u=1$, so that a parameterization in terms of just $\hat F_0$ and $\hat C_0$ captures most of the bispectrum shape.
In the right panel of Fig.~\ref{fig:source}, we display the difference between $\hat F_0$ and $\hat C_0$ as a function of $\mu$, and hence the mass of the exchange particle. As we mentioned above, this deviation from the pure contact shape is a measure of the mass of the exchange particle.

\newpage
\section{Conclusions and Outlook}
\label{sec:conclusions}

In this paper, we have presented a systematic study of
inflationary correlation functions, following a perspective
familiar from the study of scattering amplitudes. We used symmetries and
singularities to {\it uniquely fix} the correlators, rather than {\it computing} them from cosmological time evolution. 
Remarkably, the space of correlators is completely characterized by one fundamental object, the four-point function of conformally coupled scalars arising from the exchange of a massive scalar. 
Solutions with massless external and spinning internal fields are obtained simply through the application of suitable weight-shifting and spin-raising operators.

\vskip 4pt
Our findings pave the way towards a ``cosmological bootstrap", in which cosmological correlation functions are determined from consistency conditions, without recourse to a Lagrangian.  It is amusing to see the central philosophical distinction between the ``bootstrap" and
``Lagrangian" approaches manifest itself in this context. In the Lagrangian method, the correlator is the result of ``just a computation", albeit in many cases an incredibly tedious and un-illuminating one. The bootstrap method is different---instead of asking ``what is the {\it answer} for the correlator?" it asks, ``what is the {\it question} to which the correlator is the unique answer?" This second philosophy has a number of advantages. Most pragmatically,  by construction it is easy to check whether the answer is right or wrong, and hidden simplicities and structures in the final results are more transparent.  More deeply, the bootstrap approach is better suited to the ambitious goal of understanding ``time without time". Indeed, although our formalism never makes explicit reference to time evolution, the effects of the time-dependent background, such as the spontaneous production of massive particles, emerge from our solutions where they are encoded in the momentum dependence of the boundary correlators. 

\vskip 4pt
We have clearly only scratched the surface of a large and fascinating subject, which holds great promise for future developments. 
Indeed the explosion of progress in our understanding of scattering amplitudes  virtually guarantees this, since, as we have stressed repeatedly, the scattering amplitudes are contained
within the correlators. Many of the wonderful insights into the physics of scattering amplitudes must therefore have counterparts in the physics of cosmological correlators.

\vskip 4pt
We close by pointing out a few obvious avenues for future exploration: 
\begin{itemize}
\item To begin with, there are a number of points where we feel our analysis can be streamlined and improved. For example, our discussion of the polarization sums in cosmology, as well as the construction of spin-raising operators, was largely carried out by inspired guesswork; clearly the de Sitter symmetries should help to organize these computations in a more powerful way than we have been exploiting. A similar comment applies to the weight-shifting operators mapping between correlators of conformally coupled and massless scalars.

\item It would also be nice to apply the same philosophy to the computation
of correlators with external spin, especially for gravitons. Standard
Lagrangian computations for these correlators are notoriously
complicated, but we expect that the bootstrap method should yield
transparent analogs of the results we have seen here for scalars. 
In flat space, the bootstrap approach has been very powerful in determining the allowed three- and four-point graviton amplitudes (see e.g.~\cite{Benincasa:2007xk, Cheung:2017pzi}). Similarly, in de Sitter space, the graviton three-point function is highly constrained by symmetry~\cite{Maldacena:2011nz}. 

\item Moving beyond tree level, the analytic structure of one-loop
scattering amplitudes is well-understood; for instance, in four
dimensions, the final results are given as a sum over box, triangle
and bubble integrals, together with rational terms. This must have an
analog for cosmological correlators. A precise understanding of this
will allow a huge extension of ``cosmological collider physics" to encompass further plausible and potentially realistic scenarios~\cite{Chen:2016hrz, Chen:2016uwp, Chen:2016nrs}. For instance, if the Higgs field has a coupling to curvature, it can pick up a mass of order the Hubble scale during inflation, and naturally couple to the inflaton in pairs, contributing to non-Gaussianities at loop level. The same could be true for scalar partners in supersymmetric theories that might exist anywhere up to the inflationary scale.

\item More conceptually, the constraints of causality on scattering are reflected in the polynomial boundedness of the amplitudes in the Regge limit, where $t$ is held fixed and $s \to \infty$; it is important to determine the precise analog of this statement in cosmology. For amplitudes, this has led to powerful dispersion relations, which in turn result in nontrivial positivity constraints on the coefficients of higher-dimension operators in the low-energy effective field theory expansion~\cite{Adams:2006sv} (see also~\cite{EFT-hydron, deRham:2017avq, deRham:2017xox, deRham:2017imi, deRham:2017zjm, deRham:2018qqo, Bonifacio:2018vzv}). We expect entirely analogous results to hold for cosmological correlators. In particular, we expect the EFT expansion of the correlators to inherit interesting positivity properties from the positivity of the scattering amplitudes that live inside them (see e.g.~\cite{Baumann:2015nta}).

\item Some very early steps in identifying new
combinatoric/geometric structures underlying cosmological correlators,
playing a roughly analogous role to amplituhedra and generalized
associahedra for scattering amplitudes, have recently appeared in~\cite{Arkani-Hamed:2017fdk} with the
discovery of ``cosmological polytopes", at least for the case of
conformally coupled scalars with polynomial interactions. Using the
results of our paper, it would be interesting to extend
these ideas to more realistic theories of particles with general masses
and spins.

\item Finally, we have worked under the lamppost of weakly broken conformal symmetry.  This made our analysis particularly clean and precise, but also limited the strength of the couplings between the inflaton and the additional massive particles.  Obtaining larger levels of non-Gaussianity, requires interactions that break the conformal symmetry more strongly, such as in theories with a reduced sound speed for the inflationary fluctuations~\cite{Cheung:2007st, Senatore:2010wk, Lee:2016vti, Silverstein:2003hf, Delacretaz:2016nhw, Bordin:2018pca}. It would be interesting to extend our bootstrap methods to such examples where they would have immediate observational relevance.

\end{itemize}

\vspace{0.5cm}
\paragraph{Acknowledgements} 
We are grateful to Dionysios Anninos, Matteo Biagetti, Jan de Boer, Horng Sheng Chia, Paolo Creminelli, Lance Dixon, Carlos Duaso Pueyo, Garrett Goon, Daniel Green, Sadra Jazayeri, Eiichiro Komatsu, Juan Maldacena, Scott Melville, Mehrdad Mirbabayi, Enrico Pajer, Jo\~ao Penedones, Jan Pieter van der Schaar, Eva Silverstein, John Stout, and Yi Wang for helpful discussions. NA-H and GP would especially like to thank Juan Maldacena for countless stimulating discussions over many years on all aspects of cosmological correlators. DB, HL and GP~thank the Institute for Advanced Study for hospitality while parts of this work was being performed. NA-H thanks Rajesh Gopakumar and the organizers and participants of the ICTS 2018 winter school for stimulating cosmological discussions. DB thanks the Institute Henri Poincar\'e for hospitality during the completion of this work, and the organizers of the SUSY and COSMO conferences for the opportunity to present preliminary versions of this work. GP thanks the organizers of the Solvay-PI-APC workshop ``Cosmological Frontiers in Fundamental Physics" and of the XIV Modave Summer School in Mathematical Physics for the opportunity to present this work. NA-H and DB thank the
Institute D'Etudes Scientifiques de Cargese for hospitality. 
NA-H~is supported by DOE grant DE-SC0009988. DB~is supported by a Vidi grant of the Netherlands Organisation for Scientific Research~(NWO) that is funded by the Dutch Ministry of Education, Culture and Science~(OCW). HL is supported by NSF grant AST-1813694 and DOE grant DE-SC0019018. GP is supported by the European Union's Horizon 2020 research and innovation programme under the Marie-Sk\l{}odowska Curie grant agreement number 751778. The work of DB and GP is part of the Delta-ITP consortium.

\newpage
\appendix

\section{Conformal Symmetry}\label{app:ward}

In this appendix, we derive the conformal symmetry constraints on three- and four-point functions in de Sitter space. We first review the map between the semiclassical wavefunction of the universe and correlation functions, showing that the four-point functions contain a connected part and a disconnected part.
The latter is given by a product of three-point functions and  we present the explicit forms of the three-point functions that are used in this work. Finally, we show explicitly that the combined action of dilatation and special conformal transformations on the scalar four-point function leads to the conformal invariance constraint~(\ref{WI_SCT2}).

\subsection[Wavefunction $\to$ Correlators]{Wavefunction $\boldsymbol{\to}$ Correlators}
\label{sec:Psi}
The late-time wavefunction of the universe can be obtained by evaluating the on-shell action with appropriate boundary conditions, namely $\upphi(\k,\eta)\to \phi(\k)$ at late times and $\upphi \sim e^{-i k \eta}$ in the far past, selecting the Bunch-Davies vacuum.  We write the wavefunction for the fields $\phi$ and $\sigma$ as\footnote{We are dropping a phase factor that diverges at late times but does not affect correlation functions. We assume that $\sigma$ is a generic spinning field, but sometimes suppress its indices for brevity.}
\begin{align}
	\Psi[\phi,\sigma] = \exp\Bigg[\sum_{n+m\ge 2}^\infty \frac{1}{m!\hskip 1pt n!}\int\frac{\d^3 k_1\cdots \d^3 k_{m+n}}{(2\pi)^{3(m+n)}}\,\psi_{m,n}(\{\k_l\})\prod_{i=1}^{m}\prod_{j=1}^{n} \phi(\k_i)\sigma(\k_{m+j})   \Bigg]\, , \label{equ:Psi}
\end{align}
where $\{\k_l\}\equiv \{\k_1,\cdots\hskip -1pt,\k_{m+n}\}$ denotes the set of all momenta. 
Expectation values of the bulk fields are then computed as
\begin{align}
	\langle\phi_1\cdots \phi_m \sigma_{m+1}\cdots \sigma_{m+n}\rangle = \frac{1}{N}\int D\phi D\sigma\,  \phi_1\cdots \phi_m \sigma_{m+1}\cdots \sigma_{m+n}\,|\Psi[\phi,\sigma]|^2\, ,
\end{align}
where $\{\phi_n, \sigma_m\} \equiv \{\phi(\k_n), \sigma(\k_m)\}$ and $N$ is an appropriate normalization. Translation invariance implies that the wavefunction coefficients take the form
\begin{align}
	\psi_{m,n} = \langle O_1\cdots O_m \Sigma_{m+1}\cdots \Sigma_{m+n}\rangle'\times (2\pi)^3\delta^{3}(\k_1+\cdots +\k_{m+n})\, ,
\end{align}
where the prime on the expectation value indicates that the momentum-conserving delta function has been stripped. The operators $O$ and $\Sigma$ can be interpreted as primary operators in a dual conformal field theory, although this won't be needed in this work. 
The two-point function of $\Sigma$ can be written as 
\beq
	\langle\Sigma_{i_1\cdots i_S}(\k)\Sigma^{j_1\cdots j_S}(-\k)\rangle'= \Pi_{i_1\cdots i_S}^{j_1\cdots j_S}(\hat\k)\,\langle\Sigma_S(\k)\Sigma_S(-\k)\rangle'\, ,
\eeq
where $\Pi_{i_1\cdots i_S}^{j_1\cdots j_S}$ is a symmetric, traceless tensor structure.\footnote{Explicitly, it is given by~\cite{Arkani-Hamed:2015bza,Lee:2016vti}
\begin{align}
	\Pi_{i_1\cdots i_S,j_1\cdots j_S}(\hat\k) = \sum_{m=0}^S I_2^+(S,m)\hs\epsilon_{i_1\cdots i_S}^m(\hat\k)\epsilon_{j_1\cdots j_S}^{-m}(\hat\k)\, ,
\end{align}with $I_2^+(S,m)$ defined in \eqref{I2}. This captures the non-local part of the two-point function. Contracting with null momenta $\z$ and $\tilde \z$, the two-point function becomes
\begin{align}
	\langle\Sigma^{(S)}(\k;\z)\Sigma^{(S)}(-\k;\tilde\z)\rangle' &= c_\Sigma\hs k^{2\Delta-3} \big[(\z\cdot\hat\k)(\tilde\z\cdot\hat\k)\big]^S P_S^{(\Delta-S-3/2,-1/2)}\bigg(1-\frac{\z\cdot \tilde\z}{(\z\cdot\hat\k)(\tilde\z\cdot\hat\k)} \bigg)+c.c.\nn
	&= c_\Sigma\hs k^{2\Delta-3} P_S^{(\Delta-S-3/2,-1/2)}(-\cos\chi)+c.c.\, ,
\end{align}
where $\Sigma^{(S)}(\k;\z) \equiv \z^{i_1}\cdots \z^{i_S}\Sigma_{i_1\cdots i_S}(\k_3)$ and $P_S^{(a,b)}$ is the Jacobi polynomial. We used $\langle\Sigma_S(\k)\Sigma_S(-\k)\rangle'=c_\Sigma\hs k^{2\Delta-3}$ and set $\z=(\cos\psi,\sin\psi,i)$, $\tilde\z=(\cos\psi',\sin\psi',-i)$, $\k=(0,0,k)$ in the second line with $\chi\equiv \psi-\psi'$.}
 The two-point functions of $\phi$ and $\sigma$ are then given by 
\begin{align}
	\langle \phi(\k)\phi(-\k)\rangle' &= -\frac{1}{2\hskip1pt {\rm Re} \langle O(\k)O(-\k)\rangle'}\, , \\
	 \langle \sigma_{i_1\cdots i_S}(\k)\sigma^{j_1\cdots j_S}(-\k)\rangle' &= -\frac{\Pi_{i_1\cdots i_S}^{j_1\cdots j_S}(\hat\k)}{2\hskip 1pt{\rm Re} \langle \Sigma_S(\k)\Sigma_S(-\k)\rangle'} \label{Sigma2pt}\, ,
\end{align}
with
\begin{align}
	\langle O(\k)O(-\k)\rangle' &= c_O\, k^{\Delta_O-3}\, ,\\
	\langle \Sigma_S(\k)\Sigma_S(-\k)\rangle' &= c_\Sigma\, k^{\Delta_\Sigma-3}\, .
\end{align}
The four-point function of $\phi$ is
\begin{align}
	\langle \phi(\k_1)\phi(\k_2)\phi(\k_3)\phi(\k_4)\rangle'  = \frac{\langle O^4\rangle_{\rm c}'+\langle O^4\rangle_{\rm d}'}{\prod_{n=1}^4 2\,{\rm Re}\langle O(\k_n)O(-\k_n)\rangle'}\, ,
\end{align} 
where the connected and disconnected contributions are 
\begin{align}
	\langle O^4\rangle_{\rm c}' &\equiv 2\hskip 1pt {\rm Re}\langle O(\k_1)O(\k_2)O(\k_3)O(\k_4)\rangle'\label{connected}\\[5pt]
	\langle O^4\rangle_{\rm d}' &\equiv\frac{2\hskip 1pt{\rm Re}\langle O(\k_1)O(\k_2)\Sigma^{i_1\cdots i_S}(-\s)\rangle'\,\Pi_{i_1\cdots i_s}^{j_1\cdots j_S}(\hat\s)\, {\rm Re}\langle \Sigma_{j_1\cdots j_S}(\s) O(\k_3)O(\k_4)\rangle'}{{\rm Re}\langle\Sigma_S(\s)\Sigma_S(-\s)\rangle'} + \text{perms}\, .\label{disconnected}
\end{align}
We see that the disconnected part is given by the product of two three-point functions $\langle OO\Sigma\rangle$. When $\sigma$ is a generic massive particle, both the connected and disconnected parts are conformally invariant, and can be obtained by solving~(\ref{WI_SCT2}). The normalization of the connected part is fixed by the coupling constant, while the disconnected part is associated with the homogeneous solutions that we are free to add. The four-point function is then fully determined by fixing the relative coefficient between the connected and disconnected parts to be ``1''. This provides us with a purely boundary way of fixing the correlation function. 

\subsection{Three-Point Functions}
\label{app:3pt}

In this section, we collect the conformal invariance conditions for different types of three-point functions.  We will focus on the three-point functions that are most relevant for this work.

\paragraph{Three scalars}
The three-point function of generic scalar operators depends only on the magnitudes of the three momenta. Using the scaling symmetry, we can write this correlator as a function of two variables 
\begin{align}
	\langle O_1(\k_1)O_2(\k_2)O_3(\k_3)\rangle'  = k_3^{\Delta_t-6} \hat G(p,q)\, , \qquad p \equiv \frac{k_1+k_2}{k_3}\, , \quad q \equiv \frac{k_1-k_2}{k_3}\, ,
\end{align}
where $\Delta_t=\sum \Delta_n$. Writing the generator of special conformal transformations as $\K_n^i \equiv k_n^i K_n$, we find that the only nontrivial way to satisfy the condition $\sum {\cal K}^i_n(k_3^{\Delta_t-6}\hat G)=0$ is by demanding 
\beq
(K_n-K_m)(k_3^{\Delta_t-6}\hat G)=0\, ,
\eeq 
for $n,m\in\{1,2,3\}$. In terms of the variables $p$ and $q$, the $12$-equation becomes
\beq
	\Big[(p^2-q^2)\partial_{pq}+(\Delta_{12}^+-4)(q\partial_p-q\partial_q)-\Delta_{12}^-(p\partial_p-q\partial_q)\Big]\,\hat G(p,q) =0\, ,
\eeq
where $\Delta_{12}^\pm \equiv \Delta_1\pm\Delta_2$. Combining the $13$- and $23$-equations, we find
\begin{align}
	&\bigg\{(p^2-1)\partial_p^2+2pq\partial_{pq}+(q^2-1)\partial_q^2+(\Delta_t-6)(\Delta_{12}^+-\Delta_3-3)\nn
	&-2\bigg[\bigg((\Delta_{12}^+-5)p+\frac{\Delta_1-2}{p+q}+\frac{\Delta_2-2}{p-q}\bigg)\partial_p\ +\ p\leftrightarrow q\bigg]\bigg\}\, \hat G(p,q)\,=\,0\, .
\end{align}
Solutions to these equations for the special cases of interest $\Delta_1=\Delta_2\in\{2,3\}$ and $\Delta_3\notin\{2,3\}$ were obtained in~\cite{Arkani-Hamed:2015bza}. In terms of the variables $u=p^{-1}$ and $\alpha=qs$, with $s=k_3$, the function $\hat G$ satisfies
\begin{align}
\Delta_1=\Delta_2=2: \qquad	\bigg[\Delta_{u}+\bigg(\mu^2 + \frac{1}{4}\bigg)\bigg]\hat G &= 0\, , \\
\Delta_1=\Delta_2=3: \qquad	\bigg[\tilde\Delta_{u}+\bigg(\mu^2 + \frac{9}{4}\bigg)\bigg]\tilde G &= 0\, ,\quad\text{with}\quad \hat G = O_{12}\, \tilde G\, ,
\end{align} 
where $\tilde\Delta_{u}\equiv (1-u^2)u^2\partial_u^2 + 4(1-u^2)u\partial_u$. The solutions to these equations take the form of a hypergeometric function. In the limit $\Delta_3\to \{2,3\}$, the hypergeometric solutions degenerate into trivial local terms (either a constant or $k_1^3+k_2^3+k_3^3$), and the physical correlators satisfy 
{\it anomalous} conformal Ward identities, allowing for a local violation of the dilatation constraint. The solutions for these special cases $\Delta_3\in\{2,3\}$ typically contain logarithms, which arise from renormalization of the correlators, and are discussed in great detail in \cite{Bzowski:2015pba, Bzowski:2015yxv}.

\paragraph{Two scalars and one general tensor}
The three-point function of two identical scalars and one general spinning operator was derived in \cite{Arkani-Hamed:2015bza} (see also \cite{Isono:2018rrb}). The most general ansatz for this correlator is
\begin{align}
	\langle O(\k_1) O(\k_2) \Sigma^{(S)}(\k_3;\z)\rangle'  = k_3^{\Delta_t-6-S}\sum_{m=0}^S\gamma^m\delta^{S-m}\hat a_m(p,q)\, ,
\end{align}
where we have defined 
\begin{align}
	\Sigma^{(S)}(\k_3;\z) \equiv \z^{i_1}\cdots \z^{i_S}\Sigma_{i_1\cdots i_S}(\k_3)\, ,  \quad \gamma \equiv \z\cdot(\k_1-\k_2)\, , \quad \delta \equiv \z\cdot(\k_1+\k_2)\, ,
\end{align}
with $\z^{i}$ being a null vector, so that the trace of $\Sigma_{i_1\cdots i_S}$ is automatically projected out. The special conformal generator becomes
\begin{align}
	b^i{\cal K}^i = 2(3-\Delta)b^i\partial_{k^i}-b^ik^i\partial_{k_j}\partial_{k_j}+2\hs k_i\partial_{k_i}b^j\partial_{k^j}+2\hs (z^i\partial_{k^i} b^j\partial_{z^j}-b^iz^i \partial_{k^j}\partial_{z^j})\, ,
\end{align}
and the coefficients $a_m$ obey the following recursive relation:  
\begin{align}
	\hat a_{m-1} = -\frac{m\big\{ [q(1-p^2)\partial_p+p(1-q^2)\partial_q+(2\Delta_O+\Delta_\Sigma-6-S)pq]\,\hat a_m+(m+1)a_{m+1}\big\} }{(S-m+1)(\Delta_\Sigma-2+m)}\, . \label{recurs}
\end{align}
It can be shown that the highest-helicity components satisfy 
\begin{align}
	\left[\Delta_{S,u}+\bigg(\mu^2+\frac{(2S+1)^2}{4}\bigg)\right]\hat a_S &= 0\, , \\
	\left[\tilde\Delta_{S,u}+\bigg(\mu^2+\frac{(3-2S)^2}{4}\bigg)\right]\tilde a_S &= 0\, ,\quad\text{with}\quad \hat a_S = O_{12}\, \tilde a_S\, ,
\end{align}
for $\Delta=2$ and $\Delta=3$, respectively, with $\Delta_{S,u}$ and $\tilde \Delta_{S,u}$ defined in \eqref{equ:DS1} and \eqref{equ:DS2}. This allows us to determine all lower helicity components via~\eqref{recurs}.

\paragraph{Two scalars and one conserved tensor} Of particular interest for inflation are correlation functions involving the stress tensor, which are dual to correlators involving tensor metric perturbations.  In particular, the four-point function of slow-roll inflation contains a contribution from graviton exchange. Its disconnected part includes factors of $\langle OOT_{ij}\rangle$, which is not conformally invariant, but is constrained by Ward identities.

\vskip 4pt
Momentum-space Ward identities involving stress tensors and conserved currents were derived in \cite{Bzowski:2013sza} (see also \cite{Bzowski:2017poo, Bzowski:2018fql, Coriano:2013jba, Coriano:2018bbe, Coriano:2018bsy}). For example, the transverse and traceless parts of $\langle OOT_{ij}\rangle$ satisfy
\begin{align}
	k_3^j\langle O(\k_1)O(\k_2)T_{ij}(\k_3)\rangle' &= k_{1}^i\langle O(\k_1)O(-\k_1)\rangle'+k_{2}^i\langle O(\k_2)O(-\k_2)\rangle'\, ,\\
	\langle O(\k_1)O(\k_2)T_{ii}(\k_3)\rangle' &= -\Delta_O \big(\langle O(\k_2)O(-\k_1)\rangle'+\langle O(\k_2)O(-\k_2)\rangle' \big) \, .
\end{align}
 This allows us to write 
\begin{align}
	\langle O(\k_1)O(\k_2)T_{ij}(\k_3)\rangle' &= \langle O(\k_1)O(\k_2)t_{ij}(\k_3)\rangle' +\text{local}\, ,
	\end{align}
where 
$t_{ij}$ is the transverse-traceless part and ``local'' denotes terms proportional to the two-point function $\langle OO\rangle$. The transverse-traceless part is given by 
\begin{align}
	\langle O(\k_1)O(\k_2)t_{ij}(\k_3)\rangle' = -2c_OB_{OOT}(k_1,k_2,k_3)\hs  k_1^lk_2^mP_{ij,lm}(\hat\k_3)  \, ,
\end{align}
where $c_O$ is the normalization of $\langle OO\rangle$. The tensor structure $P_{ij,lm}$ follows from kinematics, while $B_{OOT}$ is fixed by conformal symmetry. For $\Delta_2=\Delta_3=3$, we have~\cite{Bzowski:2013sza, Mata:2012bx} 
\begin{align}
	B_{OOT}(k_1,k_2,k_3) = k_t-\frac{\sum_{n>m}k_n k_m}{k_t}-\frac{k_1k_2k_3}{k_t^2}\, , \label{equ:BOOT}
\end{align}
which we used in Section~\ref{sec:Massless} to determine the disconnected contribution to graviton exchange. 

\subsection{Four-Point Functions}

In this section, we derive the conformal symmetry constraint~(\ref{WI_SCT2}). We begin by collecting some basic formulas. Differentiating a momentum vector gives
\begin{align}
	\frac{\partial k_n^i}{\partial k_n} = \frac{\partial k_n}{\partial k_n^i}= \hat k_n^i\, , \quad  \frac{\partial  k_n^i}{\partial k_n^j}= \delta^{ij}\, , \quad  \frac{\partial \hat k_n^i}{\partial k_n^j}= \frac{\delta^{ij}-\hat k_n^i\hat k_n^j}{k_n}\, ,\quad \frac{\partial \hat k_n^i}{\partial k_n}=0\, .
\end{align}
The derivatives of the $s,t$ variables are
\beq
\begin{aligned}
	\frac{\partial s}{\partial k_1^i}&=\frac{\partial s}{\partial k_2^i} = \frac{k_{12}^i}{s}\, , \quad \frac{\partial s}{\partial k_3^i}=\frac{\partial s}{\partial k_4^i} = 0\, ,\\
	 \frac{\partial t}{\partial k_2^i}&=\frac{\partial t}{\partial k_3^i}= \frac{k_{23}^i}{t}\, , \quad \frac{\partial t}{\partial k_1^i}=\frac{\partial t}{\partial k_4^i} =0 \, ,
\end{aligned}
\eeq
where $k_{nm}^{i} \equiv k_n^i + k_m^i$.
Using the chain rule, the first derivatives with respect to the momentum vectors can be replaced by
\beq
\begin{aligned}
	\red{\frac{\partial}{\partial k_1^i}} &= \frac{k_1^i}{k_1}\frac{\partial}{\partial k_1}+\frac{k_{12}^i}{s}\frac{\partial}{\partial s}\, ,\quad \green{\frac{\partial}{\partial k_3^i} }=\frac{k_3^i}{k_3}\frac{\partial}{\partial k_3}+\frac{k_{23}^i}{t}\frac{\partial}{\partial t}\, ,\\[4pt]
	\orange{\frac{\partial}{\partial k_2^i} }&=\frac{k_2^i}{k_2}\frac{\partial}{\partial k_2}+\frac{k_{12}^i}{s}\frac{\partial}{\partial s}+\frac{k_{23}^i}{t}\frac{\partial}{\partial t}\, ,\quad \blue{\frac{\partial}{\partial k_4^i}} =\frac{k_4^i}{k_4}\frac{\partial}{\partial k_4}\, .
\end{aligned}
\eeq
Moreover, the second derivatives are 
\beq
\begin{aligned}
        \red{\frac{\partial^2}{\partial k_1^i\partial k_1^j} }
	&=\hat k_1^i\hat k_1^j\frac{\partial^2}{\partial k_1^2} +\frac{\delta^{ij}-\hat k_1^i \hat k_1^j}{k_1}\frac{\partial}{\partial k_1}+\frac{k_1^ik_{12}^j+k_{12}^ik_1^j}{s k_1}\frac{\partial^2}{\partial s\partial k_1} +\frac{\delta^{ij}-\hat s^{ij}}{s}\frac{\partial}{\partial s} + \hat s^{ij}\frac{\partial^2}{\partial s^2}\, ,\\[4pt]
	\orange{\frac{\partial^2}{\partial k_2^i \partial k_2^j}} &=\hat k_2^i\hat k_2^j\frac{\partial^2}{\partial k_2^2}+\frac{\delta^{ij}-\hat k_2^i\hat k_2^j}{k_2}\frac{\partial}{\partial k_2}+\frac{k_{23}^ik_{12}^j+k_{12}^ik_{23}^j}{st}\frac{\partial^2}{\partial s\partial t}\\
	&\hskip 16pt +\frac{\delta^{ij}-\hat s^{ij}}{s}\frac{\partial}{\partial s}+\frac{k_2^ik_{12}^j+k_{12}^ik_2^j}{s k_2}\frac{\partial^2}{\partial s\partial k_2}+\hat s^{ij}\frac{\partial^2}{\partial s^2}\\
	&\hskip 16pt +\frac{\delta^{ij}-\hat t^{ij}}{t}\frac{\partial}{\partial t}+\frac{k_2^ik_{23}^j+k_{23}^ik_2^j}{tk_2}\frac{\partial^2}{\partial t\partial k_2}+\hat t^{ij}\frac{\partial^2}{\partial t^2}\, ,\\[4pt]
	\green{\frac{\partial^2}{\partial k_3^i \partial k_3^j}} &=\hat k_3^i\hat k_3^j\frac{\partial^2}{\partial k_3^2} +\frac{\delta^{ij}-\hat k_3^i \hat k_3^j}{k_3}\frac{\partial}{\partial k_3}+\frac{k_3^ik_{23}^j+k_{23}^ik_3^j}{t k_3}\frac{\partial^2}{\partial t\partial k_3} +\frac{\delta^{ij}-\hat t^{ij}}{t}\frac{\partial}{\partial t} + \hat t^{ij}\frac{\partial^2}{\partial t^2}\, ,\\[4pt]
	\blue{\frac{\partial^2}{\partial k_4^i \partial k_4^j} }&=\hat k_4^i\hat k_4^j\frac{\partial^2}{\partial k_4^2} +\frac{\delta^{ij}-\hat k_4^i \hat k_4^j}{k_4}\frac{\partial}{\partial k_4}\, ,
\end{aligned}
\eeq
where we have defined 
\begin{align}
	\hat s^{ij} \equiv \frac{(k_1^i+k_2^i)(k_1^j+k_2^j)}{s^2}\, ,\quad \hat t^{ij}\equiv \frac{(k_2^i+k_3^i)(k_2^j+k_3^j)}{t^2}\, .
\end{align}
Below, we write down the constraint equations arising from dilatations and SCTs.

\paragraph{Dilatation}
Expanding \eqref{WI_D} gives
\begin{align}
	0&= \left[9-\sum_{n=1}^4 \left(\Delta_n- k_n^i\frac{\partial}{\partial k_n^i}\right)\right]F\nn	&=(9-\Delta_t)F +k_1\frac{\partial F}{\partial k_1} + k_2\frac{\partial F}{\partial k_2}+k_3\frac{\partial F}{\partial k_3}+k_4\frac{\partial F}{\partial k_4} + s\frac{\partial F}{\partial s}+t\frac{\partial F}{\partial t}\, .
\end{align}
where $\Delta_t\equiv\sum_{n=1}^4\Delta_n$ denotes the total conformal weight. Differentiating this with respect to $s$ and $t$ gives
\begin{align}
	-\frac{\partial^2 F}{\partial s^2}&=\frac{10-\Delta_t}{s}\frac{\partial F}{\partial s}+\frac{k_1}{s}\frac{\partial^2 F}{\partial s\partial k_1} + \frac{k_2}{s}\frac{\partial^2 F}{\partial s\partial k_2}+\frac{k_3}{s}\frac{\partial^2 F}{\partial s\partial k_3}+\frac{k_4}{s}\frac{\partial^2 F}{\partial s\partial k_4} +\frac{t}{s}\frac{\partial^2 F}{\partial s\partial t}\, ,\label{Fs2cond} \\
	-\frac{\partial^2 F}{\partial t^2}&=\frac{10-\Delta_t}{t}\frac{\partial F}{\partial t}+\frac{k_1}{t}\frac{\partial^2 F}{\partial t\partial k_1} + \frac{k_2}{t}\frac{\partial^2 F}{\partial t\partial k_2}+\frac{k_3}{t}\frac{\partial^2 F}{\partial t\partial k_3}+\frac{k_4}{t}\frac{\partial^2 F}{\partial t\partial k_4} +\frac{s}{t}\frac{\partial^2 F}{\partial s\partial t}\, .\label{Ft2cond}
\end{align}
We will substitute these derivatives in our derivation of the SCT constraints.

\paragraph{Linear term}
We first consider the first-order term in \eqref{WI_SCT}. We have
\begin{align}
	\sum_{n=1}^4 2(\Delta_{n}-3)\frac{\partial F}{\partial k_{n}^{i}} 
	& =\red{k_1^{i}} \left[2(\Delta_1-3)\frac{1}{k_1}\frac{\partial F}{\partial k_1}+2(\Delta_{12}-6)\frac{1}{s}\frac{\partial F}{\partial s} \right]\nn
	&+\orange{k_2^{i}}\left[2(\Delta_2-3)\frac{1}{k_2}\frac{\partial F}{\partial k_2}+2(\Delta_{12}-6)\frac{1}{s}\frac{\partial F}{\partial s}+2(\Delta_{23}-6)\frac{1}{t}\frac{\partial F}{\partial t} \right]\nn
	&+\green{k_3^{i}}\left[2(\Delta_3-3)\frac{1}{k_3}\frac{\partial F}{\partial k_3}+2(\Delta_{23}-6)\frac{1}{t}\frac{\partial F}{\partial t}\right]\nn
	&+\blue{k_4^{i}}\left[2(\Delta_4-3)\frac{1}{k_4}\frac{\partial F}{\partial k_4}\right] ,
\end{align}
where $\Delta_{nm} \equiv \Delta_n + \Delta_m$.

\paragraph{Cross term} 
There are two second-order terms in \eqref{WI_SCT}. The cross term is given by
\begin{align}
	\sum_{n=1}^4 k_{n}^{j} \frac{\partial^2 F}{\partial k_{n}^{i} \partial k_{n}^{j}} &=\red{k_1^{i}}\bigg[ \frac{\partial^2}{\partial k_1^2} +\frac{s_1^2+k_1^2}{s k_1}\frac{\partial^2 F}{\partial s\partial k_1} +\frac{t_2^2}{st}\frac{\partial^2 F}{\partial s\partial t}+\frac{k_2}{s}\frac{\partial^2F}{\partial s\partial k_2}+\frac{\partial^2F}{\partial s^2}\bigg]\nn[4pt]
	&+\orange{k_2^{i}}\bigg[ \frac{k_1^2}{s k_1}\frac{\partial^2F}{\partial s\partial k_1} + \frac{\partial^2F}{\partial s^2}+\frac{\partial^2F}{\partial k_2^2}+\frac{s_2^2+t_2^2}{st}\frac{\partial^2F}{\partial s\partial t}+\frac{s_2^2+k_2^2}{s k_2}\frac{\partial^2F}{\partial s\partial k_2}+\frac{t_2^2+k_2^2}{tk_2}\frac{\partial^2F}{\partial t\partial k_2} \nonumber \\
	&\hspace{0.9cm}+\frac{k_3^3}{t k_3}\frac{\partial^2F}{\partial t\partial k_3} + \frac{\partial^2F}{\partial t^2}\bigg]\nn[4pt]
	&+\green{k_3^{i}}\bigg[\frac{\partial^2F}{\partial k_3^2} +\frac{t_3^2+k_3^2}{t k_3}\frac{\partial^2F}{\partial t\partial k_3} +\frac{s_2^2}{st}\frac{\partial^2F}{\partial s\partial t}+\frac{k_2}{t}\frac{\partial^2F}{\partial t\partial k_2}+\frac{\partial^2F}{\partial t^2}\bigg]\nn[4pt]
	&+\blue{k_4^{i}}\bigg[\frac{\partial^2F}{\partial k_4^2}\bigg]\, ,
\end{align}
where we have defined
\begin{align}
	s_1^2\equiv k_1^{i}(k_1^{i}+k_2^{i})\, , \quad s_2^2\equiv k_2^{i}(k_1^{i}+k_2^{i})\, , \quad t_2^2\equiv k_2^{i}(k_2^{i}+k_3^{i})\, , \quad t_3^2\equiv k_3^{i}(k_2^{i}+k_3^{i})\, .
\end{align}

\paragraph{Quadratic term} 
Lastly, we consider the ``Laplacian'' term in \eqref{WI_SCT}. This is given by
\begin{align}
	\sum_{n=1}^4 k_{n}^{i} \frac{\partial^2 F}{\partial k_{n}^{j} \partial k_{n}^{j}} &=\red{k_1^{i}}\bigg[\frac{\partial^2F}{\partial k_1^2} +\frac{2}{k_1}\frac{\partial F}{\partial k_1}+\frac{2s_1^2}{s k_1}\frac{\partial^2F}{\partial s\partial k_1} +\frac{2}{s}\frac{\partial F}{\partial s} + \frac{\partial^2F}{\partial s^2}\bigg]\nn[4pt]
	&+\orange{k_2^{i}} \bigg[\frac{\partial^2 F}{\partial k_2^2}+\frac{2}{k_2}\frac{\partial F}{\partial k_2}+\frac{2s_2^2+2t_2^2 + 2k_1^{i} k_3^{i} -2k_2^2}{st}\frac{\partial^2F}{\partial s\partial t}+\frac{2}{s}\frac{\partial F}{\partial s}+\frac{2s_2^2}{s k_2}\frac{\partial^2 F}{\partial s\partial k_2} \nn &\hspace{1cm}+\frac{\partial^2F}{\partial s^2}+\frac{2}{t}\frac{\partial F}{\partial t}+\frac{2t_2^2}{tk_2}\frac{\partial^2 F}{\partial t\partial k_2}+\frac{\partial^2F}{\partial t^2}\bigg]\nn[4pt]
	&+\green{k_3^{i}} \bigg[\frac{\partial^2F}{\partial k_3^2} +\frac{2}{k_3}\frac{\partial F}{\partial k_3}+\frac{2t_3^2}{t k_3}\frac{\partial^2F}{\partial t\partial k_3} +\frac{2}{t}\frac{\partial F}{\partial t} + \frac{\partial^2F}{\partial t^2}\bigg]\nn[4pt]
	&+\blue{k_4^{i}}\bigg[\frac{\partial^2F}{\partial k_4^2} +\frac{2}{k_4}\frac{\partial F}{\partial k_4}\bigg]\, .
\end{align}

\paragraph{Total}
We now sum over the contributions from the terms above. This gives
\begin{align}
	\sum_{n=1}^4 {\cal K}_{n}^{i} F &=\red{k_1^{i}} \bigg[\frac{2(\Delta_1-2)}{k_1}\frac{\partial F}{\partial k_1}+\frac{2(\Delta_{12}-5)}{s}\frac{\partial F}{\partial s}-\frac{\partial^2F}{\partial k_1^2} -\frac{2k_1}{s}\frac{\partial^2F}{\partial s\partial k_1} -\frac{2t_2^2}{st}\frac{\partial^2F}{\partial s\partial t} \nn
	&\hspace{1cm}-\frac{2k_2}{s}\frac{\partial^2F}{\partial s\partial k_2}-\frac{\partial^2F}{\partial s^2}\bigg]\nn[4pt]
	&+\orange{k_2^{i}}\bigg[\frac{2(\Delta_2-2)}{k_2}\frac{\partial F}{\partial k_2}+\frac{2(\Delta_{12}-5)}{s}\frac{\partial F}{\partial s}+\frac{2(\Delta_{23}-5)}{t}\frac{\partial F}{\partial t}-\frac{\partial^2F}{\partial k_2^2} - \frac{\partial^2F}{\partial s^2}  - \frac{\partial^2F}{\partial t^2}\nn
	&\hspace{1cm} -\frac{2k_1}{s }\frac{\partial^2F}{\partial s\partial k_1} -\frac{2k_2}{s }\frac{\partial^2F}{\partial s\partial k_2}-\frac{2k_2}{t}\frac{\partial^2F}{\partial t\partial k_2}-\frac{2k_3}{t}\frac{\partial^2F}{\partial t\partial k_3} +\frac{2( k_1^{j} k_3^{j} -k_2^2)}{st}\frac{\partial^2F}{\partial s\partial t}\bigg]\nn[4pt]
	&+\green{k_3^{i}}\bigg[\frac{2(\Delta_3-2)}{k_3}\frac{\partial F}{\partial k_3}+\frac{2(\Delta_{23}-5)}{t}\frac{\partial F}{\partial t}-\frac{\partial^2F}{\partial k_3^2} -\frac{2k_3}{t}\frac{\partial^2F}{\partial t\partial k_3} -\frac{2s_2^2}{st}\frac{\partial^2F}{\partial s\partial t} \nn
	&\hspace{1cm}-\frac{2k_2}{t}\frac{\partial^2F}{\partial t\partial k_2} - \frac{\partial^2F}{\partial t^2}\bigg]\nn[4pt]
	&+\blue{k_4^{i}}\bigg[\frac{2(\Delta_4-2)}{k_4}\frac{\partial F}{\partial k_4}-\frac{\partial^2F}{\partial k_4^2}\bigg]\, .
\end{align}
Because of the asymmetric definitions of $s$ and $t$ in terms of the momenta $k_n$, the result is not manifestly symmetric under cyclic permutations. We can use momentum conservation to bring these into a more symmetric form.  Moreover, to make the resulting expression symmetric between $s$ and $t$, we use the constraints \eqref{Fs2cond} and \eqref{Ft2cond}. As a consequence of momentum conservation, we can drop all terms that are equal in each square brackets.
These include terms such as $\partial^2F/\partial s^2$, $\partial^2F/\partial t^2$ and $\partial F/\partial s$, $\partial F/\partial t$.
The final result is of the form (\ref{WI_SCT2}) with each coefficient given by cyclic permutations of (\ref{W1F}). This completes the derivation.

\newpage
\section{Singularity Structure}\label{app:bulk}
In this appendix, we examine the singularities of tree-exchange four-point functions  from both the bulk and boundary perspectives. 
We find agreement between both computations, providing a useful consistency check of the reasoning we advocate in the main text.

\subsection{Boundary Perspective}

We begin by analyzing the singularities of the boundary correlator for $\Delta=2$. We will examine the series solution and the homogeneous solution separately. 

\paragraph{Series solution} 
Let us analyze the leading singular behavior of the series solution \eqref{fuv} in the limit $u,v\to \pm 1$. 
Since we expect the  solution to have a logarithmic singularity as $u,v\to \pm 1$, we look at its first derivative \begin{align}
	\partial_u\hat F_<(u,v) = \sum_{m,n=0}^\infty (2m+n+1)c_{mn}\,u^{2m}(u/v)^{n}\, .
\end{align}
The sum over $n$, for general $u$ and $v$, can be expressed as 
\begin{align}
	\partial_u\hat F_<(u,v) = \sum_{m=0}^\infty \bigg[F_{1}-\frac{u}{v}(F_2+F_3)+\frac{u^2}{v^2}F_4\bigg]u^{2m}\, ,\label{SeriesDerivative}
\end{align}
where we have defined
\begin{align}
	F_1 & \equiv\frac{\Gamma(2+2m)}{4^{1+m}(\frac{1}{4}-\frac{i\mu}{2})_{1+m}(\frac{1}{4}+\frac{i\mu}{2})_{1+m}} {}_4F_3\Bigg[\begin{array}{c} \frac{1}{2}+m,1+m,\frac{1}{4}-\frac{i\mu}{2},\frac{1}{4}+\frac{i\mu}{2}\\[2pt] \frac{1}{2},\frac{5}{4}+m-\frac{i\mu}{2},\frac{5}{4}+m+\frac{i\mu}{2} \end{array}\Bigg|\, \frac{u^2}{v^2}\Bigg] \, , \\[5pt]
	F_2 &  \equiv\frac{(2)_{2m}}{4^{1+m}(\frac{3}{4}-\frac{i\mu}{2})_{1+m}(\frac{3}{4}+\frac{i\mu}{2})_{1+m}}{}_4F_3\Bigg[\begin{array}{c} 1+m,\frac{3}{2}+m,\frac{3}{4}-\frac{i\mu}{2},\frac{3}{4}+\frac{i\mu}{2}\\[2pt] \frac{1}{2},\frac{7}{4}+m-\frac{i\mu}{2},\frac{7}{4}+m+\frac{i\mu}{2} \end{array}\Bigg|\, \frac{u^2}{v^2}\Bigg]\, ,\\[5pt]
	F_3 & \equiv\frac{(2m+1)(2)_{2m}}{4^{1+m}(\frac{3}{4}-\frac{i\mu}{2})_{1+m}(\frac{3}{4}+\frac{i\mu}{2})_{1+m}} {}_4F_3\Bigg[\begin{array}{c} 1+m,\frac{3}{2}+m,\frac{3}{4}-\frac{i\mu}{2},\frac{3}{4}+\frac{i\mu}{2}\\[2pt] \frac{3}{2},\frac{7}{4}+m-\frac{i\mu}{2},\frac{7}{4}+m+\frac{i\mu}{2} \end{array}\Bigg|\, \frac{u^2}{v^2}\Bigg]\, ,\\[5pt]
	F_4 & \equiv \frac{\Gamma(3+2m)}{4^{1+m}(\frac{5}{4}-\frac{i\mu}{2})_{1+m}(\frac{5}{4}+\frac{i\mu}{2})_{1+m}}{}_4F_3\Bigg[\begin{array}{c} \frac{3}{2}+m,2+m,\frac{5}{4}-\frac{i\mu}{2},\frac{5}{4}+\frac{i\mu}{2}\\[2pt] \frac{3}{2},\frac{9}{4}+m-\frac{i\mu}{2},\frac{9}{4}+m+\frac{i\mu}{2} \end{array}\Bigg|\, \frac{u^2}{v^2}\Bigg]\, .
\end{align}
To see whether the series diverges or not, we look at the large $m$ behavior of the series coefficients.
Only the terms $F_1$ and $F_3$ are relevant in the limit $m\to\infty $, giving
\begin{align}
	\lim_{m\to \infty} F_1 & = \frac{\Gamma(\frac{1}{4}-\frac{i\mu}{2})\Gamma(\frac{1}{4}+\frac{i\mu}{2})}{2\sqrt{\pi}}{}_2F_1\Bigg[\begin{array}{c} \frac{1}{4}-\frac{i\mu}{2},\frac{1}{4}+\frac{i\mu}{2}\\[2pt] \frac{1}{2} \end{array}\Bigg|\, \frac{u^2}{v^2}\Bigg] \, ,\label{F1} \\[5pt]
	\lim_{m\to \infty} F_3 & = \frac{\Gamma(\frac{3}{4}-\frac{i\mu}{2})\Gamma(\frac{3}{4}+\frac{i\mu}{2})}{\sqrt{\pi}}{}_2F_1\Bigg[\begin{array}{c} \frac{3}{4}-\frac{i\mu}{2},\frac{3}{4}+\frac{i\mu}{2}\\[2pt] \frac{3}{2} \end{array}\Bigg|\, \frac{u^2}{v^2}\Bigg] \, .\label{F2}
\end{align}
Notice that the coefficient of the series \eqref{SeriesDerivative} becomes $m$-independent in the  limit $m\to \infty$; the series thus diverges as $u,v\to \pm 1$.
Naively, both \eqref{F1} and \eqref{F2} have $\log(u-v)$ singularities as $v\to u$. However, it turns out that these exactly cancel, so that the limit $v\to u$ is actually finite. Using the identity
\begin{align}
	\lim_{m\to \infty}\lim_{v\to u}\bigg[F_{1}-\frac{u}{v}(F_2+F_3)+\frac{u^2}{v^2}F_4\bigg] = \frac{\pi}{\cosh\pi\mu}\, ,
\end{align}
we find
\begin{align}
		\lim_{u,v\to \pm 1}\hat F_<(u,v) = -\frac{\pi}{2\cosh\pi\mu}\log(1\mp u)\, .\label{seriessingularity}
\end{align}
We notice that there is a spurious singularity as $u,v\to 1$. However, as we will show below, this is cancelled by the singularity of the homogeneous solution in the same limit.

\paragraph{Homogeneous solution} 

Next, we describe how the nonperturbative part of the boundary correlator is fixed by imposing the correct singularity structure. We start by writing down the most general solution to the differential equation \eqref{InhomogeneousFu}:
\beq
\hat F(u,v) = \begin{cases}  \displaystyle \ \sum_{m,n=0}^\infty c_{mn}\,u^{2m+1}(u/v)^{n} + \hat g(u,v)& \text{$u \le v$}\, , \\[20pt]
 \displaystyle \ \sum_{m,n=0}^\infty c_{mn}\,v^{2m+1}(v/u)^{n} + \hat h(u,v)& \text{$u \ge v$}\, , \end{cases} \label{equ:SeriesSol2}
\eeq
where the functions $\hat g$ and $\hat h$ contain the homogeneous solutions. Demanding $\hat F$ to be symmetric under the exchange $u\leftrightarrow v$ implies $\hat h(u,v)=\hat g(v,u)$, whereas the matching condition gives
\beq
	\hat g(u,v)- \hat g(v,u) = \frac{\pi}{\cosh \pi\mu}\Big(\hat F_+(u)\hat F_-(v)-\hat F_-(u)\hat F_+(v)\Big)\, .
\eeq
This fixes the function $\hat g$ up to three parameters:
\begin{align}
	\hat g(u,v) & = \frac{\pi}{2\cosh \pi\mu }\Big\{  \big[\hat {F}_+(u) \hat{F}_-(v) - \hat{F}_-(u)  \hat{F}_+(v) \big] +\beta_+\hat F_+(u)\hat F_+(v) \nn[4pt]
	&\hspace{2.5cm} +\beta_-\hat F_-(u)\hat F_-(v)+\beta_0\big[\hat F_-(u)\hat F_+(v)+\hat F_+(u)\hat F_-(v)\big]\Big\}\, .\label{generalghat}
\end{align}
We will fix the parameters $\beta_+$, $\beta_-$, $\beta_0$ in turn.

\vskip 4pt
Let us first comment on the analytic structure of the homogeneous solutions $\hat F_{\pm}$ defined in~\eqref{equ:homo}. The hypergeometric function that we are dealing with is of zero-balanced type, which has logarithmic singularities at $u^2=1$ and $v^2=1$. Moreover, there are branch points at $u=0$ and $v=0$ due to the overall non-analytic factor $u^{1/2\pm i\mu}$. On the other hand, the functions $\hat F_\pm$ are analytic as $u,v\to \infty$, so that a closed loop enclosing the point at infinity (or the three branch points at $u=\pm 1,0$) has no nontrivial monodromy. This means that we can choose the branch cut to run along the real interval $u\in[-1,1]$.

\vskip 4pt
We begin by looking at the singularities in the physical region $u,v\in[0, 1]$. Taking the limit $u \to 1$, for generic $v \in [0,1]$, we get\footnote{Note that we should be looking at $\hat h(u,v)=\hat g(v,u)$ when $u\ge v$.}
\begin{align}
	\lim_{u\to 1}\hat g(v,u) &= \Big[(\beta_- +\beta_0-1)\hat G_-(v) -(\beta_+ +\beta_0+1)\hat G_+(v)\Big] \log(1-u)\, ,
	\end{align}
where we have shown the leading singular behavior and defined 
\beq
	\hat G_{\pm}(v) \equiv \frac{\Gamma(\tfrac{1}{2}\pm i\mu)\Gamma(\mp i\mu)}{4\sqrt\pi}\left(\frac{v}{2}\right)^{\frac{1}{2}\pm i\mu} {}_2F_1\Bigg[\begin{array}{c} \frac{1}{4}\pm \frac{i\mu}{2},\frac{3}{4}\pm \frac{i\mu}{2}\\[2pt] 1 \pm i\mu \end{array}\Bigg|\, v^2\Bigg] \, .
\eeq
We see that the function $\hat g(u,v)$ has logarithmic singularities as $u,v\to 1$ in general, but, as we have argued, these singularities are unphysical. Removing theses singularities amounts to choosing
\begin{align}
	\beta_+ = -1-\beta_0\, , \quad \beta_- = 1-\beta_0\, .\label{beta}
\end{align}
The same choice also removes the singularity at $v=1$ for $u\ne 1$. For $|v|<1$, there are now two branch points at $u=-1,0$; this shrinks the branch cut down to the interval $u\in [-1,0]$. There is, however, a singularity at $u=v=1$: 
\begin{align}
	\lim_{u,v\to 1}\hat g(u,v) &= \frac{\pi}{2\cosh \pi\mu}\log(1-u)\, .
\end{align}
This singularity nicely cancels the spurious singularity of the series solution in \eqref{seriessingularity}, so that the full solution is regular at $u=v=1$. This means that the four-point function is real-valued on the real interval $u\in [0,1]$, which implies that there is no cut there. Since all the singularities in the unphysical region come from $\hat g(u,v)$, the Schwarz reflection principle implies that
\begin{align}
	\hat g^*(u,v) = \hat g(u^*,v)\, .\label{schwarz}
\end{align}
This tells us the behavior of the analytically-continued function $\hat g(u,v)$ as we approach the negative real interval $u\in [-1,0]$ above and below the branch cut.

\vskip 4pt
As we have alluded to before, there are several ways of fixing the remaining parameter $\beta_0$. One way is by analytically continuing to the complex plane and looking at the singularities in the unphysical region. The coefficients of these singularities can then be normalized through comparison with the bulk expectation. This procedure was described in Section~\ref{sec:dS-4pt}. 
Alternatively, we can fix the solution by correctly normalizing the disconnected part of the four-point function in the physical region (see \S\ref{sec:Psi}). To do this, we note that the term proportional to $\beta_0$ is just the product of two three-point functions:
\begin{align}
	\beta_0\frac{\partial}{\partial \beta_0}\hat g(u,v) = \frac{i\beta_0\sinh\pi\mu}{2\pi}\Big[\hat G_+(u)+\hat G_-(u)\Big]\Big[\hat G_+(v)+\hat G_-(v)\Big]\, ,
\end{align}
where 
\begin{align}
	\langle O(\k_1)O(\k_2) \Sigma(-\s) \rangle' &= \frac{s^{\Delta-2}\pi}{\sqrt{2}\cosh \pi\mu}\,{}_2F_1\Bigg[\begin{array}{c} \frac{1}{2}-i\mu,\frac{1}{2}+i\mu\\[2pt] 1 \end{array}\Bigg|\, \frac{u-1}{2u}\Bigg]\nn[5pt]
	&=\frac{s^{\Delta-2}}{2\sqrt{\pi}}\left[\hat G_+(u)+\hat G_-(u)\right] .\label{3pt}
\end{align}
Up to an overall scaling, the three-point function satisfies the same equation as the homogeneous piece of the four-point function, and the first line can be obtained by solving the equation in terms of the variable $u^{-1}$, for which case it is easy to impose regularity at $u=1$.
In going from the first to the second line of (\ref{3pt}) we used the identities \eqref{hyper1} and \eqref{hyper2}. We have set the normalization of the three-point function to be 1, consistent with the choice for our four-point function. Equation~\eqref{disconnected} then implies
\begin{align}
	\frac{\beta_0}{s}\frac{\partial}{\partial \beta_0}\hat g(u,v) &= \frac{2\langle O(\k_1)O(\k_2) \Sigma(-\s) \rangle'\langle \Sigma(\s)O(\k_3)O(\k_4)  \rangle'}{\langle\Sigma(\s)\Sigma(-\s)\rangle'}\, ,
\end{align}
where we have restored the mass dimension of the left-hand side. Using $\langle\Sigma(\s)\Sigma(-\s)\rangle' = s^{2\Delta-3}$, this fixes $\beta_0=1/i\sinh\pi\mu$.

\vskip 4pt
Finally, there is a third, hybrid way of fixing $\beta_0$ by sending only one of the variables $u$ and $v$ to the singularity in the unphysical region. In this limit, the four-point function factorizes into the product of a three-point correlator and a three-particle amplitude. For example, taking the limit $u \to -1$, for generic $v > 0$, we get
\beq
	\lim_{u\to -1}\hat g(u,v) = -2i\sinh \pi\mu\left[(\beta_0+1)\hat G_+(v)+(\beta_0-1)\hat G_-(v) \right]\log(1+u)\, .\label{guvlimit1}
\eeq
Comparing this to the bulk calculation in the next section, we again find~$\beta_0=1/i\sinh\pi\mu$.

\subsection{Bulk Perspective}
\label{sec:bulk}

In this section, we analyze the singularities of the four-point function of conformally coupled scalars from the bulk perspective. The standard method to compute vacuum expectation values in a time-dependent background is the Schwinger-Keldysh or in-in formalism~\cite{Maldacena:2002vr, Weinberg:2005vy} (for recent reviews see~\cite{Chen:2010xka, Wang:2013eqj, Baumann:2014nda, TASI2017}).  
In this formalism, the equal-time vacuum expectation value of some operator $Q(\eta)$ consisting of a product of quantum fields at different positions is 
\begin{align}
	\langle Q(\eta)\rangle = \big\langle 0\big|\big[\bar{\rm T}\,e^{i\int_{-\infty(1-i\epsilon)}^\eta H_{\rm int}^I(\eta')\,d\eta'}\big]Q^I(\eta)\big[{\rm T}\,e^{-i\int_{-\infty(1+i\epsilon)}^\eta H_{\rm int}^I(\eta')\,d\eta'}\big]\big|0\big\rangle\, ,\label{inin}
\end{align}
where ${\rm T}$~($\bar{\rm T}$) denotes a (anti-)time ordered product, $H_{\rm int}^I$ is the interaction Hamiltonian, and the superscript $I$ indicates that the operators are evaluated in the interaction picture. The standard $i\epsilon$ prescription is used to project the interacting vacuum to the free vacuum, $|0\rangle$.\footnote{In the following, we will suppress the extra $i\epsilon$ with the understanding that the integration contours are deformed appropriately.}

\vskip 4pt
Quantization of fields in dS proceeds in a straightforward way. We decompose the free conformally coupled field $\upvarphi$ and the massive field $\sigma$ in Fourier space as
\begin{align}
	\upvarphi^I(\k,\eta) = \upvarphi_k(\eta)	a_{\upvarphi}(\k) + \upvarphi_k^*(\eta) a_{\upvarphi}^\dagger(-\k)\, , \quad \sigma^I(\k,\eta) = \sigma_k(\eta)	a_\sigma(\k) + \sigma_k^*(\eta)	a_\sigma^\dagger(-\k)\, ,
\end{align}
where $a_{\upvarphi,\sigma}$ and $a_{\upvarphi,\sigma}^\dagger$ are the annihilation and creation operators. The mode functions are
\begin{align}
	\upvarphi_k(\eta) &= (-H\eta)\frac{e^{-ik\eta}}{\sqrt{2k}}\, ,\\
	\sigma_k(\eta) &= \frac{H\sqrt\pi}{2}e^{i\pi/4}e^{-\pi\mu/2}(-\eta)^{3/2}H_{i\mu}^{(1)}(-k\eta) \ \xrightarrow{\ \eta \to - \infty \ } \ (-H\eta)\frac{e^{-ik\eta}}{\sqrt{2k}} \, ,
\end{align}
which reduce to the Bunch-Davies vacuum~\cite{Bunch:1978yq} at early times. Demanding that $a_{\upvarphi,\sigma}$ and $a_{\upvarphi,\sigma}^\dagger$ satisfy the canonical commutation relations amounts to imposing the Wronskian normalization on the mode functions, $W[\upvarphi_k(\eta),\upvarphi_k^*(\eta)]=W[\sigma_k(\eta),\sigma_k^*(\eta)]=iH^2\eta^2$.

\vskip 4pt
The expectation value \eqref{inin} is computed by performing time integrals in an (anti-)time-ordered manner. For this purpose, it is convenient to introduce the following Green's functions
\beq
\begin{aligned}
	G_{++}(k,\eta,\eta') &= \sigma_k(\eta)\sigma_k^*(\eta')\Theta(\eta-\eta')+\sigma_k^*(\eta)\sigma_k(\eta')\Theta(\eta'-\eta)\, ,\\
	G_{+-}(k,\eta,\eta') &=\sigma_k(\eta)\sigma_k^*(\eta')\, ,\\
	G_{+-}(k,\eta,\eta') &=\sigma_k^*(\eta)\sigma_k(\eta')\, ,\\
	G_{--}(k,\eta,\eta') &= \sigma_k(\eta)\sigma_k^*(\eta')\Theta(\eta'-\eta)+\sigma_k^*(\eta)\sigma_k(\eta')\Theta(\eta-\eta')\, ,
\end{aligned}
\label{equ:Gpmpm}
\eeq
where $\pm$ indicates different parts of the integration contour and $\Theta$ is the Heaviside function. The functions $G_{\pm\pm}$ satisfy the inhomogeneous equation
\begin{align}
	(\eta^2\partial_\eta^2-2\eta\partial_\eta+k^2\eta^2+m^2/H^2) G_{\pm\pm}(k,\eta,\eta') = -i\eta^2\eta'^2\delta(\eta-\eta')\, ,\label{Gpp}
\end{align}
and a similar equation for $\eta'$, while $G_{\pm\mp}$ satisfy the corresponding the homogeneous equation.

\vskip 4pt
Using the above definitions, 
 the four-point function arising from the interaction $g \upvarphi^2\sigma$ can be written as
\begin{align}
	\langle\varphi_1\varphi_2\varphi_3\varphi_4\rangle' = \frac{\eta_0^4\hs H^8}{2k_1k_2k_3k_4}F(k_{12},k_{34},s) \ +\ \text{$t$- and $u$-channels}\, ,\label{phi4bulk}
\end{align}
where we have introduced a small late-time cutoff $\eta_0$ and defined
\begin{align}
	F & \equiv F_{++}+F_{+-}+F_{-+}+F_{--}\, ,
\end{align}	
with 
\begin{align}	
	F_{\pm\pm} &=g^2\,\frac{(\pm i)(\pm i)}{2}\int_{-\infty}^0\frac{\d\eta}{\eta^2}\,e^{\pm ik_{12}\eta}\int_{-\infty}^0\frac{\d\eta'}{\eta'^2}\,e^{\pm ik_{34}\eta'}G_{\pm\pm}(s,\eta,\eta')\, .\label{Fpmpm}
\end{align}
Using the equations of motion for the Green's functions $G_{\pm \pm}$, it can be shown that the function $F$ obeys the differential equation
\begin{align}
	\big[(k_{12}^2-s^2)\partial_{k_{12}}^2+2k_{12}\partial_{k_{12}}+m^2-2\big]F = \frac{g^2}{k_t}\, ,
\end{align}
which, in terms of $u$, $v$ and $\hat F=s F$, becomes
\begin{align}
	\bigg[\Delta_u+\bigg(\mu^2+\frac{1}{4}\bigg)\bigg]\hat F = g^2\frac{uv}{u+v}\, .
\end{align}
This is precisely the conformal invariance equation of the boundary correlator we have been using extensively in the main text.

\vskip 4pt
Let us now analyze the singularities of the integrals in (\ref{Fpmpm}). For convenience, we will set $H=1$ and $g^2=1$. In the limit $u=v \to -1$, the integral $\hat F_{-+} \equiv s F_{-+}$ picks up contributions from $\eta=\eta'=-\infty$, and, hence, we get 
\begin{align}
	\lim_{u,v\to- 1} (\hat F_{-+}+\hat F_{+-}) &= \frac{1}{4}\int_{-\infty}^0\frac{\d\eta}{\eta}\,e^{-i(k_{12}+s)\eta}\int_{-\infty}^0\frac{\d\eta'}{\eta'}\,e^{i(k_{34}+s)\eta'} + (k_{12}\leftrightarrow k_{34}) \nn[5pt]
	&= \frac{1}{2}\log(1+u)\log(1+v)\, .
\end{align}
This explains our normalization of the boundary correlator in \eqref{uvminus1}. Next, consider the integral
\begin{align}
	\hat F_{\pm\pm} &=-\frac{s}{2}\int_{-\infty}^0\frac{\d\eta}{\eta^2}\,e^{\pm ik_{12}\eta}\sigma_s(\eta)\int_{-\infty}^\eta\frac{\d\eta'}{\eta'^2}\,e^{\pm ik_{34}\eta'}\sigma_s^*(\eta') + (k_{12}\leftrightarrow k_{34})\, .\label{Fpp}
\end{align}
In the limit $u,v\to -1$, the function $\hat F_{++} $ 
picks up contributions of the inner integral from $\eta'=-\infty$. The upper limit of the inner integral then becomes irrelevant, so the inner and outer integrals factorize. The latter gives a finite contribution and can be evaluated to give the familiar cosh factor. We get the same behavior for $F_{--}$ in the limit $u,v\to - 1$. Precisely, we have
\begin{align}
	\lim_{u,v\to - 1}(\hat F_{++}+\hat F_{--}) = 
	-\frac{\pi}{2\cosh\pi\mu}\log(1+ u)\, .
\end{align}
This agrees with the behavior of \eqref{seriessingularity} in the same limits. Some useful formulas for deriving this result are presented in the insert at the end of the section.

\vskip 4pt
Next, let us examine the behavior of $F_{++}$ near $u=-1$ for $|v|<1$. First, notice that the first term in \eqref{Fpp} has no singularity at $u=-1$. This can be seen after computing the inner integral, which is non-singular for generic $v\ne \pm 1$; the only singularity of the resulting outer integral is then at $u=-v$, the usual flat-space limit. On the other hand, the inner integral of the second term picks up a singular contribution when $u=-1$, rendering the integrals in a factorized form. Evaluating the non-singular integral of the second term then gives
\begin{align}
	\lim_{u\to -1}\hat F_{++} = \frac{\pi}{4\cosh\pi\mu}\,{}_2F_1\Bigg[\begin{array}{c} \frac{1}{2}-i\mu,\frac{1}{2}+i\mu\\[2pt] 1 \end{array}\Bigg|\, \frac{v+1}{2v}\Bigg] \log(1+u)\, .
\end{align}
We see that the coefficient of the log looks almost like the three-point function (\ref{3pt}) except that it has $(1+v)/2v$ as the argument of the hypergeometric function rather than $(1-v)/2v$. In order to compare this with the boundary calculation, we use the identities \eqref{hyper1} and \eqref{hyper2} to express this as
\begin{align}
	\lim_{u\to -1}\hat F_{++} &= -i\left[e^{\pi\mu}\,\hat G_+(v)+e^{-\pi\mu}\,\hat G_-(v)\right]\log(1+u)\, ,\label{Fpp2}
\end{align}
for $v>0$. This accounts for half of the terms in \eqref{guvlimit1} not proportional to $\beta_0$. The remaining terms correspond to the singular contribution from $F_{--}$. For $v>0$, we have 
\begin{align}
	\lim_{u\to -1}\hat F_{--} &= i\left[e^{-\pi\mu}\,\hat G_+(v)+e^{\pi\mu}\,\hat G_-(v)\right]\log(1+u)\, ,
\end{align}
which is the complex conjugate of \eqref{Fpp2}.  

\vskip 4pt
The singular behavior of $F_{\mp\pm}$ near $u=-1$ can be obtained from the results for $F_{\pm\pm}$ by the analytic continuation $v\to -v$. This gives
\begin{align}
	\lim_{u\to -1}\hat F_{\mp\pm} &= -\frac{\pi}{4\cosh \pi\mu}\,{}_2F_1\Bigg[\begin{array}{c} \frac{1}{2}-i\mu,\frac{1}{2}+i\mu\\[2pt] 1 \end{array}\Bigg|\, \frac{v-1}{2v}\Bigg] \log(1+u)\, , \nonumber \\[10pt]
	&= -\Big[\hat G_+(v)+\hat G_-(v)\Big]\log(1+u)\, ,
\end{align}
where the second line holds for $v>0$. 
Combining everything, we find
\begin{align}
	\lim_{u\to -1}\hat F = -2\Big[(1+i\sinh\pi\mu)G_+(v)+(1-i\sinh\pi\mu)G_-(v)\Big]\log(1+u)\, .
\end{align}
This agrees precisely with the boundary expression \eqref{guvlimit1} with the choice $\beta_0=1/i\sinh\pi\mu $.

\begin{framed}
\small \noindent 
{\it Derivation.}---In this insert, we derive the analytic expression of the integral
\begin{align}
	{\sf I}_n(a,b) &\equiv  \int^0_{-\infty}\d \eta\, (-\eta)^{n-2}e^{-ia\eta}\sigma_{b}(\eta)= \frac{H\sqrt\pi}{2}\,e^{-\pi\mu/2}e^{-3i\pi/4}\int_0^\infty\d x\, x^{n-\frac{1}{2}}e^{iax}H_{i\mu}(b x)\, ,\label{In}
\end{align}
which is a basic element of any bulk calculation involving the exchange of a massive scalar field $\sigma$. We will do so by solving the differential equation that ${\sf I}_n$ satisfies. First, note that the equation of motion of $\sigma$ implies the following differential equation:
\begin{align}
	\bigg[\partial_\eta^2 + \frac{2-2n}{\eta}\partial_\eta + \frac{k^2\eta^2+(n-\frac{1}{2})^2+\mu^2}{\eta^2}\bigg] \big(\eta^{n-2}\sigma_k(\eta)\big) = 0\, .
\end{align}
Integrating by parts, we may replace the derivative $\partial_\eta$ with $\partial_a$ acting on the exponential function. Pulling the resulting derivative operator out of the integral, we obtain
\begin{align}
	\Big[(a^2-b^2)\partial_a^2+2a(1+2n)\partial_a + (n+\tfrac{1}{2})^2+\mu^2\Big] {\sf I}_n(a,b) =0\, .
\end{align}
The solution to the differential equation that is regular at $a=b$ is given by
\begin{align}
	{\sf I}_n(a,b) & \,\propto\, \frac{1}{(a-b)^{n}}\, {}_2\tilde F_1\Bigg[\begin{array}{c} \frac{1}{2}-i\mu,\hs \frac{1}{2}+i\mu\\[2pt] 1-n\end{array}\Bigg|\,\frac{b-a}{2b}\Bigg]\, ,
\end{align}
where $ {}_2\tilde F_1(a,b,c,z)={}_2 F_1(a,b,c,z)/\Gamma(c)$ is the regularized hypergeometric function. The normalization is fixed by looking at the limit $a+b\to 0	$ of the bulk integral:
\begin{align}
	\lim_{a+b\to 0}{\sf I}_n(a,b) = -\frac{H}{\sqrt{2b}} \int_0^\infty \d x\, x^{n-1}e^{i(a+b)x}= -\frac{H}{\sqrt{2b}}\times\begin{cases} \displaystyle \frac{\Gamma(n)}{(-i(a+b))^n}  & n> 0\,,\\[12pt] \displaystyle \frac{[i(a+b)]^{|n|}}{\Gamma(1+|n|)}\log(a+b) & n \le 0\, .
	\end{cases}
\end{align}
This fixes the solution to be
\begin{align}
	{\sf I}_n(a,b) &=-\frac{H}{\sqrt{2b}}\frac{\pi}{\cosh\pi\mu}\frac{1}{[i(a-b)]^{n}}\, {}_2\tilde F_1\Bigg[\begin{array}{c} \frac{1}{2}-i\mu,\hs \frac{1}{2}+i\mu\\[2pt] 1-n\end{array}\Bigg|\,\frac{b-a}{2b}\Bigg]\, .
\end{align}
In physical cases of interest, $n$ will be a non-negative integer. It is then more convenient to use the alternative representation~\cite{Lee:2016vti}
\begin{align}
	{\sf I}_n(a,b) &=-\frac{H}{\sqrt{2b}}\left(\frac{i}{2b}\right)^n \Gamma(\tfrac{1}{2}+n-i\mu)\Gamma(\tfrac{1}{2}+n+i\mu)\, {}_2\tilde F_1\Bigg[\begin{array}{c} \frac{1}{2}+n-i\mu,\hs \frac{1}{2}+n+i\mu\\[2pt] 1+n\end{array}\Bigg|\,\frac{b-a}{2b}\Bigg]\, .\label{In2} 
	\end{align}
The hypergeometric function becomes unity when $a=b$, but is singular when $a=-b$. When $n<-1/2$, the integral \eqref{In} diverges as $\eta^{n+1/2}$. In this case, the expression \eqref{In2} computes the finite part of the integral. In the main text, we will often drop factors of $H$ to avoid clutter.
\end{framed}

\newpage
\section{Aspects of the $\mathbf{\Delta=2}$ Solutions}
\label{app:hyper}

In this appendix, we express the scalar exchange solution for $\Delta=2$ in the canonical hypergeometric form and analyze its behavior in various limits. 
 
\subsection{Hypergeometric Form}

We first write the series solution \eqref{fuv} in terms of the Pochhammer symbol
\begin{align}
	\hat F_<(u,v)&=u\sum_{n,m=0}^\infty \frac{(\frac{5+2i\mu}{4})_{n/2}(\frac{5-2i\mu}{4})_{n/2}}{(\frac{1}{2}+n)^2+\mu^2}\frac{ (1)_m  (\frac{1}{2})_{m+n/2}(1)_{m+n/2} }{(\frac{5+2i\mu}{4})_{m+n/2}(\frac{5-2i\mu}{4})_{m+n/2}}\frac{(u^{2})^m}{m!}\frac{(-2u/v)^n}{n!}\, ,
\end{align}
where $(\lambda)_n=\Gamma(\lambda+n)/\Gamma(\lambda)$ and we used the following identities 
\begin{align}
	(\lambda)_{n+m} = (\lambda+m)_{n}(\lambda)_{m}\, , \quad (\lambda)_{2n} = 4^{n}(\tfrac{\lambda}{2})_n(\tfrac{\lambda+1}{2})_n\, .
\end{align}
We then note the following properties of the summation involving of the Pochhammer symbols with an half-integer index:
\begin{align}
	\sum_{n=0}^\infty \frac{(\blue{a})_n}{(\red{\lambda})_{\green{n/2}}} \frac{x^n}{n!} &={}_2F_2\Bigg[\begin{array}{c} \blue{\frac{a}{2}},\blue{\frac{1}{2}+\frac{a}{2}}\\ \green{\frac{1}{2}},\red{\lambda} \end{array}\Bigg|\, x^2\Bigg]+\frac{\blue{a} x}{(\red{\lambda})_{1/2}}\, {}_2F_2\Bigg[\begin{array}{c} \blue{\frac{1}{2}+\frac{a}{2}},\blue{1+\frac{a}{2}}\\ \green{\frac{3}{2}},\red{\frac{1}{2}+\lambda}\end{array}\Bigg|\, x^2\Bigg]\, ,\\[4pt]
	\sum_{n=0}^\infty \frac{(\red{\lambda})_{\green{n/2}}}{(\blue{a})_n} \frac{x^n}{n!}&={}_1F_3\Bigg[\begin{array}{c}\red{\lambda}\\ \green{\frac{1}{2}},\blue{\frac{a}{2}},\blue{\frac{1}{2}+\frac{a}{2}}\end{array}\Bigg|\,\frac{x^2}{\green{16}}\Bigg]+(\red{\lambda})_{1/2}\frac{ x}{\blue{a}}\, {}_1F_3\Bigg[\begin{array}{c} \red{\frac{1}{2}+\lambda}\\ \green{\frac{3}{2} },\blue{\frac{1}{2}+\frac{a}{2}},\blue{1+\frac{a}{2}}\end{array}\Bigg|\,\frac{x^2}{\green{16}}\Bigg]\, ,
\end{align}
which can be shown by splitting the sum into even and odd powers. A similar pattern exists for summations involving multiple Pochhammer symbols. Finally, using
\begin{align}
	\frac{1}{(\frac{1}{2}+n)^2+\mu^2} 	&=\frac{i}{\mu(1+2i\mu)}\frac{(\frac{1}{2}+i\mu)_n}{(\frac{3}{2}+i\mu)_n}-\frac{i}{\mu(1-2i\mu)}\frac{(\frac{1}{2}-i\mu)_n}{(\frac{3}{2}-i\mu)_n}\, ,
\end{align}
we can express the series in the canonical hypergeometric form as
\begin{align}
	\hat F_<(u,v) &= u\sum_{n,m=0}^\infty \left(a_{mn}+\frac{u}{v}\hs b_{mn}+c.c.\right)\frac{(u^{2})^m}{m!}\frac{(u^2/v^2)^{n}}{n!}\nn
	&=\frac{i}{\mu(1+2i\mu)}F^{2|1|3}_{\,2|0|1}\Bigg[\begin{array}{c} \frac{1}{2},1\\[2pt]\frac{5+2i\mu}{4},\frac{5-2i\mu}{4}\end{array}\Bigg| \begin{array}{c} 1\\[2pt] - \, \end{array} \Bigg| \begin{array}{c} \frac{5+2i\mu}{4},\frac{5-2i\mu}{4},\frac{1}{2}+i\mu \\[2pt] \frac{3}{2}+i\mu \end{array}\Bigg| \, u^2,\frac{u^2}{v^2}\Bigg]\nn
	&\hskip 12pt -\frac{i}{\mu(1+2i\mu)}\frac{u}{v}F^{2|1|3}_{\,2|0|1}\Bigg[\begin{array}{c} \frac{3}{2},1\\[2pt] \frac{7+2i\mu}{4},\frac{7-2i\mu}{4}\end{array}\Bigg| \begin{array}{c} 1\\[2pt] - \, \end{array} \Bigg| \begin{array}{c} \frac{7+2i\mu}{4},\frac{7-2i\mu}{4},\frac{1}{2}+i\mu \\[2pt] \frac{3}{2}+i\mu \end{array}\Bigg| \, u^2,\frac{u^2}{v^2}\Bigg]+c.c.\, ,
\end{align}
where
\begin{align}
	a_{mn} &\equiv \frac{   (\frac{1}{2})_{m+n}(1)_{m+n} }{(\frac{5+2i\mu}{4})_{m+n}(\frac{5-2i\mu}{4})_{m+n}}\frac{i(1)_m}{\mu(1+2i\mu)}\frac{(\frac{5+2i\mu}{4})_{n}(\frac{5-2i\mu}{4})_{n}(\frac{1}{2}+i\mu)_n}{(\frac{3}{2}+i\mu)_n}\, ,\\
	b_{mn} &\equiv \frac{   (1)_{m+n}(\frac{3}{2})_{m+n} }{(\frac{7+2i\mu}{4})_{m+n}(\frac{7-2i\mu}{4})_{m+n}}\frac{(1)_m}{i\mu(1+2i\mu)}\frac{(\frac{7+2i\mu}{4})_{n}(\frac{7-2i\mu}{4})_{n}(\frac{1}{2}+i\mu)_n}{(\frac{3}{2}+i\mu)_n}\, .
\end{align}
The generalized hypergeometric function $F^{a|b|c}_{\,d|e|f}$ is known as the Kamp\'e de F\'eriet function; the first two indices $a$, $d$ denote the mixed terms in the double sum. 

\subsection{Limiting Behaviors}

It turns out that the solution in the limit $u\ll v$ takes a much simpler form
\begin{align}
	\hat F_<(u,v\gg u) = \frac{1}{\mu^2+\frac{1}{4}}\, {}_3 \hskip -0.5pt F_2\Bigg[\begin{array}{c}\frac{1}{2},1,1\\[2pt] \frac{5}{4}+\frac{i\mu}{2},\frac{5}{4}-\frac{i\mu}{2} \end{array}\Bigg|\, u^2\Bigg]\, .\label{3F2}
\end{align}
Consider the behavior of this function as we approach the singularities near $u\to \pm 1$. In our previous discussion, with $|v|<1$, we had to switch from $\hat F_<$ to $\hat F_>$ at $u=v\in (-1,1)$, so we didn't encounter this singularity, but we do for $|v|>1$. Using the formula for the behavior of the generalized hypergeometric function near the boundary of the disk of convergence \eqref{GenHyperBoundary}, we can determine the leading singular behavior of $\hat F_<$ as $u\to \pm 1$ for large $v$:
\begin{align}
	\lim_{u\to \pm 1}\hat F_<(u,v\gg u) = \frac{\Gamma(\frac{1}{4}-\frac{i\mu}{2})\Gamma(\frac{1}{4}-\frac{i\mu}{2})}{4\sqrt{\pi}}\log(1\mp u)\, .\label{3F2behavior}
\end{align}
Of course, this can be directly determined from the series expansion as well. Let us first take the term with $n=0$, and look at $\partial_u\hat F_<(u,v)$: 
\begin{align}
	\sum_{m=0}^\infty \frac{(2m+1)!}{(\frac{1}{4}+\mu^2)(\frac{25}{4}+\mu^2)\cdots ((\frac{1}{2}+2m)^2+\mu^2)}u^{2m}\, .
\end{align}
Now, we can see why the series diverges as $u\to 1$: for large $m$, the coefficient of $u^{2m}$ becomes $m$-independent; indeed, by again examining residues in $\mu$, we can easily deduce that
\begin{align}
	\frac{1}{2}\lim_{m\to \infty}\frac{(2m+1)!}{(\frac{1}{4}+\mu^2)(\frac{25}{4}+\mu^2)\cdots ((\frac{1}{2}+2m)^2+\mu^2)} = \frac{\Gamma(\frac{1}{4}-\frac{i\mu}{2})\Gamma(\frac{1}{4}-\frac{i\mu}{2})}{4\sqrt{\pi}}\, ,
\end{align}
reproducing the coefficient in \eqref{3F2behavior}.

\vskip 4pt
Finally, we note that in the limit $\mu \to 0$, which lies on the boundary of the principal and complementary series, the homogeneous solutions to our differential equation become elliptic integrals:
\begin{align}
	\hat F_\pm(u)\, \propto\, u^{\frac{1}{2}\pm i\mu}\, {}_2F_1\Bigg[\begin{array}{c} \frac{1}{4}\pm i\mu,\hs \frac{3}{4}\pm i\mu\\[2pt] 1 \pm i\mu \end{array}\Bigg|\,u^2\Bigg] \ &\xrightarrow{\mu\to 0}\ \frac{2}{\pi}\sqrt{\frac{u}{1+u}}\,K\bigg(\frac{2u}{1+u}\bigg)\, ,
\end{align}
where $K$ 
is the complete elliptic integral of the first kind. Because the two homogeneous solutions $\hat F_\pm$ degenerate into a single solution for $\mu \to 0$, we must find a second linearly independent homogeneous solution in this limit. 
Looking for a series solution around $u=0$, the general homogeneous solution can be written as
\begin{align}
	(A_1+A_2\log u)\sum_{n=0}^\infty r_n u^{\frac{1}{2}+2n}+A_2\sum_{n=0}^\infty  \tilde r_n u^{\frac{1}{2}+2n}\, ,\label{mu0homogeneous}
\end{align}
with free coefficients $A_1$, $A_2$, and
\begin{align}
	r_n \equiv \frac{3\sqrt 2\pi}{32}\frac{(\frac{5}{4})_{n-1}(\frac{7}{4})_{n-1}}{(n!)^2}\, , \quad \tilde r_n \equiv (H_{2n-1/2}-H_n)r_n\, ,
\end{align}
where $H_n$ is the $n$-th harmonic number, $H_n\equiv \sum_{k=1}^n k^{-1}$. We see that the leading behavior of this solution as $u\to 0$ is $\sqrt{u}\log u$, as expected from our discussion in \S\ref{sec:exchange}.

\vskip 4pt
It will be convenient to take the two linearly independent homogeneous solutions to be
\begin{align}
	\hat F_K(u) &\equiv \sum_{n=0}^\infty r_n u^{\frac{1}{2}+2n}\, ,\label{FKseries}\\
	\hat F_G(u) &\equiv\frac{2}{\pi^2}(\log 2-\log u)\sum_{n=0}^\infty r_n u^{\frac{1}{2}+2n}-\frac{2}{\pi^2}\sum_{n=0}^\infty \tilde r_n u^{\frac{1}{2}+2n}\, ,\label{FGseries}
\end{align} 
where $\hat F_K$ is just an expansion of the complete elliptic integral around $u=0$.
These solutions were normalized such that their Wronskian is given by $W[\hat F_G,\hat F_K]=1/(1-u^2)$. Moreover, the coefficient for $\hat F_G$ has been chosen so that it is regular at $u=1$. This can be seen from the large-$n$ behavior of the series coefficients: 
\begin{align}
	\lim_{n\to\infty} r_n\log 2-\tilde r_n = O(1/n^2)\, ,
\end{align} 
which implies that the sum is finite at $u=1$.  An important feature of the function $\hat F_G$ is that it is {\it not} symmetric in $u\leftrightarrow -u$ due to the presence of the logarithmic term. Indeed, it is easily checked that $\hat F_G$ behaves as $\log(1+u)$ when $u\to -1$, whilst being regular at $u=1$ by construction. Having said that, considering it as a function of $u^2$ for a moment would allow us to write 
\begin{align}
	\hat F_G(u) = G^{2,0}_{2,0}\Bigg[\begin{array}{c}\frac{1}{2},\, 1\\[2pt] \frac{1}{4},\, \frac{1}{4} \end{array}\Bigg|\, u^2\Bigg]\, ,\label{meijer}
\end{align}
where $G^{m,n}_{p,q}$ is	 the Meijer G-function. However, this representation {\it cannot} be extended beyond the physical interval $u\in[0,1]$. We will therefore find it more useful to use the series representation of the solutions \eqref{FKseries} and \eqref{FGseries} for analyzing singularities.

\vskip 4pt
The particular solution we obtained for generic $\mu$ in \S\ref{sec:exchange} does not have a singularity at $\mu=0$, so the limit $\mu\to 0$ is perfectly well defined. The procedure of fixing the homogeneous solutions is the same as for general values of $\mu$. Our Wronskian normalization implies that the matching condition is given by
\begin{align}
	\hat F_{<}(u,v)-\hat F_{>}(u,v) = \pi\Big(\hat F_G(u)\hat F_K(v)-\hat F_K(u)\hat F_G(v)\Big)\, .
\end{align}
The most general solution that is symmetric in $u\leftrightarrow v$ is then
\begin{align}
	\hat F(u,v)&= \hat F_<(u,v) +  \frac{\pi}{2}\Big\{\big[\hat F_K(u)\hat F_G(v)-\hat F_K(v)\hat F_G(u)\big]  \nn 
	&\quad + \beta_K \hat F_K(u)\hat F_K(v) + \beta_G \hat F_G(u)\hat F_G(v)+ \beta_{KG}\big[\hat F_K(u)\hat F_G(v)+\hat F_K(v)\hat F_G(u)\big]\Big\}\, .
\end{align}
Demanding the absence of a folded singularity at $u=1$ now uniquely fixes $\beta_K=0$ and $\beta_{KG}=1$. The remaining parameter $\beta_G$ is again fixed by going to the factorization channel
\begin{align}
	\lim_{u,v\to -1} \hat F(u,v) = \frac{\beta_G}{2\pi}\hs \log(1+u)\log(1+v)\, .
\end{align}
Fixing the coefficient to be $\frac{1}{2}$, as before, we get $\beta_G=\pi$. The leading behavior of this solution in the collapsed limit is then
\begin{align}
	\lim_{u,v\to 0} \hat F(u,v) = \sqrt{uv}\hs\log u\log v\, .
\end{align}
This agrees with the leading behavior of the result \eqref{collapsedmu0} as $\mu\to 0$.

\newpage
\section{Details of the $\mathbf{\Delta=3}$ Solutions}
\label{app:D3}

In this appendix, we present important technical details of the $\Delta =3$ solutions of Section~\ref{sec:Massless}. We begin 
by analyzing the conformal constraint equations for the four-point functions of massless scalar fields in de Sitter space.  This is the case that is most relevant to inflation. The analysis of these equations is very similar to that performed in Section~\ref{sec:dS-4pt}, but some details are different.
We pay particular attention to the shapes that break the shift symmetry of the inflaton, which includes the famous example of local non-Gaussianity. After classifying the possible contact terms in \S\ref{sec:contactD3}, we derive the solution for scalar exchange in \S\ref{sec:treeD3}. In \S\ref{sec:flat}, we confirm that the flat-space limit of the solution has the expected singularity.  
Finally, in \S\ref{sec:Inf3}, we derive the inflationary three-point functions due to scalar exchange using only the boundary perspective.

\subsection{Contact Interactions}
\label{sec:contactD3}

 The simplest solution of the conformal invariance equations doesn't depend on $s$. Writing 
\beq
{\cal F}=O_{12}O_{34}\hs h(k_{12},k_{34})\,,\label{O12O34h}
\eeq  
the $(D_1-D_3){\cal F}=0$ constraint takes the form
\beq
\left[\left(\partial_{k_{12}}^2 - \frac{2}{k_{12}} \partial_{k_{12}}\right)  - \left(\partial_{k_{34}}^2 - \frac{2}{k_{34}} \partial_{k_{34}}\right)  \right] h = 0\, .
\eeq
By the same logic as in \eqref{equ:anssol1234}, this equation is solved by 
\beq
h(k_{12},k_{34}) = \left(1- \frac{k_{12} k_{34}}{k_t}\partial_{k_t}\right) \tilde h(k_t)\, , \label{equ:f}
\eeq
where the function $\tilde h(k_t)$ will be fixed by its scaling dimension. 
The function $\tilde h$ has mass dimension $+3$, so the simplest choice is $\tilde h =k_t^3$.  This leads to $h = (k_1+k_2)^3 + (k_3+k_4)^3$, and hence 
\beq
{\cal F}_{\rm loc} = O_{12}O_{34} \left[ (k_1+k_2)^3 + (k_3+k_4)^3\right] = \sum_{n=1}^4 k_n^3 \, ,\label{Delta3localF}
\eeq
which corresponds to local non-Gaussianity. 
As we explained in Section~\ref{sec:Massless}, the solution (\ref{Delta3localF}) is somewhat trivial as it arising from a zero mode of the operator enforcing conformal invariance.  Indeed local non-Gaussianity doesn't contain a nontrivial dependence on any sum of the momenta~$k_n$, and can be removed by a field redefinition. 

\vskip 4pt
The reason we haven't found a more nontrivial solution is that the choice
 $\tilde h=k_t^3$ determined by scale invariance was too restrictive. The simplest nontrivial
solution follows from the ansatz
\beq
\tilde h = -\frac{1}{3} \,k_t^3\log k_t\, , \label{equ:g}
\eeq 
where the factor of $-1/3$ was introduced for later convenience.
This solution mildly breaks the scaling symmetry by local terms of the form above. To capture this we should have allowed for local terms on the right-hand side of (\ref{WI_D}). When transformed to position space these terms become delta functions which only have support at coincident points.  This ansatz for $g$ leads to the four-point function  $\C_0$ corresponding to a $\upphi^4$ interaction in the bulk, which indeed has a logarithmic infrared singularity.  
Substituting (\ref{equ:g}) into (\ref{equ:f}) and \eqref{O12O34h}, the four-point function is given by $\C_0 = s^3O_{12}O_{34}\hs\hat\C_0$ with
\beq
\hat {\cal C}_0(u,v)\equiv  \frac{1}{3} \left[\left(\frac{1}{u^3} + \frac{1}{v^3}\right)\log\left(\frac{ u v}{u+v}\right)+\left(\frac{1}{u}+\frac{1}{v}\right) \frac{1}{u v} \right]  , \label{equ:C0}
\eeq
where the argument of the logarithm was made dimensionless by introducing of the momentum $s$ as an IR cutoff. Strictly speaking, one should use a late-time cutoff to make the logarithm dimensionless, but we will ignore this subtlety for now. 

\vskip 4pt
Higher-derivative contact interactions give us rational functions, which depend nontrivially on~$s$. We don't expect the higher contact terms to have a logarithmic dependence on the momenta, so it is now easier to solve for the dimensionless function $\hat \F_c$. 
The simplest rational solution of (\ref{equ:d3cons}) with a pole as $k_t \to 0$, or $u+v \to 0$, is
\beq
\hat {\cal C}_1(u,v)\equiv\frac{u^2+v^2+u v-1}{u v(u+v)}\, .
\eeq
Higher-order contact interactions, corresponding to higher-order poles in $k_t$, are generated by applying $\tilde\Delta_u$ to this result, i.e.~$\hat {\cal C}_n = \tilde\Delta_u^{n-1}\hat {\cal C}_1$. Acting with $\tilde\Delta_u$ on $\hat {\cal C}_0$, however, we obtain 
\beq
\tilde\Delta_u \hat {\cal C}_0 = \hat {\cal C}_1 -\frac{1}{u^3}\, , \label{equ:rec}
\eeq
where the extra local term arises from the logarithmic term in $\hat {\cal C}_0$. This is because $\tilde\Delta_u$ is a combination of 
dilatation and special conformal transformations, and the dilatation operator doesn't quite annihilate $\hat {\cal C}_0$. However, as we have discussed above, the term $1/u^3$ is local and can be removed by a field redefinition. In summary, the most general
contact interactions are of the form 
\begin{align}
\hat {\cal F}_c(u,v) &\,=\, c_0\hs \hat {\cal C}_0(u,v) +
 \sum_{n=1}^\infty c_n \tilde\Delta_u^{n-1} \hat {\cal C}_1(u,v)\, .
\end{align}
As before, the solution is symmetric in $u \leftrightarrow v$. Notice that, starting from $\hat {\cal C}_2$, all contact interactions vanish in the soft limit $k_4\to 0$ (or $v\to 1$), because the massless scalar has vanishing gradients in the soft limit and is coupled in a shift-symmetric fashion.\footnote{We will show in the next section that the shift-symmetric solutions to the $\Delta=3$ equation can also be obtained directly from solutions of the $\Delta=2$ equation.} In fact, even $\hat {\cal C}_1$ trivializes into a local term in this limit. The only contact interactions that do not vanish in the soft limit are local non-Gaussianity and $\hat {\cal C}_0$. 

\vskip 4pt
In the flat-space limit, $k_t\to 0$, the physical contact terms, ${\cal C}_n \equiv  s^3 O_{12}O_{34}\, \hat {\cal C}_n$,  satisfy 
\beq\label{equ:Cnflatd3}
\lim_{k_t \to 0} \frac{ {\cal C}_n}{\prod_m k_m} = (2n)!  \,\frac{s_{\rm flat}^n}{k_t^{2n+1}}\, .
\eeq
This is the same as (\ref{equ:Cnflat}), after taking into account the factor of $\Pi_m k_m$ arising from the different normalization of the mode functions of massless scalars, $\upphi \propto k^{-1}(1+i k\eta)e^{-i k \eta}$, and conformally coupled scalars, $\upvarphi  \propto \eta e^{-i k\eta}$. 
 
\subsection{Tree-Level Exchange}
\label{sec:treeD3}

As before, tree exchange corresponds to an inhomogeneous equation, whose source term is given by one of the contact interactions:  
\beq
\Big[\tilde\Delta_u + \tilde M^2 \Big] \hat {\cal F}_n = (-1)^n\hs\hat {\cal C}_n\, , 
\label{equ:INH3}
\eeq
where $\tilde M^2 \equiv \mu^2+\frac{9}{4}$. Again, the solutions $\hat {\cal F}_n$ associated with the different contact interactions~$\hat {\cal C}_n$ are related to each other by a recursive relation:
\begin{align}
\hat {\cal F}_{n+1} &= \tilde M^2 \hat {\cal F}_{n} - \hat {\cal C}_{n}  + \frac{1}{\tilde M^2} \frac{1}{u^3} \hskip 1pt \delta_{n0}\, .\label{FnRecur}
\end{align}
This shows that all solutions can be inferred from the simplest exchange solution $\hat {\cal F}_1$ and knowledge of the contact terms $\hat {\cal C}_n$.
We will therefore solve for $\hat {\cal F}_1$ explicitly, and then use the above relationship to infer $\hat {\cal F}_{n}$.

\vskip 4pt
The homogeneous solutions of (\ref{equ:INH3}) are
\beq
\hat {\cal F}_\pm(u)  =  \frac{1}{4\mu^2} \left(\frac{i u}{2\mu}\right)^{-\frac{3}{2} \pm i\mu }  {}_2F_1\Bigg[\begin{array}{c} -\frac{3}{4}\pm\frac{i\mu}{2},\hs \frac{3}{4}\pm\frac{i\mu}{2}\\[2pt] 1\pm i\mu \end{array}\Bigg|\, u^2\Bigg]\, , \label{equ:homo2}
\eeq
for which the Wronskian is $W(\hat {\cal F}_+,\hat {\cal F}_-)=1/u^4$. In the limit $u\to 1$, these solutions become 
\begin{align}
	\lim_{u\to 1}\hat \F_\pm(u) = \tilde \alpha_\pm (1-u)\log(1-u)\, , \quad \tilde \alpha_\pm \equiv \frac{1}{2\mu^2}\left(\frac{i}{2\mu}\right)^{-\frac{3}{2}\pm i\mu} \frac{\Gamma(1\pm i\mu)}{\Gamma(\frac{3}{4}\pm \frac{i\mu}{2})\Gamma(-\frac{3}{4}\pm\frac{i\mu}{2})}\, .
\end{align}
This is similar to the behavior of the homogeneous solutions \eqref{equ:Shomo} for $\Delta=2$, but this time the leading singularity is given by $(1-u)\log(1-u)$. The expected $\log(1-u)$ singularity is reproduced for the physical four-point function once we act on this solution with the operator~$O_{12}$.

\vskip 4pt
Using the contact term $\hat\C_1$ as the source, the exchange solution $\hat \F_1$ satisfies 
\beq
\left[ u^2(1-u^2) \partial_u^2 +4 u(1-u^2)\partial_u + \left(\mu^2 + \frac{9}{4}\right) \right] \hat {\cal F}_1 = \frac{1-(u^2+v^2 + uv)}{uv(u+v)}\, . \label{equ:F1}
\eeq
As discussed in \S\ref{sec:exchange}, the series expansion of $\hat {\cal F}_{1,<}(u,v)$ is uniquely fixed once we demand analyticity at the origin, since the homogeneous solutions have branch points at $u=0$. This time, however, the solution will be meromorphic in $u$, having a single pole rather than being analytic at $u=0$.  
We therefore consider the following ansatz
\beq
\hat {\cal F}_{1,<}(u,v)=\sum_{m,n=0}^\infty d_{mn}\,u^{2m-3}(u/v)^n \, . \label{equ:F1<}
\eeq
This form of the series solution is motivated by the series expansion of the source term in (\ref{equ:F1}). 
The coefficients $d_{mn}$ are given by 
\begin{align}
d_{mn}&=\begin{cases}\displaystyle \ \frac{n-1}{2m+n-3}\frac{\frac{1}{4}+\mu^2}{(n-\frac{3}{2})^2+\mu^2}\,c_{m-1,n} & m\ne 0,\, n\ne 1\,,\\[5pt]\ \displaystyle c_{0,n-2} & m=0,\, n\ne\{0,1\}\,, \\[5pt] \ 0 & \text{otherwise}\, ,
\end{cases}
\end{align}
where the coefficients $c_{mn}$ are the same as in~\eqref{cmn}:
\beq
c_{mn}  = \frac{(-1)^n(n+1)(n+2)\cdots(n+2m)}{[(n+\frac{1}{2})^2+\mu^2][(n+\frac{5}{2})^2+\mu^2]\cdots [(n+\frac{1}{2}+2m)^2+\mu^2]}\, .\label{cmn2}
\eeq
 We see that the series has a simple pole at $u=0$, but no higher-order poles or branch cuts. As before, the solution which is meromorphic around $u=\infty$, denoted $\hat {\cal F}_{1,>}(u,v)$, can be obtained by demanding symmetry under the exchange of $u$ and $v$, so that $\hat {\cal F}_{1,>}(u,v) = \hat {\cal F}_{1,<}(v,u)$. 

\vskip 4pt
Once again, the difference between these particular solutions gives a homogeneous solution, so we can write 
\beq
\hat {\cal F}_{1,<}(u,v) - \hat {\cal F}_{1,>}(u,v)= \sum_\pm \A_\pm(v;\mu) \hat {\cal F}_\pm(u) \, .
\eeq
Evaluating this at $u= v$ gives $\A_\pm (v;\mu)= \mp \hs \tilde a(v;\mu) \hat {\cal F}_\mp (v)$, for some function $a(v;\mu)$. Matching the $u$-derivative at $u = v$ fixes the function in terms of the Wronskian of the homogeneous solutions, namely
\begin{align}
\left.\left(\partial_u \hat {\cal F}_{1,<} - \partial_u \hat {\cal F}_{1,>}\right)\right|_{u \to v} &= \tilde a(v;\mu)\Big(\hat {\cal F}_+(v) \hat {\cal F}_-^{\hskip 1pt\prime}(v) - \hat {\cal F}_+^{\hskip 1pt \prime}(v)  \hat {\cal F}_-(v) \Big) = \frac{\tilde a(v;\mu)}{v^4} \, .
\end{align}
It remains to evaluate the left-hand side
\begin{align}
\left.\left(\partial_u \hat {\cal F}_{1,<} - \partial_u \hat {\cal F}_{1,>}\right)\right|_{u \to v} = \sum_{m,n=0}^\infty (2m+2n-3)d_{mn}v^{2m-4}\, .
\end{align}
Somewhat remarkably, the sum over $n$ for all $m\ne0$ is zero, while, for $m=0$, we get
\begin{align}
	\sum_{n=0}^\infty (2n-3)d_{0n} =\frac{\pi}{\cosh\pi\mu}\, .\label{D3cosh}
\end{align}
This allows us to fix $\tilde a(v;\mu)={\pi/\cosh\pi\mu}$ independent of $v$, i.e.~we have the same matching condition as in the $\Delta=2$ case:
\beq
\hat {\cal F}_{1,<}(u,v) - \hat {\cal F}_{1,>}(u,v)= \frac{\pi}{\cosh\pi\mu}\Big(\hat\F_+(u)\hat\F_-(v)-\hat\F_-(u)\hat\F_+(v)\Big)\equiv \hat \F_h \, .\label{equ:match}
\eeq
The solution of the differential equation which is meromorphic around $u=0$ therefore is
\beq
\hspace{-0.35cm}\hat {\cal F}_{1,<}(u,v) = \begin{cases}  \displaystyle\, \sum_{m,n=0}^\infty d_{mn}\,u^{2m-3}(u/v)^{n} & \text{$u \le v$}\,, \\[20pt]
 \displaystyle \, \sum_{m,n=0}^\infty d_{mn}\,v^{2m-3}(v/u)^{n} + \hat {\cal F}_h(v,u) & \text{$u \ge v$}\,. \end{cases}
 \label{equ:Fsmall}
\eeq
As before, the physical four-point function is obtained after symmetrizing $u$ and $v$, and removing unphysical singularities. We first write the most general homogeneous solution in terms of $\hat\F_\pm$ as in \eqref{generalF}, and then demand the absence of the singularity at $u=1$. This again fixes the coefficients to be $\beta_\pm = -(\beta_0\pm 1)\tilde \alpha_\mp/\tilde \alpha_\pm$. To determine the remaining parameter $\beta_0$, we look at the singular behavior in the limit $u=v\to -1$. Isolating the physically relevant term, we obtain
\begin{align}
\lim_{u,v\to -1} \hat {\cal F}_1(u,v) 
 &\,=\, \frac{i\beta_0(\mu^2+\frac{9}{4})}{2}   \sinh\pi\mu \,(1+u)(1+v)\log(1+u)\log(1+v)\, . \label{F1uvminus1}
\end{align}
This is similar to the result in \eqref{uvminus1}. The extra factors of $(1+u)(1+v)$ are present in $\hat {\cal F}_1$, but disappear in the physical four-point function ${\cal F}_1$ upon the action of $O_{12}O_{34}$. To understand the appearance of the  prefactor $\mu^2+\frac{9}{4}$, recall that we are solving for the four-point function $\hat {\cal F}_{1}$ with $\hat {\cal C}_1$ as the source term. This is related to the solution $\hat {\cal F}_{0}$ with the source term~$\hat {\cal C}_0$ by $\hat {\cal F}_{1}= (\mu^2+\frac{9}{4})\hat {\cal F}_{0}+\cdots$, cf.~\eqref{FnRecur}. Since we have implicitly set the coupling constant for the contact term $\hat {\cal C}_0$ to be unity, we expect  the solution $\hat {\cal F}_1$ to have an overall factor of $\mu^2+\frac{9}{4}$. The correct normalization of the singularity in the limit $u,v \to -1$ therefore requires $\beta_0=1/i\sinh\pi\mu$. 
As a consistency check, we look at the non-analytic contribution to the collapsed limit: 
\begin{align}
	\lim_{u,v\to 0}O_{12}O_{34}\hat {\cal F}_1(u,v) \,=\ & \frac{1}{512\pi}\left(\frac{uv}{4}\right)^{-\frac{3}{2}+i\mu}(1+i\sinh \pi\mu)\frac{(\frac{3}{2}+i\mu)^2(\frac{5}{2}+i\mu)^2\Gamma(\frac{1}{2}+i\mu)^2\Gamma(-i\mu)^2}{\mu^2+\frac{9}{4}} \nonumber \\ &+c.c.\, ,
\end{align}
which agrees with equation (5.104) in~\cite{Arkani-Hamed:2015bza}.

\subsection{Flat-Space Limit} 
\label{sec:flat}

To analyze the flat-space limit, it is convenient to look at the conformal invariance equation with the source term given by (\ref{equ:C0}): 
\beq
\left[u^2(1-u^2) \partial_u^2 +4 u(1-u^2)\partial_u + \left(\mu^2+\frac{9}{4}\right)\right] \hat {\cal F}_0 = \frac{u+v}{3(uv)^2} - \frac{u^3+v^3}{3(uv)^3} \log\left(\frac{u+v}{uv}\right) . 
\label{equ:INH4}
\eeq
In the limit $u \to -v$, the leading singularity is 
\beq
\lim_{u \to -v} \frac{\partial^2 \hat {\cal F}_0}{\partial u^2} = \frac{(u+v)}{v^6(1-v^2)} \log(u+v)  \quad \Rightarrow \quad \lim_{u \to -v} \hat {\cal F}_0 =  \frac{1}{6} \frac{(u+v)^3}{v^6(1-v^2)} \log(u+v)  \, .
\label{equ:F0}
\eeq
For the physical correlator, this implies
\beq
\lim_{k_t\to 0}\,\frac{{\cal F}_0}{\prod_m k_m} = \lim_{k_t\to 0}\,\frac{s^3 O_{12}O_{34} \hat {\cal F}_0}{\prod_m k_m} =  \frac{1}{s_{\rm flat}}\, k_t \log k_t \,.
\eeq
We see that the coefficient of the $k_t \log k_t$ singularity is again given by the high-energy limit of the flat-space amplitude, $A_{\rm flat} = 1/s_{\rm flat}$.
Moreover, equation (\ref{equ:F0}) implies that $A_{\rm flat}$ can be related to the discontinuity of the third derivative of $\hat {\cal F}_0$: 
\beq
A_{\rm flat} = \frac{v^8}{s^2} \lim_{u \to -v} \frac{{\rm Disc}[\hat {\cal F}_0^{\hskip 1pt\prime \prime \prime}]}{2\pi i} \, . \label{equ:AFLAT}
\eeq
As before, we would like to relate this to the discontinuity of the homogeneous solution in the same limit.  

\vskip 6pt
We first recall that
$\hat {\cal F}_0$ and $\hat {\cal F}_1$ are related, up to a local term, by 
\beq
\hat {\cal F}_1= \tilde M^2 \hat {\cal F}_0 - \hat {\cal C}_0\, . \label{equ:FF1}
\eeq
The non-analytic behavior of $\hat {\cal C}_0$ in the limit $u \to -v$ is 
\begin{align}
	\lim_{u\to -v}\hat {\cal C}_0 &=\left(\frac{u+v}{v^4}+\frac{2(u+v)^2}{v^5}+\frac{10(u+v)^3}{3v^6}+\cdots \right) \log(u+v)+\cdots\, . \label{equ:C0exp}
\end{align}
As before, we will use the matching condition (\ref{equ:match}) to relate the discontinuities of  $\hat {\cal F}_0$  and $\hat {\cal C}_0$ to that of $\hat {\cal F}_h(u,v)$.
We expect the homogeneous solution to have discontinuities at every derivative in $u$, but only the discontinuity in the third derivative is related to the pole in the flat-space amplitude. The discontinuities in the first and second derivatives are related to $\hat {\cal C}_0$, which contributes a constant term in the flat-space limit \eqref{equ:Cnflatd3}.
Using (\ref{equ:FF1}), we can write (\ref{equ:AFLAT}) as 
\beq
A_{\rm flat} = \frac{v^8}{s^2} \frac{1}{\tilde M^2} \lim_{u \to -v} \frac{{\rm Disc}[\hat {\cal F}_1^{\hskip 1pt\prime \prime \prime}] \,+\, {\rm Disc}[\hat {\cal C}_0^{\hskip 1pt\prime \prime \prime}]}{2\pi i} \,. \label{equ:AFLAT2}
\eeq
The discontinuity of $\hat {\cal F}_1^{\hskip 1pt\prime \prime \prime}$ can be related to that of $\hat {\cal F}_h^{\hskip 1pt\prime \prime \prime}$ via the matching condition (\ref{equ:match}), while that of $\hat {\cal C}_0^{\hskip 1pt\prime \prime \prime}$ can be extracted from (\ref{equ:C0exp}).  
 This leads us to 
 \beq
 A_{\rm flat} =  \frac{v^8}{s^2} \frac{1}{\tilde M^2} \left(\lim_{u \to -v}  \frac{{\rm Disc}[\hat {\cal F}_h^{\hskip 1pt\prime \prime \prime}]}{2\pi i}  + \frac{20}{v^6} \right) . \label{equ:AFLAT3}
 \eeq
 Using the differential equation,  we get
 \beq
\hat {\cal F}_\pm^{\hskip 1pt\prime \prime \prime} = \left(\frac{20}{u^2} - \frac{\tilde M^2}{u^2(1-u^2)}\right) \hat {\cal F}_\pm^{\hskip 1pt\prime} - \frac{2 \tilde M^2(4u^2-3)}{u^3(1-u^2)^2 } \hat {\cal F}_\pm\, ,
\eeq
 which implies that
 the discontinuity of the third derivative of $\hat {\cal F}_h$ can be related to the Wronskian of the homogeneous solution: 
 \begin{align}
\lim_{u \to -v} \frac{{\rm Disc}[\hat {\cal F}_h^{\hskip 1pt\prime \prime \prime} (u,v)]}{2\pi i } &\,=\, \frac{\pi}{\cosh \pi \mu}  \lim_{u \to -v}  \frac{{\rm Disc}[\hat {\cal F}_+^{\hskip 1pt\prime \prime \prime} (u) \hat {\cal F}_-(v) - \hat {\cal F}_-^{\hskip 1pt\prime \prime \prime} (u) \hat {\cal F}_+(v)]}{2\pi i } \nonumber \\[4pt]
&= \left(\frac{20}{v^2} - \frac{\tilde M^2}{v^2(1-v^2)}\right) W[\hat {\cal F}_-, \hat {\cal F}_+] \nn
&= \frac{\tilde M^2}{v^6(1-v^2)}-\frac{20}{v^6}\, .
\end{align}
We therefore get 
\beq
A_{\rm flat} = \frac{v^2}{s^2} \frac{1}{1-v^2} = \frac{1}{s_{\rm flat}}\, ,
\eeq
as expected.

\subsection{Inflationary Bispectra}
\label{sec:Inf3}

In Section~\ref{sec:inflation}, we used bulk arguments to find a prescription to obtain the three-point functions from the $\Delta=2$ de Sitter four-point functions. We will now show that the same results can be obtained purely from the boundary perspective by analyzing the constraint equations. For concreteness, we will restrict the presentation to scalar exchange. 

\vskip 4pt
Consider the de Sitter four-point function for $\Delta_4=3-\epsilon$ and  $\Delta_n=3$, $n\ne 4$, and write it~as 
\beq
{\cal F}= \bar {\cal F}+\epsilon \,{\cal I} +\cdots\, , \label{equ:I}
\eeq
where $\bar {\cal F}$ is the four-point function for $\Delta=3$ and ${\cal I}$ is its (slow-roll) correction for $\epsilon \ne 0$.  For derivatively-coupled interactions, the $k_4 \to 0$ limit of  
$\bar\F$ is trivial, so the interesting part of the inflationary three-point function is given by\footnote{Recall that this is just the $s$-channel contribution. The full inflationary three-point function should include the $t$- and $u$-channel contributions. We add these by symmetrizing the final answer in the momenta $k_1$, $k_2$, $k_3$.} 
\beq
B = \epsilon \lim_{k_4 \to 0} {\cal I}+ \text{perms}\, .
\eeq
As we will see, ${\cal I}$ will be determined by the properties of ${\cal F}$ away from the soft limit. For scalar exchange, the four-point function is independent of $t$ and 
the conformal symmetry constraints 
$(D_m-D_n)\F=0$, with $D_m =  \bar D_m + \epsilon\hskip 1pt d_m$, at order $\epsilon$, become 
\begin{align}
(\bar D_1- \bar D_2 )\,{\cal I} &= 0\, , \label{equ:12}\\[4pt]
(\bar D_1- \bar D_3 )\,{\cal I} &= \frac{1}{s}\partial_s {\cal F}\, ,  \label{equ:13}\\
(\bar D_1- \bar D_4 )\,{\cal I} &=  \frac{1}{s}\partial_s {\cal F} \,+\, \frac{2}{k_4}\partial_{k_4} {\cal  F} \, . \label{equ:14}
\end{align}
All other equations are linearly dependent on these equations. 

\vskip 4pt
Equation~(\ref{equ:12}) is solved by the ansatz\hs\footnote{Notice that the source function $b$ in (\ref{equ:III})  is strictly speaking {\it not} the same as that used in Section~\ref{sec:inflation}. We will choose to use the same symbol, however, to highlight that it plays the same role of generating the inflationary bispectrum.}  
\beq
{\cal I} = O_{12}\,b(k_{12},k_3,k_4,s)\, .\label{equ:III}
\eeq
To determine the function $j$ in the soft limit $k_4\to0$, we only need to solve equation (\ref{equ:13}).
To see this, we note that the small $k_4$ expansion of ${\cal I}$ can be written as ${\cal I}= {\cal I}_0+k_4^2 \,{\cal I}_1+\cdots$, where the absence of a term linear in $k_4$ follows from $k_4^{-1}\partial_{k_4} {\cal F}$ having no pole in $k_4$. Equations (\ref{equ:13}) and (\ref{equ:14}) determine ${\cal I}_0$ and ${\cal I}_1$, respectively. In the soft limit $k_4\to0$, we therefore only need to solve equation (\ref{equ:13}). 

\vskip 4pt
To analyze (\ref{equ:13}), we must find $s^{-1} \partial_s {\cal F}$ and evaluate it in the limit $k_4 \to 0$ (and hence $s \to k_3$). A quick calculation gives
\beq
\lim_{k_4 \to 0}\frac{1}{s}\partial_s {\cal F} \,=\,  k_3 O_{12} \lim_{v\to 1} \left [ 3 \hat {\cal F}(u,v)+ v \partial_v \hat {\cal F}(u,v)\right] , \label{equ:FFF}
\eeq
where we have used that $\partial_u \hat {\cal F} = 0$ in the limit $v \to 1$. We can also discard the $\hat {\cal F}(u,1)$ term, as it will always be proportional to a zero mode of $\tilde \Delta_u$ and can therefore be removed by the addition of a suitable local term.
The differential operator on the left-hand side of (\ref{equ:13}), $( D_1 - D_3 )\,{\cal I}$, 
can be analyzed in the same way as in our discussion of the de Sitter four-point function with $\Delta=3$. In particular, we write  
\beq
  \lim_{k_4 \to 0} {\cal I} =k_3^3\, O_{12} \,\hat b(k_3/{k_{12}})\, . \label{equ:bbbb}
\eeq This ansatz is compatible with the momentum dependence of the source term, so we can discard the $s$ dependence in the $ D_n$'s. Moreover, by scaling, we can pull out the factor of $k_3^3$ and make $\hat b$ dimensionless, with functional dependence on the ratio of momenta. We will  write $u\equiv k_3/{k_{12}}$, but it is important to keep in mind that there is no $s$ dependence in $\lim_{k_4 \to 0}{\cal I}$. We then get
\begin{align}
( D_1 - D_3)   \lim_{k_4 \to 0} {\cal I}  &=-k_3 O_{12}\left[u^2(1-u^2) \partial^2_u+4u(1-u^2) \partial_u  \right] \hat b \nonumber \\[4pt]
&\equiv -k_3 O_{12} \tilde \Delta_u \hs\hat b \, , \label{equ:bbb}
\end{align}
where $\tilde \Delta_u$ is the same operator as before, although $u$ has nothing to do with $s$ anymore. Finally, comparing (\ref{equ:bbb}) and (\ref{equ:FFF}), we get  
\begin{eBox}
\beq\label{equ:infl3pt}
\tilde \Delta_u \hs\hat b = -  \lim_{v\to 1}   \partial_v \hat {\cal F}(u,v)\, .
\eeq
\end{eBox}
We see that the momentum dependence of the inflationary three-point function is determined by the 
differential operator $\tilde \Delta_u$, with a source term given by the $v \to 1$ limit of the de Sitter four-point function. 
 This equation is valid for functions $\hat \F$ corresponding to contact terms and exchange diagrams. Our task now is to solve this equation for the various possible sources in de Sitter. 

\paragraph{Contact interactions} We begin by considering contact terms as sources in \eqref{equ:infl3pt}:
\beq
\tilde \Delta_u \hat b_{c,n} = -\partial_v\hat {\cal C}_n(u,1)\, ,
\eeq
where we have introduced the subscript $n$ in $\hat b_{c,n}$ to label the solutions corresponding to the different contact terms $\hat {\cal C}_n$.
 We note that, except for the non-derivatively coupled $\hat {\cal C}_0$, all $\hat  {\cal C}_n$'s will have trivial soft limits. In particular, $\hat  {\cal C}_1(u,1)=1$ and $\hat  {\cal C}_{n>1}(u,1)=0$. 

\vskip 4pt
The analysis for the first contact term $\hat  {\cal C}_1$ is a bit tricky due to the appearance of a logarithm in the answer for the inflationary three-point function. 
We will therefore treat it separately, starting from the ansatz (\ref{equ:III}) and the constraint equation (\ref{equ:13}). 
 Evaluating the $s$-derivative of $\hat {\cal C}_1$ in the soft limit, we obtain 
\beq
\lim_{k_4 \to 0}\frac{1}{s}\partial_s  {\cal C}_1 = 
O_{12} \left[2\left(k_3+\frac{k_{12}^2}{k_{12}+k_3}\right) \right] .
\eeq
The operator $\bar D_3$ commutes with $O_{12}$,
while for $\bar D_1$ we have
\beq
\bar D_1
O_{12}  \,b_{c,1} = 
O_{12}
 \left(\partial^2_{k_{12}}-\frac{2}{k_{12}}\partial_{k_{12}}\right)  b_{c,1}\, .
\eeq
Finally, we must therefore solve 
\beq
\left[\left(\partial^2_{k_{12}}-\frac{2}{k_{12}}\partial_{k_{12}}\right) - \left(\partial^2_{k_{3}}-\frac{2}{k_{3}}\partial_{k_{3}}\right)\right] b_{c,1}=-2\left(k_3+\frac{k_{12}^2}{k_{12}+k_3}\right) .
\eeq
The solution can be written in the form
\beq
b_{c,1} = -k_{12} k_3^2 +k_3^3 \log(k_{12}+k_3) +A k_3^3+ B k_{12}^3\, ,
\eeq
and the corresponding inflationary bispectrum is
\begin{align}\label{infcont1}
B_{c,1}(k_1,k_2,k_3) 
&=\epsilon \left[A k_3^3+B(k_1^3+k_2^3)-(k_1+k_2)k_3^2+k_3^3 \log k_t+\frac{k_1k_2k_3^2}{k_t}\right] + {\rm perms}\,.
\end{align}
Summing over permutations, the pole in $k_t$ disappears, and we get a milder $\log k_t$ singularity. This is reminiscent of the fact that the second contact term of the de Sitter four-point function is  equivalent to the first contact term, after summing over the permutations of all legs. Finally, the appearance of the $\log$ term is what prevents us from using the ansatz (\ref{equ:bbbb}) in a straightforward fashion. A naive use of this ansatz would lead to a
solution for $\hat b_{c,1}(u)$ with a spurious singularity at $k_{12}=k_3$. Finally, the three-point function (\ref{infcont1}) matches the result for the bulk interaction $(\partial_\mu \upphi)^2 \upphi^2$, after replacing one of the derivatively coupled legs with $\dot{\bar\upphi}$.\footnote{One must choose specific values of $A$ and $B$ to match the bulk calculation, but as we have argued these values can be adjusted by field redefinitions.} 

\vskip 4pt
Next, we consider 
\begin{align}
\tilde \Delta_u  \hat b_{c,2} &= \partial_v \hat {\cal C}_2(u,1) \nonumber \\[4pt]
 &= \frac{4(u-1)(1+3u+u^2)}{u(1+u^2)} \, .
\end{align}
It is easily verified that this equation is solved by 
\beq
\hat b_{c,2} =-\frac{2}{u(1+u)}+Z(u)\, , 
\eeq
where $Z(u)=A+B/u^3$ is the zero mode of the operator $\tilde\Delta_u$.
Acting with $O_{12}$ on $Z(u)$ leads to the local non-Gaussianity $B_{\rm loc}=Ak_3^3+B(k_1^3+k_2^3)$. 
The inflationary three-point function for $n=2$ is 
\beq
B_{c,2}= \epsilon\left[A k_3^3+B(k_1^3+k_2^3)-2k_3^2\left(\frac{k_1^2+k_1 k_2+k_2^2}{k_t}-\frac{k_1k_2k_3}{k_t^2}\right) \right] +{\rm perms}\, . \label{Ic2}
\eeq
This time, symmetrization will not remove the pole.
Notice that the highest-order pole is second order rather than cubic, so the singularity structure of the inflationary contact term is milder than its de Sitter counterpart.
From the bulk perspective, $B_{c,2}$ is obtained from the higher-derivative interaction $(\partial_\mu \upphi)^4$ 
and evaluating the 
one of the derivative legs on the background solution. 
Upon fixing $A=-1-2B$ and symmetrizing the momenta, \eqref{Ic2} indeed reproduces the three-point function associated with $(\partial_\mu \upphi)^4$; i.e.~we obtain precisely equation~(14) in~\cite{Creminelli:2003iq}.

\vskip 4pt
Since $\hat {\cal C}_{n>2} = \tilde \Delta_u^{n-2} \hat  {\cal C}_2$, the general solution for $n >2$ can be written as
\begin{align}
\hat b_{c,n} &= \tilde \Delta_u^{n-2} \hat b_{c,2}+Z(u)\, .
\end{align}
This concludes our analysis of  inflationary three-point functions of the contact type. 

\paragraph{Exchange diagrams}

The discussion of exchange diagrams turns out to be even simpler. To solve equation (\ref{equ:infl3pt}), we now consider the following ansatz
\beq
\hat b = -\frac{1}{M^2} \left(\partial_v \hat {\cal F}(u,1)-\hat b_c(u)\right) , \label{equ:ansatz}
\eeq
where $\hat b_c(u)$ is (for now) an arbitrary function.
Substituting (\ref{equ:ansatz}) into the left-hand side of~(\ref{equ:infl3pt}), we get
\begin{align}
\tilde \Delta_u \hat b &\ =\ -\frac{1}{M^2} \tilde \Delta_u \left( \partial_v \hat  {\cal F}(u,1)-\hat b_c(u)\right) \nonumber \\
&\ =\ \frac{1}{M^2}\left(M^2 \partial_v \hat  {\cal F}- \partial_v \hat  {\cal C}+ \tilde \Delta_u \hat b_c\right)  \ =\ \partial_v \hat  {\cal F}- \frac{1}{M^2}\left( \- \partial_v \hat  {\cal C}- \tilde \Delta_u \hat b_c\right) ,
\end{align}
where we have used that $\hat  {\cal F}$ (and hence $\partial_v \hat  {\cal F}$) satisfies (\ref{equ:INH3}).
We see that (\ref{equ:ansatz}) solves (\ref{equ:infl3pt}) if the function $\hat b_c$ obeys 
\beq
\tilde \Delta_u \hat b_c=\partial_v \hat  {\cal C}(u,1)\, .
\eeq
This is nothing but the equation we solved above; i.e.~the function $\hat b_c$ describes a contact inflationary three-point function. This means that we are done. The total inflationary three-point function will be given by
\begin{eBox}
\beq\label{equ:inflexc}
B_n = \epsilon\, k_3^3 \,O_{12}\, \hat b_n(k_3/k_{12}) + \text{perms}\, , 
\eeq
\vskip 1pt
\end{eBox}
where we restored the index $n$ and defined the source function
\begin{eBox}
\beq
\hat b_n(u)=-\frac{1}{\mu^2+\frac{9}{4}} \,\left(\partial_v \hat  {\cal F}_n(u,1)- \hat b_{c,n}(u)\right) .
\eeq
\end{eBox}

\vskip 4pt
As a consistency check, we compare the squeezed limit of our result to the known result in the literature; cf.~Section 6.1 in \cite{Arkani-Hamed:2015bza}.
For concreteness, let us specialize to the case of $\hat b_1$. 
The derivative of the corresponding four-point function $\hat {\cal F}_1$ can be written in the form
\beq
\partial_v \hat  {\cal F}_1(u,1)= \sum_{n=0}^\infty a_n u^{n-1} + \frac{\pi}{2 \cosh \pi \mu}\Big(A_+\tilde {\cal F}_+(u)+A_- \tilde {\cal F}_-(u) \Big) \,,  \label{equ:dvF1}
\eeq
where
\begin{align}
A_\pm = \left(\frac{3}{2} \pm i \mu\right)\sqrt{\pi}(1\mp\beta_0) \frac{\Gamma\left(\frac{1}{2}\pm i \mu \right)}{\Gamma\left(1\pm i \mu\right)}\left(\frac{i}{\mu}\right)^{-\frac{1}{2}\mp i \mu}\, .
\end{align}
The coefficients $a_n$ of the particular solution could either be written in terms of the coefficients $d_{mn}$ of the solution (\ref{equ:F1<}), or determined directly by solving the differential equation again, this time for $\partial_v \hat  {\cal F}_1(u,1)$.  Having said that, since the squeezed limit is dominated by the homogeneous solution, we won't need the precise form of the particular solution for our consistency check. 
Acting with $O_{12}$ on (\ref{equ:dvF1}), we indeed obtain 
the same squeezed limit as equation~(6.131) of \cite{Arkani-Hamed:2015bza}.

\newpage
\section{Weight-Shifting Operators}
\label{sec:ladder}

\begin{figure}[h!]
    \centering
      \includegraphics[scale=1]{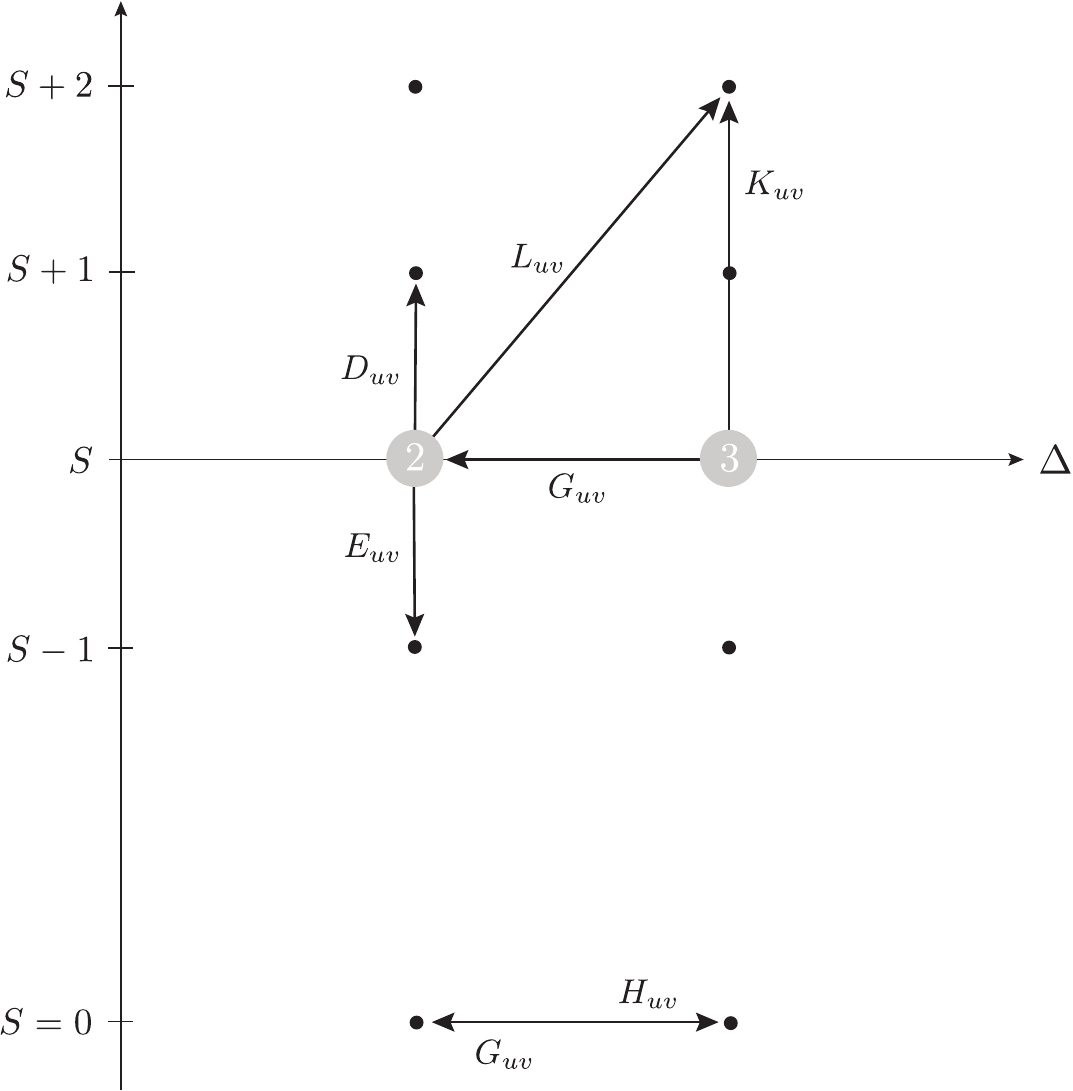}
           \caption{Illustration of the web of relations between $\Delta=2$ and $\Delta=3$ solutions corresponding to the exchange of particles with spin $S$. Only the operators $D_{uv}$, $G_{uv}$, $L_{uv}$, and $K_{uv}$ maintain the number of derivatives of the solution (from the bulk point of view), while $E_{uv}$ and $H_{uv}$ increase the number of derivatives.}
    \label{fig:web}
\end{figure}

In this appendix, we introduce a web of relations between the $\Delta =2$ and $\Delta=3$ solutions (see Fig.~\ref{fig:web}). Through this web, we will show that most solutions of the $\Delta=3$ constraint equations can be obtained directly from solutions of the $\Delta=2$ equations, a fact we use extensively in the main text.
This provides a powerful way of bootstrapping the $\Delta=3$ solutions from the $\Delta=2$ solutions.

\vskip 4pt
For notational clarity, we will use the following definitions
\begin{align}
	\Delta_{S,u} &\equiv (1-u^2)u^2\partial_u^2 - 2(u^2+S) u\partial_u\, , \label{equ:DS1}\\
	\tilde\Delta_{S,u} &\equiv (1-u^2) u^2\partial_u^2 -2(2u^2+S-2)u\partial_u\, , \label{equ:DS2}
\end{align}
which denote the differential operators for $\Delta=2$ (without tilde) and $\Delta=3$ (with tilde) respectively. We have not introduced a separate ansatz for the spinning $\Delta=3$ solutions, but we will always refer to its top helicity component. If we write ${\cal F}_S=s^3 \bar \Pi_S \tilde \Pi_{S,S} \hat{\cal A}_{S,S}+\cdots$, then $\hat{\cal A}_{S,S}$ will satisfy the formulas we present below. For $\Delta=3$, the lower-helicity components become more complicated, and we rely on the bootstrap operators determined in Section~\ref{sec:Massless}. The coefficient functions satisfy  
\begin{align}
(\Delta_{m,u}-\Delta_{m,v})\hat A_{S,m} &=0\, , \\
(\tilde\Delta_{S,u}-\tilde\Delta_{S,v})\hat {\cal A}_{S,S}&=0\, , 
\end{align}
for $\Delta=2$ and $\Delta = 3$, respectively. Notice that the $\Delta_{m,u}$ operators act in the same way on all helicity-$m$ components of the $\Delta=2$ solutions, regardless of the total spin $S$. This is not the case for $\Delta=3$ coefficient functions.

\begin{itemize}
\item For $\Delta = 2$, we define the following {\it spin-raising} and {\it spin-lowering} operators 
\begin{align}
D_{uv} &\equiv (uv)^2 \partial_u \partial_v\, ,  \label{equ:Duv}\\
E_{uv} &\equiv r^{S-1} D_{uv}(r^{-S}\ \cdot \ )\, , \quad {\rm with} \quad r \equiv \frac{u^2}{1-u^2} \frac{v^2}{1-v^2}\, , \label{equ:Euv}
\end{align}
These operators raise/lower the spin of the solution by one unit: 
\begin{align}
\hat A_{S+1} &= D_{uv}\hat A_S\, , \\
\hat A_{S-1} &= E_{uv}\hat A_{S} =  r^{S-1} D_{uv}(r^{-S}\hat A_{S}) \, . 
\end{align}
In Section~\ref{sec:spin}, we used $D_{uv}$ to build the spin-$S$ solution from the scalar solution. Notice that we did not specify the total spin of the solution, just the helicity, as the constraint equations are the same for a given helicity, regardless of the total spin.

\item The operator
\begin{align}
	L_{uv}\equiv \partial_u \partial_{v}(uv \ \cdot) \label{equ:Luv}
\end{align} 
 takes the spin-$S$, $\Delta=2$ solution to the spin-$(S+2)$, $\Delta=3$ solution. 
\item
For $\Delta=3$, the operator
\begin{align}
	K_{uv}\equiv (uv)^2(4 u\partial_u+u^2\partial_u^2)(4v\partial_v+v^2\partial_v^2) \label{equ:Kuv}
\end{align}
 raises the spin of the solution by two units. 

\item For $S=0$, the operators
\begin{align}
G_{uv} &\equiv (uv)^3 \partial_u \partial_v\, , \label{equ:Guv}\\
H_{uv} &\equiv r^{-1} \partial_u \partial_v(uv\ \cdot\ )\, , \label{equ:Huv}
\end{align}
relate the scalar solutions for $\Delta=2$ and $\Delta=3$:
\begin{align}\label{3to2}
\hat F  &= G_{uv} \hat {\cal F}\, ,\\
\label{2to3}
\hat{\cal F} &= H_{uv} \hat F\, .
\end{align}
Given a solution for the massless scalar equation, we can obtain a solution for the conformally coupled scalar by applying \eqref{3to2}, and vice versa for \eqref{2to3}.  
Of course, the function $\hat F$ in \eqref{3to2} and \eqref{2to3} are not related to each other. Namely, applying the equations twice, we do {\it not} get the original $\hat F$ we started with. 
Instead, we get
\begin{align}
G_{uv} H_{uv} \hat F&= (\Delta_u-2)(\Delta_v-2) \hat F\, , \\
H_{u v} G_{uv}{\cal \hat F}&= \tilde\Delta_u \tilde\Delta_v {\cal \hat F}\, .
\end{align}
Finally, it turns out that the formula 
\beq
\hat F_S= G_{u v} {\cal \hat F}_S
\eeq 
is valid for any spin. However, the map from $\Delta=2$ to $\Delta=3$ is not the same.
\end{itemize}
 We have not found a use for all of the relations presented in Fig.~\ref{fig:web}. Moreover, we don't claim that the set of displayed relations is complete. In fact, we suspect that it is not.
 
\newpage
\section{Useful Identities}
\label{app:identities}

In this appendix, we collect some useful identities of the (generalized) hypergeometric function. The following formulas relate hypergeometric functions at different values of the argument~\cite{Abramowitz:1974}:
\begin{align}
	{}_2F_1\bigg[\begin{array}{c} a,\hs b\\ \frac{a+b+1}{2}\end{array}\bigg|\,z\bigg] &= \frac{\Gamma(\frac{1}{2})\Gamma(\frac{a+b+1}{2})}{\Gamma(\frac{a+1}{2})\Gamma(\frac{b+1}{2})}{}_2F_1\bigg[\begin{array}{c} \frac{a}{2},\hs \frac{b}{2}\\ \frac{1}{2}\end{array}\bigg|\,(1-2z)^2\bigg]\nn
	&\quad +(1-2z)\frac{\Gamma(-\frac{1}{2})\Gamma(\frac{a+b+1}{2})}{\Gamma(\frac{a}{2})\Gamma(\frac{b}{2})}{}_2F_1\bigg[\begin{array}{c} \frac{a+1}{2},\hs \frac{b+1}{2}\\ \frac{3}{2}\end{array}\bigg|\,(1-2z)^2\bigg]\, ,\label{hyper1}\\[6pt]
	{}_2F_1\bigg[\begin{array}{c} a,\hs b\\ c\end{array}\bigg|\,z\bigg] &=\frac{1}{(-z)^{a}}\frac{\Gamma(b-a)\Gamma(c)}{\Gamma(b)\Gamma(c-a)}{}_2F_1\bigg[\begin{array}{c} a,\hs a-c+1\\ a-b+1\end{array}\bigg|\,\frac{1}{z}\bigg]\nn
	&\quad +\frac{1}{(-z)^{b}}\frac{\Gamma(a-b)\Gamma(c)}{\Gamma(a)\Gamma(c-b)}{}_2F_1\bigg[\begin{array}{c} b,\hs b-c+1\\ -a+b+1\end{array}\bigg|\,\frac{1}{z}\bigg]\, .\label{hyper2}
\end{align}
These two transformation formulas can be combined to express the hypergeometric function with the argument $(u\pm 1)/2u$ in terms of that of $u^2$. One needs to be a bit careful in manipulating these formulas for complex parameters $a$ and $b$ due to the presence of the factors $(-z)^{-a}$ and $(-z)^{-b}$.

\vskip 4pt
The series expansion of the hypergeometric function around $z=0$ has a radius of convergence of 1. The behavior of the hypergeometric function near $z=1$ is
\begin{align}
	\lim_{z\to 1}\,{}_2F_1\bigg[\begin{array}{c} a,\hs b\\ c\end{array}\bigg|\,z\bigg] = \begin{cases}\displaystyle \ \frac{\Gamma(c)\Gamma(c-a-b)}{\Gamma(c-a)\Gamma(c-b)} & {\rm Re}(a+b-c)<0\, ,\ c\notin\mathbb{Z}^-\, , \\[13pt] \displaystyle \ -\frac{\log(1-z)+R(a,b)}{B(a,b)}+O(1) & {\rm Re}(a+b-c)=0\, ,\\[10pt] \displaystyle \ (1-z)^{c-a-b}\, {}_2F_1\bigg[\begin{array}{c} c-a,\hs c-b\\ c\end{array}\bigg|\,z\bigg] & {\rm Re}(a+b-c)>0\, ,\end{cases}\label{2F1behavior}
\end{align}
where $\mathbb{Z}^-=\{0,-1,-2,\cdots\}$,
 \begin{align}
 	B(a,b)=\frac{\Gamma(a)\Gamma(b)}{\Gamma(a+b)}\, , \quad \psi(x)=\frac{\Gamma'(x)}{\Gamma(x)}\, ,
 \end{align}
 are the beta function and the digamma function, respectively, and $R(a,b)\equiv \psi(a)+\psi(b)+2\gamma_E$, with $\gamma_E$ the Euler-Mascheroni constant. The special case ${\rm Re}(a+b-c)=0$ is of particular interest due to the presence of a logarithmic singularity.  Hypergeometric functions with this property are called {\it zero-balanced}, and have a branch point at $z=1$. As we have seen, all solutions of the conformal invariance equation turn out to be of this type.

\vskip 4pt
Similarly, the near $z=1$ behavior of the generalized hypergeometric function
\begin{align}
	{}_{p+1}F_p\bigg[\begin{array}{c} a_1,\cdots,a_{p+1}\\ b_1,\cdots,b_p\end{array}\bigg|\,z\bigg] = \sum_{n=0}^\infty\frac{(a_1)_n\cdots (a_{p+1})_n}{(b_1)_n\cdots (b_{p})_n}\frac{z^n}{n!}\, ,
\end{align}
is characterized by the parameter
\begin{align}
	\omega =\sum_{j=1}^p b_j -\sum_{j=1}^{p+1}a_j\, .
\end{align}
The series converges absolutely at $|z|=1$ when ${\rm Re}(\omega)>0$ and diverges otherwise. We will be particularly interested in the zero-balanced case, for which $\omega=0$. Its behavior near $z=1$ is given by~\cite{kampe}
\begin{align}
	\lim_{z\to 1}\, {}_{p+1}F_p\bigg[\begin{array}{c} a_1,\cdots,a_{p+1}\\ b_1,\cdots,b_p\end{array}\bigg|\,z\bigg] = -\frac{\Gamma(b_1)\cdots \Gamma(b_p)}{\Gamma(a_1)\cdots \Gamma(a_{p+1})} \big[\log(1-z)+R(a,b)\big] +\cdots\, ,\label{GenHyperBoundary}
\end{align}
where the ellipsis denotes a finite remainder.

\newpage
\section{Notation and Conventions}
\label{app:notation}

\begin{center}
\renewcommand*{\arraystretch}{1.08}
\begin{longtable}{c p{10cm} c}
\toprule
\multicolumn{1}{c}{\textbf{Symbol}} &
\multicolumn{1}{l}{\textbf{Meaning}} &
\multicolumn{1}{c}{\textbf{Reference}} \\
\midrule
\endfirsthead
\multicolumn{3}{c}
{} \\
\toprule
\multicolumn{1}{c}{\textbf{Symbol}} &
\multicolumn{1}{l}{\textbf{Meaning}} &
\multicolumn{1}{c}{\textbf{Reference}} \\
\midrule
\endhead
\bottomrule
\endfoot
\bottomrule
\endlastfoot
$p_\mu$	  & 	Four-momentum	&	 \S\ref{sec:FlatSpace} 	\\ 
$s$	  & Mandelstam variable, $s \equiv -(p_1+p_2)^2$ 	&  \S\ref{sec:FlatSpace}	\\  
$t$	  & Mandelstam variable, $t \equiv -(p_2+p_3)^2$ 	& \S\ref{sec:FlatSpace}	\\  
$u$	  & Mandelstam variable, $u \equiv -(p_2+p_4)^2$	& \S\ref{sec:FlatSpace}	\\ 
$A_4$ & Four-particle amplitude & \S\ref{sec:FlatSpace}  \\
$A_3^\lambda$ & Three-particle amplitude & \S\ref{sec:FlatSpace}  \\
$m$ & Mass of external particle &  \S\ref{sec:FlatSpace} \\
$M$ & Mass of exchange particle &  \S\ref{sec:FlatSpace} \\
$S$ & Spin of exchange particle & \S\ref{sec:FlatSpace} \\ 
$\lambda$ & Helicity of exchange particle & \S\ref{sec:FlatSpace} \\ 
$g$ & Coupling contant & \S\ref{sec:FlatSpace} \\ 
\midrule
$\sigma$ & Generic bulk scalar field & \S\ref{sec:dS} \\
$\upvarphi$ & Conformally-coupled scalar field & \S\ref{sec:dS-4pt}  \\
$\upphi$ & Massless scalar field & \S\ref{sec:Massless} \\
$O$ & Operator dual to $\sigma$ & \eqref{equ:dual}\\ 
$\varphi$ & Operator dual to $\upvarphi$ $(\Delta = 2)$ & \S\ref{sec:dS}\\ 
$\phi$	  & Operator dual to $\upphi$ ($\Delta = 3$)	&	\S\ref{sec:dS}	\\  
$H$ & Hubble parameter, $H \equiv \dot a/a$ & \S\ref{sec:dS} \\ 
$\eta$ & Conformal time, $\d \eta \equiv \d t/a(t)$ & \S\ref{sec:dS} \\ 
${\cal D}$ & Dilatation operator & (\ref{equ:D}) \\ 
${\cal K}^i$ & Special conformal transformation operator &  (\ref{equ:SCT}) \\ 
$\mu$ &  
Mass parameter, $\mu^2\equiv m^2-\frac{9}{4}$ & (\ref{equ:mdimsc}) \\ 
$\Delta$	  & Scaling dimension (conformal weight), $\Delta \equiv \frac{3}{2}\pm i \mu$	&	(\ref{equ:mdimsc})	\\  
$\Delta_t$	  & Total conformal weight, $\Delta_t \equiv \sum_n \Delta_n$ &	 \S\ref{sec:dS}		\\ 
$M$ & 
Mass parameter (for $\Delta=2$), $M^2\equiv \mu^2+\frac{1}{4}$ & \S\ref{sec:dS} \\ 
$\k$ & Three-momentum & \S\ref{sec:dS} \\
$k_i$	  & 	Spatial component of $\k$	&	\S\ref{sec:dS}	\\ 
$\k_n$	 	&Momentum of the $n$-th leg	&	\S\ref{sec:dS}	\\ 
$k_n$	 	& Magnitude of $\k_n$, $k_n \equiv |\k_n|$	&	\S\ref{sec:dS}	\\ 
$k_t$	  & Total energy, $k_t \equiv \sum_n k_n$	&	\S\ref{sec:dS}	\\ 
$s$	  &  Exchange momentum, $s \equiv |\k_1+\k_2|$  	&	\S\ref{sec:dS}	\\ 
$t$	 	& Exchange momentum, $t \equiv |\k_2+\k_3|$  	&\S\ref{sec:dS}		\\ 
$k_{nm}$ & Sum of $k_n$ and $k_m$, $k_{nm} \equiv k_n+k_m$ & \S\ref{sec:dS}	\\ 
$u$ & Ratio of $s$ and $k_{12}$, $u \equiv s/k_{12}$& \eqref{equ:p} \\ 
$v$ & Ratio of $s$ and $k_{34}$, $v \equiv s/k_{34}$& \eqref{equ:p} \\ 
$\Psi$ & Wavefunction of the universe & (\ref{equ:Psi}) \\ 
$B$ & Scalar three-point function &  (\ref{equ:scalarB}) \\ 
$F$ & Scalar four-point function & (\ref{equ:scalarF}) \\ 
$\hat F$ & Dimensionless four-point function, $\hat F \equiv s^{9-\Delta_t} F$ & \S\ref{sec:dS}	\\ 
$D_n$	  & Differential operator	&	(\ref{W1F})	\\  
$D_{nm}$	  & Difference of $D_n$ and $D_m$, $D_{nm} \equiv D_n - D_m$	& \S\ref{sec:dS}\\  
$\Delta_u$	  & Differential operator (for $\Delta=2$)	&(\ref{equ:Du})	\\  
$F_c$ & Contact four-point function & \S\ref{sec:dS} \\ 
$C$ & General contact term & \S\ref{sec:dS} \\ 
$\hat C_0$ & Lowest-order contact term, $\hat C_0 \equiv uv/(u+v)$ & (\ref{equ:C00}) \\ 
$\hat C_n$ & $n$-th order contact term, $C_n \equiv \Delta_u^n C_0$ & (\ref{equ:Cn}) \\ 
$G$ & Bulk-to-bulk propagator & (\ref{propagator}) \\ 
$F_{\pm \pm}$ & In-in integral  & (\ref{Fpmpm}) \\ 
$G_{\pm \pm}$ & In-in propagators  & (\ref{equ:Gpmpm}) \\ 
$A_{\rm flat}$	  & Flat-space scattering amplitude	&  \S\ref{sec:dS}	\\  
$s_{\rm flat}$	  & Mandelstam variable, $s_{\rm flat} \equiv -(p_1+p_2)^2$ 	&  \S\ref{sec:dS}	\\  
$t_{\rm flat}$	  & Mandelstam variable, $t_{\rm flat} \equiv -(p_2+p_3)^2$ 	& \S\ref{sec:dS}	\\  
\midrule
$\hat F_\pm$ & Homogeneous solutions & (\ref{equ:homo}) \\ 
$\hat F_h$ & Matched homogeneous solution &  (\ref{equ:Hom}) \\ 
$\hat F_K$ & Homogeneous solution for $\mu=0$ & \eqref{FKseries} \\ 
$\hat F_G$ & Homogeneous solution for $\mu=0$ & \eqref{FGseries}  \\
$\hat F_<$ & Series solution around $u=0$ & \eqref{equ:SOL1} \\ 
$\hat F_>$ & Series solution around $u=\infty$, $\hat F_>(u,v) = \hat F_<(v,u)$ & \S\ref{sec:exchange} \\ 
$\hat F_n$ & $n$-th order exchange solution & (\ref{highercontact})\\
\midrule
$m$ & Helicity of exchange particle & \S\ref{sec:prelim}\\ 
$F_S$ & Four-point function from spin-$S$ exchange & (\ref{equ:Ansatz})\\ 
$A_{S,m}$ & Coefficient function of $F_S$ & (\ref{equ:Ansatz})\\ 
$x$ & Sum of $k_1$ and $k_2$, $x \equiv k_1+k_2$ & \eqref{equ:xy} \\ 
$y$ & Sum of $k_3$ and $k_4$, $y \equiv k_3+k_4$& \eqref{equ:xy} \\ 
$\alpha$ & Difference of $k_1$ and $k_2$, $\alpha \equiv k_1-k_2$ & \eqref{equ:tau} \\ 
$\beta$ & Difference of $k_3$ and $k_4$, $\beta \equiv k_3-k_4$  & \eqref{equ:tau} \\ 
$\tau$	  & 	Angular variable	&	(\ref{equ:tau})	\\ 
$\hat T$ & Angular variable &  (\ref{equ:T}) \\ 
$\hat L$ & Angular variable & (\ref{equ:Z})\\ 
$P_{ij}$ & Spin-1 projection tensor & (\ref{equ:S1P}) \\
$P_{i_1\cdots i_S}^{i_j \cdots j_S}$ & Spin-S projection tensor & (\ref{projtensor}) \\
$\bar \epsilon^\lambda_{i_1\cdots i_S}$	  & Transverse polarization tensor	& \S\ref{sec:prelim}	\\  
$\tilde \epsilon_{i_1\cdots i_S}$	  & Longitudinal polarization tensor	& \S\ref{sec:prelim}	\\  
$\bar \Pi_m$	  & Transverse polarization sum	& (\ref{equ:trans})	\\  
$\tilde \Pi_{S,m}$	  & Longitudial polarization sum	& (\ref{equ:long})	\\  
$\Delta_{m,u}$ & Differential operator (for helicity $m$)	&(\ref{equ:Dum})	\\ 
$D_{uv}$	  &  Spin-raising operator, $D_{uv} \equiv (uv)^2 \partial_u \partial_v$	& \eqref{equ:SpinRaising}	\\  
$E_{uv}$	  &   Spin-lowering operator	& (\ref{equ:Euv})	\\  
$L_{uv}$	  &   Spin-raising operator	&  (\ref{equ:Luv})	\\  
$K_{uv}$	  &   Spin-raising operator	&  (\ref{equ:Kuv})	\\ 
$f$ & Scalar exchange solution for $\Delta=2$, $f\equiv F_{S=0}$ & \S\ref{sec:results} \\ 
$\sigma$ & Exchange particle &  \S\ref{sec:PM}  \\ 
$M_\sigma$ & 
Mass of exchange particle & \S\ref{sec:PM} \\ 
$\Delta_\sigma$	  & Scaling dimension of exchange field	&	(\ref{equ:Deltas})	\\  
\midrule
$\tilde M$ &  
Mass parameter (for $\Delta =3$), $\tilde M^2\equiv \mu^2+\frac{9}{4}$ & \S\ref{app:D3} \\ 
${\cal F}$ & Scalar four-point function for $\Delta=3$ & \S\ref{sec:Massless} \\ 
$\hat {\cal F}$ & Dimensionless four-point function for $\Delta=3$, $\hat {\cal F} = s^{-3} {\cal F}$ & \S\ref{app:D3} \\ 
$\hat {\cal F}_\pm$ & Homogeneous solutions for $\Delta=3$ & \eqref{equ:homo2} \\ 
 $\hat {\cal F}_<$ & Series solution around $u=0$ & \eqref{equ:Fsmall}\\
  $\hat {\cal F}_>$ & Series solution around $u=\infty$ & \S\ref{app:D3}\\
${\cal I}$ & Slow-roll correction to ${\cal F}$ &  (\ref{equ:I}) \\ 
${\cal F}_S$ & Spin-S exchange solution for $\Delta=3$ & \S\ref{sec:Massless} \\ 
${\cal C}$ & Contact term for $\Delta=3$ &  \S\ref{sec:Massless} \\ 
$O_{12}$	  & Differential operator & (\ref{equ:Oij2})	\\  
$\tilde \Delta_u$	  & Differential operator (for $\Delta=3$)	&(\ref{equ:tildeD})	\\   
$U_{12}$	  & Weight-shifting operator & (\ref{equ:U12})	\\   
$U_{12}^{S,m}$	  & Weight-shifting operator (for helicity $m$) & 	\eqref{equ:USm} \\ 
$G_{uv}$	  &   Weight-shifting operator	&  (\ref{equ:Guv})	\\  
$H_{uv}$	  &   Weight-shifting operator	&  (\ref{equ:Huv})	\\
$J_\alpha^\upvarphi$ & Spin-1 current for $\upvarphi$ & (\ref{J1}) \\ 
$J_\alpha^\upphi$ & Spin-1 current for $\upphi$ & (\ref{J2}) \\ 
$T_{\alpha \beta}^\upvarphi$ & Spin-2 tensor for $\upvarphi$ & (\ref{equ:tensors0}) \\ 
$T_{\alpha\beta}^\upphi$ & Spin-2 tensor for $\upphi$ & (\ref{equ:tensors}) \\ 
${\cal F}_{\rm inf}$ & Inflationary trispectrum & (\ref{inf4pt}) \\ 
${\cal F}_{\rm GE}$ & Graviton exchange part of ${\cal F}_{\rm inf}$ & \S\ref{sec:mapping} \\ 
${\cal F}_{\rm CT}$ & Contact part of ${\cal F}_{\rm inf}$ & \S\ref{sec:mapping} \\ 
${\cal F}_{\rm GE,c}$ & Connected part of ${\cal F}_{\rm GE}$ &   \eqref{ConnectedPart}\\
${\cal F}_{\rm GE,d}$ & Disconnected part of ${\cal F}_{\rm GE}$ &   \eqref{DisconnectedPart}\\
${\cal F}_{\rm loc}$ & Local trispectrum, ${\cal F}_{\rm loc} \equiv \sum_n k_n^3$ &   \eqref{Delta3localF}\\
%
$\epsilon$ & Slow-roll parameter &  (\ref{epsilon}) \\ 
$B_{\rm inf}$ & Bispectrum of slow-roll inflation & \eqref{equ:Binf}\\ 
$b_S$ & Source function of the inflationary bispectrum & (\ref{equ:bs}) \\ 
$B_{\rm loc}$ & Local bispectrum, $B_{\rm loc} \equiv \sum_n k_n^3$& \eqref{sec:InfBi} \\ 
\midrule
$P_S$ & Legendre polynomial of order $S$ & \cite{MathWorld} \\ 
$P_S^m$ & Associated Legendre polynomial & \cite{MathWorld} \\ 
$P_{d,S}$ & Gegenbauer polynomial & \cite{MathWorld} \\ 
${}_a F_b$ & Generalized hypergeometric function & \cite{MathWorld}\\ 
$F^{a|b|c}_{\,d|e|f}$ &  Kamp\'e de F\'eriet function & \cite{MathWorld}\\ 
$G^{a,b}_{ c,d}$ &  Meijer G-function & \cite{MathWorld}\\
$\Gamma$ & Gamma function & \cite{MathWorld} \\ 
$(\cdot)_n$ & Pochhammer symbol & \cite{MathWorld} \\ 
${\rm Li}_2$ & Dilogarithm & \cite{MathWorld} \\ 
$\gamma_E$ & Euler-Mascheroni constant & \cite{MathWorld} \\ 
$\Theta$ & Heaviside function  & \cite{MathWorld} \\ 
$H_{i\mu}^{(1)}$ & Hankel function of the first kind  & \cite{MathWorld} \\ 
$K$ & Complete elliptic integral of the first kind& \cite{MathWorld} \\
$H_n$ & $n$-th harmonic number & \cite{MathWorld} \\
$W$ & Wronskian & \cite{MathWorld}  \\ 
\end{longtable}
\end{center}

\clearpage
\phantomsection
\addcontentsline{toc}{section}{References}
\bibliographystyle{utphys}
\bibliography{4pt}

\end{document}